\documentclass[12pt]{article}
\usepackage{youngtab}
\makeatletter \@addtoreset{equation}{section} \makeatother
\renewcommand{\theequation}{\thesection.\arabic{equation}}
\newcommand{\ba}{\begin{array}}
\newcommand{\ea}{\end{array}}
\newcommand{\beq}{\begin{equation}}
\newcommand{\eeq}{\end{equation}}
\newcommand{\bea}{\begin{eqnarray}}
\newcommand{\eea}{\end{eqnarray}}
\def\bce{\begin{center}}
\def\ece{\end{center}}

\def\nonu{\nonumber}

\def\pa{\partial}

\def\be{\beta}

\def\de{\delta}

\def\la{\lambda}
\def\La{\Lambda}

\def\si{\sigma}

\def\eps6{{\displaystyle \mathop{\epsilon}^{6}}{}}
\def\g6{{\displaystyle \mathop{g}^{6}}{}}

\def\nab6{{\displaystyle \mathop{\nabla}^{6}}{}}
\def\0{{\sst{(0)}}}
\def\1{{\sst{(1)}}}
\def\2{{\sst{(2)}}}
\def\3{{\sst{(3)}}}
\def\4{{\sst{(4)}}}
\def\5{{\sst{(5)}}}
\def\6{{\sst{(6)}}}
\def\7{{\sst{(7)}}}
\def\8{{\sst{(8)}}}

\def\ba{\begin{array}}
\def\ea{\end{array}}
\def\beq{\begin{equation}}
\def\eeq{\end{equation}}
\def\be{\begin{equation}}
\def\ee{\end{equation}}

\def\la{\lambda}
\def\eps{\epsilon}

\def\ba{\begin{array}}
\def\ea{\end{array}}
\def\beq{\begin{equation}}
\def\eeq{\end{equation}}
\def\be{\begin{equation}}
\def\ee{\end{equation}}

\def\la{\lambda}
\def\eps{\epsilon}

\def\eps6{{\displaystyle \mathop{\epsilon}^{6}}{}}

\def\nab6{{\displaystyle \mathop{\nabla}^{6}}{}}

\newcommand{\bean}{\begin{eqnarray*}}
\newcommand{\eean}{\end{eqnarray*}}

\begin{document}
\thispagestyle{empty} \addtocounter{page}{-1}
   \begin{flushright}
%PUPT-2395 \\
%CALT-68-nnnn \\
%{\tt hep-th/yymmnnn}\\
\end{flushright}

\vspace*{1.3cm}
  
\centerline{ \Large \bf   
Wolf Space Coset Spectrum
% } 
%\vspace*{0.3cm}
%\centerline{ \Large \bf 
 in the Large ${\cal N}=4$ Holography} 
%and }
\vspace*{1.5cm}
\centerline{{\bf Changhyun Ahn}
\footnote{On leave from the Department of Physics, Kyungpook National University, Taegu
  41566, Korea and 
address until Aug. 31, 2018:
C.N. Yang Institute for Theoretical Physics,
Stony Brook University,
Stony Brook, NY 11794-3840, USA
%Department of Physics, Princeton University, Jadwin Hall, 
%Princeton, NJ 08544, USA
}
} 
\vspace*{1.0cm} 
\centerline{\it
C.N. Yang Institute for Theoretical Physics,
Stony Brook University,
Stony Brook, NY 11794-3840, USA}
 \centerline{\it 
Department of Physics, Kyungpook National University, Taegu
41566, Korea} 
%\centerline{\it 
%Department of Physics, Princeton University, Jadwin Hall, 
%Princeton, NJ 08544, USA}
\vspace*{0.8cm} 
\centerline{\tt ahn@knu.ac.kr
%, \qquad kimhyun@knu.ac.kr 
} 
\vskip2cm

\centerline{\bf Abstract}
\vspace*{0.5cm}

After reviewing the four eigenvalues (the conformal dimension,
two $SU(2)$ quantum number, and $U(1)$ charge)
in the minimal (and higher) representations 
in the Wolf space coset where the ${\cal N}=4$ superconformal algebra
is realized by $11$ currents in nonlinear way,
these four eigenvalues in the higher representations up to three boxes
(of Young tableaux) are examined in detail. 
The eigenvalues associated with the higher spin-$1,2,3$ currents in the
(minimal and) higher representations up to two boxes
are studied. They are expressed in terms of the two finite parameters
$(N, k)$ where the Wolf space coset contains the group $SU(N+2)$
and the affine Kac-Moody spin $1$ current has the level $k$.
Under the large $(N,k)$ 't Hooft-like limit, they are simply
linear combinations
of the eigenvalues in the minimal representations.
For the linear case where the ${\cal N}=4$ superconformal algebra
is realized by $16$ currents in linear way,
the eigenvalues, corresponding to
the spin $2$ current and the higher spin $3$ current,
which are the only different quantities (compared to the
nonlinear case), are also obtained
at finite $(N,k)$. They coincide with the results for the nonlinear
case above after the large $(N,k)$ 't Hooft-like limit is taken.
As a by product, the three-point functions of the higher spin currents
with two scalar operators can be determined at finite $(N,k)$.

\baselineskip=18pt
\newpage
\renewcommand{\theequation}
{\arabic{section}\mbox{.}\arabic{equation}}

\tableofcontents

%%%%%%%%%%%%%%%%%%%%%%%%%%%%%%%%%%%%%%%%%%%%%%%%%%%%%%%%%%%%%%%%%%%%%
%%%%%%%%%%%%%%%%%%%%%%%%%%%%%%%%%%%%%%%%%%%%%%%%%%%%%%%%%%%%%%%%%%%%%%
\section{Introduction}
%section1%%%%%%%%%%%%%%%%%%%%%%%%%%%%%%%%%%%%%%%%%%%%%%%%%%%%%%%%%%%%%%%%%%%%%
%%%%%%%%%%%%%%%%%%%%%%%%%%%%%%%%%%%%%%%%%%%%%%%%%%%%%%%%%%%%%%%%%%%%%

It is known in \cite{GG1305}
that the conformal dimension of a coset $\frac{G}{H}$ primary
can be described as
\bea
h(\Lambda_{+}; \Lambda_{-})= \frac{C^{(N+2)}(\Lambda_{+})}{(k+N+2)} -
\frac{C^{(N)}(\Lambda_{-})}{(k+N+2)}- \frac{\hat{u}^2}{N(N+2)(k+N+2)} +n.
\label{conformaldimension}
\eea
The quantity $C^{(N+2)}(\Lambda_{+})$ is the quadratic Casimir of
$SU(N+2)$ of $G$ on the representation $\La_+$.
Similarly, the quantity $C^{(N)}(\Lambda_{-})$ is the quadratic Casimir of
$SU(N)$ of $H$ on the representation $\La_-$.
These Casimirs depend on $N$ and have explicit forms for any
representations.
The $\hat{u}$ charge is related to the $U(1)$ (of $H$) current of
large ${\cal N}=4$ ``linear'' superconformal algebra
\cite{STVplb,npb1988,Schoutens88,ST,Saulina}.
Finally, the last quantity $n$ is known as the excitation number
which can be positive integer or half-integer. That is,
$n=\frac{1}{2}, 1, \frac{3}{2}, \cdots $.
When the representation $\La_-$ appears in the branching of
$\La_+$, then this excitation number vanishes.
Otherwise (when the representation occurs in the product of
$(\La_+;0)$ and $(0;\La_-)$),
in general, the excitation number is nonzero.
Because the $\hat{u}$ charge behaves as $N$ under the large $(N,k)$
't Hooft-like limit and due to the fact that the denominator
behaves as cubic term, the third term in (\ref{conformaldimension})
behaves differently when one compares to the first two terms of
(\ref{conformaldimension}). Note that the quadratic Casimir
behaves as $N$ under  the large $(N,k)$
't Hooft-like limit.
The quantity $k$ appearing in (\ref{conformaldimension})
is the level of spin $1$ current having the adjoint index of $SU(N+2)$.

In this paper, we would like to examine 
the above conformal dimensions closely, obtain them at finite $(N,k)$,
identify the BPS representations (as well as non BPS representations),
and observe some relation between the above formula
(\ref{conformaldimension}) and the BPS bound \cite{GPTV,PT1,PT2},
up to three boxes of Young tableaux.
The BPS bound is described in terms of two $SU(2)$ quantum numbers
$l^{\pm}$ corresponding to
six spin $1$ currents of
the large ${\cal N}=4$ ``nonlinear'' superconformal algebra
\cite{GS,VP,GPTV,GK}.
Therefore, the four quantum numbers,
$h$ corresponding to the stress energy tensor
of the large ${\cal N}=4$ nonlinear superconformal algebra,
$l^{\pm}$, and $\hat{u}$ will be presented explicitly.
One of the reasons why one should reexamine these quantum numbers
is that there is no known formula like as (\ref{conformaldimension})
for the higher spin currents. In particular, when the excitation number
is nonzero (the representation appears in the product of
$(\La_+;0)$ and $(0;\La_-)$), then the conformal dimensions
and other eigenvalues corresponding to the higher spin currents
are not sum of the ones in $(\La_+;0)$ and the ones in
$(0;\La_-)$. Instead, there is a contribution, at
finite $(N,k)$, from the commutator
between the zeromode of spin $2$ current (or
the zeromode of higher spin current) and other mode from
the multiple product of spin $\frac{1}{2}$ curents
having the adjoint index of $SU(N+2)$.
After one makes sure that the two approaches, from the conformal formula
(\ref{conformaldimension}) directly and from the collection of
three contributions above,
give the same answer, one can go to the eigenvalues for the
higher spin currents.

Because there is a higher spin symmetry in the Wolf space
coset \cite{Wolf,Alek,Salamon,BW}, it is natural to ask what are the eigenvalues
for the higher spin currents on the (minimal and) higher representations.
How one can obtain the eigenvalues of the higher spin
zero modes
acting on the higher representations?
Recall that in the above eigenvalues,
$h$, $l^{\pm}$ and $\hat{u}$,
the first three quantum numbers for the any representation
remains the same by going to its complex conjugated representation.
However, the last one, $\hat{u}$ charge can have an extra
minus sign under this process.
In particular, when one considers the vanishing excitation number
(when the representation $\La_-$ appears in the branching of $\La_+$),
the $SU(N+2)$ generators play the crucial role in the calculations
of these eigenvalues.
One can associate the quadratic $SU(N+2)$ generators with the
zeromodes of the three quantities (corresponding to $h$ and $l^{\pm}$)
while linear $SU(N+2)$ generators with the zeromode
of spin $1$ current corresponding to $\hat{u}$.
Then it is obvious to see that the former
does not change the sign and the latter
does change the sign when one changes the sign of
$SU(N+2)$ generators because the complex conjugated representation
is the minus of the original representation.
We will see that some states contain the same quantum number $h$
but they will have different quantum numbers from the higher spin currents.

Recall that the lowest $16$ currents of the ${\cal N}=4$ multiplet
have the spin contents, ($1$, $(\frac{3}{2})^4$, $2^6$,
$(\frac{5}{2})^4$, $3$). For the higher spin $1$ and $3$ currents
which transform as $SO(4)$ singlet,
one expects that there should be minus sign between the
representation and its complex conjugated representation.
On the other hand, there is no sign change for the higher spin $2$
currents, which transform as two $SU(2)$ adjoints,
between these two representations. One can make two $SU(2)$ singlets
by squaring of each higher spin $2$ current and summing over the
adjoint indices, in analogy of two quantities corresponding to
the quantum numbers $l^{\pm}$. Of course, their conformal dimensions
are $4$. For the minimal representations, $(f;0)$ (or $(\overline{f};0)$)
and $(0;f)$ (or $(0;\overline{f})$), the corresponding
eigenvalues for the higher spin currents were obtained in \cite{AK1506}.
In this paper, the eigenvalues for the higher spin currents
in the higher representations, which are obtained from the
various products of the minimal representations (which will
correspond to single or multi particle states in the $AdS_3$ bulk
theory in the large $(N,k)$ 't Hooft like limit), are determined
explicitly by considering the eigenvalues for each minimal representation
(and some additional eigenvalue contributions). 

One of the reasons why one should
obtain the eigenvalues for the higher spin currents is
that one can determine
the three-point functions (in the higher representations) for the higher
spin (which is fixed) current with two scalar operators at finite $(N,k)$. 
It will turn out that although the various three-point functions
for the higher spin currents with two scalar operators depend on
$(N,k)$ explicitly (so far, it is not known how to write them down
in terms of the data of the representations, contrary to
(\ref{conformaldimension})),
the large $(N,k)$ 't Hooft like limit of them
leads to very simple form. That is, the 't Hooft coupling dependent piece
of the three-point functions in the higher
representations (the coefficient of the two point function of two scalar
operators)
is a multiple of the ones in the minimal representations. 
For example, for the higher spin $3$ current, the fundamental quantities
are given by the eigenvalues of the representations, $(0;f)$
(or $(0;\overline{f})$), $(f;0)$ (or $(\overline{f};0)$)
and $(f;f)$ (or $(\overline{f};\overline{f})$).
The first two of these behave as the functions of 't Hooft coupling
while the last one behaves as $\frac{1}{N}$. Although we
consider the two boxes (of Young tableaux) in this paper, we expect 
that the above behavior holds for any boxes of the Young Tableaux
\footnote{
Recently \cite{EGGL}, by analyzing the BPS spectrum of string theory
and supergravity theory on $AdS_3 \times {\bf S^3} \times {\bf S^3}
\times {\bf S^1}$, it has been found that the BPS spectra of both
descriptions agree (where the world sheet approach is used).
See also \cite{FGJ,EGL,Gopakumar17}
for further studies along this direction. It would
be interesting to see how the large ${\cal N}=4$ superconformal higher spin
and CFT duality arises in the context of these world sheet approaches.}.

In section $2$, the work of Gaberdiel and Gopakumar \cite{GG1305} is reviewed.
The four eigenvalues mentioned in the abstract are obtained.
There are also new observations for the eigenvalues with
the mixed higher representation.

In section $3$, further eigenvalues are determined for the other
higher representations \footnote{The higher representations can be obtained
  from the minimal representations and have more than
  two boxes of Young tablueaux. The currents of the
  ${\cal N}=4$ nonlinear superconformal algebra
  can act on the
minimal representations as well as the higher representations.} up to three boxes of Young tableaux.
In particular, we summarize the conformal dimensions we have found
under the large $(N,k)$ 't Hooft like limit with Tables.

In section $4$, the eigenvalues for the higher spin currents in the
minimal representations \cite{AK1506} are reviewed.

In section $5$, as in sections $2$ and $3$,
the eigenvalues for the higher spin currents in the higher representations
up to two boxes of Young tableaux are obtained.
This section is one of the main results of this paper.
Some Tables under the large $(N,k)$ 't Hooft like limit
summarize this section.

In section $6$, some results \cite{AKK1703}
corresponding to sections $2$ and $3$
in the linear version are reviewed.

In section $7$,  some results \cite{AK1506} corresponding to section $4$ 
in the linear version are reviewed.

In section $8$,  some results corresponding to section $5$ 
in the linear version are obtained.
This section is also one of the main results of this paper.
Section $5$ is necessary to understand the results of this section.

In section $9$, some open problems related to this paper are presented.  
Some expectations for the particular eigenvalues for the
higher spin currents are given.

In Appendices, the $SU(N+2=5)$ generators in various higher representations
are given explicitly. Although the similar $SU(N+2=7,9,11,13)$
generators in higher representations (with two boxes) can be presented,
but they are omitted here.

The mathematica \cite{mathematica} package by Thielemans \cite{Thielemans}
is used all the time.

Although some presentations are repetitive, most of the subsections (for
example, sections $3$ or $5$) can
be read independently and are  self-contained. 

%%%%%%%%%%%%%%%%%%%%%%%%%%%%%%%%%%%%%%%%%%%%%%%%%%%%%%%%%%%%%%%%%%%%%
%%%%%%%%%%%%%%%%%%%%%%%%%%%%%%%%%%%%%%%%%%%%%%%%%%%%%%%%%%%%%%%%%%%%%%
\section{Review of the work of Gaberdiel and Gopakumar}
%section2%%%%%%%%%%%%%%%%%%%%%%%%%%%%%%%%%%%%%%%%%%%%%%%%%%%%%%%%%%%%%%%%%%%%%
%%%%%%%%%%%%%%%%%%%%%%%%%%%%%%%%%%%%%%%%%%%%%%%%%%%%%%%%%%%%%%%%%%%%%

The unitary Wolf space coset \cite{Wolf,Alek,Salamon,BW} is given by 
\bea
\mbox{Wolf}= \frac{G}{H} = 
\frac{SU(N+2)}{SU(N) \times SU(2) \times U(1)}.
%\label{coset}
\nonu
\eea
The adjoint group indices in the complex basis are divided into 
\bea 
G \quad \mbox{indices} &:& a, b, c, \cdots = 1, 2, \cdots,
\frac{1}{2} [(N+2)^2-1],  1^{\ast}, 2^{\ast}, \cdots,
\frac{1}{2} [(N+2)^2-1]^{\ast},
\nonu \\
%H \quad \mbox{indices} &:& a',b',c',\cdots
%\nonu \\
\frac{G}{H} \quad \mbox{indices} &:& \bar{a},\bar{b},\bar{c},\cdots
=1, 2, \cdots, 2N, 1^{\ast}, 2^{\ast}, \cdots, 2N^{\ast}.
\label{abnotation}
\eea
For given $(N+2) \times (N+2)$ unitary matrix, 
one can associate the above $4N$ Wolf space coset indices
as follows \cite{GG1305}:
\bea
\left(\begin{array}{rrrrr|rr}
 &&&&&{\ast} & {\ast}\\
 &&&&&{\ast} & {\ast}\\
 &&&&&\vdots & \vdots \\
 &&&&&{\ast} & {\ast}\\
 &&&&&{\ast} & {\ast}\\ \hline
 {\ast} & {\ast} & \cdots & {\ast} & {\ast}&& \\ 
 {\ast} & {\ast} & \cdots & {\ast} & {\ast}&&  \\ 
\end{array}\right)_{(N+2) \times (N+2)}.
\label{4nmatrix}
%\nonu
\eea
Note that the adjoint subgroup $H$ indices
run over $1, 2,
\cdots, \frac{1}{2} [N^2+3],1^{\ast}, 2^{\ast},
\cdots, \frac{1}{2} [N^2+3]^{\ast}$.

The operator product expansions (OPEs) between the spin-$1$
current $V^a(z)$ and the spin-$\frac{1}{2}$
current $Q^a(z)$  are described as  \cite{KT1985}
\bea
V^a(z) \, V^b(w) & = & \frac{1}{(z-w)^2} \, k \, g^{ab}
-\frac{1}{(z-w)} \, f^{ab}_{\,\,\,\,\,\,c} \, V^c(w) 
+\cdots,
\nonu \\
Q^a(z) \, Q^b(w) & = & -\frac{1}{(z-w)} \, (k+N+2) \, g^{ab} + \cdots,
\nonu \\
V^a(z) \, Q^b(w) & = & + \cdots.
\label{opevq}
\eea
The positive integer $k$ is the level of the spin $1$ current.
The metric $g_{ab}$ in (\ref{opevq})
is given by $g_{ab} = \mbox{Tr} (T_a T_b)$ and
the structure constant $f_{abc}$ is given by $f_{abc} = \mbox{Tr}
(T_c [T_a, T_b])$ where $T_a$ is the $SU(N+2)$ generator.
Note that the nonvanishing metric components are given by
$g_{A A^{\ast}} = g_{A^{\ast} A} =1$ where $A = 1, 2, \cdots,
\frac{1}{2}[(N+2)^2-1]$. Then by raising the $SU(N+2)$
adjoint lower index $A$, one has the $SU(N+2)$ adjoint upper
index $A^{\ast}$ and vice versa.
Note that the above adjoint indices $a,b, \cdots$ for the group $G$
are further divided into index $A$ and index $A^{\ast}$.

The explicit 
$11$ currents of large $\mathcal N = 4$ nonlinear superconformal algebra 
with (\ref{abnotation}) are given by
\bea
G^{0}(z) &  = &   \frac{i}{(k+N+2)}  \, Q_{\bar{a}} \, V^{\bar{a}}(z),
\qquad
G^{i}(z)  =  \frac{i}{(k+N+2)} 
\, h^{i}_{\bar{a} \bar{b}} \, Q^{\bar{a}} \, V^{\bar{b}}(z),
\nonu \\
A^{+i}(z) &  = & 
-\frac{1}{4N} \, f^{\bar{a} \bar{b}}_{\,\,\,\,\,\, c} \, h^i_{\bar{a} \bar{b}} \, V^c(z), 
\qquad
A^{-i}(z)  =  
-\frac{1}{4(k+N+2)} \, h^i_{\bar{a} \bar{b}} \, Q^{\bar{a}} \, Q^{\bar{b}}(z),
\nonu \\
T(z)  & = & 
\frac{1}{2(k+N+2)^2} \left[ (k+N+2) \, V_{\bar{a}} \, V^{\bar{a}} 
+k \, Q_{\bar{a}} \, \pa \, Q^{\bar{a}} 
+f_{\bar{a} \bar{b} c} \, Q^{\bar{a}} \, Q^{\bar{b}} \, V^c  \right] (z)
\nonu \\
&&- \frac{1}{(k+N+2)} \sum_{i=1}^3 ( A^{+i}+A^{-i} )^2 (z),
\label{11currents}
\eea
where $i=1,2,3$. The $G^{\mu}(z)$ with $\mu =(0, i)$
currents are four supersymmetry currents,
$A^{\pm i}(z)$ are six spin-$1$ generators of $SU(2)_{k} \times SU(2)_{N}$
and 
$T(z)$ is the spin-$2$ stress energy tensor. 
The three almost complex structures $ h^i_{\bar{a} \bar{b}}$ are given by
$4N \times 4N$ matrices as in \cite{AK1506}.
Note that the spin-$1$ current $A^{+i}(z)$ with level $k$ depends on
the spin-$1$ current $V^a(z)$ only while
 the spin-$1$ current $A^{-i}(z)$ with level $N$ depends on
the spin-$\frac{1}{2}$ current $Q^{\bar{a}}(z)$ only.
Then it is obvious that the OPEs between them are regular.

%%%%%%%%%%%%%%%%%%%%%%%%%%%%%%%%%%%%%%%%
\subsection{The minimal representations
 in the $\frac{SU(N+2)}{SU(N)
    \times SU(2)
  \times U(1)}$ Wolf space coset
}
%%%%%%%%%%%%%%%%%%%%%%%%%%%%%%%%%%%%%%%%

Let us describe the eigenvalues of
\bea
&& 1) \,\, \mbox{the zero mode of
the stress energy tensor spin-$2$ current}: \, \, T_0,
\nonu \\
&& 2) \,\, \mbox{the zero mode of sum of the square of spin-$1$ current
  }: \,\,  -\left[\sum_{i=1}^3 (A^{+i})^2 \right]_0,
\nonu \\
&& 3) \,\,
\mbox{the zero mode of sum of the square of  spin-$1$ current
  }: \,\,
-\left[\sum_{i=1}^3 ( A^{-i} )^2 \right]_0,
%\,\, \mbox{ and}
\nonu \\
&& 4) \,\, \mbox{the zero mode of other spin-$1$ current}: \,\,
\frac{1}{2} U_0,
\nonu
\eea
acting on the minimal representations \cite{GG1305}.
The corresponding eigenvalues are denoted by $h$, $l^{+}(l^{+}+1)$,
$l^-(l^-+1)$ and $\hat{u}$ respectively.
Note that the $U(1)$ current $U(z)$
is equivalent to the one ${\bf U}(z)$ of
the large ${\cal N}=4$ linear superconformal algebra.
That is, $U(z) = 2 i \sqrt{N(N+2)} \,{\bf U}(z)$ \cite{AK1506}
appearing in section $6$.

%%%%%%%%%%%%%%%%%%%%%%%%%%%%%%%%%%%%%%%
\subsubsection{ The $(0;f)$ representation
\label{0frep}}
%%%%%%%%%%%%%%%%%%%%%%%%%%%%%%%%%%%%%%%

One of the minimal representations is given by 
the fundamental representation $f$ (or $ \tiny\yng(1)$)
of $SU(N)$ living in the denominator of the Wolf space
coset. 
The corresponding state is given by \cite{GG1305}
\bea
|(0;f)>=\frac{1}{\sqrt{k+N+2}} \, Q_{-\frac{1}{2}}^{\bar{A}^{\ast}}|0>,
\qquad \bar{A}^{\ast} = 1^{\ast}, 2^{\ast}, \cdots, 2N^{\ast}. 
\label{state0f}
\eea
They correspond to the rectangular $N \times 2$ unitary matrix
inside of $(N+2) \times (N+2)$ unitary matrix in (\ref{4nmatrix}).
Under the decomposition of $SU(N+2)$ into the $SU(N) \times SU(2)$,
the adjoint representation of $SU(N+2)$ contains
$({\bf N}, {\bf 2})$ which corresponds to this $(0;f)$ representation.
Furthermore, the representation $(0;\bar{f})$ corresponds to
$(\overline{{\bf N}}, {\bf 2})$ associated with $
\frac{1}{\sqrt{k+N+2}} \, Q_{-\frac{1}{2}}^{\bar{A}}|0>$
with $\bar{A} = 1, 2, \cdots, 2N $ in the other rectangular $2 \times N$
unitary matrix in (\ref{4nmatrix}). 

The four eigenvalues are summarized by \cite{GG1305}
\bea
h (0;
  \tiny\yng(1)
) & = & \frac{(2k+3)}{4(N+k+2)},
\nonu \\
%l^{+} (0;{ \tiny\yng(1)}) & = & 0, \qquad
l^{+} (l^{+}+1)(0;{ \tiny\yng(1)})  & = &  0,
\nonu \\
%l^{-} (0;{ \tiny\yng(1)}) & = & \frac{1}{2},\qquad
l^{-} (l^{-}+1)  (0;{ \tiny\yng(1)})  & = &  \frac{3}{4},
\nonu \\
\hat{u}(0;  \tiny\yng(1) ) & = & \frac{(N+2)}{2}.
\label{0ffoureigen}
\eea
The eigenvalue $h$ can be obtained by using the relation
(\ref{conformaldimension}). The excitation number $n$ is given
by $\frac{1}{2}$. Note that
the excitation number for the fermion is given by
$\frac{1}{2}$ and for the multiple fermions, one can have
the excitation number which is given by the number of fermions divided
by two. 
Or one can calculate the OPE between
the ``reduced'' $T(z)$, where the spin-$1$ current $V^a(z)$ dependence
is ignored completely, and $Q^{\bar{A}^{\ast}}(w)$
and read off the second-order pole.

For the $l^-$ quantum number, the eigenvalue is obtained from the
OPE between the corresponding operator $
-\sum_{i=1}^3 ( A^{-i} )^2(z)$
and $Q^{\bar{A}^{\ast}}(w)$
and read off the second-order pole.
Therefore, one obtains $l^-=\frac{1}{2}$ which is consistent with the fact
that the above state transforms as a doublet under the $SU(2)_N$. 

For the $l^+$ quantum number, the relevant spin-$1$ current is given by
$A^{+i}(z)$ which contains only the spin-$1$ current $V^a(z)$. Therefore, 
the corresponding eigenvalue vanishes and $l^+=0$.

Similarly, the $U(1)$ charge $\hat{u}$ can be determined by
calculating the first order pole in the OPE between the $U(1)$ current
$\frac{1}{2}  U(z)$
and $Q^{\bar{A}^{\ast}}(w)$.

It is known that the BPS bound
for the conformal dimension \cite{GPTV,PT1,PT2} is
\bea
\frac{1}{(N+k+2)} \Bigg[ (k+1)l^{-} +(N+1) l^{+} +(l^{+}-l^{-})^2 \Bigg].
\label{BPS}
\eea
One can easily check that by substituting the quantum numbers
$l^+=0$ and $l^-=\frac{1}{2}$, the above conformal dimension
becomes the BPS bound \footnote{
\label{zerofbarexplanation}
  For the state $|(0;\overline{f})>$,
one can obtain similar eigenvalues where
the only difference appears in the eigenvalue $\hat{u}$. That is,
$\hat{u}(0;  \overline{\tiny\yng(1)} )  =  -\frac{(N+2)}{2}$.
Because the conformal dimension depends on $\hat{u}^2$
in (\ref{conformaldimension}),
the sign change for the $\hat{u}$ does not change the conformal dimension.}.
The large $(N,k)$ 't Hooft like limit for the BPS bound can be
obtained.

%%%%%%%%%%%%%%%%%%%%%%%%%%%%%%%%%%%%%%%
\subsubsection{ The $(f;0)$ representation
\label{f0subsectionname}}
%%%%%%%%%%%%%%%%%%%%%%%%%%%%%%%%%%%%%%%

Other minimal representation
is given by $(f;0)$ representation. In other words,
the singlets with respect to the $SU(N)$ can be obtained from
the fundamental representation in the $SU(N+2)$.
It is known that the branching rule for the fundamental representation
in the $SU(N+2)_k$ with respect to the $SU(N)_k \times SU(2)_k \times U(1)$
is characterized by
\bea
    {\tiny\yng(1)} \rightarrow ({\tiny\yng(1)},{\bf 1})_1
    + ({\bf 1},{\bf 2})_{-\frac{N}{2}}.
\label{f0branching}
\eea
The indices $1$ and $-\frac{N}{2}$ are the $U(1)$ charges $\hat{u}$
\cite{Slansky,Feger}.
Then the state $|(f;0)>$ corresponds to
$ ({\bf 1},{\bf 2})_{-\frac{N}{2}}$.

The four eigenvalues are described as \cite{GG1305}
\bea
h ({\tiny\yng(1)};0) & = & \frac{(2N+3)}{4(N+k+2)},
\nonu \\
%l^{+} ({ \tiny\yng(1)};0) & = & \frac{1}{2}, \qquad
l^{+} (l^{+}+1)({ \tiny\yng(1)};0)  & = &  \frac{3}{4},
\nonu \\
%l^{-} ({ \tiny\yng(1)};0) & = & 0,\qquad
l^{-} (l^{-}+1)  ({ \tiny\yng(1)};0)  & = &  0,
\nonu \\
\hat{u}(  \tiny\yng(1);0 ) & = & -\frac{N}{2}.
\label{boxzeroresult}
\eea
One can obtain the conformal dimension using the formula
(\ref{conformaldimension}). Or
after substituting the $SU(N+2)$ generator $T_{a^{\ast}}$ into the
zero mode of spin-$1$ current $V_0^a$ in the quadratic of the
reduced spin-$2$
current $T(z)$ where all the $Q^a$ dependent terms are ignored,
one obtains $(N+2) \times (N+2)$ unitary matrix acting on the state
$|(f;0)>$.
Then the conformal dimension for this state can be determined by
the diagonal elements of the last $2 \times 2$ subdiagonal matrix. 

For the $l^+$ quantum number, as described in the conformal dimension,
one can find $(N+2) \times (N+2)$ unitary matrix for the zero mode
of $-\sum_{i=1}^3 (A^{+i})^2$ which contains only the spin-$1$ current
$V^a(z)$. It turns out that the diagonal element of the last
$2 \times 2$ subdiagonal matrix  is given by $\frac{3}{4}$ which implies
$l^+=\frac{1}{2}$.  
This is consistent with the fact that this state
is a doublet under the $SU(2)_k$ $ ({\bf 1},{\bf 2})_{-\frac{N}{2}}$.

For the $l^-$ quantum number, because the spin-$\frac{1}{2}$ current
$Q^a(z)$ does not contribute to the eigenvalue equation associated with
this state, one has $l^-=0$.

For the $U(1)$ charge, one can construct
 $(N+2) \times (N+2)$ unitary matrix for the zero mode
of the reduced $\frac{1}{2} U(z)$ where the spin-$\frac{1}{2}$ current
$Q^a(z)$ dependence is removed. Then
 the diagonal element of the last
$2 \times 2$ subdiagonal matrix  is given by $-\frac{N}{2}$.

One can check the above conformal dimension satisfies the BPS bound
by substituting $l^+=\frac{1}{2}$ and $l^-=0$.

For the state $|(\overline{f};0)>$,
one can obtain similar eigenvalues where
the only difference appears in the eigenvalue $\hat{u}$. That is,
$\hat{u}(\overline{\tiny\yng(1)};0 )  =  \frac{N}{2}$.

%%%%%%%%%%%%%%%%%%%%%%%%%%%%%%%%%%%%%%%%
\subsection{The higher representations}
%%%%%%%%%%%%%%%%%%%%%%%%%%%%%%%%%%%%%%%%

Let us describe the four eigenvalues for the higher representations
which arise in the various products of the above minimal representations
$(0;f)$, $(f;0)$, $(0;\overline{f})$ and $(\overline{f};0)$.

%%%%%%%%%%%%%%%%%%%%%%%%%%%%%%%%%%%%%%%
\subsubsection{ The $(f;f)$ representation}
%%%%%%%%%%%%%%%%%%%%%%%%%%%%%%%%%%%%%%%

According to the previous branching rule in (\ref{f0branching}),
this higher representation
corresponds to $
({\tiny\yng(1)},{\bf 1})_1$ transforming as a fundamental representation
$f$ (or ${\tiny\yng(1)}$)
under the $SU(N)$ and a singlet under the $SU(2)_k$.
The nonzero $\hat{u}$ charge is given by $1$ from the subscript.

One can summarize the following eigenvalues \cite{GG1305}
\bea
h ({\tiny\yng(1)};{\tiny\yng(1)}) & = & \frac{1}{(N+k+2)},
\nonu \\
%l^{+} ({\tiny\yng(1)};{ \tiny\yng(1)}) & = & 0, \qquad
l^{+} (l^{+}+1)({\tiny\yng(1)};{ \tiny\yng(1)})  & = &  0,
\nonu \\
%l^{-} ({\tiny\yng(1)};{ \tiny\yng(1)}) & = & 0,\qquad
l^{-} (l^{-}+1)  ({\tiny\yng(1)};{ \tiny\yng(1)})  & = &  0,
\nonu \\
\hat{u}({\tiny\yng(1)};  \tiny\yng(1) ) & = & 1.
\label{ffeigenvalues}
\eea

In this case, the conformal dimension can be determined by
reading off the diagonal elements in the
$N \times N$ subdiagonal unitary matrix inside of
$(N+2) \times (N+2)$ unitary matrix obtained in previous
subsection. Or the formula in (\ref{conformaldimension}) can be used also.
From   the diagonal elements in the
$N \times N$ subdiagonal unitary matrix inside of
 $(N+2) \times (N+2)$ unitary matrix for the zero mode
of $-\sum_{i=1}^3 (A^{+i})^2$, one can determine the $l^+$ quantum number
which is equal to $0$. This is consistent with the singlet under the
$SU(2)_k$ in $
({\tiny\yng(1)},{\bf 1})_1$. For the $l^-$ quantum number,
it is the same as before and $l^-=0$.
From   the diagonal elements in the
$N \times N$ subdiagonal unitary matrix inside of
 $(N+2) \times (N+2)$ unitary matrix for the zero mode
of the reduced $\frac{1}{2} U(z)$, the above
$\hat{u}$ charge can be obtained.
It is easy to see that the above representation does not satisfy
the BPS bound because the conformal dimension for the BPS bound
is equal to $0$ by substituting $l^+=l^-=0$.

%%%%%%%%%%%%%%%%%%%%%%%%%%%%%%%%%%%%%%%
\subsubsection{ The $(f;\overline{f})$ representation
\label{ffbarrep}}
%%%%%%%%%%%%%%%%%%%%%%%%%%%%%%%%%%%%%%%

This representation
can be obtained from the product
between the minimal representations $(f;0)$ and $(0;\overline{f})$.
Then one can write down
the state as
$\frac{1}{\sqrt{k+N+2}} \, Q_{-\frac{1}{2}}^{\bar{A}}|(f;0)>$
where $\bar{A} = 1, 2, \cdots, 2N$ (\ref{state0f}).
The four eigenvalues are given by \cite{GG1305}
\bea
h ({\tiny\yng(1)}; \overline{{\tiny\yng(1)}}) & = & \frac{1}{2},
\nonu \\
%l^{+} ({\tiny\yng(1)};\overline{{ \tiny\yng(1)}}) & = & \frac{1}{2}, \qquad
l^{+} (l^{+}+1)({\tiny\yng(1)};\overline{{ \tiny\yng(1)}})  & = &
\frac{3}{4},
\nonu \\
%l^{-} ({\tiny\yng(1)}; \overline{{ \tiny\yng(1)}}) & = & \frac{1}{2},\qquad
l^{-} (l^{-}+1)  ({\tiny\yng(1)};\overline{{ \tiny\yng(1)}})  & = &
\frac{3}{4},
\nonu \\
\hat{u}({\tiny\yng(1)};  \overline{\tiny\yng(1)} ) & = & -N-1.
\label{ffbareigenvalues}
\eea

For the conformal dimension, there exists other contribution
in addition to the sum of $h({\tiny\yng(1)} ;0)$ and $h(0;
\overline{{\tiny\yng(1)}})$ (which is equal to
 $h(0;{\tiny\yng(1)})$).
The extra contribution coming from the lower order pole in the commutator
$[T_0, Q_{-\frac{1}{2}}^{\bar{A}}]$ takes the form 
\bea
\left(
\begin{array}{cc}
 \frac{1}{3(N+k+2)}{\bf 1}_{N \times N}  & 0 
  \\
0  & -\frac{1}{2(N+k+2)}{\bf 1}_{2 \times 2} 
\end{array}
\right).
\nonu
\eea
By realizing the diagonal elements in
the lower $2 \times 2$ subdiagonal matrix as the above extra contribution,
one can write down the final conformal dimension
from (\ref{boxzeroresult}) and (\ref{0ffoureigen}) as
\bea
\frac{(2N+3)}{4(N+k+2)}+ 
\frac{(2k+3)}{4(N+k+2)}-
\frac{1}{2(N+k+2)} = \frac{1}{2}.
\nonu
\eea
Of course, this counting can be seen from the formula
(\ref{conformaldimension}).
For the $l^{\pm}$ quantum numbers,
one can add each contribution from (\ref{boxzeroresult}) and
(\ref{0ffoureigen}).
For the $\hat{u}$ charge, one can add
each contribution and it turns out that
$-\frac{N}{2} -\frac{N+2}{2}=-N-1$.
By substituting $l^{\pm} =\frac{1}{2}$, the above
conformal dimension satisfies the 
BPS bound.

%%%%%%%%%%%%%%%%%%%%%%%%%%%%%%%%%%%%%%%
\subsubsection{ The $(\rm{symm};0)$ representation
\label{twosymmetric}}
%%%%%%%%%%%%%%%%%%%%%%%%%%%%%%%%%%%%%%%

This representation can be obtained from the product of
minimal representations $(f;0)$ and $(f;0)$.
By taking the product between the two identical branching rules
in (\ref{f0branching}), we obtain \cite{GG1305,Slansky,Feger}
\bea
    {\tiny\yng(1)} \otimes  {\tiny\yng(1)}
= {\tiny\yng(2)} + {\tiny\yng(1,1)}
    & \rightarrow & \Bigg[ ({\tiny\yng(1)},{\bf 1})_1
      + ({\bf 1},{\bf 2})_{-\frac{N}{2}} \Bigg] \otimes
    \Bigg[ ({\tiny\yng(1)},{\bf 1})_1
      + ({\bf 1},{\bf 2})_{-\frac{N}{2}} \Bigg]
    \nonu \\
    & = &  \Bigg[ ({\tiny\yng(2)},{\bf 1})_2 +
     ({\tiny\yng(1)},{\bf 2})_{1-\frac{N}{2}}
      +  ({\bf 1},{\bf 3})_{-N} \Bigg] \nonu \\
    & + & \Bigg[  ({\tiny\yng(1,1)},{\bf 1})_2 +
      ({\tiny\yng(1)},{\bf 2})_{1-\frac{N}{2}} +
      ({\bf 1},{\bf 1})_{-N} \Bigg].
    \nonu 
\eea
Then one can identify the following branching rules
under the $SU(N)_k \times SU(2)_k \times U(1)$
\bea
    {\tiny\yng(2)} & \rightarrow &
     ({\tiny\yng(2)},{\bf 1})_2 +
     ({\tiny\yng(1)},{\bf 2})_{1-\frac{N}{2}}
      +  ({\bf 1},{\bf 3})_{-N},
      \nonu \\
\tiny\yng(1,1) & \rightarrow &
({\tiny\yng(1,1)},{\bf 1})_2 +
      ({\tiny\yng(1)},{\bf 2})_{1-\frac{N}{2}} +
      ({\bf 1},{\bf 1})_{-N}.
  \label{Branching}
\eea

The 
$(\rm{symm};0)$ representation,
which is the singlet in $SU(N)_k$, corresponds to
$ ({\bf 1},{\bf 3})_{-N}$.
Then one can describe the four eigenvalues as follows:
\bea
h ({\tiny\yng(2)};0) & = & \frac{(N+2)}{(N+k+2)},
\nonu \\
%l^{+} ({ \tiny\yng(2)};0) & = & 1, \qquad
l^{+} (l^{+}+1)({ \tiny\yng(2)};0)  & = &  2,
\nonu \\
%l^{-} ({ \tiny\yng(2)};0) & = & 0,\qquad
l^{-} (l^{-}+1)  ({ \tiny\yng(2)};0)  & = &  0,
\nonu \\
\hat{u}(  \tiny\yng(2);0 ) & = & -N.
\label{eigenvaluestwosymm}
\eea
One can calculate the conformal dimension by using the formula
or by substituting the $SU(N+2)$ generator $T_{a^{\ast}}$ into the
zero mode of spin-$1$ current $V_0^a$ in the
reduced spin-$2$
current $T(z)$,
one obtains $\frac{1}{2} (N+2)(N+3) \times \frac{1}{2} (N+2)(N+3)$
unitary matrix acting on the state
$|(\mbox{symm};0)>$.
See Appendix $A$ for the generators for $N=3$. 
Then the conformal dimension for this state can be determined by
the diagonal elements of the last $3 \times 3$ subdiagonal matrix. 
The role of remaining diagonal elements in the
 $\frac{1}{2} N(N+1) \times \frac{1}{2} N(N+1)$
unitary matrix or
 $ 2N \times 2N$
unitary matrix will be explained in next subsection.

From the  whole
unitary matrix for the zero mode
of $-\sum_{i=1}^3 (A^{+i})^2$,
the diagonal element of the last
$3 \times 3$ subdiagonal matrix  (triplet under $SU(2)_k$)  implies
$l^+=1$.  
This is consistent with the fact that this state
is a triplet under the $SU(2)_k$ $({\bf 1},{\bf 3})_{-N}$.
For the $l^-$ quantum number, one has trivial $l^-=0$.
For the $U(1)$ charge, one can construct
the whole  unitary matrix for the zero mode
of the reduced $\frac{1}{2} U(z)$. Then
 the diagonal element of the last
$3 \times 3$ subdiagonal matrix  is given by $-N$
 which is simply the sum of two $U(1)$ charge $-\frac{N}{2}$
 for $\tiny\yng(1)$.
One can check the above conformal dimension satisfies the BPS bound
by substituting $l^+=1$ and $l^-=0$.
For the state $|(\overline{\mbox{symm}};0)>$,
one can obtain similar eigenvalues where
the only difference appears in the eigenvalue $\hat{u}$. That is,
$\hat{u}(\overline{\tiny\yng(2)};0 )  =  N$.

%%%%%%%%%%%%%%%%%%%%%%%%%%%%%%%%%%%%%%%
\subsubsection{The $(\rm{antisymm};0)$ representation
\label{twoantisymmetric}}
%%%%%%%%%%%%%%%%%%%%%%%%%%%%%%%%%%%%%%%

The other representation can arise from
 the product of
minimal representations $(f;0)$ and $(f;0)$.
The 
$(\rm{antisymm};0)$ representation,
which is the singlet in $SU(N)_k$, corresponds to
$ ({\bf 1},{\bf 1})_{-N}$ in the branching rule (\ref{Branching}).

Then the four eigenvalues are given as follows:
\bea
h ({\tiny\yng(1,1)};0) & = & \frac{N}{(N+k+2)},
\nonu \\
%l^{+} ({ \tiny\yng(1,1)};0) & = & 0, \qquad
l^{+} (l^{+}+1)({ \tiny\yng(1,1)};0)  & = &  0,
\nonu \\
%l^{-} ({ \tiny\yng(1,1)};0) & = & 0,\qquad
l^{-} (l^{-}+1)  ({ \tiny\yng(1,1)};0)  & = &  0,
\nonu \\
\hat{u}( \tiny \yng(1,1);0 ) & = & -N.
\label{antidoubleboxzero}
\eea
The conformal dimension formula
can be used here
or by substituting the $SU(N+2)$ generator $T_{a^{\ast}}$ into the
zero mode of spin-$1$ current $V_0^a$ in the
reduced spin-$2$
current $T(z)$,
one obtains $\frac{1}{2} (N+2)(N+1) \times \frac{1}{2} (N+2)(N+1)$
unitary matrix acting on the state
$|(\mbox{antisymm};0)>$.
See also Appendix $B$ for the generators with  $N=3$.
The conformal dimension for this state can be obtained by
the last diagonal element (singlet of $SU(2)_k$). 
The detailed descriptions  of remaining diagonal elements in the
 $\frac{1}{2} N(N-1) \times \frac{1}{2} N(N-1)$
unitary matrix or
 $ 2N \times 2N$
unitary matrix will be given in next subsection.

By reading off the last diagonal element in the whole matrix
for the zero mode of the operator corresponding to $l^+$ quantum number,
one obtains $l^+=0$ which is a singlet under the $SU(2)_k$
$({\bf 1},{\bf 1})_{-N}$.
As before, one has a trivial $l^-=0$ quantum number.
The similar analysis for the $\hat{u}$ charge can be done.
One can check the above conformal dimension does not satisfy
the BPS bound
by substituting $l^{\pm}=0$.
For the state $|(\overline{\mbox{antisymm}};0)>$,
one has
$\hat{u}(\overline{\tiny\yng(1,1)};0 )  =  N$.

%%%%%%%%%%%%%%%%%%%%%%%%%%%%%%%%%%%%%%%
\subsubsection{The $(0;\rm{symm})$ representation
\label{0symmtwo}}
%%%%%%%%%%%%%%%%%%%%%%%%%%%%%%%%%%%%%%%

One can also construct the product of
two minimal representations
$(0;f)$ and $(0;f)$.
The $(0;\rm{symm})$ representation can arise.
Let us focus on $N=3$ case for simplicity.
One can visualize the spin-$\frac{1}{2}$ current
$Q^{\bar{a}}(z)$ in the following $5 \times 5$ unitary matrix
\bea
\left(
\begin{array}{ccc|cc}
 0 & 0 & 0 & Q^{13} \equiv Q^{1^{\ast}} & Q^{16} \equiv Q^{4^{\ast}}  \\
 0 & 0 & 0 & Q^{14}  \equiv Q^{2^{\ast}} & Q^{17} \equiv Q^{5^{\ast}} \\
 0 & 0 & 0 & Q^{15} \equiv Q^{3^{\ast}} & Q^{18} \equiv Q^{6^{\ast}} \\
\hline
 Q^1 & Q^2 & Q^3 & 0 & 0 \\
 Q^4 & Q^5 & Q^6 & 0 & 0 \\
\end{array}
\right).
\label{5matrix}
\eea
The $Q^{1^{\ast}}, \cdots, Q^{6^{\ast}}$ appear in (\ref{state0f}). 

Let us construct the symmetric combinations
between the spin-$\frac{1}{2}$ currents
$Q^{\bar{A}^{\ast}}(z)$.
There are six states and the corresponding operators are
as follows:
\bea
&&  Q^{13} Q^{16}(z), \qquad
(Q^{13} Q^{17} + Q^{14} Q^{16})(z),
\qquad
(Q^{13} Q^{18} + Q^{15} Q^{16})(z),
\nonu \\
&&  Q^{14} Q^{17}(z),
\qquad
 (Q^{14} Q^{18} + Q^{15} Q^{17})(z),
\qquad  Q^{15} Q^{18}(z),
\label{symmetriclist}
\eea
up to some overall normalizations.
First of all,  the two $SU(2)$ indices should be
different from each other because of the fermionic property of
$Q^{\bar{a}}(z)$.
One should have 
one factor from the elements, $Q^{13}(z)$, $Q^{14}(z)$ or $Q^{15}(z)$ 
and the other factor from the elements,
$Q^{16}(z)$, $Q^{17}(z)$ or $Q^{18}(z)$.
When the $SU(N=3)$ indices are equal (i.e., if one takes
two operators in the same row of the matrix (\ref{5matrix})),
the interchange of these
indices gives the original term. Then one obtains
the single terms in (\ref{symmetriclist}).
When the $SU(N=3)$ indices are not equal to each other,
then there are extra terms. Then we have the remaining terms in
(\ref{symmetriclist}).   

The four eigenvalues are given by
\bea
h ({0;\tiny\yng(2)}) & = & \frac{k}{(N+k+2)},
\nonu \\
%l^{+} (0;{ \tiny\yng(2)}) & = & 0, \qquad
l^{+} (l^{+}+1)(0;{ \tiny\yng(2)})  & = &  0,
\nonu \\
%l^{-} (0;{ \tiny\yng(2)}) & = & 0,\qquad
l^{-} (l^{-}+1)  (0;{ \tiny\yng(2)})  & = &  0,
\nonu \\
\hat{u}(  0;\tiny\yng(2) ) & = & N+2.
\label{eigenvaluessymmtwo}
\eea

One can interpret the conformal dimension by
calculating the following second order pole of the OPE
\bea
T(z) \, Q^{13} Q^{16}(w)\Bigg|_{\frac{1}{(z-w)^2}} = 
\frac{k}{(N+k+2)} \, 
Q^{13} Q^{16}(w). \nonu
\eea
Of course, the stress energy tensor spin-$2$ current
does not contain the spin-$1$ current $V^a(z)$.
The $N$-dependence appearing in the
denominator can be easily generalized from $N=3$ result.
It is easy to check that the similar calculations for other
symmetric combinations in (\ref{symmetriclist})
lead to the same results.

For the $l^+$ quantum number, we have trivial $l^+=0$.
For the $l^-$ quantum number, one can compute the following
OPE and read off the coefficient of second order pole 
\bea
-\sum_{i=1}^{3} (A^{-i})^2 (z) \,
 Q^{13} Q^{16}(w) \Bigg|_{\frac{1}{(z-w)^2}} =0.
\nonu
\eea
Similarly, the $\hat{u}$ charge can be added
and leads to
\bea
i \sqrt{N(N+2)} \, {\bf U}(z) \, 
 Q^{13} Q^{16} (w)\Bigg|_{\frac{1}{(z-w)}} =
 (N+2) \, Q^{13} Q^{16}(w).
\nonu
\eea
One can check the above conformal dimension does not satisfy
the BPS bound
by substituting $l^{\pm}=0$ \footnote{
\label{0symmbarexplanation}
  One has
$\hat{u}(0;\overline{\tiny\yng(2)} )  = -( N+2)$.
We have also checked that 
 other
symmetric combinations in (\ref{symmetriclist})
lead to the corresponding same quantum numbers.}.

%%%%%%%%%%%%%%%%%%%%%%%%%%%%%%%%%%%%%%%
\subsubsection{The $(0;\rm{antisymm})$ representation
\label{0antisymm}}
%%%%%%%%%%%%%%%%%%%%%%%%%%%%%%%%%%%%%%%

Let us construct the antisymmetric combinations
between the spin-$\frac{1}{2}$ currents
$Q^{\bar{a}}(z)$.
There are six states and the corresponding operators are
as follows:
\bea
&&
 Q^{13} Q^{14}(z),
\qquad
Q^{13} Q^{15}(z),
\qquad
Q^{14} Q^{15}(z),
\nonu \\
&& Q^{16} Q^{17}(z)
\qquad
Q^{16} Q^{18}(z),
\qquad
Q^{17} Q^{18}(z).
\label{antisymmetriclist}
\eea
The two $SU(2)$ indices should be
equal to each other because of the fermionic property of
$Q^{\bar{a}}(z)$.
One should take the two operators in (\ref{5matrix})
from the same column.

The four eigenvalues are summarized by
\bea
h ({0;\tiny\yng(1,1)}) & = & \frac{(k+2)}{(N+k+2)},
\nonu \\
%l^{+} (0;{ \tiny\yng(1,1)}) & = & 0, \qquad
l^{+} (l^{+}+1)(0;{ \tiny\yng(1,1)})  & = &  0,
\nonu \\
%l^{-} (0;{ \tiny\yng(1,1)}) & = & 1,\qquad
l^{-} (l^{-}+1)  (0;{ \tiny\yng(1,1)})  & = &  2,
\nonu \\
\hat{u}( 0; \tiny\yng(1,1) ) & = & N+2.
\label{eigenvaluestwoantisymm}
\eea

For the conformal dimension, one can calculate the following
OPE
\bea
T(z) \, Q^{13} Q^{14}(w)\Bigg|_{\frac{1}{(z-w)^2}} = 
\frac{(k+2)}{(N+k+2)} \, 
Q^{13} Q^{14}(w). \nonu
\eea
Similarly, one can check that
the conformal dimension for other operators in (\ref{antisymmetriclist})
leads to the same quantum number.
The denominator for general $N$ can be expected.

For $l^+$ quantum number, we have trivial $l^+=0$.
For the $l^-$ quantum number, one can compute the following OPE
and read off the second order pole
\bea
-\sum_{i=1}^{3} (A^{-i})^2 (z) \,
Q^{13} Q^{14}(w) \Bigg|_{\frac{1}{(z-w)^2}} =
2 \, Q^{13} Q^{14}(w). 
\nonu
\eea
This implies that the $l^-$ quantum number is
$l^-=1$.
Similarly, the $\hat{u}$ charge can be obtained from the following
OPE
\bea
i \sqrt{N(N+2)} \, {\bf U}(z) \, 
 Q^{13} Q^{14} (w)\Bigg|_{\frac{1}{(z-w)}} =
 (N+2) \, Q^{13} Q^{14}(w).
\nonu
\eea
By substituting $l^+=0$ and $l^-=1$ into the 
BPS bound for the conformal dimension, one sees that
this state satisfies the BPS bound \footnote{
  We see that 
$\hat{u}(0;\overline{\tiny\yng(1,1)} )  = -( N+2)$.}.

%%%%%%%%%%%%%%%%%%%%%%%%%%%%%%%%%%%%%%%
\subsubsection{The $(0;\rm{antisymm})$ representation with three boxes
\label{0antisymmthree}}
%%%%%%%%%%%%%%%%%%%%%%%%%%%%%%%%%%%%%%%

For the antisymmetric representation with three boxes,
one has the following operators
\bea
Q^{13} Q^{14} Q^{15}(z),
\qquad
Q^{16} Q^{17} Q^{18}(z).
\nonu
\eea

The four eigenvalues
are
\bea
h ({0;\tiny\yng(1,1,1)}) & = & \frac{3(2k+5)}{4(N+k+2)},
\nonu \\
%l^{+} (0;{ \tiny\yng(1,1,1)}) & = & 0, \qquad
l^{+} (l^{+}+1)(0;{ \tiny\yng(1,1,1)})  & = &  0,
\nonu \\
%l^{-} (0;{ \tiny\yng(1,1,1)}) & = & \frac{3}{2},\qquad
l^{-} (l^{-}+1)  (0;{ \tiny\yng(1,1,1)})  & = &  \frac{15}{4},
\nonu \\
\hat{u}( 0; \tiny\yng(1,1,1) ) & = & \frac{3}{2}(N+2).
\label{eigenvaluesthreeantisymm}
\eea

The conformal dimension can be obtained from the following
OPE and the coefficient of second order pole 
leads to
\bea
T(z) \, Q^{13} Q^{14} Q^{15}(w)\Bigg|_{\frac{1}{(z-w)^2}} = 
\frac{3(2k+5)}{4(N+k+2)} \, 
Q^{13} Q^{14} Q^{15}(w). \nonu
\eea
Or
the conformal dimension formula (\ref{conformaldimension})
is given by
\bea
-\frac{3(N-3)(N+1)}{ 2N(k+N+2)}-
\frac{(\frac{3 }{2} (N+2) )^2}{N (N+2) (k+N+2)}+\frac{3}{2} =
\frac{3 (2 k+5)}{4 (k+N+2)}.
\nonu
\eea

As before, the $l^+$ quantum number is trivial and $l^+=0$.
For the $l^-$ quantum number, one can calculate the following
OPE
\bea
-\sum_{i=1}^{3} (A^{-i})^2 (z) \,
Q^{13} Q^{14} Q^{15}(w)\Bigg|_{\frac{1}{(z-w)^2}} = 
\frac{15}{4} \, 
Q^{13} Q^{14} Q^{15}(w).
\nonu
\eea
This implies that the $l^-$ quantum number
is given by $l^-=\frac{3}{2}$.
For the $\hat{u}$ charge, the following result holds
\bea
i \sqrt{N(N+2)} \, {\bf U}(z) \, 
Q^{13} Q^{14} Q^{15}(w)\Bigg|_{\frac{1}{(z-w)}} =
\frac{3}{2} (N+2) \, Q^{13} Q^{14} Q^{15}(w).
\nonu
\eea
With $l^+=0$ and $l^-=\frac{3}{2}$, the above conformal
dimension satisfies the BPS bound \footnote{ The
following quantum number $
\hat{u}( 0; \overline{\tiny\yng(1,1,1)} )  =  -\frac{3}{2}(N+2)
$ can be checked also.}.

%%%%%%%%%%%%%%%%%%%%%%%%%%
\subsubsection{New result: the $(0;\rm{mixed})$ representation
\label{0mixed}}
%%%%%%%%%%%%%%%%%%%%%%%%%%

In the  triple product of
three minimal representations
$(0;f)$, $(0;f)$ and $(0;f)$, one has
also 
the $(0;\rm{mixed})$ representation.
There are two fundamental representations,
$(Q^{13}, Q^{14}, Q^{15})$ for fixed $SU(2)$ index and
$(Q^{16}, Q^{17}, Q^{18})$ for fixed other $SU(2)$ index,
under the $SU(N=3)$.
Let us select one of the three quantities in
$(Q^{13}, Q^{14}, Q^{15})$. Then the next
quantity is chosen from the one of the
three quantities 
$(Q^{16}, Q^{17}, Q^{18})$.
Then the final quantity is selected from the first
$SU(3)$ fundamental $(Q^{13}, Q^{14}, Q^{15})$.
The mixed representation is either the one that is antisymmetric
in the first two indices or the other one
that is symmetric in the first two indices.
For the first choice, one has
the following states (with the conventions in Appendices) 
\bea
\hat{u}_1 &= & \frac{1}{\sqrt{2}}(E_{2 1 1}-E_{1 2 1}), \qquad
\hat{u}_2 = \frac{1}{\sqrt{2}} (E_{3 1 1} - E_{1 3 1}),
\nonu \\
\hat{u}_3 & = &  \frac{1}{2} (E_{3 2 1}-E_{2 3 1}+ E_{3 1 2}- E_{1 3 2}),
\qquad
\hat{u}_4  =   \frac{1}{\sqrt{2}} (E_{3 1 3}- E_{1 3 3}),
\nonu \\
\hat{u}_5 &= & \frac{1}{\sqrt{2}} (E_{3 2 3}- E_{2 3 3}),
\qquad
\hat{u}_6 = \frac{1}{\sqrt{2}} (E_{2 1 2} - E_{1 2 2}),
\nonu \\
\hat{u}_7 & = & \frac{1}{2} (E_{3 1 2} - E_{1 3 2}+ E_{2 1 3}- E_{1 2 3}), \qquad
\hat{u}_8 =  \frac{1}{\sqrt{2}} (E_{3 2 2}- E_{2 3 2}),
\label{eightcases}
\eea
up to some overall normalizations.
The three numbers stand for each fundamentals. 
For example,
the operator correpsonding to $(2 1 1- 1 2 1)$
is given by
$(Q^{14} Q^{16} Q^{13}-Q^{13} Q^{17} Q^{13})(z)$.
Of course the second term is identically zero by moving
$Q^{17}$ to the left.
See also Appendix $D$ for $SU(5)$ generators in mixed
representation.

Then after rewriting (\ref{eightcases}) with the help of
adjoint spin-$\frac{1}{2}$ current,
one has the following operators corresponding to mixed representations
\bea
&&
Q^{14} Q^{16} Q^{13} (z),
\qquad Q^{15} Q^{16} Q^{13} (z),
\nonu \\
&&
(Q^{15} Q^{17} Q^{13}-Q^{14} Q^{18} Q^{13}+
Q^{15} Q^{16} Q^{14}-Q^{13} Q^{18} Q^{14})(z),
\qquad
Q^{13} Q^{18} Q^{15} (z),
\nonu \\
&& Q^{14} Q^{18} Q^{15} (z),
\qquad
Q^{13} Q^{17} Q^{14} (z),
\label{mixedlist}
\\
&&
(Q^{15} Q^{16} Q^{14}-Q^{13} Q^{18} Q^{14}+
Q^{14} Q^{16} Q^{15}-Q^{13} Q^{17} Q^{15})(z),
\qquad
Q^{14} Q^{18} Q^{15} (z).
\nonu
\eea

The four eigenvalues are described as
\bea
h ({0;\tiny\yng(2,1)}) & = & \frac{3(1+2k)}{4(N+k+2)},
\nonu \\
%l^{+} (0;{ \tiny\yng(2,1)}) & = & 0, \qquad
l^{+} (l^{+}+1)(0;{ \tiny\yng(2,1)})  & = &  0,
\nonu \\
%l^{-} (0;{ \tiny\yng(2,1)}) & = & \frac{1}{2},\qquad
l^{-} (l^{-}+1)  (0;{ \tiny\yng(2,1)})  & = &  \frac{3}{4},
\nonu \\
\hat{u}( 0; \tiny\yng(2,1) ) & = & \frac{3}{2}(N+2).
\label{eigenvaluesmixed}
\eea

By using  the quadratic Casimir for the mixed representation
\cite{GG1011}
\bea
C^{(N)}({ \tiny\yng(2,1)}) = C^{(N)}([1,1,0, \cdots, 0])
=\frac{3 (N^2-3)}{2 N},
\nonu
\eea
and inserting the $\hat{u}$ charge $
\hat{u}( 0; \tiny\yng(2,1) )  =  \frac{3}{2}(N+2)$,
one can calculate the conformal dimension formula
and it is given by
\bea
-\frac{3 (N^2-3)}{2 N (k+N+2)}-
\frac{(\frac{3 }{2} (N+2) )^2}{N (N+2) (k+N+2)}+\frac{3}{2} =
\frac{3 (2 k+1)}{4 (k+N+2)}.
\nonu
\eea
Here the numerical value $\frac{3}{2}$ comes from 
the three products of spin-$\frac{1}{2}$ current $Q^{\bar{a}}(z)$.
Or one can calculate the following OPE
and focus on the second order pole
\bea
T(z) \, Q^{14} Q^{16} Q^{13} (w)\Bigg|_{\frac{1}{(z-w)^2}} = 
\frac{3(2k+1)}{4(N+k+2)} \, 
Q^{14} Q^{16} Q^{13} (w), \nonu
\eea
where we also put the $N$ dependence in the denominator.
For the $l^+$ quantum number, we have $l^+=0$
and for the $l^-$ quantum number, one can calculate
the following OPE and read off the second order pole 
\bea
-\sum_{i=1}^{3} (A^{-i})^2 (z) \,
Q^{14} Q^{16} Q^{13} (w) \Bigg|_{\frac{1}{(z-w)^2}} = 
\frac{3}{4} \, 
Q^{14} Q^{16} Q^{13} (w). \nonu
\eea
This implies that we have $l^-=\frac{1}{2}$.
For the $\hat{u}$ charge, the similar OPE
calculation leads to 
\bea
i \sqrt{N(N+2)} \, {\bf U}(z) \, 
Q^{14} Q^{16} Q^{13} (w)\Bigg|_{\frac{1}{(z-w)}} =
\frac{3}{2} (N+2) \,
Q^{14} Q^{16} Q^{13} (w),
\nonu
\eea
where
also the general $N$ dependence is included.
With $l^+=0$ and $l^-=\frac{1}{2}$, the above conformal
dimension does not satisfy the BPS bound
\footnote{For the higher representation
$( 0; \overline{\tiny\yng(2,1)} )$, the similar analysis can be done.}.
It is easy to check that the above four eigenvalues
can be obtained for other mixed representations in (\ref{mixedlist})
as follows:
\bea
\hat{u}_1 &= & \frac{1}{\sqrt{6}}(2 E_{112} -E_{2 1 1}-E_{1 2 1}), \qquad
\hat{u}_2 = \frac{1}{\sqrt{6}} (2 E_{113}- E_{3 1 1} - E_{1 3 1}),
\nonu \\
\hat{u}_3 & = &  \frac{1}{\sqrt{12}}
(2 E_{123}+ 2 E_{213}- E_{3 2 1}-E_{2 3 1}- E_{3 1 2}- E_{1 3 2}),
\qquad
\hat{u}_4  =   \frac{1}{\sqrt{6}} ( E_{3 1 3}-2 E_{331} + E_{1 3 3}),
\nonu \\
\hat{u}_5 &= & \frac{1}{\sqrt{6}} (E_{212} -2 E_{2 2 1}+ E_{1 2 2}),
\qquad
\hat{u}_6 = \frac{1}{2}
(E_{132}-  E_{231}+ E_{3 1 2}-E_{3 2 1}),
\nonu \\
\hat{u}_7 & = & \frac{1}{\sqrt{6}}
(-E_{3 2 2} +2 E_{2 2 3}- E_{2 3 2}), \qquad
\hat{u}_8 =  \frac{1}{\sqrt{6}} (-2 E_{3 3 2}+ E_{ 3 2 3}+E_{2 3 3}).
\nonu
\eea
These are symmetric in the first two indices.

%%%%%%%%%%%%%%%%%%%%%%%%%%
\subsubsection{The $(0;\rm{symm})$ representation with three boxes}
%%%%%%%%%%%%%%%%%%%%%%%%%%

Are there any states?
There are no such states because of the property of the fermions
$Q^{\bar{a}}$. There is a minus sign when two of them are interchanged
each other.

%%%%%%%%%%%%%%%%%%%%%%%%%%
\subsubsection{Summary of this section}
%%%%%%%%%%%%%%%%%%%%%%%%%%

Let us summarize what has been done in this section.
The conformal dimension for any representation can be
encoded in the formula (\ref{conformaldimension}).
For the representation $(\La_+;0)$, the $\hat{u}$ charge
is additive and is given by $-\frac{N}{2}$ (which is the
$\hat{u}$ charge in $(\tiny\yng(1);0)$) times the number of boxes. 
Its $l^-$ quantum number is trivial $l^-=0$. For the $l^+$ quantum number,
the maximum number, which is the $\frac{1}{2}$ times the
number of boxes, can arise in the symmetric representation. 
Other $l^+$ quantum numbers arise in the mixed representation
which will appear in next section.

For the  representation $(0;\La_-)$, the $\hat{u}$ charge
is additive and is given by $\frac{1}{2}(N+2)$ (which is the
$\hat{u}$ charge in $(0;\tiny\yng(1))$) times the number of boxes. 
Its $l^+$ quantum number is trivial $l^+=0$.
 For the $l^-$ quantum number,
the maximum number, which is the $\frac{1}{2}$ times the
number of boxes, can arise in the antisymmetric representation. 
Other $l^-$ quantum numbers arise in the mixed representation
which will appear in next section.

For the representation $(\La_+;\La_-)$, there are two cases.

1) The case where the representation $\La_-$ appears in the
branching of $\La_+$ under the $SU(N)_k \times SU(2)_k \times U(1)$.
There is a trivial $l^-=0$ quantum number.
For the $l^+$ quantum number and $\hat{u}$ charge
can be read off from the multiple product of
$  ({\tiny\yng(1)},{\bf 1})_1
    + ({\bf 1},{\bf 2})_{-\frac{N}{2}}$ in (\ref{f0branching}).

2) The case where the representation arises in the product of
$(\La_+;0)$ and $(0;\La_-)$.
The $l^{\pm}$ and $\hat{u}$ quantum numbers are additive.
In other words, the $l^+$ quantum number of $(\La_+;\La_-)$
comes from the one of $(\La_+;0)$ while
 the $l^-$ quantum number of $(\La_+;\La_-)$
 comes from the one of $(0;\La_-)$.
 The $\hat{u}$ charge of  $(\La_+;\La_-)$ is the sum of
 the one in  $(\La_+;0)$ and the one in  $(0;\La_-)$.
 Note that this is not true for the conformal dimension
because there exists the extra contribution.

%%%%%%%%%%%%%%%%%%%%%%%%%%%%%%%%%%%%%%%%%%%%%%%%%%%%%%%%%%%%%%%%%%%%%
%%%%%%%%%%%%%%%%%%%%%%%%%%%%%%%%%%%%%%%%%%%%%%%%%%%%%%%%%%%%%%%%%%%%%%
\section{ More eigenvalues for the
  higher representations in the $\frac{SU(N+2)}{SU(N)
    \times SU(2)
  \times U(1)}$ Wolf space coset}
%section3%%%%%%%%%%%%%%%%%%%%%%%%%%%%%%%%%%%%%%%%%%%%%%%%%%%%%%%%%%%%%%%%%%%%%
%%%%%%%%%%%%%%%%%%%%%%%%%%%%%%%%%%%%%%%%%%%%%%%%%%%%%%%%%%%%%%%%%%%%%

In this section, 
we would like to construct the four eigenvalues for the zero modes
of previous stress energy tensor spin-$2$ current, the
sum of square of spin-$1$ current,
the sum of square of other spin-$1$ current and
other spin-$1$ current, acting on
other higher representations
by considering the multiple
products of $(0;f)$, $(f;0)$, $(0;\overline{f})$
or $(\overline{f};0)$.

%%%%%%%%%%%%%%%%%%%%%%%%%%%%%%%%%%%%
%%%%%%%%%%%%%%%%%%%
\subsection{
The symmetric  representations $\La_+$ with two boxes
%$(0;\yng(3))$
}
%%%%%%%%%%%%%%%%%%%
%%%%%%%%%%%%%%%%%%%%%%%%%%%%%%%%%%%%%

One of the simplest higher representation is
given by the symmetric representation
$\mbox{symm}$ or 
$
\tiny\yng(2)
$.
It is known that
the product of
 the minimal representation $(f;0)$ and  itself $(f;0)$
 implies
 that there are symmetric and antisymmetric representations.
 From the branching rule for the $SU(N+2)$ under the
 $SU(N)_k \times SU(2)_k \times U(1)$, one can identify
 the following branching rules
 \bea
    {\tiny\yng(1)} \otimes  {\tiny\yng(1)}
= {\tiny\yng(2)} + {\tiny\yng(1,1)}
    & \rightarrow & \Bigg[ ({\tiny\yng(1)},{\bf 1})_1
      + ({\bf 1},{\bf 2})_{-\frac{N}{2}} \Bigg] \otimes
    \Bigg[ ({\tiny\yng(1)},{\bf 1})_1
      + ({\bf 1},{\bf 2})_{-\frac{N}{2}} \Bigg]
    \nonu \\
    & = &  \Bigg[ ({\tiny\yng(2)},{\bf 1})_2 +
     ({\tiny\yng(1)},{\bf 2})_{1-\frac{N}{2}}
      +  ({\bf 1},{\bf 3})_{-N} \Bigg] \nonu \\
    & + & \Bigg[  ({\tiny\yng(1,1)},{\bf 1})_2 +
      ({\tiny\yng(1)},{\bf 2})_{1-\frac{N}{2}} +
      ({\bf 1},{\bf 1})_{-N} \Bigg].
    \nonu 
\eea
The subscript stands for the $U(1)$ charge $\hat{u}$.
The final $\hat{u}$ charge is obtained by adding each
$\hat{u}$ charge.
Then one obtains the following branching rules
for the symmetric and antisymmetric representations
under the  $SU(N)_k \times SU(2)_k \times U(1)$ (again from
(\ref{Branching}))
\bea
    {\tiny\yng(2)} & \rightarrow &
     ({\tiny\yng(2)},{\bf 1})_2 +
     ({\tiny\yng(1)},{\bf 2})_{1-\frac{N}{2}}
      +  ({\bf 1},{\bf 3})_{-N},
      \nonu \\
\tiny\yng(1,1) & \rightarrow &
({\tiny\yng(1,1)},{\bf 1})_2 +
      ({\tiny\yng(1)},{\bf 2})_{1-\frac{N}{2}} +
      ({\bf 1},{\bf 1})_{-N}.
  \label{transtrans}
\eea
Note that the last representations in (\ref{transtrans}) correspond to
the symmetric and antisymmetric representations $|(\tiny\yng(2) ;0) >$
and  $|(\tiny\yng(1,1) ;0) >$ repspectively.

The two-index symmetric parts of the $SU(N+2)$ representation
can be obtained from the generators of the fundamental representation
of $SU(N+2)$ by using the projection operator $\frac{1}{2} (\de_{ik}
\de_{jl}+\de_{jk} \de_{il})$ where $i \leq j$ and $ k \leq l$
and $i, j, k, l = 1, 2, \cdots, (N+2) $ \cite{Cvitanovic,Joseph}.
Then by acting on
the space
$T_a \otimes {\bf 1}_{(N+2) \times (N+2)} + {\bf 1}_{(N+2) \times (N+2)}
\otimes T_a$,
one has
the generators for the symmetric representation for the $SU(N+2)$
\bea
(T_a)_{ik} \, \delta_{jl} +
(T_a)_{jk} \, \delta_{il} +
 \delta_{ik} \, (T_a)_{jl}  +
\delta_{jk} \, (T_a)_{il}.  
\nonu
\eea

For $N=3$, one has
$\frac{1}{2}(N+2)(N+3) \times \frac{1}{2}(N+2)(N+3)
= 15 \times 15$ unitary matrix, 
and the row and columns are characterized by
the following double index notations
\bea
11, \, 12, \, 13, \, 22, \, 23, \, 33; \,
14, \, 15, \, 24, \, 25, \, 34, \, 35; \, 44, \, 45, \, 55, 
\nonu
\eea
corresponding to $\hat{u}_1, \cdots, \hat{u}_{15}$ in Appendix $A$.
The first six elements correspond to the
symmetric representation for $SU(3)$.
The next six elements correspond to
the fundamental representation of $SU(3)$
with $SU(2)_k$ doublet.
The last three elements correspond to
the singlet of $SU(3)$ with $SU(2)_k$ triplet
according to (\ref{transtrans}).
The explicit form for the $24$ generators of
$SU(5)$ is presented in Appendix $A$.

Let us calculate the zero mode for the reduced stress energy tensor
spin-$2$ current acting on the state $|(\tiny\yng(2) ;0) >$. 
It turns out that the $15 \times 15$ matrix
is given by
\bea
\left(
\begin{array}{ccc}
  \frac{2}{(5+k)} {\bf 1}_{6 \times 6}& 0 & 0
  \\
0  & \frac{17}{4(5+k)}{\bf 1}_{6 \times 6} &  0\\
 0 &0 & \frac{5}{(5+k)}{\bf 1}_{3 \times 3}
\end{array}
\right).
\label{spin2symmetric}
\eea
There are three block diagonal elements. 
The last block diagonal elements correspond to
the eigenvalue on the  state $|(\tiny\yng(2) ;0) >$.
We will describe the detailed quantum numbers
for the other eigenvalues soon.

One can also calculate the zero mode for the sum of the square for
the spin-$1$ current with minus sign acting on the above
state $|(\tiny\yng(2) ;0) >$
and the explicit result is given by 
\bea
\left(
\begin{array}{ccc}
  0 {\bf 1}_{6 \times 6} & 0 & 0
  \\
0  & \frac{3}{4}{\bf 1}_{6 \times 6} &  0\\
 0 &0 & 2 {\bf 1}_{3 \times 3}
\end{array}
\right).
\label{su2symmetric}
\eea
The diagonal elements correspond to the eigenvalues
and in particular, the last one is the eigenvalue for the
state  $|(\tiny\yng(2) ;0) >$ which behaves as a singlet under the
$SU(3)$. 

Similarly, one can also compute the
zero mode for the spin-$1$ current
acting on the 
state $|(\tiny\yng(2) ;0) >$
and one obtains
\bea
\left(
\begin{array}{ccc}
 2{\bf 1}_{6 \times 6}  & 0 & 0
  \\
0  & -\frac{1}{2}{\bf 1}_{6 \times 6} &  0\\
 0 &0 & -3 {\bf 1}_{3 \times 3}
\end{array}
\right).
\label{u1chargesymm}
\eea
In this case, also the last elements are the eigenvalues
for the state  $|(\tiny\yng(2) ;0) >$.

%%%%%%%%%%%%%%%%%%%%%%%%%%%%%%%%%%%%%%%
\subsubsection{The $(\rm{symm};\rm{symm})$ representation}
%%%%%%%%%%%%%%%%%%%%%%%%%%%%%%%%%%%%%%%

Let us consider the higher representation 
where the symmetric representation in $SU(N)$ $ ({\tiny\yng(2)},{\bf 1})_2$
survives
in the branching of (\ref{transtrans}).
The four eigenvalues are given by
\bea
h ({\tiny\yng(2);\tiny\yng(2)}) & = & \frac{2}{(N+k+2)},
\nonu \\
%l^{+} (\tiny\yng(2);{ \tiny\yng(2)}) & = & 0, \qquad
l^{+} (l^{+}+1)(\tiny\yng(2);{ \tiny\yng(2)})  & = &  0,
\nonu \\
%l^{-} (\tiny\yng(2);{ \tiny\yng(2)}) & = & 0,\qquad
l^{-} (l^{-}+1)  (\tiny\yng(2);{ \tiny\yng(2)})  & = &  0,
\nonu \\
\hat{u}( \tiny\yng(2); \tiny\yng(2) ) & = & 2.
\label{doubleboxsymm}
\eea
One can calculate the conformal dimension for this
representation using the previous formula or
one performs the explicit form for the
matrix in this particular representation
as in (\ref{spin2symmetric}). 
Therefore, one obtains
$\frac{2}{(k+5)}$ in the first diagonal elements.
One can apply for other $N$ values where $N=5,7,9, 11, 13, \cdots$. 
It turns out that the numerator of the above quantity  does not
depend on $N$ and takes the common value and the denominator
is generalized to $(k+N+2)$.
By realizing that the quadratic Casimir of $SU(N+2)$ for the
symmetric representation $C^{(N+2)}(\tiny\yng(2))=\frac{(N+1)(N+4)}{(N+2)}$
(and the one for the symmetric representation
$C^{(N)}(\tiny\yng(2))=\frac{(N-1)(N+2)}{N}$)
and the correct $\hat{u}$ charge is given by $2$,
the following relation can be obtained
\bea
\frac{2 (N+1) (N+2+2)}{2 (N+2) (k+N+2)}
-\frac{2 (N-1) (N+2)}{2 N (k+N+2)}-\frac{4}{N (N+2) (k+N+2)}=
\frac{2}{(N+k+2)}.
\nonu
\eea

Similarly, the quantum number for the
$l^{+}$ can be determined by 
the above matrix calculation in (\ref{su2symmetric}).
From the zero eigenvalues appearing in the first block diagonal matrix
in (\ref{su2symmetric}), one can see the above
$l^{+}$ quantum number, a singlet under the $SU(2)_k$.
This also canbe seen from the previous expression
$ ({\tiny\yng(2)},{\bf 1})_2$.
The trivial $l^-$ quantum number $l^-=0$ arises.
For the last eigenvalue corresponding to $\hat{u}$ charge,
one uses the previous matrix calculation given in (\ref{u1chargesymm}).
The eigenvalues  
appearing in the first block diagonal matrix
in (\ref{u1chargesymm}) imply that the $\hat{u}$ charge is given by $2$.
This is also consistent with the representation 
$ ({\tiny\yng(2)},{\bf 1})_2$ where the subscript denotes
the $\hat{u}$ charge.
One observes that the above conformal dimension does not satisfy
the vanishing BPS bound with $l^{\pm} =0$.

%%%%%%%%%%%%%%%%%%%%%%%%%%%%%%%%%%%%%%%
\subsubsection{The $({\rm symm};f)$ representation}
%%%%%%%%%%%%%%%%%%%%%%%%%%%%%%%%%%%%%%%

Let us consider the
 higher representation 
 where the fundamental representation in
 $SU(N)$ $ ({\tiny\yng(1)},{\bf 2})_{1-\frac{N}{2}}$
survives
in the branching of (\ref{transtrans}).
The four eigenvalues can be summarized by
\bea
h ({\tiny\yng(2);\tiny\yng(1)}) & = & \frac{(2N+11)}{4(N+k+2)},
\nonu \\
%l^{+} (\tiny\yng(2);{ \tiny\yng(1)}) & = & \frac{1}{2}, \qquad
l^{+} (l^{+}+1)(\tiny\yng(2);{ \tiny\yng(1)})  & = &  \frac{3}{4},
\nonu \\
%l^{-} (\tiny\yng(2);{ \tiny\yng(1)}) & = & 0,\qquad
l^{-} (l^{-}+1)  (\tiny\yng(2);{ \tiny\yng(1)})  & = &  0,
\nonu \\
\hat{u}( \tiny\yng(2); \tiny\yng(1) ) & = & -\frac{N}{2}+1.
\label{symmfeigen}
\eea
The explicit form for the
matrix in this particular representation
of conformal dimension is given by (\ref{spin2symmetric}). 
Then, one obtains
$\frac{17}{4(k+5)}$ in the second block diagonal elements.
One can apply for other $N$ values  
and it turns out that the numerator of the above quantity  does 
depend on $N$ linearly as well as the constant term
while the denominator
is generalized to $4(k+N+2)$.
One can also use the formula with the correct $\hat{u}$ charge
\bea
\frac{2 (N+1) (N+2+2)}{2 (N+2) (k+N+2)}-\frac{(\frac{N}{2}
  -\frac{1}{2 N})}{(k+N+2)}-\frac{(1-\frac{N}{2})^2}{N (N+2) (k+N+2)}
=\frac{(2N+11)}{4(N+k+2)},
\nonu
\eea
where the quadratic Casimir 
$C^{(N)}(\tiny\yng(1))=(\frac{N}{2}-\frac{1}{2N})$ is used.
The quantum number for the
$l^{+}$ can be obtained by 
the above matrix calculation in (\ref{su2symmetric}).
From the  eigenvalues $\frac{3}{4}$
appearing in the second block diagonal matrix
in (\ref{su2symmetric}), one can see the above
$l^{+}$ quantum number $l^{+} =\frac{1}{2}$, a doublet under the $SU(2)_k$.
This also canbe seen from the previous expression
$ ({\tiny\yng(1)},{\bf 2})_{1-\frac{N}{2}}$.
The trivial $l^-=0$ quantum number holds in this representation.
For the last eigenvalue corresponding to $\hat{u}$ charge,
the previous matrix calculation given in (\ref{u1chargesymm})
can be used.
The eigenvalues  
appearing in the second block diagonal matrix
in (\ref{u1chargesymm}) imply that the $\hat{u}$ charge is given by
$-\frac{1}{2}$.
Varying the $N$ values,
one finds that the $\hat{u}$ charge is linear in $N$
as well as the constant term.
This is also consistent with the representation 
$ ({\tiny\yng(1)},{\bf 2})_{1-\frac{N}{2}}$ where the subscript denotes
the $\hat{u}$ charge.
One observes that the above conformal dimension does not satisfy
the  BPS bound.

%%%%%%%%%%%%%%%%%%%%%%%%%%%%%%%%%%%%%%%
\subsubsection{The $({\rm symm};0)$ representation}
%%%%%%%%%%%%%%%%%%%%%%%%%%%%%%%%%%%%%%%

The eigenvalues in this higher representation
are given in the subsection \ref{twosymmetric}.
One sees that there are eigenvalues in (\ref{eigenvaluestwosymm}).
From the three matrices
in (\ref{spin2symmetric}), (\ref{su2symmetric}) and (\ref{u1chargesymm}),
the relevant eigenvalues are given by
$\frac{5}{(k+5)}$, $2$ and $-3$ for the nontrivial quantum numbers.
They are generalized to $\frac{(N+2)}{(k+N+2)}$, $2$ and $-N$
respectively. The conformal dimension can be checked from
$\frac{2(N+1)(N+2+2)}{2(N+2)(k+N+2)}-\frac{N^2}{N(N+2)(k+N+2)}$ also.

%%%%%%%%%%%%%%%%%%%%%%%%%%%%%%%%%%%%%%%
\subsubsection{The $({\rm symm};\overline{f})$ representation
\label{previoussubsection}}
%%%%%%%%%%%%%%%%%%%%%%%%%%%%%%%%%%%%%%%

Let us consider the higher representation
which arises from the product of $(\tiny\yng(2);0)$
and $(0;\overline{\tiny\yng(1)})$.
The former occurs in the subsection 
\ref{twosymmetric} and the latter occurs in the subsection
\ref{0frep} together with the complex conjugation in the footnote
\ref{zerofbarexplanation}.

In this case, the corresponding
four eigenvalues are described by
\bea
h ({\tiny\yng(2);\overline{\tiny\yng(1)}}) & = & \frac{(4N+2k+7)}{4(N+k+2)},
\nonu \\
%l^{+} (\tiny\yng(2);\overline{ \tiny\yng(1)}) & = & 1, \qquad
l^{+} (l^{+}+1)(\tiny\yng(2);\overline{ \tiny\yng(1)})  & = &  2,
\nonu \\
%l^{-} (\tiny\yng(2);\overline{ \tiny\yng(1)}) & = & \frac{1}{2},\qquad
l^{-} (l^{-}+1)  (\tiny\yng(2);\overline{ \tiny\yng(1)})  & = &  \frac{3}{4},
\nonu \\
\hat{u}( \tiny\yng(2); \overline{\tiny\yng(1)} ) & = & -\frac{3N}{2}-1.
\label{eigenforsymmf}
\eea

First of all, one can obtain the following $15 \times 15$ matrix
by calculating the commutator
$[T_0, Q_{-\frac{1}{2}}^{\bar{A}}]$ as in the subsection
\ref{ffbarrep}
\bea
\left(
\begin{array}{ccc}
 \frac{2}{3(5+k)}{\bf 1}_{6 \times 6}  & 0 & 0
  \\
0  & -\frac{1}{6(5+k)}{\bf 1}_{6 \times 6} &  0\\
 0 &0 & -\frac{1}{(5+k)} {\bf 1}_{3 \times 3}
\end{array}
\right).
\label{extramatrix}
\eea
The last three eigenvalues (the $N$ generalization is
straightforward to obtain) appearing in the last block diagonal
matrix in (\ref{extramatrix})
provide the extra contribution as well as the sum of conformal dimensions
of  $(\tiny\yng(2);0)$
and $(0;\overline{\tiny\yng(1)})$.
They are given in (\ref{eigenvaluestwosymm}) and (\ref{0ffoureigen})
respectively.
Then one obtains the final conformal dimension by adding the above
contribution appearing in (\ref{extramatrix}) as follows
\bea
\frac{(N+2)}{(N+k+2)}+ 
\frac{(2k+3)}{4(N+k+2)}-
\frac{1}{(N+k+2)} = \frac{(4N+2k+7)}{4(N+k+2)},
\nonu
\eea
as in (\ref{eigenforsymmf}).
It is also useful to interpret the above result
from the conformal dimension formula.
One determines the following result
\bea
\frac{2 (N+1) (N+2+2)}{2 (N+2) (k+N+2)}
-\frac{(\frac{N}{2}-\frac{1}{2 N})}{(k+N+2)}
-\frac{(-\frac{1}{2} (N+2)-N)^2}{N (N+2) (k+N+2)}
+\frac{1}{2} =
\frac{(4N+2k+7)}{4(N+k+2)}.
\nonu
\eea
Here we used the quadratic Casimirs for $C^{(N+2)}(\tiny\yng(2))$
and $C^{(N)}(\overline{\tiny\yng(1)})$. The correct $\hat{u}$ charge is inserted. The excitation number is given by $\frac{1}{2}$.

For the $l^+$ quantum number, due to the vanishing of $l^+$ in
$(0;\overline{\tiny\yng(1)})$, it turns out that
the $l^+$ is the same as the one($l^+=1$) in $(\tiny\yng(2);0)$. 
For the $l^-$ quantum number, due to the vanishing of $l^-$ in
 $(\tiny\yng(2);0)$, it turns out that
the $l^-$ is the same as the one($l^-=\frac{1}{2}$) in
$(0;\overline{\tiny\yng(1)})$. 
It is easy to see that
the above conformal dimension satisfies the BPS bound by substituting
$l^+=1$ and $l^-=\frac{1}{2}$.
One can add each $\hat{u}$ charge and it is obvious that
the total $\hat{u}$ charge is given by $-N -\frac{(N+2)}{2}$
which leads to the above result. Note that
the $\hat{u}$ charge for $(0;\overline{\tiny\yng(1)})$ is opposite to
the one for $(0;\tiny\yng(1))$. 

%%%%%%%%%%%%%%%%%%%%%%%%%%%%%%%%%%%%%%%
\subsubsection{The $(\rm{symm};\rm{antisymm})$ representation}
%%%%%%%%%%%%%%%%%%%%%%%%%%%%%%%%%%%%%%%

Let us consider the higher representation
which arises from the product of $(\tiny\yng(2);0)$
and $(0;\tiny\yng(1,1))$.
The former occurs in the subsection 
\ref{twosymmetric} and the latter occurs in the subsection
\ref{0antisymm}.

The four eigenvalues can be summarized by
\bea
h ({\tiny\yng(2);\tiny\yng(1,1)}) & = & \frac{(N+k+6)}{(N+k+2)},
\nonu \\
%l^{+} (\tiny\yng(2);{ \tiny\yng(1,1)}) & = & 1, \qquad
l^{+} (l^{+}+1)(\tiny\yng(2);{ \tiny\yng(1,1)})  & = &  2,
\nonu \\
%l^{-} (\tiny\yng(2);{ \tiny\yng(1,1)}) & = & 1,\qquad
l^{-} (l^{-}+1)  (\tiny\yng(2);{ \tiny\yng(1,1)})  & = &  2,
\nonu \\
\hat{u}( \tiny\yng(2); \tiny\yng(1,1) ) & = & 2.
\label{symmantisymmexp}
\eea

The following $15 \times 15$ matrix
by calculating the commutator
$[T_0,  Q_{-\frac{1}{2}}^{13} Q_{-\frac{1}{2}}^{14}]$ as in the subsection
\ref{previoussubsection} 
can be obtained
\bea
\left(
\begin{array}{ccc}
 -\frac{4}{3(5+k)}{\bf 1}_{6 \times 6}  & 0 & 0
  \\
0  & \frac{1}{3(5+k)}{\bf 1}_{6 \times 6} &  0\\
 0 &0 & \frac{2}{(5+k)} {\bf 1}_{3 \times 3}
\end{array}
\right).
\label{extramatrix1}
\eea
The last three eigenvalues (the $N$ generalization is
simply $ \frac{2}{(N+k+2)}$) appearing in the last block diagonal
matrix in (\ref{extramatrix1})
give the extra contribution as well as the sum of conformal dimensions
of  $(\tiny\yng(2);0)$
and $(0;\tiny\yng(1,1))$.
They are given in (\ref{eigenvaluestwosymm}) and
(\ref{eigenvaluestwoantisymm})
respectively.
Then one obtains the final conformal dimension by adding the above
contribution appearing in (\ref{extramatrix1}) as follows
\bea
\frac{(N+2)}{(N+k+2)}+ 
\frac{(k+2)}{(N+k+2)}+
\frac{2}{(N+k+2)} = \frac{(N+k+6)}{(N+k+2)}.
\nonu
\eea
Furthermore,
this analysis can be seen from the conformal dimension formula
\bea
\frac{2 (N+1) (N+2+2)}{2 (N+2) (k+N+2)}
-\frac{2 (N-2) (N+1)}{2 N (k+N+2)}
-\frac{((N+2)-N)^2}{N (N+2) (k+N+2)}
+1=\frac{(N+k+6)}{(N+k+2)}.
\nonu
\eea
Here we also use the quadratic Casimir
$C^{(N)}(\tiny\yng(1,1))=\frac{(N-2)(N+1)}{N}$ and the excitation number
is equal to $1$. The correct $\hat{u}$ charge is inserted.
For the $l^+$ quantum number, due to the vanishing of $l^+$ in
$(0;\tiny\yng(1,1))$, 
the $l^+$ is the same as the one($l^+=1$) in $(\tiny\yng(2);0)$. 
For the $l^-$ quantum number, due to the vanishing of $l^-$ in
 $(\tiny\yng(2);0)$, 
the $l^-$ is the same as the one($l^-=1$) in
$(0;\tiny\yng(1,1))$. 
The total $\hat{u}$ charge is given by $-N+(N+2)$
which leads to the above result.
One can easily see that the above conformal dimension
does not lead to the BPS bound with $l^{\pm} =1$ \footnote{For the
  higher representation  $(\tiny\yng(2);\overline{\tiny\yng(1,1)} )$,
  the similar analysis can be done.
We have
\bea
h (\tiny\yng(2);\overline{\tiny\yng(1,1)}) & = & 1,
\qquad
%l^{+} (\tiny\yng(3);\tiny\yng(1,1,1)) & = &  ,\qquad
l^{+} (l^{+}+1)(\tiny\yng(2);\overline{\tiny\yng(1,1)})   =  2,  
\nonu \\
%l^{-} (\tiny\yng(3);\tiny\yng(1,1,1)) & = & , \qquad
l^{-} (l^{-}+1)  (\tiny\yng(2);\overline{\tiny\yng(1,1)}) & = & 2, 
\qquad
\hat{u}( \tiny\yng(2);\overline{\tiny\yng(1,1)} )  =  -2N-2.
\nonu
\eea
This conformal dimension
does lead to the BPS bound with $l^{\pm} =1$. Note that
different $\hat{u}$ charge leads to the different
$h$ value.}.

%%%%%%%%%%%%%%%%%%%%%%%%%%%%%%%%%%%%%%%
\subsubsection{The $(\rm{symm};\overline{\rm{symm}})$ representation}
%%%%%%%%%%%%%%%%%%%%%%%%%%%%%%%%%%%%%%%

Let us consider the higher representation
which arises from the product of $(\tiny\yng(2);0)$
and $(0;\overline{\tiny\yng(2)})$.
The former occurs in the subsection 
\ref{twosymmetric} while the latter occurs in the subsection
\ref{0symmtwo} with complex conjugation in the footnote
\ref{0symmbarexplanation}.

The four eigenvalues are given by 
\bea
h (\tiny\yng(2);\overline{\tiny\yng(2)}) & = & \frac{(k+N)}{(k+N+2)},
\nonu \\
%l^{+} (\tiny\yng(2);\overline{ \tiny\yng(2)}) & = & 1, \qquad
l^{+} (l^{+}+1)(\tiny\yng(2);\overline{ \tiny\yng(2)})  & = &  2,
\nonu \\
%l^{-} (\tiny\yng(2);\overline{ \tiny\yng(2)}) & = & 0,\qquad
l^{-} (l^{-}+1)  (\tiny\yng(2);\overline{ \tiny\yng(2)}) & = &  0,
\nonu \\
\hat{u}( \tiny\yng(2); \overline{\tiny\yng(2)} ) & = & -2N-2.
\label{symmsymmbarexpression}
\eea

One should calculate the commutator
$[T_0,  Q_{-\frac{1}{2}}^{1} Q_{-\frac{1}{2}}^{4}]$
and it turns out that
\bea
\left(
\begin{array}{ccc}
 \frac{4}{3(5+k)}{\bf 1}_{6 \times 6}  & 0 & 0
  \\
0  & -\frac{1}{3(5+k)}{\bf 1}_{6 \times 6} &  0\\
 0 &0 & -\frac{2}{(5+k)} {\bf 1}_{3 \times 3}
\end{array}
\right).
\label{extramatrix2}
\eea
The last three eigenvalues (the $N$ generalization is
simply $ -\frac{2}{(N+k+2)}$) appearing in the last block diagonal
matrix in (\ref{extramatrix2})
give the extra contribution as well as the sum of conformal dimensions
of  $(\tiny\yng(2);0)$
and $(0;\overline{\tiny\yng(2)})$.
They are given in (\ref{eigenvaluestwosymm}) and
(\ref{eigenvaluessymmtwo})
respectively.
Then one obtains the final conformal dimension by adding the above
contribution appearing in (\ref{extramatrix2}) as follows
\bea
\frac{(N+2)}{(k+N+2)}+\frac{k}{(k+N+2)}-\frac{2}{(k+N+2)}
= \frac{(k+N)}{(k+N+2)}.
\nonu
\eea
The conformal dimension formula implies that
\bea
\frac{2 (N+1) (N+2+2)}{2 (N+2) (k+N+2)}-
\frac{2 (N-1) (N+2)}{(2 N) (k+N+2)}
-\frac{(-(N+2)-N)^2}{N (N+2) (k+N+2)}
+1 = \frac{(k+N)}{(k+N+2)}.
\nonu
\eea
The quadratic Casimir
$C^{(N)}(\overline{\tiny\yng(2)})$ is the same as
$C^{(N)}(\tiny\yng(2))$ and the excitation number
is equal to $1$. The correct $\hat{u}$ charge is inserted.
For the $l^+$ quantum number, due to the vanishing of $l^+$ in
$(0;\overline{\tiny\yng(2)})$, 
the $l^+$ is the same as the one($l^+=1$) in $(\tiny\yng(2);0)$. 
For the $l^-$ quantum number, due to the vanishing of $l^-$ in
 $(\tiny\yng(2);0)$ and $(0;\overline{\tiny\yng(2)})$, 
the total $l^-$ is given by $l^-=0$, a singlet. 
The total $\hat{u}$ charge is given by $-N-(N+2)$
which leads to the above result. Again,
the $\hat{u}$ charge for $(0;\overline{\tiny\yng(2)})$ is opposite to
the one for $(0;\tiny\yng(2))$.
One can easily see that the above conformal dimension
does not lead to the BPS bound with $l^{+} =1$ and $l^-=0$.

%%%%%%%%%%%%%%%%%%%%%%%%%%%%%%%%%%%%%%%%%%%%%%%%%%%%%%%
%%%%%%%%%%%%%%%%%%%
\subsection{ The antisymmetric representations
  $\La_+$ with two boxes}
%%%%%%%%%%%%%%%%%%%
%%%%%%%%%%%%%%%%%%%%%%%%%%%%%%%%%%%%%%%%%%%%%%%%%%%%%%%

%{\tiny
%  \bea
%\tiny\yng(1,1)
%\nonu
%\eea}

The two-index antisymmetric parts of the $SU(N+2)$ representation
can be obtained from the generators of the fundamental representation
of $SU(N+2)$ by using the projection operator $\frac{1}{2} (\de_{ik}
\de_{jl}-\de_{jk} \de_{il})$ where $i < j$ and $ k < l$
and $i, j, k, l = 1, 2, \cdots, (N+2) $ \cite{Cvitanovic,Joseph}.
Then by acting on
the space
$T_a \otimes {\bf 1}_{(N+2) \times (N+2)} +
{\bf 1}_{(N+2) \times (N+2)} \otimes T_a$,
one has
the generators for the antisymmetric representation for the $SU(N+2)$
\bea
(T_a)_{ik} \, \delta_{jl} -
(T_a)_{jk} \, \delta_{il} +
\delta_{ik} \, (T_a)_{jl}   -
 \delta_{jk} \, (T_a)_{il}. 
\nonu
\eea

For $N=3$, one has
$\frac{1}{2}(N+2)(N+1) \times \frac{1}{2}(N+2)(N+1)
= 10 \times 10$ unitary matrix, 
and the row and columns are characterized by
the following double index notations
\bea
12, \, 13,  \, 23;  \,
14, \, 15, \, 24, \, 25, \, 34, \, 35;  \, 45,
\nonu
\eea
corresponding to $\hat{u}_1, \hat{u}_2, \cdots, \hat{u}_{10}$
in Appendix $B$.
The first three elements correspond to the
antisymmetric representation for $SU(3)$.
The next six elements correspond to
the fundamental representation of $SU(3)$
with $SU(2)_k$ doublet.
The last element corresponds to
the singlet of $SU(3)$ with $SU(2)_k$ singlet.
The explicit form for the $24$ generators of
$SU(5)$ is presented in Appendix $B$.
See also (\ref{transtrans}).

Let us calculate the zero mode for the reduced stress energy tensor
spin-$2$ current acting on the state $|(\tiny\yng(1,1) ;0) >$. 
It turns out that the $10 \times 10$ matrix
is given by
\bea
\left(
\begin{array}{ccc}
  \frac{2}{(5+k)} {\bf 1}_{3 \times 3}& 0 & 0
  \\
0  & \frac{9}{4(5+k)}{\bf 1}_{6 \times 6} &  0\\
 0 &0 & \frac{3}{(5+k)}
\end{array}
\right).
\label{spin2antisymmetric}
\eea
The last block diagonal element gives us the eigenvalue
on the state 
$|(\tiny\yng(1,1);0)>$ and its $N$ generalization is straightforward.

One can also calculate the
zero mode for the spin-$1$ current
acting on the 
state $|(\tiny\yng(1,1) ;0) >$
and one obtains
\bea
\left(
\begin{array}{ccc}
  0  {\bf 1}_{3 \times 3} & 0 & 0
  \\
0  & \frac{3}{4}{\bf 1}_{6 \times 6} &  0\\
 0 &0 & 0
\end{array}
\right).
\label{su2antisymmetric}
\eea
Again, 
the last block diagonal element gives us the eigenvalue
associated with the above operator
on the state 
$|(\tiny\yng(1,1);0)>$ and its $N$ generalization is straightforward.

Similarly, one can also compute the
zero mode for the spin-$1$ current
acting on the 
state $|(\tiny\yng(1,1) ;0) >$
and it turns out that
\bea
\left(
\begin{array}{ccc}
 2{\bf 1}_{3 \times 3}  & 0 & 0
  \\
0  & -\frac{1}{2}{\bf 1}_{6 \times 6} &  0\\
 0 &0 & -3 
\end{array}
\right).
\label{u1chargeantisymm}
\eea
The last block diagonal element gives us the eigenvalue
associated with the above spin-$1$ operator
on the state 
$|(\tiny\yng(1,1);0)>$ and its $N$ generalization is straightforward.

%%%%%%%%%%%%%%%%%%%%%%%%%%%%%%%%%%%%%%%
\subsubsection{The $(\rm{antisymm};\rm{antisymm})$ representation}
%%%%%%%%%%%%%%%%%%%%%%%%%%%%%%%%%%%%%%%

Let us consider the higher representation 
where the antisymmetric
representation in $SU(N)$ $ ({\tiny\yng(1,1)},{\bf 1})_2$
survives
in the branching of (\ref{transtrans}).
The four eigenvalues are given by
\bea
h ({\tiny\yng(1,1);\tiny\yng(1,1)}) & = & \frac{2}{(N+k+2)},
\nonu \\
%l^{+} (\tiny\yng(1,1);{ \tiny\yng(1,1)}) & = & 0, \qquad
l^{+} (l^{+}+1)(\tiny\yng(1,1);{ \tiny\yng(1,1)})  & = &  0,
\nonu \\
%l^{-} (\tiny\yng(1,1);{ \tiny\yng(1,1)}) & = & 0,\qquad
l^{-} (l^{-}+1)  (\tiny\yng(1,1);{ \tiny\yng(1,1)}) & = & 0,
\nonu \\
\hat{u}( \tiny\yng(1,1); \tiny\yng(1,1) ) & = & 2.
\label{antisymmantisymmresult}
\eea

One can calculate the conformal dimension for this
representation using the previous formula or
one performs the explicit form for the
matrix in this particular representation
as in (\ref{spin2antisymmetric}). 
Therefore, one obtains
$\frac{2}{(k+5)}$ in the first block diagonal elements.
It turns out that the numerator of the above quantity  does not
depend on $N$ and takes the common value and the denominator
is generalized to $(k+N+2)$.
By realizing that the quadratic Casimir of $SU(N+2)$ for the
antisymmetric representation $C^{(N+2)}(\tiny\yng(1,1))=
\frac{N(N+3)}{(N+2)}$
(and the one for the antisymmetric representation
$C^{(N)}(\tiny\yng(1,1))=\frac{(N-2)(N+1)}{N}$)
and the correct $\hat{u}$ charge is given by $2$,
the following relation can be obtained
\bea
\frac{2 (N+2-2) (N+2+1)}{2 (N+2) (k+N+2)}-\frac{2 (N-2) (N+1)}{2 N (k+N+2)}-\frac{4}{N (N+2) (k+N+2)}=
\frac{2}{(N+k+2)}.
\nonu
\eea

Similarly, the quantum number for the
$l^{+}$ can be determined by 
the above matrix calculation in (\ref{su2antisymmetric}).
From the zero eigenvalues appearing in the first block diagonal matrix
in (\ref{su2antisymmetric}), one can see the above
$l^{+}$ quantum number which is  a singlet under the $SU(2)_k$.
This also can be seen from the previous expression
$ ({\tiny\yng(1,1)},{\bf 1})_2$.
The trivial $l^-$ quantum number $l^-=0$ occurs.
For the last eigenvalue $\hat{u}$ charge,
one uses the previous matrix calculation given in (\ref{u1chargeantisymm}).
The eigenvalue  
appearing in the first block diagonal matrix
in (\ref{u1chargeantisymm})
implies that the $\hat{u}$ charge is given by $2$.
This is also consistent with the representation 
$ ({\tiny\yng(1,1)},{\bf 1})_2$.
The above conformal dimension does not satisfy
the vanishing BPS bound with $l^{\pm} =0$.

%%%%%%%%%%%%%%%%%%%%%%%%%%%%%%%%%%%%%%%
\subsubsection{The $({\rm antisymm};f)$ representation}
%%%%%%%%%%%%%%%%%%%%%%%%%%%%%%%%%%%%%%%

Let us consider the
 higher representation 
 where the fundamental representation in
 $SU(N)$ $ ({\tiny\yng(1)},{\bf 2})_{1-\frac{N}{2}}$
survives
in the branching of (\ref{transtrans}).
The four eigenvalues can be summarized by
\bea
h ({\tiny\yng(1,1);\tiny\yng(1)}) & = & \frac{(2N+3)}{4(N+k+2)},
\nonu \\
%l^{+} (\tiny\yng(1,1);{ \tiny\yng(1)}) & = & \frac{1}{2}, \qquad
l^{+} (l^{+}+1)(\tiny\yng(1,1);{ \tiny\yng(1)})  & = &  \frac{3}{4},
\nonu \\
%l^{-} (\tiny\yng(1,1);{ \tiny\yng(1)}) & = & 0,\qquad
l^{-} (l^{-}+1)  (\tiny\yng(1,1);{ \tiny\yng(1)})  & = &  0,
\nonu \\
\hat{u}( \tiny\yng(1,1); \tiny\yng(1) ) & = & -\frac{N}{2}+1.
\label{antifexpression}
\eea

The explicit form for the
matrix in this particular representation
is given by (\ref{spin2antisymmetric}). 
Then, one obtains
$\frac{9}{4(k+5)}$ in the second block diagonal elements.
One can apply for other $N$ values  
and it turns out that the numerator of the above quantity  does 
depend on $N$ linearly as well as the constant term
while the denominator
is generalized to $4(k+N+2)$.
One can also use the formula with the correct $\hat{u}$ charge
as follows:
\bea
\frac{2 (N+2-2) (N+2+1)}{2 (N+2) (k+N+2)}
-\frac{(\frac{N}{2}-\frac{1}{2 N})}{(k+N+2)}
-\frac{(1-\frac{N}{2})^2}{N (N+2) (k+N+2)}
=
\frac{(2N+3)}{4(N+k+2)}.
\nonu
\eea
where the quadratic Casimir 
$C^{(N)}(\tiny\yng(1))$ is used.
The quantum number for the
$l^{+}$ can be obtained by 
the above matrix calculation in (\ref{su2antisymmetric}).
From the  eigenvalues $\frac{3}{4}$
appearing in the second block diagonal matrix
in (\ref{su2antisymmetric}), one can see the above
$l^{+}$ quantum number $l^{+} =\frac{1}{2}$, a doublet under the $SU(2)_k$.
This also can be seen from the previous expression
$ ({\tiny\yng(1)},{\bf 2})_{1-\frac{N}{2}}$.
The trivial $l^-=0$ quantum number holds in this representation.
For the last eigenvalue  $\hat{u}$ charge,
the previous matrix calculation given in (\ref{u1chargeantisymm})
can be used.
The eigenvalues  
appearing in the second block diagonal matrix
in (\ref{u1chargeantisymm}) imply that the $\hat{u}$ charge is given by
$-\frac{1}{2}$.
One finds that the $\hat{u}$ charge is linear in $N$
as well as the constant term.
This is also consistent with the representation 
$ ({\tiny\yng(1)},{\bf 2})_{1-\frac{N}{2}}$.
One observes that the above conformal dimension does  satisfy
the  BPS bound.

%%%%%%%%%%%%%%%%%%%%%%%%%%%%%%%%%%%%%%%
\subsubsection{The $({\rm antisymm};0)$ representation}
%%%%%%%%%%%%%%%%%%%%%%%%%%%%%%%%%%%%%%%

The eigenvalues in this higher representation
are given in the subsection \ref{twoantisymmetric}.
One sees that there are eigenvalues in (\ref{antidoubleboxzero}).
From the three matrices
in (\ref{spin2antisymmetric}), (\ref{su2antisymmetric}) and
(\ref{u1chargeantisymm}),
the relevant eigenvalues are given by
$\frac{3}{(k+5)}$, $0$ and $-3$ for the nontrivial quantum numbers.
They are generalized to $\frac{N}{(k+N+2)}$, $0$ and $-N$
respectively.
The conformal dimension can be checked from
$\frac{2N(N+2+1)}{2(N+2)(k+N+2)}-\frac{N^2}{N(N+2)(k+N+2)}$ also.

%%%%%%%%%%%%%%%%%%%%%%%%%%%%%%%%%%%%%%%
\subsubsection{The $({\rm antisymm};\overline{f})$ representation}
%%%%%%%%%%%%%%%%%%%%%%%%%%%%%%%%%%%%%%%

Let us consider the higher representation
which arises from the product of $(\tiny\yng(1,1);0)$
and $(0;\overline{\tiny\yng(1)})$.
The former occurs in the subsection 
\ref{twoantisymmetric} and the latter occurs in the subsection
\ref{0frep} together with the complex conjugation in the
footnote \ref{zerofbarexplanation}.

In this case, the corresponding
four eigenvalues are described by
\bea
h (\tiny\yng(1,1);\overline{\tiny\yng(1)}) & = &
  \frac{(4N+2k-1)}{4(N+k+2)},
\nonu \\
%l^{+} (\tiny\yng(1,1);\overline{ \tiny\yng(1)}) & = & 0, \qquad
l^{+} (l^{+}+1)(\tiny\yng(1,1);\overline{ \tiny\yng(1)})  & = &  0,
\nonu \\
%l^{-} (\tiny\yng(1,1);\overline{ \tiny\yng(1)}) & = & \frac{1}{2},\qquad
l^{-} (l^{-}+1)  (\tiny\yng(1,1);\overline{ \tiny\yng(1)})  & = &
\frac{3}{4},
\nonu \\
\hat{u}( \tiny\yng(1,1); \overline{\tiny\yng(1)} ) & = & -\frac{3N}{2}-1.
\label{antisymmfbarexpression}
\eea

One can obtain the following $10 \times 10$ matrix
by calculating the commutator
$[T_0, Q_{-\frac{1}{2}}^{\bar{A}}]$ as in the subsection
\ref{ffbarrep}
\bea
\left(
\begin{array}{ccc}
  \frac{2}{3(5+k)} {\bf 1}_{3 \times 3}& 0 & 0
  \\
0  & -\frac{1}{6(5+k)}{\bf 1}_{6 \times 6} &  0\\
 0 &0 & -\frac{1}{(5+k)}
\end{array}
\right).
\label{extramatrix3}
\eea
The last eigenvalue (the $N$ generalization is
straightforward to obtain) appearing in the last  diagonal
element in (\ref{extramatrix3})
provides the extra contribution
as well as the sum of conformal dimensions
of  $(\tiny\yng(1,1);0)$
and $(0;\overline{\tiny\yng(1)})$.
They are given in (\ref{antidoubleboxzero}) and
(\ref{0ffoureigen})
respectively.
Then one obtains the final conformal dimension by adding the above
contribution appearing in (\ref{extramatrix3}) as follows
\bea
\frac{N}{(N+k+2)}+ 
\frac{(2k+3)}{4(N+k+2)}-
\frac{1}{(N+k+2)} = \frac{(4N+2k-1)}{4(N+k+2)},
\nonu
\eea
as described in (\ref{eigenforsymmf}).
It is also useful to interpret the above result
from the conformal dimension formula.
One determines the following result
\bea
\frac{2 (N+2-2) (N+2+1)}{2 (N+2) (k+N+2)}
-\frac{(\frac{N}{2}-\frac{1}{2 N})}{(k+N+2)}
-\frac{(-\frac{1}{2} (N+2)-N)^2}{N (N+2) (k+N+2)}
+\frac{1}{2} =
\frac{(4N+2k-1)}{4(N+k+2)}.
\nonu
\eea
Here we used the quadratic Casimirs for $C^{(N+2)}(\tiny\yng(1,1))$
and $C^{(N)}(\overline{\tiny\yng(1)})$.
The correct $\hat{u}$ charge is inserted.
The excitation number is given by $\frac{1}{2}$.

For the $l^+$ quantum number, due to the vanishing of $l^+$ in
$(0;\overline{\tiny\yng(1)})$ and
$(\tiny\yng(1,1);0)$, it turns out that
the total $l^+$ is trivial. 
For the $l^-$ quantum number, due to the vanishing of $l^-$ in
 $(\tiny\yng(1,1);0)$, it turns out that
the $l^-$ is the same as the one($l^-=\frac{1}{2}$) in
$(0;\overline{\tiny\yng(1)})$. 
It is easy to see that
the above conformal dimension does not satisfy
the BPS bound by substituting
$l^+=0$ and $l^-=\frac{1}{2}$.
One can add each $\hat{u}$ charge and it is obvious that
the total $\hat{u}$ charge is given by $-N -\frac{(N+2)}{2}$
which leads to the above result. Note that
the $\hat{u}$ charge for $(0;\overline{\tiny\yng(1)})$ is opposite to
the one for $(0;\tiny\yng(1))$. 

%%%%%%%%%%%%%%%%%%%%%%%%%%%%%%%%%%%%%%%
\subsubsection{The $(\rm{antisymm};\rm{symm})$ representation}
%%%%%%%%%%%%%%%%%%%%%%%%%%%%%%%%%%%%%%%

Let us consider the higher representation
which arises from the product of $(\tiny\yng(1,1);0)$
and $(0;\tiny\yng(2))$.
The former occurs in the subsection 
\ref{twoantisymmetric} and the latter occurs in the subsection
\ref{0symmtwo}.

The four eigenvalues can be summarized by
\bea
h ({\tiny\yng(1,1);\tiny\yng(2)}) & = & 1,
\nonu \\
%l^{+} (\tiny\yng(1,1);{ \tiny\yng(2)}) & = & 0, \qquad
l^{+} (l^{+}+1)(\tiny\yng(1,1);{ \tiny\yng(2)})  & = &  0,
\nonu \\
%l^{-} (\tiny\yng(1,1);{ \tiny\yng(2)}) & = & 0,\qquad
l^{-} (l^{-}+1)  (\tiny\yng(1,1);{ \tiny\yng(2)})  & = &  0,
\nonu \\
\hat{u}( \tiny\yng(1,1); \tiny\yng(2) ) & = & 2.
\label{antisymmsymmeigen}
\eea

The following $15 \times 15$ matrix
by calculating the commutator
$[T_0,  Q_{-\frac{1}{2}}^{13} Q_{-\frac{1}{2}}^{16}]$ as in the
previous subsections
can be obtained
\bea
\left(
\begin{array}{ccc}
  -\frac{4}{3(5+k)} {\bf 1}_{3 \times 3}& 0 & 0
  \\
0  & \frac{1}{3(5+k)}{\bf 1}_{6 \times 6} &  0\\
 0 &0 & \frac{2}{(5+k)}
\end{array}
\right)
\label{extramatrix4}
\eea
The last eigenvalue (the $N$ generalization is
simply $ \frac{2}{(N+k+2)}$) appearing in the last element in
(\ref{extramatrix4})
gives the extra contribution as well as the sum of conformal dimensions
of  $(\tiny\yng(1,1);0)$
and $(0;\tiny\yng(2))$.
They are given in (\ref{antidoubleboxzero}) and
(\ref{eigenvaluessymmtwo})
respectively.
Then one obtains the final conformal dimension by adding the above
contribution appearing in (\ref{extramatrix4}) as follows
\bea
\frac{N}{(N+k+2)} +\frac{k}{(N+k+2)} +\frac{2}{(N+k+2)}
= 1.
\nonu
\eea
Furthermore,
this analysis can be seen from the conformal dimension formula
\bea
\frac{2 N (N+2+1)}{2 (N+2) (k+N+2)}
-\frac{2 (N-1) (N+2)}{2 N (k+N+2)}
-\frac{((N+2)-N)^2}{N (N+2) (k+N+2)}
+1=1.
\nonu
\eea
Here we also use the quadratic Casimirs for
$C^{(N+2)}(\tiny\yng(1,1))$ and $C^{(N)}(\tiny\yng(2))$
    and the excitation number
is equal to $1$. The correct $\hat{u}$ charge is inserted.
For the $l^{\pm}$ quantum number, due to the vanishing of $l^{\pm}$ in
$(\tiny\yng(1,1);0)$ and $(0;\tiny\yng(2))$, 
the total $l^{\pm}$ is trivial. 
The total $\hat{u}$ charge is given by $-N+(N+2)$
which leads to the above result.
One can easily see that the above conformal dimension
does not lead to the BPS bound with $l^{\pm} =0$ \footnote{For the
  higher representation  $(\tiny\yng(1,1);\overline{\tiny\yng(2)} )$,
  the similar analysis can be done by changing
  the $\hat{u}$ charge correctly.}.

%%%%%%%%%%%%%%%%%%%%%%%%%%%%%%%%%%%%%%%
\subsubsection{The $(\rm{antisymm};\overline{\rm{antisymm}})$ representation}
%%%%%%%%%%%%%%%%%%%%%%%%%%%%%%%%%%%%%%%

Let us consider the higher representation
which arises from the product of $(\tiny\yng(1,1);0)$
and $(0;\overline{\tiny\yng(1,1)})$.
The former occurs in the subsection 
\ref{twoantisymmetric} while the latter occurs in the subsection
\ref{0antisymm} with complex conjugation.

The four eigenvalues are given by 
\bea
h (\tiny\yng(1,1);\overline{\tiny\yng(1,1)}) & = &
  \frac{(k+N)}{(k+N+2)},
\nonu \\
%l^{+} (\tiny\yng(1,1);\overline{ \tiny\yng(1,1)}) & = & 0, \qquad
l^{+} (l^{+}+1)(\tiny\yng(1,1);\overline{ \tiny\yng(1,1)})  & = &  0,
\nonu \\
%l^{-} (\tiny\yng(1,1);\overline{ \tiny\yng(1,1)}) & = & 1,\qquad
l^{-} (l^{-}+1)  (\tiny\yng(1,1);\overline{ \tiny\yng(1,1)})  & = &  2,
\nonu \\
\hat{u}( \tiny\yng(1,1); \overline{\tiny\yng(1,1)} ) & = & -2N-2.
\label{antisymmantisymmbarresult}
\eea

One should calculate the commutator
$[T_0,  Q_{-\frac{1}{2}}^{1} Q_{-\frac{1}{2}}^{2}]$
and it turns out that
\bea
\left(
\begin{array}{ccc}
  \frac{4}{3(5+k)} {\bf 1}_{3 \times 3}& 0 & 0
  \\
0  & -\frac{1}{3(5+k)}{\bf 1}_{6 \times 6} &  0\\
 0 &0 & -\frac{2}{(5+k)}
\end{array}
\right)
\label{Extramatrix5}
\eea
The last eigenvalue (the $N$ generalization is
simply $ -\frac{2}{(N+k+2)}$) appearing in the last element
 in (\ref{Extramatrix5})
gives the extra contribution as well as the sum of conformal dimensions
of  $(\tiny\yng(1,1);0)$
and $(0;\overline{\tiny\yng(1,1)})$.
They are given in (\ref{antidoubleboxzero}) and
(\ref{eigenvaluestwoantisymm})
respectively.
Then one obtains the final conformal dimension by adding the above
contribution appearing in (\ref{Extramatrix5}) as follows
\bea
\frac{N}{(k+N+2)} +\frac{k+2}{(k+N+2)}-\frac{2}{(k+N+2)}
=\frac{(k+N)}{(k+N+2)}.
\nonu
\eea
The conformal dimension formula implies that
{\small
  \bea
\frac{2 (N+2-2) (N+2+1)}{2 (N+2) (k+N+2)}
-\frac{2 (N-2) (N+1)}{(2 N) (k+N+2)}
-\frac{(-(N+2)-N)^2}{N (N+2) (k+N+2)}
+1 =
 \frac{(k+N)}{(k+N+2)}.
\nonu
\eea}
The quadratic Casimir
$C^{(N)}(\overline{\tiny\yng(1,1)})$ is the same as
$C^{(N)}(\tiny\yng(1,1))$ and the excitation number
is equal to $1$. The correct $\hat{u}$ charge is inserted.
For the $l^+$ quantum number, due to the vanishing of $l^+$ in
$(0;\overline{\tiny\yng(1,1)})$ and $(\tiny\yng(1,1);0)$, 
the total $l^+$ is trivial. 
For the $l^-$ quantum number, due to the vanishing of $l^-$ in
 $(\tiny\yng(1,1);0)$, 
the total $l^-$ is given by $l^-=1$ in $(0;\overline{\tiny\yng(1,1)})$. 
The total $\hat{u}$ charge is given by $-N-(N+2)$
which leads to the above result. Again,
the $\hat{u}$ charge for
the representation $(0;\overline{\tiny\yng(1,1)})$ is opposite to
the one for the representation $(0;\tiny\yng(1,1))$.
One can easily see that the above conformal dimension
does not lead to the BPS bound with $l^{+} =0$ and $l^-=1$.

%%%%%%%%%%%%%%%%%%%
%\subsection{ The higher representation}
%%%%%%%%%%%%%%%%%%%

%\bea
%\overline{\tiny\yng(2)}
%\nonu
%\eea

%%%%%%%%%%%%%%%%%%%
%\subsection{ The higher representation}
%%%%%%%%%%%%%%%%%%%

%\bea
%\overline{
%\tiny\yng(1,1)}
%\nonu
%\eea

%%%%%%%%%%%%%%%%%%%%%%%%%%%%%%%%%%%%%%%%%%%%%%%%%%%%%%%%%%%%%%
%%%%%%%%%%%%%%%%%%%
\subsection{
The symmetric  representations $\La_+$ with three boxes
%$(0;\yng(3))$
}
%%%%%%%%%%%%%%%%%%%
%%%%%%%%%%%%%%%%%%%%%%%%%%%%%%%%%%%%%%%%%%%%%%%%%%%%%%%%%%%%%%

%\bea
%{\tiny\yng(3)}
%\nonu
%\eea
From the triple product of the fundamental representation of
$SU(N+2)$, one obtains the following branching under the
$SU(N) \times SU(2) \times U(1)$
\bea
    {\tiny\yng(1)} \otimes  {\tiny\yng(1)} \otimes
    {\tiny\yng(1)} 
     & = &  \Bigg[ {\tiny\yng(2)} + {\tiny\yng(1,1)} \Bigg]
    \otimes
        {\tiny\yng(1)} =
        {\tiny\yng(3)} + 2 {\tiny\yng(2,1)} +
         {\tiny\yng(1,1,1)} \nonu \\
& \rightarrow &
\Bigg[ ({\tiny\yng(1)},{\bf 1})_1
      + ({\bf 1},{\bf 2})_{-\frac{N}{2}} \Bigg] \otimes
    \Bigg[ ({\tiny\yng(1)},{\bf 1})_1
      + ({\bf 1},{\bf 2})_{-\frac{N}{2}} \Bigg]
     \otimes
    \Bigg[ ({\tiny\yng(1)},{\bf 1})_1
      + ({\bf 1},{\bf 2})_{-\frac{N}{2}} \Bigg]
    \nonu \\
    &= & \Bigg[  ({\tiny\yng(3)},{\bf 1})_3+
      ({\tiny\yng(2)},{\bf 2})_{2-\frac{N}{2}} +
      ({\tiny\yng(1)},{\bf 3})_{1-N}
+  ({\bf 1},{\bf 4})_{-\frac{3N}{2}}
      \Bigg] \nonu \\
    &+& 2 \Bigg[({\tiny\yng(2,1)},{\bf 1})_3+
      ({\tiny\yng(1,1)},{\bf 2})_{2-\frac{N}{2}} +
       ({\tiny\yng(2)},{\bf 2})_{2-\frac{N}{2}} +
      ({\tiny\yng(1)},{\bf 3})_{1-N} +({\tiny\yng(1)},{\bf 1})_{1-N}
+  ({\bf 1},{\bf 2})_{-\frac{3N}{2}}
      \Bigg] \nonu \\
         &+& \Bigg[  ({\tiny\yng(1,1,1)},{\bf 1})_3+
      ({\tiny\yng(1,1)},{\bf 2})_{2-\frac{N}{2}} +
      ({\tiny\yng(1)},{\bf 1})_{1-N} \Bigg].
    \nonu 
    \eea
    Furthermore \footnote{The higher representations on the
      remaining three subsections,
      $3.3$, $3.4$ and $3.5$ will not be considered for the eigenvalues
      of the higher spin currents in the
      remaining sections because it is rather complicated to
      construct the $SU(N+2)$ generators with these particular
      representations having three boxes for several $N$ values.},
    the three representations,
    symmetric, mixed and antisymmetric  ones
    have the following decompositions under
    the $SU(N) \times SU(2) \times U(1)$ \cite{Slansky,Feger}
    \bea
        {\tiny\yng(3)} & \rightarrow &
          ({\tiny\yng(3)},{\bf 1})_3+
      ({\tiny\yng(2)},{\bf 2})_{2-\frac{N}{2}} +
      ({\tiny\yng(1)},{\bf 3})_{1-N}
+  ({\bf 1},{\bf 4})_{-\frac{3N}{2}},
        \nonu \\
              {\tiny\yng(2,1)} & \rightarrow &
({\tiny\yng(2,1)},{\bf 1})_3+
      ({\tiny\yng(1,1)},{\bf 2})_{2-\frac{N}{2}} +
       ({\tiny\yng(2)},{\bf 2})_{2-\frac{N}{2}} +
      ({\tiny\yng(1)},{\bf 3})_{1-N} +({\tiny\yng(1)},{\bf 1})_{1-N}
+  ({\bf 1},{\bf 2})_{-\frac{3N}{2}},              
\nonu \\
              {\tiny\yng(1,1,1)}
              & \rightarrow &
({\tiny\yng(1,1,1)},{\bf 1})_3+
      ({\tiny\yng(1,1)},{\bf 2})_{2-\frac{N}{2}} +
      ({\tiny\yng(1)},{\bf 1})_{1-N}.              
              \label{transtranstrans}
    \eea
    Note that
    there is no $SU(N)$ singlet in the antisymmetric representation
    contrary to the symmetric and mixed ones.    

The three-index symmetric parts of the $SU(N+2)$ representation
can be obtained from the generators of the fundamental representation
of $SU(N+2)$ by using the projection operator $\frac{1}{6} (\de_{il}
\de_{jm} \de_{kn} + \de_{il} \de_{km} \de_{jn} +\de_{kl}
\de_{im} \de_{jn} +\de_{jl} \de_{im} \de_{kn}+\de_{jl}
\de_{km} \de_{in} + \de_{kl} \de_{jm} \de_{in} )$ where
$i \leq j \leq k$ and $ l \leq m \leq n$
and $i, j, k, l,m,n = 1, 2, \cdots, (N+2) $ \cite{Cvitanovicbook}.
Then by acting on
the space
$T_a \otimes {\bf 1}_{(N+2) \times (N+2)} \otimes {\bf 1}_{(N+2) \times (N+2)}
+ {\bf 1}_{(N+2) \times (N+2)} \otimes T_a \otimes {\bf 1}_{(N+2) \times (N+2)} +
{\bf 1}_{(N+2) \times (N+2)} \otimes {\bf 1}_{(N+2) \times (N+2)} \otimes T_a$,
one has
the generators for the symmetric representation for the $SU(N+2)$
as follows:
{\small
  \bea
&& (T_a)_{il} \, \delta_{jm} \, \delta_{kn} +
(T_a)_{il} \, \delta_{km} \, \delta_{jn} +
(T_a)_{kl} \, \delta_{im} \, \delta_{jn} +
(T_a)_{jl} \, \delta_{im} \, \delta_{kn} +
(T_a)_{jl} \, \delta_{km} \, \delta_{in} +
(T_a)_{kl} \, \delta_{jm} \, \delta_{in}  \nonu\\
&& +
\delta_{il} \, (T_a)_{jm} \, \delta_{kn}+
\delta_{il} \, (T_a)_{km} \, \delta_{jn}+
\delta_{kl} \, (T_a)_{im} \, \delta_{jn}+
\delta_{jl} \, (T_a)_{im} \, \delta_{kn}+
\delta_{jl} \, (T_a)_{km} \, \delta_{in}+
\delta_{kl} \, (T_a)_{jm} \, \delta_{in}
\nonu \\
&& +
\delta_{il} \, \delta_{jm} \, (T_a)_{kn}
+
\delta_{il} \, \delta_{km} \, (T_a)_{jn}
+
\delta_{kl} \, \delta_{im} \, (T_a)_{jn}
+
\delta_{jl} \, \delta_{im} \, (T_a)_{kn}
+
\delta_{jl} \, \delta_{km} \, (T_a)_{in}
+
\delta_{kl} \, \delta_{jm} \, (T_a)_{in}. 
\nonu
\eea}
For $N=3$, one has
$\frac{1}{6}(N+2)(N+3)(N+4) \times \frac{1}{6}(N+2)(N+3)(N+4)
= 35 \times 35$ unitary matrix, 
the row and columns are characterized by
the following triple index notations
\bea
&& 111,\, 112, \, 113, \, 122, \, 123, \, 133, \, 222, \, 223, \,
233, \, 333; 
\nonu \\
&& 114, \, 115, \, 124, \, 125, \, 134, \, 135, \, 224, \,
225, \, 234, \, 235, \, 334, \, 335;
\nonu \\
&& 144, \, 145, \, 155, \,  244, \, 245, \, 255, \, 344, \,
345, \, 355;
\nonu \\
&& 444, \, 445, \, 455, \, 555. 
\label{35cases}
\eea
See also Appendix $C$ in $\hat{u}_1, \cdots, \hat{u}_{35}$.
The first ten elements correspond to the
symmetric representation for $SU(3)$.
The next twelve elements correspond to
the symmetric representation of $SU(3)$
with $SU(2)_k$ doublet.
The next nine elements correspond to
the fundamental representation of $SU(3)$
with $SU(2)_k$ triplet.
The last four elements correspond to
the singlet of $SU(3)$ with $SU(2)_k$ quartet.
The explicit form for the $24$ generators of
$SU(5)$ is presented in Appendix $C$.
Note that the orderings given in (\ref{35cases})
are different from the ones in Appendix $C$.

Let us calculate the zero mode for the reduced stress energy tensor
spin-$2$ current acting on the state $|(\tiny\yng(3) ;0) >$. 
It turns out that the $35 \times 35$ matrix
is given by
%{ \small
%  \bea
%\left(
%\begin{array}{cccccccc}
%  \frac{3}{(5+k)} {\bf 1}_{10 \times 10}& 0 & 0
% & 0 &0 & 0 & 0 &0  \\
%0  & \frac{25}{4(5+k)}{\bf 1}_{6 \times 6} &  0 & 0 & 0 & 0 & 0 & 0   \\
%0 &0 & \frac{8}{(5+k)} {\bf 1}_{3 \times 3} & 0  & 0 & 0 & 0 & 0 \\
%0  & 0 & 0 & \frac{25}{4(5+k)}{\bf 1}_{4 \times 4}  & 0 & 0 & 0 & 0   \\
%0 &0 &  0  & 0 & \frac{8}{(5+k)} {\bf 1}_{3 \times 3} & 0 & 0 & 0 \\
%0  &  0 & 0 & 0 & 0 &
%\frac{25}{4(5+k)}{\bf 1}_{2 \times 2} &  0 & 0    \\
%0 &0 & 0 & 0 & 0 & 0 &
%\frac{8}{(5+k)} {\bf 1}_{3 \times 3} & 0 \\
%0& 0& 0&  0 & 0 & 0 & 0 &   \frac{33}{4(5+k)} {\bf 1}_{4\times 4}
%\end{array}
%\right)
%_{35 \times 35}
%\label{spin2symmetricthree}
%\eea }
\bea
&& \frac{1}{(5+k)} \mbox{diag}  (3,3,3,3,3,3,3,3,3,3,\frac{25}{4},
\frac{25}{4},\frac{25}{4},\frac{25}{4},\frac{25}{4},\frac{25}{4},
\nonu \\
&& 8,8,8,\frac{25}{4},\frac{25}{4},\frac{25}{4},\frac{25}{4},8,8,8,\frac{25}{4},\frac{25}{4},8,8,8,\frac{33}{4},\frac{33}{4},\frac{33}{4},\frac{33}{4}).
\label{spin2symmetricthree}
\eea
If one takes the ordering from (\ref{35cases}),
then all the diagonal elements will look nicer.
There are four block diagonal elements. 
The last block diagonal elements correspond to
the eigenvalue on the  state $|(\tiny\yng(3) ;0) >$.
We will describe the detailed quantum numbers
for the other eigenvalues soon.

One can also calculate the zero mode for the sum of the square for
the spin-$1$ current with minus sign acting on the above
state $|(\tiny\yng(3) ;0) >$
and the explicit result is given by 
%\bea
%\left(
%\begin{array}{cccc}
%  0 {\bf 1}_{10\times 10}& 0 & 0
% & 0   \\
%0  & \frac{3}{4}{\bf 1}_{12\times 12} &  0 & 0   \\
%0 &0 & 2 {\bf 1}_{9\times 9} & 0   \\
%0 &0 &0 &  \frac{15}{4} {\bf 1}_{4\times 4}
%\end{array}
%\right)
%_{35 \times 35}
%\label{su2symmetricthree}
%\eea
%\bea
\bea
&& \mbox{diag}  (
0,0,0,0,0,0,0,0,0,0,\frac{3}{4},\frac{3}{4},\frac{3}{4},\frac{3}{4},
\frac{3}{4},\frac{3}{4},\nonu \\
&& 2,2,2,\frac{3}{4},\frac{3}{4},\frac{3}{4},\frac{3}{4},2,2,2,\frac{3}{4},\frac{3}{4},2,2,2,\frac{15}{4},\frac{15}{4},\frac{15}{4},\frac{15}{4}).
\label{su2symmetricthree}
\eea
The diagonal elements correspond to the eigenvalues
and in particular, the last one $\frac{15}{4}$ is the eigenvalue for the
state  $|(\tiny\yng(3) ;0) >$ which behaves as a quartet under the
$SU(2)$. 

Similarly, one can also compute the
zero mode for the spin-$1$ current
acting on the 
state $|(\tiny\yng(3) ;0) >$
and one obtains (twice of the $\hat{u}$)
%\bea
%\left(
%\begin{array}{cccc}
%  3 {\bf 1}_{10\times 10}& 0 & 0
% & 0  \\
%0  & \frac{1}{2}{\bf 1}_{12\times 12} &  0 & 0   \\
%0 &0 & -2 {\bf 1}_{9\times 9} & 0   \\
%0 &0 &0 &   -\frac{9}{2} {\bf 1}_{4\times 4}
%\end{array}
%\right)
%_{35 \times 35}
%\label{u1chargesymmthree}
%\eea
\bea
&& \mbox{diag}  (
6,6,6,6,6,6,6,6,6,6,1,1,1,1,1,1,\nonu \\
&& -4,-4,-4,1,1,1,1,
 -4,-4,-4,1,1,-4,-4,-4,-9,-9,-9,-9).
\label{u1chargesymmthree}
\eea
In this case, also the last elements are the eigenvalues
for the state  $|(\tiny\yng(3) ;0) >$.

%%%%%%%%%%%%%%%%%%%%%%%%%%%%%%%%%%%%%%%
\subsubsection{The $(\rm{symm};\rm{symm})$ representation
with three boxes}
%%%%%%%%%%%%%%%%%%%%%%%%%%%%%%%%%%%%%%%

Let us consider the higher representation 
where the symmetric representation in $SU(N)$ $ ({\tiny\yng(3)},{\bf 1})_3$
survives
in the branching of (\ref{transtranstrans}).
The four eigenvalues are given by
\bea
h ({\tiny\yng(3);\tiny\yng(3)}) & = & \frac{3}{(N+k+2)},
\nonu \\
%l^{+} ({\tiny \yng(3)};{ \tiny\yng(3)}) & = & 0, \qquad
l^{+} (l^{+}+1)(\tiny\yng(3);{ \tiny\yng(3)})  & = &  0,
\nonu \\
%l^{-} (\tiny\yng(3);{ \tiny\yng(3)}) & = & 0,\qquad
l^{-} (l^{-}+1)  (\tiny\yng(3);{ \tiny\yng(3)})  & = &  0,
\nonu \\
\hat{u}( \tiny\yng(3); \tiny\yng(3) ) & = & 3.
\nonu
\eea
One can calculate the conformal dimension for this
representation using the previous formula or
one performs the explicit form for the
matrix in this particular representation
as in (\ref{spin2symmetricthree}). 
Therefore, one obtains
$\frac{3}{(k+5)}$ in the first block diagonal elements.
One can apply for other $N$ values where $N=5,7,9, 11, 13, \cdots$. 
It will turn out that the numerator of the above quantity  does not
depend on $N$ and takes the common value and the denominator
is generalized to $(k+N+2)$.
By realizing that the quadratic Casimir of $SU(N+2)$ for the
symmetric representation $C^{(N+2)}(\tiny\yng(3))=\frac{3(N+1)(N+5)}{2(N+2)}$
(and the one for the symmetric representation
$C^{(N)}(\tiny\yng(3))=\frac{3(N-1)(N+3)}{2N}$)
and the correct $\hat{u}$ charge is given by $3$,
the following relation can be obtained
\bea
\frac{3 (N+1) (N+3+2)}{2 (N+2) (k+N+2)}-
\frac{3 (N-1) (N+3)}{2 N (k+N+2)}-\frac{9}{N (N+2) (k+N+2)}
=\frac{3}{(N+k+2)}.
\nonu
\eea

Similarly, the quantum number for the
$l^{+}$ can be determined by 
the above matrix calculation in (\ref{su2symmetricthree}).
From the zero eigenvalues appearing in the first block diagonal matrix
in (\ref{su2symmetricthree}), one can see the above
$l^{+}$ quantum number, a singlet under the $SU(2)_k$.
This also can be seen from the previous expression
$ ({\tiny\yng(3)},{\bf 1})_3$.
The trivial $l^-$ quantum number $l^-=0$ arises.
For the last eigenvalue corresponding to $\hat{u}$ charge,
one uses the previous matrix calculation given in (\ref{u1chargesymmthree}).
The eigenvalues  
appearing in the first block diagonal matrix
in (\ref{u1chargesymmthree}) imply that the $\hat{u}$ charge is given by $3$.
This is also consistent with the representation 
$ ({\tiny\yng(3)},{\bf 1})_3$ where the subscript denotes
the $\hat{u}$ charge.
One observes that the above conformal dimension does not satisfy
the vanishing BPS bound with $l^{\pm} =0$.

%%%%%%%%%%%%%%%%%%%%%%%%%%%%%%%%%%%%%%%
\subsubsection{The $(\rm{symm};\rm{symm})$ representation
with three and two boxes}
%%%%%%%%%%%%%%%%%%%%%%%%%%%%%%%%%%%%%%%

Let us consider the
 higher representation 
 where the fundamental representation in
 $SU(N)$ $ ({\tiny\yng(2)},{\bf 2})_{2-\frac{N}{2}}$
survives
in the branching of (\ref{transtranstrans}).
The four eigenvalues can be summarized by
\bea
h ({\tiny\yng(3);\tiny\yng(2)}) & = & \frac{(2N+19)}{4(N+k+2)},
\nonu \\
%l^{+} (\tiny\yng(3);{ \tiny\yng(2)}) & = & \frac{1}{2}, \qquad
l^{+} (l^{+}+1)(\tiny\yng(3);{ \tiny\yng(2)})  & = &  \frac{3}{4},
\nonu \\
%l^{-} (\tiny\yng(3);{ \tiny\yng(2)}) & = & 0,\qquad
l^{-} (l^{-}+1)  (\tiny\yng(3);{ \tiny\yng(2)})  & = &  0,
\nonu \\
\hat{u}( \tiny\yng(3); \tiny\yng(2) ) & = & -\frac{N}{2}+2.
%\label{symmsymmeigen}
\nonu
\eea
The explicit form for the
matrix in this particular representation
is given by (\ref{spin2symmetricthree}). 
Then, one obtains
$\frac{25}{4(k+5)}$ in the second block diagonal elements.
One can apply for other $N$ values  
and it will turn out that the numerator of the above quantity  does 
depend on $N$ linearly as well as the constant term
while the denominator
is generalized to $4(k+N+2)$.
One can also use the formula with the correct $\hat{u}$ charge
\bea
\frac{3 (N+1) (N+3+2)}{2 (N+2) (k+N+2)}-\frac{2 (N-1) (N+2)}{2 N (k+N+2)}
-\frac{(2-\frac{N}{2})^2}{N (N+2) (k+N+2)}
=\frac{(2N+19)}{4(N+k+2)},
\nonu
\eea
where the quadratic Casimir 
$C^{(N+2)}(\tiny\yng(3))$ and $C^{(N)}(\tiny\yng(2))$ are used.
The quantum number for the
$l^{+}$ can be obtained by 
the above matrix calculation in (\ref{su2symmetricthree}).
From the  eigenvalues $\frac{3}{4}$
appearing in the second block diagonal matrix
in (\ref{su2symmetricthree}), one can see the above
$l^{+}$ quantum number $l^{+} =\frac{1}{2}$, a doublet under the $SU(2)_k$.
This also can be seen from the previous expression
$ ({\tiny\yng(2)},{\bf 2})_{2-\frac{N}{2}}$.
The trivial $l^-=0$ quantum number holds in this representation.
For the last eigenvalue corresponding to $\hat{u}$ charge,
the previous matrix calculation given in (\ref{u1chargesymmthree})
can be used.
The eigenvalues  
appearing in the second block diagonal matrix
in (\ref{u1chargesymmthree}) imply that the $\hat{u}$ charge is given by
$\frac{1}{2}$.
Varying the $N$ values,
one finds that the $\hat{u}$ charge is linear in $N$
as well as the constant term.
This is also consistent with the representation 
$ ({\tiny\yng(2)},{\bf 2})_{2-\frac{N}{2}}$ where the subscript denotes
the $\hat{u}$ charge.
One observes that the above conformal dimension does not satisfy
the BPS bound.

%%%%%%%%%%%%%%%%%%%%%%%%%%%%%%%%%%%%%%%
\subsubsection{The $({\rm symm};f)$ representation with three boxes}
%%%%%%%%%%%%%%%%%%%%%%%%%%%%%%%%%%%%%%%

Let us consider the higher representation 
where the fundamental representation in $SU(N)$
$ ({\tiny\yng(1)},{\bf 3})_{1-N}$
survives
in the branching of (\ref{transtranstrans}).
The four eigenvalues are given by
\bea
h ({\tiny\yng(3);\tiny\yng(1)}) & = & \frac{(N+5)}{(N+k+2)},
\nonu \\
%l^{+} (\tiny\yng(3);{ \tiny\yng(1)}) & = & 1, \qquad
l^{+} (l^{+}+1)(\tiny\yng(3);{ \tiny\yng(1)})  & = &  2,
\nonu \\
%l^{-} (\tiny\yng(3);{ \tiny\yng(1)}) & = & 0,\qquad
l^{-} (l^{-}+1)  (\tiny\yng(3);{ \tiny\yng(1)})  & = & 0,
\nonu \\
\hat{u}( \tiny\yng(3); \tiny\yng(1) ) & = & -N+1.
\nonu
\eea

The explicit form for the
matrix in this particular representation
is given by (\ref{spin2symmetricthree}). 
Then, one obtains
$\frac{8}{(k+5)}$ in the third block diagonal elements.
One can apply for other $N$ values  
and it will turn out that the numerator of the above quantity  does 
depend on $N$ linearly as well as the constant term
while the denominator
is generalized to $(k+N+2)$.
One can also use the formula with the correct $\hat{u}$ charge
\bea
\frac{3 (N+1) (N+3+2)}{2 (N+2) (k+N+2)}
-\frac{(\frac{N}{2}-\frac{1}{2 N})}{(k+N+2)}
-\frac{(1-N)^2}{N (N+2) (k+N+2)}
=
\frac{(N+5)}{(N+k+2)}.
\nonu
\eea
where the quadratic Casimir 
$C^{(N+2)}(\tiny\yng(3))$ and $C^{(N)}(\tiny\yng(1))$ are used.
The quantum number for the
$l^{+}$ can be obtained by 
the above matrix calculation in (\ref{su2symmetricthree}).
From the  eigenvalues $2$
appearing in the third block diagonal matrix
in (\ref{su2symmetricthree}), one can see the above
$l^{+}$ quantum number $l^{+} =1$, a triplet under the $SU(2)_k$.
This also can be seen from the previous expression
$ ({\tiny\yng(1)},{\bf 3})_{1-N}$.
The trivial $l^-=0$ quantum number holds in this representation.
For the last eigenvalue corresponding to $\hat{u}$ charge,
the previous matrix calculation given in (\ref{u1chargesymmthree})
can be used.
The eigenvalues  
appearing in the third block diagonal matrix
in (\ref{u1chargesymmthree}) imply that the $\hat{u}$ charge is given by
$-2$.
Varying the $N$ values,
one finds that the $\hat{u}$ charge is linear in $N$
as well as the constant term.
This is also consistent with the representation 
$ ({\tiny\yng(2)},{\bf 3})_{1-N}$ where the subscript denotes
the $\hat{u}$ charge.
One observes that the above conformal dimension does not satisfy
the BPS bound.

%%%%%%%%%%%%%%%%%%%%%%%%%%%%%%%%%%%%%%%
\subsubsection{The $(\rm{symm};0)$ representation with three boxes
\label{threesymmzero}}
%%%%%%%%%%%%%%%%%%%%%%%%%%%%%%%%%%%%%%%

Let us consider the higher representation 
where the singlet representation in $SU(N)$
$ ({\bf 1},{\bf 4})_{-\frac{3N}{2}}$
survives
in the branching of (\ref{transtranstrans}).
The four eigenvalues are given by
\bea
h (\tiny\yng(3);0) & = & \frac{3(2N+5)}{4(N+k+2)},
\nonu \\
%l^{+} (\tiny\yng(3);0) & = & \frac{3}{2}, \qquad
l^{+} (l^{+}+1)(\tiny\yng(3);0)  & = &  \frac{15}{4},
\nonu \\
%l^{-} (\tiny\yng(3);0) & = & 0,\qquad
l^{-} (l^{-}+1)  (\tiny\yng(3);0)  & = &  0,
\nonu \\
\hat{u}( \tiny\yng(3);0 ) & = & -\frac{3N}{2}.
\label{eigenvaluesthreesymm}
\eea

The explicit form for the
matrix in this particular representation
is given by (\ref{spin2symmetricthree}). 
Then, one obtains
$\frac{33}{4(k+5)}$ in the last block diagonal elements.
One can apply for other $N$ values  
and it will turn out that the numerator of the above quantity  does 
depend on $N$ linearly as well as the constant term
while the denominator
is generalized to $4(k+N+2)$.
One can also use the formula with the correct $\hat{u}$ charge
\bea
\frac{3 (N+1) (N+3+2)}{2 (N+2) (k+N+2)}-\frac{(-\frac{3 N}{2})^2}{N (N+2)
  (k+N+2)}=
\frac{3(2N+5)}{4(N+k+2)},
\nonu
\eea
where the quadratic Casimir 
$C^{(N+2)}(\tiny\yng(3))$ is used.
The quantum number for the
$l^{+}$ can be obtained by 
the above matrix calculation in (\ref{su2symmetricthree}).
From the  eigenvalues $\frac{15}{4}$
appearing in the last block diagonal matrix
in (\ref{su2symmetricthree}), one can see the above
$l^{+}$ quantum number $l^{+} =\frac{3}{2}$, a quartet under the $SU(2)_k$.
This also can be seen from the previous expression
$ ({\bf 1},{\bf 4})_{-\frac{3N}{2}}$.
The trivial $l^-=0$ quantum number holds in this representation.
For the last eigenvalue corresponding to $\hat{u}$ charge,
the previous matrix calculation given in (\ref{u1chargesymmthree})
can be used.
The eigenvalues  
appearing in the last block diagonal matrix
in (\ref{u1chargesymmthree}) imply that the $\hat{u}$ charge is given by
$-\frac{9}{2}$.
Varying the $N$ values,
one finds that the $\hat{u}$ charge is linear in $N$.
This is also consistent with the representation 
$ ({\bf 1},{\bf 3})_{-\frac{3N}{2}}$ where the subscript denotes
the $\hat{u}$ charge.
One observes that the above conformal dimension does satisfy
the BPS bound.

%%%%%%%%%%%%%%%%%%%%%%%%%%%%%%%%%%%%%%%
\subsubsection{The $({\rm symm};\overline{f})$ representation with three
boxes}
%%%%%%%%%%%%%%%%%%%%%%%%%%%%%%%%%%%%%%%

Let us consider the higher representation
which arises from the product of $(\tiny\yng(3);0)$
and $(0;\overline{\tiny\yng(1)})$.
The former occurs in the subsection 
\ref{threesymmzero} and the latter occurs in the subsection
\ref{0frep} together with the complex conjugation.

In this case, the corresponding
four eigenvalues are described by
\bea
h (\tiny\yng(3);\overline{\tiny\yng(1)}) & = & \frac{(k+3 N+6)}{2 (k+N+2)},
\nonu \\
%l^{+} (\tiny\yng(3);\overline{\tiny\yng(1)}) & = & , \qquad
l^{+} (l^{+}+1)(\tiny\yng(3);\overline{\tiny\yng(1)})  & = & \frac{15}{4},  
\nonu \\
%l^{-} (\tiny\yng(3);\overline{\tiny\yng(1)}) & = & , \qquad
l^{-} (l^{-}+1)  (\tiny\yng(3);\overline{\tiny\yng(1)})  & = & \frac{3}{4},  
\nonu \\
\hat{u}( \tiny\yng(3); \overline{\tiny\yng(1)}) & = & -2N-1.
\nonu
\eea

First of all, one can obtain the following $35 \times 35$ matrix
by calculating the commutator
$[T_0, Q_{-\frac{1}{2}}^{\bar{A}}]$ as in the subsection
\ref{ffbarrep}
%\bea
%\left(
%\begin{array}{cccc}
%  \frac{1}{(5+k)} {\bf 1}_{10\times 10}& 0 & 0
% & 0  \\
%0  & \frac{1}{6(5+k)}{\bf 1}_{12\times 12} &  0 & 0   \\
%0 &0 & - \frac{2}{3(5+k)} {\bf 1}_{9\times 9} & 0   \\
%0 &0 &0 &   -\frac{3}{2(5+k)} {\bf 1}_{4\times 4}
%\end{array}
%\right)
%_{35 \times 35}
%\label{Extramatrix}
%\eea
\bea
&& \frac{1}{(5+k)} \mbox{diag}  (
1,1,1,1,1,1,1,1,1,1,\frac{1}{6},\frac{1}{6},\frac{1}{6},\frac{1}{6},
\frac{1}{6},\frac{1}{6},
\nonu \\
&& -\frac{2}{3},-\frac{2}{3},-\frac{2}{3},\frac{1}{6},\frac{1}{6},\frac{1}{6},\frac{1}{6},-\frac{2}{3},-\frac{2}{3},-\frac{2}{3},\frac{1}{6},\frac{1}{6},-\frac{2}{3},-\frac{2}{3},-\frac{2}{3},-\frac{3}{2},-\frac{3}{2},-\frac{3}{2},-\frac{3}{2}).
\label{Extramatrix}
\eea
The last four eigenvalues (the $N$ generalization is
straightforward to obtain) appearing in the last block diagonal
matrix in (\ref{Extramatrix})
provide the extra contribution as well as the sum of conformal dimensions
of  $(\tiny\yng(3);0)$
and $(0;\overline{\tiny\yng(1)})$.
They are given in (\ref{eigenvaluesthreesymm}) and (\ref{0ffoureigen})
respectively.
Then one obtains the final conformal dimension by adding the above
contribution appearing in (\ref{Extramatrix}) as follows
\bea
\frac{3 (2 N+5)}{4 (k+N+2)}
+\frac{(2 k+3)}{4 (k+N+2)}-\frac{3}{2 (k+N+2)}=\frac{(k+3 N+6)}{2 (k+N+2)}
\nonu
\eea
as in (\ref{eigenforsymmf}).

It is also useful to interpret the above result
from the conformal dimension formula.
One determines the following result
\bea
\frac{3 (N+1) (N+3+2)}{2 (N+2) (k+N+2)}
-\frac{(\frac{N}{2}-\frac{1}{2 N})}{(k+N+2)}
-\frac{(-\frac{1}{2} (N+2)-\frac{3 N}{2})^2}{N (N+2) (k+N+2)}
+\frac{1}{2}=\frac{(k+3 N+6)}{2 (k+N+2)}.
\nonu
\eea
Here we used the quadratic Casimirs for $C^{(N+2)}(\tiny\yng(3))$
and $C^{(N)}(\overline{\tiny\yng(1)})$. The correct $\hat{u}$
charge is inserted. The excitation number is given by $\frac{1}{2}$.

For the $l^+$ quantum number, due to the vanishing of $l^+$ in
$(0;\overline{\tiny\yng(1)})$, it turns out that
the $l^+$ is the same as the one($l^+=\frac{3}{2}$) in $(\tiny\yng(3);0)$. 
For the $l^-$ quantum number, due to the vanishing of $l^-$ in
 $(\tiny\yng(3);0)$, it turns out that
the $l^-$ is the same as the one($l^-=\frac{1}{2}$) in
$(0;\overline{\tiny\yng(1)})$. 
It is easy to see that
the above conformal dimension satisfies the BPS bound by substituting
$l^+=\frac{3}{2}$ and $l^-=\frac{1}{2}$.
One can add each $\hat{u}$ charge and it is obvious that
the total $\hat{u}$ charge is given by $-\frac{3N}{2} -\frac{(N+2)}{2}$
which leads to the above result. Note that
the $\hat{u}$ charge for $(0;\overline{\tiny\yng(1)})$ is opposite to
the one for $(0;\tiny\yng(1))$.

%%%%%%%%%%%%%%%%%%%%%%%%%%%%%%%%%%%%%%%
\subsubsection{The $(\rm{symm};\rm{antisymm})$ representation
with three and two boxes}
%%%%%%%%%%%%%%%%%%%%%%%%%%%%%%%%%%%%%%%
Let us consider the higher representation
which arises from the product of $(\tiny\yng(3);0)$
and $(0;\tiny\yng(1,1))$.
The former occurs in the subsection 
\ref{threesymmzero} and the latter occurs in the subsection
\ref{0antisymm}.

The four eigenvalues can be summarized by
\bea
h (\tiny\yng(3);\tiny\yng(1,1)) & = & \frac{(4 k+6 N+35)}{4 (k+N+2)},
\nonu \\
%l^{+} (\tiny\yng(3);\tiny\yng(1,1)) & = & , \qquad
l^{+} (l^{+}+1)(\tiny\yng(3);\tiny\yng(1,1))  & = & \frac{15}{4},  
\nonu \\
%l^{-} (\tiny\yng(3);\tiny\yng(1,1)) & = & , \qquad
l^{-} (l^{-}+1)  (\tiny\yng(3);\tiny\yng(1,1)) & =& 2,  
\nonu \\
\hat{u}( \tiny\yng(3);\tiny\yng(1,1) ) & = & -\frac{N}{2}+2.
\nonu
\eea

The following $35 \times 35$ matrix
by calculating the commutator
$[T_0,  Q_{-\frac{1}{2}}^{13} Q_{-\frac{1}{2}}^{14}]$ as in the
previous subsection
can be obtained
%\bea
%\left(
%\begin{array}{cccc}
%  -\frac{2}{(5+k)} {\bf 1}_{10\times 10}& 0 & 0
% & 0  \\
%0  & -\frac{1}{3(5+k)}{\bf 1}_{12\times 12} &  0 & 0   \\
%0 &0 &  \frac{4}{3(5+k)} {\bf 1}_{9\times 9} & 0   \\
%0 &0 &0 &   \frac{3}{(5+k)} {\bf 1}_{4\times 4}
%\end{array}
%\right)
%_{35 \times 35}
%\label{Extramatrix1}
%\eea
\bea
&& \frac{1}{(5+k)} \mbox{diag}  (
-2,-2,-2,-2,-2,-2,-2,-2,-2,-2,
-\frac{1}{3},-\frac{1}{3},-\frac{1}{3},-\frac{1}{3},-\frac{1}{3},
-\frac{1}{3},
\nonu \\
&& \frac{4}{3},\frac{4}{3},\frac{4}{3},-\frac{1}{3},-\frac{1}{3},-\frac{1}{3},-\frac{1}{3},\frac{4}{3},\frac{4}{3},\frac{4}{3},-\frac{1}{3},-\frac{1}{3},\frac{4}{3},\frac{4}{3},\frac{4}{3},3,3,3,3).
\label{Extramatrix1} 
\eea
The last four eigenvalues (the $N$ generalization is
simply $ \frac{3}{(N+k+2)}$) appearing in the last block diagonal
matrix in (\ref{Extramatrix1})
give the extra contribution as well as the sum of conformal dimensions
of  $(\tiny\yng(3);0)$
and $(0;\tiny\yng(1,1))$.
They are given in (\ref{eigenvaluesthreesymm}) and
(\ref{eigenvaluestwoantisymm})
respectively.
Then one obtains the final conformal dimension by adding the above
contribution appearing in (\ref{Extramatrix1}) as follows
\bea
\frac{3 (2 N+5)}{4 (k+N+2)}+
\frac{(k+2)}{(k+N+2)}+\frac{3}{(k+N+2)}=
\frac{(4 k+6 N+35)}{4 (k+N+2)}.
\nonu
\eea
Furthermore,
this analysis can be seen from the conformal dimension formula
{\small
  \bea
\frac{3 (N+1) (N+3+2)}{2 (N+2) (k+N+2)}-
\frac{2 (N-2) (N+1)}{2 N (k+N+2)}
-\frac{((N+2)-\frac{3 N}{2})^2}{N (N+2) (k+N+2)}
+1
=
\frac{(4 k+6 N+35)}{4 (k+N+2)}.
\nonu
\eea}
Here we also use the quadratic Casimir
$C^{(N)}(\tiny\yng(1,1))=\frac{(N-2)(N+1)}{N}$ and the excitation number
is equal to $1$. The correct $\hat{u}$ charge is inserted.
For the $l^+$ quantum number, due to the vanishing of $l^+$ in
$(0;\tiny\yng(1,1))$, 
the $l^+$ is the same as the one($l^+=\frac{3}{2}$) in $(\tiny\yng(3);0)$. 
For the $l^-$ quantum number, due to the vanishing of $l^-$ in
 $(\tiny\yng(3);0)$, 
the $l^-$ is the same as the one($l^-=1$) in
$(0;\tiny\yng(1,1))$. 
The total $\hat{u}$ charge is given by $-\frac{3N}{2}+(N+2)$
which leads to the above result.
One can easily see that the above conformal dimension
does not lead to the BPS bound with $l^{+} =\frac{3}{2}$ and $l^-=1$.
%\footnote{For the
%  higher representation  $(\tiny\yng(3);\overline{\tiny\yng(1,1)} )$,
%  the similar analysis can be done.}.

%%%%%%%%%%%%%%%%%%%%%%%%%%%%%%%%%%%%%%%
\subsubsection{The $(\rm{symm};\overline{\rm{symm}})$ representation
with three and two boxes}
%%%%%%%%%%%%%%%%%%%%%%%%%%%%%%%%%%%%%%%

Let us consider the higher representation
which arises from the product of $(\tiny\yng(3);0)$
and $(0;\overline{\tiny\yng(2)})$.
The former occurs in the subsection 
\ref{threesymmzero} while the latter occurs in the subsection
\ref{0symmtwo} with complex conjugation.

The four eigenvalues are given by 
\bea
h (\tiny\yng(3);\overline{\tiny\yng(2)}) & = &
\frac{(4 k+6 N+3)}{4 (k+N+2)},
\nonu \\
%l^{+} (\tiny\yng(3);\overline{\tiny\yng(2)}) & = & , \qquad
l^{+} (l^{+}+1)(\tiny\yng(3);\overline{\tiny\yng(2)})  & = & \frac{15}{4},  
\nonu \\
%l^{-} (\tiny\yng(3);\overline{\tiny\yng(2)}) & = & , \qquad
l^{-} (l^{-}+1)  (\tiny\yng(3);\overline{\tiny\yng(2)}) & = & 0,  
\nonu \\
\hat{u}( \tiny\yng(3);\overline{\tiny\yng(2)} ) & = & -\frac{5N}{2}-2.
\nonu
\eea

One should calculate the commutator
$[T_0,  Q_{-\frac{1}{2}}^{1} Q_{-\frac{1}{2}}^{4}]$
and it turns out that
%\bea
%\left(
%\begin{array}{cccc}
%  \frac{2}{(5+k)} {\bf 1}_{10\times 10}& 0 & 0
% & 0  \\
%0  & \frac{1}{3(5+k)}{\bf 1}_{12\times 12} &  0 & 0   \\
%0 &0 &  -\frac{4}{3(5+k)} {\bf 1}_{9\times 9} & 0   \\
%0 &0 &0 &   -\frac{3}{(5+k)} {\bf 1}_{4\times 4}
%\end{array}
%\right)
%_{35 \times 35}
%\label{Extramatrix2}
%\eea
\bea
&& \frac{1}{(5+k)} \mbox{diag}  (
2,2,2,2,2,2,2,2,2,2,\frac{1}{3},\frac{1}{3},\frac{1}{3},\frac{1}{3},\frac{1}{3},\frac{1}{3},
\nonu \\
&& -\frac{4}{3},-\frac{4}{3},-\frac{4}{3},\frac{1}{3},\frac{1}{3},\frac{1}{3},\frac{1}{3},-\frac{4}{3},-\frac{4}{3},-\frac{4}{3},\frac{1}{3},\frac{1}{3},-\frac{4}{3},-\frac{4}{3},-\frac{4}{3},-3,-3,-3,-3).
\label{Extramatrix2}
\eea
The last four eigenvalues (the $N$ generalization is
simply $ -\frac{3}{(N+k+2)}$) appearing in the last block diagonal
matrix in (\ref{Extramatrix2})
give the extra contribution as well as the sum of conformal dimensions
of  $(\tiny\yng(3);0)$
and $(0;\overline{\tiny\yng(2)})$.
They are given in (\ref{eigenvaluesthreesymm}) and
(\ref{eigenvaluessymmtwo})
respectively.
Then one obtains the final conformal dimension by adding the above
contribution appearing in (\ref{Extramatrix2}) as follows
\bea
\frac{3 (2 N+5)}{4 (k+N+2)}
+\frac{k}{(k+N+2)}-\frac{3}{(k+N+2)}
=\frac{(4 k+6 N+3)}{4 (k+N+2)}.
\nonu
\eea
The conformal dimension formula implies that
{\small
  \bea
\frac{3 (N+1) (N+3+2)}{2 (N+2) (k+N+2)}
-
\frac{2 (N-1) (N+2)}{(2 N) (k+N+2)}
-\frac{(-(N+2)-\frac{3 N}{2})^2}{N (N+2) (k+N+2)}+1
=\frac{(4 k+6 N+3)}{4 (k+N+2)}.
\nonu
\eea}
The quadratic Casimir
$C^{(N)}(\overline{\tiny\yng(2)})$ is the same as
$C^{(N)}(\tiny\yng(2))$ and the excitation number
is equal to $1$. The correct $\hat{u}$ charge is inserted.
For the $l^+$ quantum number, due to the vanishing of $l^+$ in
$(0;\overline{\tiny\yng(2)})$, 
the $l^+$ is the same as the one($l^+=\frac{3}{2}$) in $(\tiny\yng(3);0)$. 
For the $l^-$ quantum number, due to the vanishing of $l^-$ in
 $(\tiny\yng(3);0)$ and $(0;\overline{\tiny\yng(2)})$, 
the total $l^-$ is given by $l^-=0$, a singlet. 
The total $\hat{u}$ charge is given by $-\frac{3N}{2}-(N+2)$
which leads to the above result. Again,
the $\hat{u}$ charge for $(0;\overline{\tiny\yng(2)})$ is opposite to
the one for $(0;\tiny\yng(2))$.
One can easily see that the above conformal dimension
does not lead to the BPS bound with $l^{+} =\frac{3}{2}$ and $l^-=0$.

%%%%%%%%%%%%%%%%%%%%%%%%%%%%%%%%%%%%%%%
\subsubsection{The $(\rm{symm};\overline{\rm{antisymm}})$ representation
with three and two boxes}
%%%%%%%%%%%%%%%%%%%%%%%%%%%%%%%%%%%%%%%

Let us consider the higher representation
which arises from the product of $(\tiny\yng(3);0)$
and $(0;\overline{\tiny\yng(1,1)})$.
The former occurs in the subsection 
\ref{threesymmzero} while the latter occurs in the subsection
\ref{0antisymm} with complex conjugation.

The four eigenvalues are given by
\bea
h (\tiny\yng(3);\overline{\tiny\yng(1,1)}) & = &
\frac{(4 k+6 N+11)}{4 (k+N+2)},
\nonu \\
%l^{+} (\tiny\yng(3);\overline{\tiny\yng(2)}) & = & , \qquad
l^{+} (l^{+}+1)(\tiny\yng(3);\overline{\tiny\yng(1,1)})  & = & \frac{15}{4},  
\nonu \\
%l^{-} (\tiny\yng(3);\overline{\tiny\yng(2)}) & = & , \qquad
l^{-} (l^{-}+1)  (\tiny\yng(3);\overline{\tiny\yng(1,1)}) & = & 2,  
\nonu \\
\hat{u}( \tiny\yng(3);\overline{\tiny\yng(1,1)} ) & = & -\frac{5N}{2}-2.
\nonu
\eea

One should calculate the commutator
$[T_0,  Q_{-\frac{1}{2}}^{1} Q_{-\frac{1}{2}}^{2}]$
and it turns out that
%\bea
%\left(
%\begin{array}{cccc}
%  \frac{2}{(5+k)} {\bf 1}_{10\times 10}& 0 & 0
% & 0  \\
%0  & \frac{1}{3(5+k)}{\bf 1}_{12\times 12} &  0 & 0   \\
%0 &0 &  -\frac{4}{3(5+k)} {\bf 1}_{9\times 9} & 0   \\
%0 &0 &0 &   -\frac{3}{(5+k)} {\bf 1}_{4\times 4}
%\end{array}
%\right)
%_{35 \times 35}
%\label{Extramatrix3}
%\eea
\bea
&& \frac{1}{(5+k)} \mbox{diag}  (
2,2,2,2,2,2,2,2,2,2,\frac{1}{3},\frac{1}{3},\frac{1}{3},\frac{1}{3},
\frac{1}{3},\frac{1}{3},\nonu \\
&&
-\frac{4}{3},-\frac{4}{3},-\frac{4}{3},\frac{1}{3},\frac{1}{3},\frac{1}{3},\frac{1}{3},-\frac{4}{3},-\frac{4}{3},-\frac{4}{3},\frac{1}{3},\frac{1}{3},-\frac{4}{3},-\frac{4}{3},-\frac{4}{3},-3,-3,-3,-3).
\label{Extramatrix3}
\eea
The last four eigenvalues (the $N$ generalization is
simply $ -\frac{3}{(N+k+2)}$) appearing in the last block diagonal
matrix in (\ref{Extramatrix3})
give the extra contribution as well as the sum of conformal dimensions
of  $(\tiny\yng(3);0)$
and $(0;\overline{\tiny\yng(1,1)})$.
They are given in (\ref{eigenvaluesthreesymm}) and
(\ref{eigenvaluestwoantisymm})
respectively.
Then one obtains the final conformal dimension by adding the above
contribution appearing in (\ref{Extramatrix3}) as follows
\bea
\frac{3 (2 N+5)}{4 (k+N+2)}
+\frac{(k+2)}{(k+N+2)}-\frac{3}{(k+N+2)}
=\frac{(4 k+6 N+11)}{4 (k+N+2)}.
\nonu
\eea

The conformal dimension formula implies that
{\small
  \bea
\frac{3 (N+1) (N+3+2)}{2 (N+2) (k+N+2)}
-\frac{2 (N-2) (N+1)}{(2 N) (k+N+2)}
-\frac{(-(N+2)-\frac{3 N}{2})^2}{N (N+2) (k+N+2)}
+1=
\frac{(4 k+6 N+11)}{4 (k+N+2)}.
\nonu
\eea}
The quadratic Casimir
$C^{(N)}(\overline{\tiny\yng(1,1)})$ is the same as
$C^{(N)}(\tiny\yng(1,1))$ and the excitation number
is equal to $1$. The correct $\hat{u}$ charge is inserted.
For the $l^+$ quantum number, due to the vanishing of $l^+$ in
$(0;\overline{\tiny\yng(1,1)})$, 
the $l^+$ is the same as the one($l^+=\frac{3}{2}$) in $(\tiny\yng(3);0)$. 
For the $l^-$ quantum number, due to the vanishing of $l^-$ in
 $(\tiny\yng(3);0)$, 
the total $l^-$ is given by $l^-=1$ in
$(\overline{\tiny\yng(1,1)})$, a triplet. 
The total $\hat{u}$ charge is given by $-\frac{3N}{2}-(N+2)$
which leads to the above result. Again,
the $\hat{u}$ charge for $(0;\overline{\tiny\yng(1,1)})$ is opposite to
the one for $(0;\tiny\yng(1,1))$.
One can easily see that the above conformal dimension
does lead to the BPS bound with $l^{+} =\frac{3}{2}$ and $l^-=1$.

%%%%%%%%%%%%%%%%%%%%%%%%%%%%%%%%%%%%%%%
\subsubsection{The $(\rm{symm};\rm{mixed})$ representation
with three boxes}
%%%%%%%%%%%%%%%%%%%%%%%%%%%%%%%%%%%%%%%

Let us consider the higher representation
which arises from the product of $(\tiny\yng(3);0)$
and $(0;\tiny\yng(2,1))$.
The former occurs in the subsection 
\ref{threesymmzero} while the latter occurs in the subsection
\ref{0mixed}.

The four eigenvalues are given by
\bea
h (\tiny\yng(3);\tiny\yng(2,1)) & = & \frac{3 (k+N+6)}{2 (k+N+2)},
\nonu \\
%l^{+} (\tiny\yng(3);\tiny\yng(2,1)) & = &  ,\qquad
l^{+} (l^{+}+1)(\tiny\yng(3);\tiny\yng(2,1))  & = & \frac{15}{4}, 
\nonu \\
%l^{-} (\tiny\yng(3);\tiny\yng(2,1)) & = & ,\qquad
l^{-} (l^{-}+1)  (\tiny\yng(3);\tiny\yng(2,1))  & = & \frac{3}{4},  
\nonu \\
\hat{u}( \tiny\yng(3);\tiny\yng(2,1) ) & = & 3.
\nonu
\eea

One should calculate the commutator
$[T_0,  Q_{-\frac{1}{2}}^{14} Q_{-\frac{1}{2}}^{16}  Q_{-\frac{1}{2}}^{13} ]$
and it turns out that
%\bea
%\left(
%\begin{array}{cccc}
%  -\frac{3}{(5+k)} {\bf 1}_{10\times 10}& 0 & 0
% & 0  \\
%0  & -\frac{1}{2(5+k)}{\bf 1}_{12\times 12} &  0 & 0   \\
%0 &0 &  \frac{2}{(5+k)} {\bf 1}_{9\times 9} & 0   \\
%0 &0 &0 &   \frac{9}{2(5+k)} {\bf 1}_{4\times 4}
%\end{array}
%\right)
%_{35 \times 35}
%\label{Extramatrix4}
%\eea
\bea
&& \frac{1}{(5+k)} \mbox{diag}  (
-3,-3,-3,-3,-3,-3,-3,-3,-3,-3,
-\frac{1}{2},-\frac{1}{2},-\frac{1}{2},-\frac{1}{2},-\frac{1}{2},
-\frac{1}{2}, \nonu \\
&& 2,2,2,-\frac{1}{2},-\frac{1}{2},-\frac{1}{2},-\frac{1}{2},2,2,2,-\frac{1}{2},-\frac{1}{2},2,2,2,\frac{9}{2},\frac{9}{2},\frac{9}{2},\frac{9}{2}).
\label{Extramatrix4}
\eea
The last four eigenvalues (the $N$ generalization is
simply $ \frac{9}{2(N+k+2)}$) appearing in the last block diagonal
matrix in (\ref{Extramatrix4})
give the extra contribution as well as the sum of conformal dimensions
of  $(\tiny\yng(3);0)$
and $(0;\tiny\yng(2,1))$.
They are given in (\ref{eigenvaluesthreesymm}) and
(\ref{eigenvaluesmixed})
respectively.
Then one obtains the final conformal dimension by adding the above
contribution appearing in (\ref{Extramatrix4}) as follows
\bea
\frac{3 (2 N+5)}{4 (k+N+2)}+
\frac{3 (2 k+1)}{4 (k+N+2)}
+
\frac{9}{2 (k+N+2)} =\frac{3 (k+N+6)}{2 (k+N+2)}.
\nonu
\eea

The conformal dimension formula implies that
\bea
\frac{3 (N+1) (N+3+2)}{2 (N+2) (k+N+2)}
-\frac{3 (N^2-3)}{(2 N) (k+N+2)}
-\frac{(\frac{3 (N+2)}{2}-\frac{3 N}{2})^2}{N (N+2) (k+N+2)}
+\frac{3}{2}
=\frac{3 (k+N+6)}{2 (k+N+2)}.
\nonu
\eea
The quadratic Casimir
$C^{(N)}(\tiny\yng(2,1))$ is used and the excitation number
is equal to $\frac{3}{2}$.
The correct $\hat{u}$ charge is inserted.
For the $l^+$ quantum number, due to the vanishing of $l^+$ in
$(0;\tiny\yng(2,1))$, 
the $l^+$ is the same as the one($l^+=\frac{3}{2}$) in $(\tiny\yng(3);0)$. 
For the $l^-$ quantum number, due to the vanishing of $l^-$ in
 $(\tiny\yng(3);0)$, 
the total $l^-$ is given by $l^-=\frac{1}{2}$ in
$(\tiny\yng(2,1))$, a doublet. 
The total $\hat{u}$ charge is given by $-\frac{3N}{2}+\frac{3}{2}(N+2)$
which leads to the above result.
%Again,
%the $\hat{u}$ charge for $(0;\overline{\tiny\yng(1,1)})$ is opposite to
%the one for $(0;\tiny\yng(1,1))$.
One can easily see that the above conformal dimension
does not lead to the BPS bound with $l^{+} =\frac{3}{2}$ and
$l^-=\frac{1}{2}$ \footnote{ For the higher representation
  $ (\tiny\yng(3);\overline{\tiny\yng(2,1)})$, one can do similar
  analysis with an appropriate $\hat{u}$ charge.}.

%%%%%%%%%%%%%%%%%%%%%%%%%%%%%%%%%%%%%%%
\subsubsection{The $(\rm{symm};\rm{antisymm})$ representation
with three boxes}
%%%%%%%%%%%%%%%%%%%%%%%%%%%%%%%%%%%%%%%

Let us consider the higher representation
which arises from the product of $(\tiny\yng(3);0)$
and $(0;\tiny\yng(1,1,1))$.
The former occurs in the subsection 
\ref{threesymmzero} while the latter occurs in the subsection
\ref{0antisymmthree}.

The four eigenvalues are given by
\bea
h (\tiny\yng(3);\tiny\yng(1,1,1)) & = & \frac{3 (k+N+8)}{2 (k+N+2)},
\nonu \\
%l^{+} (\tiny\yng(3);\tiny\yng(1,1,1)) & = &  ,\qquad
l^{+} (l^{+}+1)(\tiny\yng(3);\tiny\yng(1,1,1))  & = & \frac{15}{4},  
\nonu \\
%l^{-} (\tiny\yng(3);\tiny\yng(1,1,1)) & = & , \qquad
l^{-} (l^{-}+1)  (\tiny\yng(3);\tiny\yng(1,1,1)) & = & \frac{15}{4}, 
\nonu \\
\hat{u}( \tiny\yng(3);\tiny\yng(1,1,1) ) & = & 3.
\label{symmantithreethree}
\eea

One should calculate the commutator
$[T_0,  Q_{-\frac{1}{2}}^{13} Q_{-\frac{1}{2}}^{14}  Q_{-\frac{1}{2}}^{15} ]$
and it turns out that
%\bea
%\left(
%\begin{array}{cccc}
%  -\frac{3}{(5+k)} {\bf 1}_{10\times 10}& 0 & 0
% & 0  \\
%0  & -\frac{1}{2(5+k)}{\bf 1}_{12\times 12} &  0 & 0   \\
%0 &0 &  \frac{2}{(5+k)} {\bf 1}_{9\times 9} & 0   \\
%0 &0 &0 &   \frac{9}{2(5+k)} {\bf 1}_{4\times 4}
%\end{array}
%\right)
%_{35 \times 35}
%\eea
\bea
&& \frac{1}{(5+k)} \mbox{diag}  (
-3,-3,-3,-3,-3,-3,-3,-3,-3,-3,-\frac{1}{2},
-\frac{1}{2},-\frac{1}{2},-\frac{1}{2},-\frac{1}{2},-\frac{1}{2},
\nonu \\
&& 2,2,2,-\frac{1}{2},-\frac{1}{2},-\frac{1}{2},-\frac{1}{2},2,2,2,-\frac{1}{2},-\frac{1}{2},2,2,2,\frac{9}{2},\frac{9}{2},\frac{9}{2},\frac{9}{2}).
\label{Extramatrix5-1}
\eea
The last four eigenvalues (the $N$ generalization is
simply $ \frac{9}{2(N+k+2)}$) appearing in the last block diagonal
matrix in (\ref{Extramatrix5-1})
give the extra contribution as well as the sum of conformal dimensions
of  $(\tiny\yng(3);0)$
and $(0;\tiny\yng(1,1,1))$.
They are given in (\ref{eigenvaluesthreesymm}) and
(\ref{eigenvaluesthreeantisymm})
respectively.
Then one obtains the final conformal dimension by adding the above
contribution appearing in (\ref{Extramatrix5-1}) as follows
\bea
\frac{3 (2 N+5)}{4 (k+N+2)}
+\frac{6 k+15}{4 (k+N+2)}
+\frac{9}{2 (k+N+2)}
=\frac{3 (k+N+8)}{2 (k+N+2)}.
\nonu
\eea

The conformal dimension formula implies that
\bea
\frac{3 (N+1) (N+3+2)}{2 (N+2) (k+N+2)}
-\frac{3 (N-3) (N+1)}{(2 N) (k+N+2)}
-\frac{(\frac{3 (N+2)}{2}-\frac{3 N}{2})^2}{N (N+2) (k+N+2)}
+\frac{3}{2}=
\frac{3 (k+N+8)}{2 (k+N+2)}.
\nonu
\eea
The quadratic Casimir
$C^{(N)}(\tiny\yng(1,1,1))$ is used and the excitation number
is equal to $\frac{3}{2}$.
The correct $\hat{u}$ charge is inserted.
For the $l^+$ quantum number, due to the vanishing of $l^+$ in
$(0;\tiny\yng(1,1,1))$, 
the $l^+$ is the same as the one($l^+=\frac{3}{2}$) in $(\tiny\yng(3);0)$. 
For the $l^-$ quantum number, due to the vanishing of $l^-$ in
 $(\tiny\yng(3);0)$, 
the total $l^-$ is given by $l^-=\frac{3}{2}$ in
$(\tiny\yng(1,1,1))$, a quartet. 
The total $\hat{u}$ charge is given by $-\frac{3N}{2}+\frac{3}{2}(N+2)$
which leads to the above result.
%Again,
%the $\hat{u}$ charge for $(0;\overline{\tiny\yng(1,1)})$ is opposite to
%the one for $(0;\tiny\yng(1,1))$.
One can easily see that the above conformal dimension
does not lead to the BPS bound with $l^{\pm} =\frac{3}{2}$
\footnote{One can do the similar analysis for the higher representation
  $(\tiny\yng(3);\overline{\tiny\yng(1,1,1)})$. We have
\bea
h (\tiny\yng(3);\overline{\tiny\yng(1,1,1)}) & = & \frac{3}{2},
\qquad
%l^{+} (\tiny\yng(3);\tiny\yng(1,1,1)) & = &  ,\qquad
l^{+} (l^{+}+1)(\tiny\yng(3);\overline{\tiny\yng(1,1,1)})   =  \frac{15}{4},  
\nonu \\
%l^{-} (\tiny\yng(3);\tiny\yng(1,1,1)) & = & , \qquad
l^{-} (l^{-}+1)  (\tiny\yng(3);\overline{\tiny\yng(1,1,1)}) & = & \frac{15}{4}, 
\qquad
\hat{u}( \tiny\yng(3);\overline{\tiny\yng(1,1,1)} )  =  -3N-3.
\nonu
\eea
This conformal dimension
does lead to the BPS bound with $l^{\pm} =\frac{3}{2}$.
Due to the different $\hat{u}$ charge, the conformal dimension
is different from the one in (\ref{symmantithreethree})}.

%%%%%%%%%%%%%%%%%%%%%%%%%%%%%%%%%%%%%%%%%%%%%%%%%%%%%%%
%%%%%%%%%%%%%%%%%%%
\subsection{ The antisymmetric representations
  $\La_+$ with three boxes}
%%%%%%%%%%%%%%%%%%%
%%%%%%%%%%%%%%%%%%%%%%%%%%%%%%%%%%%%%%%%%%%%%%%%%%%%%%%

%\bea
%\tiny\yng(1,1,1)
%\nonu
%\eea

The three-index antisymmetric parts of the $SU(N+2)$ representation
can be obtained from the generators of the fundamental representation
of $SU(N+2)$ by using the projection operator $\frac{1}{6} (\de_{il}
\de_{jm} \de_{kn} - \de_{il} \de_{km} \de_{jn} +\de_{kl}
\de_{im} \de_{jn} -\de_{jl} \de_{im} \de_{kn}+\de_{jl}
\de_{km} \de_{in} - \de_{kl} \de_{jm} \de_{in} )$ where
$i < j < k$ and $ l < m < n$
and $i, j, k, l,m,n = 1, 2, \cdots, (N+2) $ \cite{Cvitanovicbook}.
Then by acting on
the space
$T_a \otimes {\bf 1}_{(N+2) \times (N+2)} \otimes {\bf 1}_{(N+2) \times (N+2)}
+ {\bf 1}_{(N+2) \times (N+2)} \otimes T_a \otimes {\bf 1}_{(N+2) \times (N+2)} +
{\bf 1}_{(N+2) \times (N+2)} \otimes {\bf 1}_{(N+2) \times (N+2)} \otimes T_a$,
one has
the generators for the antisymmetric representation for the $SU(N+2)$
as follows:
{\small
\bea
&& (T_a)_{il} \, \delta_{jm} \, \delta_{kn} -
(T_a)_{il} \, \delta_{km} \, \delta_{jn} +
(T_a)_{kl} \, \delta_{im} \, \delta_{jn} -
(T_a)_{jl} \, \delta_{im} \, \delta_{kn} +
(T_a)_{jl} \, \delta_{km} \, \delta_{in} -
(T_a)_{kl} \, \delta_{jm} \, \delta_{in}  \nonu\\
&& +
\delta_{il} \, (T_a)_{jm} \, \delta_{kn}-
\delta_{il} \, (T_a)_{km} \, \delta_{jn}+
\delta_{kl} \, (T_a)_{im} \, \delta_{jn}-
\delta_{jl} \, (T_a)_{im} \, \delta_{kn}+
\delta_{jl} \, (T_a)_{km} \, \delta_{in}-
\delta_{kl} \, (T_a)_{jm} \, \delta_{in}
\nonu \\
&& +
\delta_{il} \, \delta_{jm} \, (T_a)_{kn}
-
\delta_{il} \, \delta_{km} \, (T_a)_{jn}
+
\delta_{kl} \, \delta_{im} \, (T_a)_{jn}
-
\delta_{jl} \, \delta_{im} \, (T_a)_{kn}
+
\delta_{jl} \, \delta_{km} \, (T_a)_{in}
-
\delta_{kl} \, \delta_{jm} \, (T_a)_{in}. 
\nonu
\eea}
For $N=3$, one has
$\frac{1}{6}(N+2)(N+1)(N) \times \frac{1}{6}(N+2)(N+1)(N)
= 10 \times 10$ unitary matrix, 
and the row and columns are characterized by
the following triple index notations
\bea
&& 123; \,  124, \, 125, \, 134, \, 135,  \, 234, \,  235;
\, 145, \, 245, \, 345. 
\label{10cases}
\eea
See also Appendix $E$ for $\hat{u}_1, \cdots, \hat{u}_{10}$.
See also (\ref{transtranstrans}).
The first element corresponds to the
antisymmetric (singlet) representation for $SU(3)$.
The next six elements correspond to
the antisymmetric representation of $SU(3)$
with $SU(2)_k$ doublet.
The last three elements correspond to
the fundamental representation of $SU(3)$
with $SU(2)_k$ singlet.
The explicit form for the $24$ generators of
$SU(5)$ is presented in Appendix $E$.
Note that the ordering in (\ref{10cases})
is different from the ones in Appendix $E$.

Let us calculate the zero mode for the reduced stress energy tensor
spin-$2$ current acting on the state $|(\tiny\yng(1,1,1) ;0) >$. 
It turns out that the $10 \times 10$ matrix
is given by
\bea
\left(
\begin{array}{ccccc}
  \frac{3}{(5+k)}  & 0 & 0 & 0 & 0
  \\
  0  & \frac{9}{4(5+k)}{\bf 1}_{4 \times 4} & 0 & 0 &  0 \\
0 & 0 &  \frac{2}{(5+k)} & 0 & 0   \\
   0  &  0 & 0 & \frac{9}{4(5+k)}{\bf 1}_{2 \times 2}  &  0 \\
0 &0 &  0 & 0 & \frac{2}{(5+k)} {\bf 1}_{2 \times 2}    
\end{array}
\right).
%_{35 \times 35}
\label{spin2antisymmetricthree}
\eea
If one takes the ordering in (\ref{10cases}), then
the diagonal elements will look nicer.
There are five block diagonal elements. 
The last block diagonal elements (together with the third one)
correspond to
the eigenvalue on the  state $|(\tiny\yng(1,1,1) ;\tiny\yng(1)) >$.
We will describe the detailed quantum numbers
for the other eigenvalues soon.

One can also calculate the zero mode for the sum of the square for
the spin-$1$ current with minus sign acting on the above
state $|(\tiny\yng(1,1,1) ;\tiny\yng(1)) >$
and the explicit result is given by 
\bea
\left(
\begin{array}{ccccc}
  0  & 0 & 0 & 0 & 0
  \\
0  & \frac{3}{4}{\bf 1}_{4 \times 4} &  0 & 0 & 0 \\
0 &0 &  0  & 0 & 0  \\
0  & 0  & 0 & \frac{3}{4}{\bf 1}_{2 \times 2}   & 0 \\
0 &0 &  0  &  0 & 0 {\bf 1}_{2 \times 2}     
\end{array}
\right).
%_{35 \times 35}
\label{su2antisymmetricthree}
\eea
The diagonal elements correspond to the eigenvalues
and in particular, the last one (together with the third one)
is the eigenvalue for the
state  $|(\tiny\yng(1,1,1) ;\tiny\yng(1)) >$
which behaves as a triplet (fundamental) under the
$SU(3)$. 

Similarly, one can also compute the
zero mode for the spin-$1$ current
acting on the 
state $|(\tiny\yng(1,1,1) ;\tiny\yng(1)) >$
and one obtains (twice of $\hat{u}$)
\bea
\left(
\begin{array}{ccccc}
  6  & 0 & 0 & 0 & 0
  \\
0  & 1 {\bf 1}_{4 \times 4} &  0 & 0 & 0 \\
0 &0 &  -4 & 0 & 0  \\
0  & 0 & 0 & 1 {\bf 1}_{2 \times 2}  & 0 \\
0 &0 &  0 & 0 & -4 {\bf 1}_{2 \times 2} 
\end{array}
\right).
%_{35 \times 35}
\label{u1chargeantisymmthree}
\eea
In this case, also the last elements (together with the third one)
are the eigenvalues
for the state  $|(\tiny\yng(1,1,1) ;\tiny\yng(1)) >$.

Note that there is no $SU(N)$ singlet in the antisymmetric representation
in (\ref{transtranstrans}).

%%%%%%%%%%%%%%%%%%%%%%%%%%%
\subsubsection{The $(\rm{antisymm};{\rm antisymm})$ representation
with three boxes}
%%%%%%%%%%%%%%%%%%%%%%%%%%%%

Let us consider the higher representation 
where the antisymmetric representation in $SU(N)$ $ ({\tiny\yng(1,1,1)},
{\bf 1})_3$
survives
in the branching of (\ref{transtranstrans}).
The four eigenvalues are given by
\bea
h (\tiny\yng(1,1,1);\tiny\yng(1,1,1)) & = & \frac{3}{(k+N+2)},
\nonu \\
%l^{+} (\tiny\yng(3);\tiny\yng(1,1,1)) & = &  ,\qquad
l^{+} (l^{+}+1)(\tiny\yng(1,1,1);\tiny\yng(1,1,1))  & = & 0,  
\nonu \\
%l^{-} (\tiny\yng(3);\tiny\yng(1,1,1)) & = & , \qquad
l^{-} (l^{-}+1)  (\tiny\yng(1,1,1);\tiny\yng(1,1,1)) & = & 0, 
\nonu \\
\hat{u}( \tiny\yng(1,1,1);\tiny\yng(1,1,1) ) & = & 3.
\nonu
\eea
One can calculate the conformal dimension for this
representation using the previous formula or
one performs the explicit form for the
matrix in this particular representation
as in (\ref{spin2symmetricthree}). 
Therefore, one obtains
$\frac{3}{(k+5)}$ in the first diagonal elements in
(\ref{spin2antisymmetricthree}).
One can apply for other $N$ values where $N=5,7,9, 11, 13, \cdots$. 
It will turn out that the numerator of the above quantity  does not
depend on $N$ and takes the common value and the denominator
is generalized to $(k+N+2)$.
By realizing that the quadratic Casimir of $SU(N+2)$ for the
antisymmetric representation
$C^{(N+2)}(\tiny\yng(1,1,1))=\frac{3(N-1)(N+3)}{2(N+2)}$
(and the one for the antisymmetric representation
$C^{(N)}(\tiny\yng(1,1,1))=\frac{3(N-3)(N+1)}{2N}$)
and the correct $\hat{u}$ charge is given by $3$,
the following relation can be obtained
\bea
\frac{3 (N+2-3) (N+2+1)}{2 (N+2) (k+N+2)}
-\frac{3 (N-3) (N+1)}{2 N (k+N+2)}
-\frac{9}{N (N+2) (k+N+2)}=
\frac{3}{(k+N+2)}.
\nonu
\eea

Similarly, the quantum number for the
$l^{+}$ can be determined by 
the above matrix calculation in (\ref{su2antisymmetricthree}).
From the zero eigenvalues appearing in the first block diagonal matrix
in (\ref{su2antisymmetricthree}), one can see the above
$l^{+}$ quantum number, a singlet under the $SU(2)_k$.
This also can be seen from the previous expression
$ ({\tiny\yng(1,1,1)},{\bf 1})_3$.
The trivial $l^-$ quantum number $l^-=0$ arises.
For the last eigenvalue corresponding to $\hat{u}$ charge,
one uses the previous matrix calculation given in
(\ref{u1chargeantisymmthree}).
The eigenvalues  
appearing in the first block diagonal matrix
in (\ref{u1chargeantisymmthree}) imply that the $\hat{u}$ charge is given by $3$.
This is also consistent with the representation 
$ ({\tiny\yng(1,1,1)},{\bf 1})_3$ where the subscript denotes
the $\hat{u}$ charge.
One observes that the above conformal dimension does not satisfy
the vanishing BPS bound with $l^{\pm} =0$.

%%%%%%%%%%%%%%%%%%%%%%%%%%%
\subsubsection{The $(\rm{antisymm};{\rm antisymm})$ representation
with three and two boxes}
%%%%%%%%%%%%%%%%%%%%%%%%%%%%

Let us consider the
 higher representation 
 where the antisymmetric representation in
 $SU(N)$ $ ({\tiny\yng(1,1)},{\bf 2})_{2-\frac{N}{2}}$
survives
in the branching of (\ref{transtranstrans}).
The four eigenvalues can be summarized by
\bea
h (\tiny\yng(1,1,1);\tiny\yng(1,1)) & = &
\frac{(2 N+3)}{4 (k+N+2)},
\nonu \\
%l^{+} (\tiny\yng(3);\tiny\yng(1,1,1)) & = &  ,\qquad
l^{+} (l^{+}+1)(\tiny\yng(1,1,1);\tiny\yng(1,1))  & = & \frac{3}{4},  
\nonu \\
%l^{-} (\tiny\yng(3);\tiny\yng(1,1,1)) & = & , \qquad
l^{-} (l^{-}+1)  (\tiny\yng(1,1,1);\tiny\yng(1,1)) & = & 0, 
\nonu \\
\hat{u}( \tiny\yng(1,1,1);\tiny\yng(1,1) ) & = & 2-\frac{N}{2}.
\nonu
\eea

The explicit form for the
matrix in this particular representation
is given by (\ref{spin2antisymmetricthree}). 
Then, one obtains
$\frac{9}{4(k+5)}$ in the second (and fourth one) diagonal elements.
One can apply for other $N$ values  
and it will turn out that the numerator of the above quantity  does 
depend on $N$ linearly as well as the constant term
while the denominator
is generalized to $4(k+N+2)$.
One can also use the formula with the correct $\hat{u}$ charge
\bea
\frac{3 (N+2-3) (N+2+1)}{2 (N+2) (k+N+2)}
-\frac{2 (N-2) (N+1)}{2 N (k+N+2)}
-\frac{(2-\frac{N}{2})^2}{N (N+2) (k+N+2)}
=\frac{(2 N+3)}{4 (k+N+2)}
\nonu
\eea
where the quadratic Casimir 
$C^{(N+2)}(\tiny\yng(1,1,1))$ and $C^{(N)}(\tiny\yng(1,1))$ are used.
The quantum number for the
$l^{+}$ can be obtained by 
the above matrix calculation in (\ref{su2antisymmetricthree}).
From the  eigenvalues $\frac{3}{4}$
appearing in the second (and fourth one) block diagonal matrix
in (\ref{su2antisymmetricthree}), one can see the above
$l^{+}$ quantum number $l^{+} =\frac{1}{2}$, a doublet under the $SU(2)_k$.
This also can be seen from the previous expression
$ ({\tiny\yng(1,1)},{\bf 2})_{2-\frac{N}{2}}$.
The trivial $l^-=0$ quantum number holds in this representation.
For the last eigenvalue corresponding to $\hat{u}$ charge,
the previous matrix calculation given in (\ref{u1chargeantisymmthree})
can be used.
The eigenvalues  
appearing in the second block diagonal matrix
in (\ref{u1chargeantisymmthree}) imply that the $\hat{u}$ charge is given by
$\frac{1}{2}$.
Varying the $N$ values,
one finds that the $\hat{u}$ charge is linear in $N$
as well as the constant term.
This is also consistent with the representation 
$ ({\tiny\yng(1,1)},{\bf 2})_{2-\frac{N}{2}}$ where the subscript denotes
the $\hat{u}$ charge.
One observes that the above conformal dimension does satisfy
the BPS bound.

%%%%%%%%%%%%%%%%%%%%%%%%%%%
\subsubsection{The $({\rm antisymm};f)$ representation
with three boxes}
%%%%%%%%%%%%%%%%%%%%%%%%%%%%

Let us consider the higher representation 
where the fundamental representation in $SU(N)$
$ ({\tiny\yng(1)},{\bf 1})_{1-N}$
survives
in the branching of (\ref{transtranstrans}).
The four eigenvalues are given by
\bea
h (\tiny\yng(1,1,1);\tiny\yng(1)) & = & \frac{(N-1)}{(k+N+2)},
\nonu \\
%l^{+} (\tiny\yng(3);\tiny\yng(1,1,1)) & = &  ,\qquad
l^{+} (l^{+}+1)(\tiny\yng(1,1,1);\tiny\yng(1))  & = & 0,  
\nonu \\
%l^{-} (\tiny\yng(3);\tiny\yng(1,1,1)) & = & , \qquad
l^{-} (l^{-}+1)  (\tiny\yng(1,1,1);\tiny\yng(1)) & = & 0, 
\nonu \\
\hat{u}( \tiny\yng(1,1,1);\tiny\yng(1) ) & = & 1-N.
\nonu
\eea

The explicit form for the
matrix in this particular representation
is given by (\ref{spin2antisymmetricthree}). 
Then, one obtains
$\frac{2}{(k+5)}$ in the third (and the last) diagonal elements.
One can apply for other $N$ values  
and it will turn out that the numerator of the above quantity  does 
depend on $N$ linearly as well as the constant term
while the denominator
is generalized to $(k+N+2)$.
One can also use the formula with the correct $\hat{u}$ charge
\bea
\frac{3 (N+2-3) (N+2+1)}{2 (N+2) (k+N+2)}
-\frac{(\frac{N}{2}-\frac{1}{2 N})}{(k+N+2)}
-\frac{(1-N)^2}{N (N+2) (k+N+2)}
=\frac{(N-1)}{(k+N+2)}
\nonu
\eea
where the quadratic Casimir 
$C^{(N+2)}(\tiny\yng(1,1,1))$ and $C^{(N)}(\tiny\yng(1))$ are used.
The quantum number for the
$l^{+}$ can be obtained by 
the above matrix calculation in (\ref{su2antisymmetricthree}).
From the  eigenvalue $0$
appearing in the third (and the last) block diagonal matrix
in (\ref{su2antisymmetricthree}), one can see the above
$l^{+}$ quantum number $l^{+} =0$, a singlet under the $SU(2)_k$.
This also can be seen from the previous expression
$ ({\tiny\yng(1)},{\bf 1})_{1-N}$.
The trivial $l^-=0$ quantum number holds in this representation.
For the last eigenvalue corresponding to $\hat{u}$ charge,
the previous matrix calculation given in (\ref{u1chargeantisymmthree})
can be used.
The eigenvalues  
appearing in the third (and the last) block diagonal matrix
in (\ref{u1chargeantisymmthree}) imply that the $\hat{u}$ charge is given by
$-2$.
Varying the $N$ values,
one finds that the $\hat{u}$ charge is linear in $N$
as well as the constant term.
This is also consistent with the representation 
$ ({\tiny\yng(1)},{\bf 1})_{1-N}$ where the subscript denotes
the $\hat{u}$ charge.
One observes that the above conformal dimension does not satisfy
the BPS bound.

\subsection{ The mixed representations $\La_+$ with three boxes}
%%%%%%%%%%%%%%%%%%%
%%%%%%%%%%%%%%%%%%%%%%%%%%%%%%%%%%%%%%%%%%%%%%%%%%%

%\bea
%\tiny\yng(2,1)
%\nonu
%\eea
Let us
describe the appropriate projection operators
which pick up the mixed parts of the
third-rank tensor representation of $SU(N+2)$.
The symmetric projection operator $S_{12}$ acting on the first
two indices of the general rank $3$ tensor
$X_{abc}$
is described as
\bea
S_{12} X_{abc}  & = & \frac{1}{2} ( X_{abc} +X_{bac}).
\label{twoterms}
\eea
Then let us introduce the antisymmetric projection
operator $A_{23}$ acting on the last two indices
of the above quantity (\ref{twoterms}) 
as follows:
\bea
A_{23} S_{12} X_{abc} & = &
\frac{1}{4} (X_{abc} + X_{bac}-X_{acb}-X_{bca}).
\label{fourterms}
\eea
Furthermore, 
one projects (\ref{fourterms}) by using the
symmetric projection operator $S_{12}$.
Then one obtains, with appropriate an overall normalization, 
\bea
\frac{4}{3} S_{12} A_{23} S_{12} X_{abc} & = & \frac{1}{6}
(2 X_{abc}+2X_{bac}-X_{bca}-X_{cba}
-X_{acb}-X_{cab}).
\label{sixterms}
\eea
Note that 
the quantity (\ref{sixterms}) is symmetric under the
interchange of the first two indices.
This representation comes from the product of the representation
$\tiny\yng(2)$ and the representation $\tiny\yng(1)$.

Similarly, one can try to do other mixed projection operator.
First of all, by acting the antisymmetric projection operator
on the rank $3$ tensor with the first two indices
\bea
A_{12} X_{abc}  & = & \frac{1}{2} ( X_{abc} -X_{bac}).
\label{twotermsother}
\eea
Then one can act  the symmetric projection operator
on (\ref{twotermsother})
with the last two indices
\bea
S_{23} A_{12} X_{abc} & = &
\frac{1}{4} (X_{abc} - X_{bac}+X_{acb}-X_{bca}).
\label{fourtermsother}
\eea
Finally, one acts
the antisymmetric projection operator on
(\ref{fourtermsother}) with the first two indices
as follows:
\bea
\frac{4}{3} A_{12} S_{23} A_{12} X_{abc} & = & \frac{1}{6}
(2 X_{abc}-2X_{bac}-X_{bca}+X_{cba}
+X_{acb}-X_{cab}).
\label{sixtermsother}
\eea
Note that 
the quantity (\ref{sixtermsother}) is antisymmetric under the
interchange of the first two indices.
This representation comes from the product of the representation
$\tiny\yng(1,1)$ and the representation $\tiny\yng(1)$.

One can easily see that the sum of the two projection operators,
(\ref{sixterms}) and (\ref{sixtermsother}), leads to
\bea
\frac{1}{6} (4X_{abc}-2 X_{cab}-2X_{bca}).
\label{twomixedsum}
\eea
It is known that the totally symmetric and antisymmetric
projection operators acting on the $X_{abc}$ are given by
\bea
S_{123}  X_{abc} & = &
\frac{1}{6} (X_{abc}+X_{acb}+X_{cab}+ X_{bac}+X_{bca}+X_{cba}),
\nonu \\
A_{123} X_{abc} & = &
\frac{1}{6} (X_{abc}-X_{acb}+X_{cab}- X_{bac}+X_{bca}-X_{cba}).
\label{symmantisymm}
\eea
By adding these two operators in (\ref{symmantisymm}), one obtains
\bea
\frac{1}{6} (2X_{abc}+2 X_{cab}+2X_{bca}).
\label{sumsymmantisymm}
\eea
As we expect,
one obtains $X_{abc}$ by adding (\ref{sumsymmantisymm}) to
(\ref{twomixedsum}).

The three-index mixed parts of the $SU(N+2)$ representation
corresponding to (\ref{sixterms})
can be obtained from the generators of the fundamental representation
of $SU(N+2)$ by using the projection operator $\frac{1}{6} (2 \de_{il}
\de_{jm} \de_{kn} +2 \de_{jl} \de_{im} \de_{kn} -\de_{jl}
\de_{km} \de_{in} -\de_{kl} \de_{jm} \de_{in}-\de_{il}
\de_{km} \de_{jn} - \de_{kl} \de_{im} \de_{jn} )$ where
$i \leq j \leq k$ and $ l \leq m \leq n$
and $i, j, k, l,m,n = 1, 2, \cdots, (N+2) $ \cite{Cvitanovicbook}.
Then by acting on
the space
$T_a \otimes {\bf 1}_{(N+2) \times (N+2)} \otimes {\bf 1}_{(N+2) \times (N+2)}
+ {\bf 1}_{(N+2) \times (N+2)} \otimes T_a \otimes {\bf 1}_{(N+2) \times (N+2)} +
{\bf 1}_{(N+2) \times (N+2)} \otimes {\bf 1}_{(N+2) \times (N+2)} \otimes T_a$,
one has
the generators for the mixed representation for the $SU(N+2)$
as follows:
{\small
  \bea
&& 2(T_a)_{il} \, \delta_{jm} \, \delta_{kn} +
2 (T_a)_{jl} \, \delta_{im} \, \delta_{kn} -
(T_a)_{jl} \, \delta_{km} \, \delta_{in} -
(T_a)_{kl} \, \delta_{jm} \, \delta_{in} -
(T_a)_{il} \, \delta_{km} \, \delta_{jn} -
(T_a)_{kl} \, \delta_{im} \, \delta_{jn}  \nonu\\
&& +
2 \delta_{il} \, (T_a)_{jm} \, \delta_{kn}+
2 \delta_{jl} \, (T_a)_{im} \, \delta_{kn}-
\delta_{jl} \, (T_a)_{km} \, \delta_{in}-
\delta_{kl} \, (T_a)_{jm} \, \delta_{in}-
\delta_{il} \, (T_a)_{km} \, \delta_{jn}-
\delta_{kl} \, (T_a)_{im} \, \delta_{jn}
\nonu \\
&& + 2
\delta_{il} \, \delta_{jm} \, (T_a)_{kn}
+ 2
\delta_{jl} \, \delta_{im} \, (T_a)_{kn}
-
\delta_{jl} \, \delta_{km} \, (T_a)_{in}
-
\delta_{kl} \, \delta_{jm} \, (T_a)_{in}
-
\delta_{il} \, \delta_{km} \, (T_a)_{jn}
-
\delta_{kl} \, \delta_{im} \, (T_a)_{jn}. 
\nonu
\eea}
See also Appendix $D$.
Let us calculate the zero mode for the reduced stress energy tensor
spin-$2$ current acting on the state $|(\tiny\yng(2,1) ;0) >$. 
It turns out that the $\frac{(N+2)(N+3)(N+1)}{3}
\times \frac{(N+2)(N+3)(N+1)}{3} =40 \times 40$ matrix (see also
\cite{Eichmann})
is given by
{\small
  \bea
&& \frac{1}{(5+k)} \mbox{diag}
(3,3,\frac{13}{4},\frac{13}{4},3,3,5,5,3,\frac{13}{4},
\frac{13}{4},3,5,5,\frac{13}{4},\frac{13}{4},5,5,\frac{21}{4},\frac{21}{4},
\label{spin2mixedthree}
 \\
&& 3,3,\frac{13}{4},\frac{21}{4},\frac{13}{4},\frac{21}{4},\frac{13}{4},
\frac{21}{4},\frac{13}{4},\frac{21}{4},
5,5,5,5,
%\frac{19}{4},\frac{15}{4},\frac{19}{4},\frac{15}{4},
\left(
\begin{array}{cccc}
 \frac{19}{4 } & -\frac{\sqrt{3}}{2 } & 0 & 0 \\
 -\frac{\sqrt{3}}{2 } & \frac{15}{4 } & 0 & 0 \\
 0 & 0 & \frac{19}{4 } & -\frac{\sqrt{3}}{2 } \\
 0 & 0 & -\frac{\sqrt{3}}{2 } & \frac{15}{4 } \\
\end{array}
\right),
5,5). 
\nonu
\eea}
There are four block diagonal elements after diagonalizing
the $35,36,37,38$ matrix elements (the eigenvalues are
given by $\frac{21}{4 (k+5)}$, $\frac{21}{4 (k+5)}$,
$\frac{13}{4 (k+5)}$ and $\frac{13}{4 (k+5)}$ and the eigenfunctions are
\bea
(0,0,-\sqrt{3},1), \qquad (-\sqrt{3},1,0,0),
\qquad (0,0,\frac{1}{\sqrt{3}},1), \qquad (\frac{1}{\sqrt{3}},1,0,0)).
\nonu
\eea
The $19$, $20$ elements  of (\ref{spin2mixedthree}),
$\frac{21}{4(k+5)}$ and $\frac{21}{4(k+5)}$, correspond to
the eigenvalue on the  state $|(\tiny\yng(2,1) ;0) >$ corresponding to
$({\bf 1},{\bf 2})_{-\frac{3N}{2}}$.
See also the $\hat{u}_{19}$ and $\hat{u}_{20}$ in Appendix $D$.
We will describe the detailed quantum numbers
for the other eigenvalues soon.

One can also calculate the zero mode for the sum of the square for
the spin-$1$ current with minus sign acting on the above
state $|(\tiny\yng(2,1) ;0) >$
and the explicit result is given by 
\bea
&& \mbox{diag} (0,0,\frac{3}{4},\frac{3}{4},0,0,2,2,0,\frac{3}{4},\frac{3}{4},0,2,2,\frac{3}{4},\frac{3}{4},2,2,\frac{3}{4},\frac{3}{4},
\nonu \\
&&  0,0,
\frac{3}{4},\frac{3}{4},\frac{3}{4},\frac{3}{4},\frac{3}{4},\frac{3}{4},\frac{3}{4},\frac{3}{4},
%\frac{1}{2},\frac{3}{2},\frac{1}{2},\frac{3}{2},
\left(
\begin{array}{cccc}
 \frac{1}{2} & \frac{\sqrt{3}}{2} & 0 & 0 \\
 \frac{\sqrt{3}}{2} & \frac{3}{2} & 0 & 0 \\
 0 & 0 & \frac{1}{2} & \frac{\sqrt{3}}{2} \\
 0 & 0 & \frac{\sqrt{3}}{2} & \frac{3}{2} \\
\end{array}
\right),
\frac{3}{4},\frac{3}{4},\frac{3}{4},\frac{3}{4},2,0).
\label{su2mixedthree}
\eea
The eigenvalues for the
$31$, $32$, $33$ and $34$ elements 
are given by $2$, $2$, $0$ and $0$.
The corresponding eigenfunctions are
$(0,0,\frac{1}{\sqrt{3}},1)$, $(\frac{1}{\sqrt{3}},1,0,0)$
$(0,0,-\sqrt{3},1)$, and $(-\sqrt{3},1,0,0)$
 as before.
 The $19$, $20$ elements  of (\ref{su2mixedthree}), $\frac{3}{4}$
 and $\frac{3}{4}$, correspond to
the eigenvalue on the  state $|(\tiny\yng(2,1) ;0) >$.
which behaves as a doublet under the
$SU(2)$. 

Similarly, one can also compute the
zero mode for the spin-$1$ current
acting on the 
state $|(\tiny\yng(2,1) ;0) >$  corresponding to
$({\bf 1},{\bf 2})_{-\frac{3N}{2}}$
and one obtains (twice of $\hat{u}$)
\bea
&& \mbox{diag}  (6,6,1,1,6,6,-4,-4,6,1,1,6,-4,-4,1,1,-4,-4,-9,-9,
\nonu \\ & & 6,6,
1,1,1,1,1,1,1,1,
-4,-4,-4,-4,1,1,1,1,-4,-4).
\label{u1chargemixedthree}
\eea
In this case, also the $19$, $20$ elements are the eigenvalues
for the state  $|(\tiny\yng(2,1) ;0) >$.
Then one can generalize the above eigenvalues
for general $N$ as follows:
\bea
-9 & \rightarrow & 2 (-\frac{3N}{2}):  ({\bf 1},{\bf 2})_{-\frac{3N}{2}},
\qquad -4 \rightarrow 2 (1-N):
({\tiny\yng(1)},{\bf 3})_{1-N} +({\tiny\yng(1)},{\bf 1})_{1-N},
\nonu \\
1  & \rightarrow &
2(2-\frac{N}{2}):  ({\tiny\yng(1,1)},{\bf 2})_{2-\frac{N}{2}} +
       ({\tiny\yng(2)},{\bf 2})_{2-\frac{N}{2}},
\qquad 6 \rightarrow 2 \times 3: ({\tiny\yng(2,1)},{\bf 1})_3,
\nonu
\eea
by carefully analyzing (\ref{transtranstrans})
\footnote{
 The row and columns are characterized by
the following triple index notations
\bea
&& 112, \,  113, \, 114, \, 115, \, 122,  \, 133, \,  144,
\, 155, \, 223, \, 224, \,
 225, \,  233, \, 244, \, 255, \, 334,  \, 335, \,  344,
\, 355, \, 445, \, 455,
\nonu \\
&&
 123, \,  132, \, 124, \, 142, \, 125,  \, 152, \,  134,
 \, 143, \, 135, \, 153, \,
  145, \,  154, \, 245, \, 254, \, 234,  \, 243, \,  235,
\, 253, \, 345, \, 354.
 \nonu
\eea
See also Appendix $D$ for $\hat{u}_1, \cdots, \hat{u}_{40}$.
For example, the eight
$({\tiny\yng(2,1)},{\bf 1})_3$
comes from $112, 113, 122, 133, 223,
233, 123$ and $132$ which do not contain the indices $4$ or $5$.
Similarly, the twelve $({\tiny\yng(1)},{\bf 3})_{1-N} +({\tiny\yng(1)},{\bf 1})_{1-N}$
comes from  $144,155,244,255,344,355,145,154,245,254,345$ and $354$
which contain a single $SU(3)$ fundamentals ($1$ or $2$ or $3$).
The two $ ({\bf 1},{\bf 2})_{-\frac{3N}{2}}$ comes from $445$ and $455$
which do not contain the indices $1$ or $2$ or $3$.
Finally, the eighteen $ ({\tiny\yng(1,1)},{\bf 2})_{2-\frac{N}{2}} +
({\tiny\yng(2)},{\bf 2})_{2-\frac{N}{2}}$ comes from
the remaining ones which contain two $SU(3)$ indices. }.

%%%%%%%%%%%%%%%%%%%%%%%%%%%
\subsubsection{The $(\rm{mixed};\rm{mixed})$ representation
with three boxes}
%%%%%%%%%%%%%%%%%%%%%%%%%%%%

Let us consider the higher representation 
where the mixed representation in $SU(N)$ $ ({\tiny\yng(2,1)},
{\bf 1})_3$
survives
in the branching of (\ref{transtranstrans}).
The four eigenvalues are given by
\bea
h (\tiny\yng(2,1);\tiny\yng(2,1)) & = & \frac{3}{(k+N+2)},
\nonu \\
%l^{+} (\tiny\yng(3);\tiny\yng(1,1,1)) & = &  ,\qquad
l^{+} (l^{+}+1)(\tiny\yng(2,1);\tiny\yng(2,1))  & = & 0,  
\nonu \\
%l^{-} (\tiny\yng(3);\tiny\yng(1,1,1)) & = & , \qquad
l^{-} (l^{-}+1)  (\tiny\yng(2,1);\tiny\yng(2,1)) & = & 0, 
\nonu \\
\hat{u}( \tiny\yng(2,1);\tiny\yng(2,1) ) & = & 3.
\nonu
\eea
One can calculate the conformal dimension for this
representation using the previous formula or
one performs the explicit form for the
matrix in this particular representation
as in (\ref{spin2symmetricthree}). 
Therefore, one obtains
$\frac{3}{(k+5)}$ in the first diagonal elements in (\ref{spin2mixedthree}).
One can apply for other $N$ values where $N=5,7,9, 11, 13, \cdots$. 
It will turn out that the numerator of the above quantity  does not
depend on $N$ and takes the common value and the denominator
is generalized to $(k+N+2)$.
By realizing that the quadratic Casimir of $SU(N+2)$ for the
mixed representation
$C^{(N+2)}(\tiny\yng(2,1))=\frac{3((N+2)^2-3)}{2(N+2)}$
(and the one for the mixed representation
$C^{(N)}(\tiny\yng(2,1))=\frac{3(N^2-3)}{2N}$)
and the correct $\hat{u}$ charge is given by $3$,
the following relation can be obtained
\bea
\frac{3 (N^2+4 N+1)}{2 (N+2)(k+N+2)}-
\frac{3 (N^2-3)}{2 N (k+N+2)}-
\frac{9}{N (N+2) (k+N+2)} =\frac{3}{(k+N+2)}.
\nonu
\eea

Similarly, the quantum number for the
$l^{+}$ can be determined by 
the above matrix calculation in (\ref{su2mixedthree}).
From the zero eigenvalues appearing in the first diagonal matrix elements
in (\ref{su2mixedthree}), one can see the above
$l^{+}$ quantum number, a singlet under the $SU(2)_k$.
This also can be seen from the previous expression
$ ({\tiny\yng(2,1)},{\bf 1})_3$.
The trivial $l^-$ quantum number $l^-=0$ arises.
For the last eigenvalue corresponding to $\hat{u}$ charge,
one uses the previous matrix calculation given in
(\ref{u1chargemixedthree}).
The eigenvalues  
appearing in the first diagonal matrix elements
in (\ref{u1chargemixedthree})
imply that the $\hat{u}$ charge is given by $3$.
This is also consistent with the representation 
$ ({\tiny\yng(2,1)},{\bf 1})_3$ where the subscript denotes
the $\hat{u}$ charge.
One observes that the above conformal dimension does not satisfy
the vanishing BPS bound with $l^{\pm} =0$.

%%%%%%%%%%%%%%%%%%%%%%%%%%%
\subsubsection{The $(\rm{mixed};\rm{antisymm})$ representation
with three and two boxes}
%%%%%%%%%%%%%%%%%%%%%%%%%%%%

Let us consider the
 higher representation 
 where the antisymmetric representation in
 $SU(N)$ $ ({\tiny\yng(1,1)},{\bf 2})_{2-\frac{N}{2}}$
survives
in the branching of (\ref{transtranstrans}).
The four eigenvalues can be summarized by
\bea
h (\tiny\yng(2,1);\tiny\yng(1,1)) & = & \frac{(2 N+15)}{4 (k+N+2)},
\nonu \\
%l^{+} (\tiny\yng(3);\tiny\yng(1,1,1)) & = &  ,\qquad
l^{+} (l^{+}+1)(\tiny\yng(2,1);\tiny\yng(1,1))  & = & \frac{3}{4},  
\nonu \\
%l^{-} (\tiny\yng(3);\tiny\yng(1,1,1)) & = & , \qquad
l^{-} (l^{-}+1)  (\tiny\yng(2,1);\tiny\yng(1,1)) & = & 0, 
\nonu \\
\hat{u}( \tiny\yng(2,1);\tiny\yng(1,1) ) & = & 2-\frac{N}{2}.
\nonu
\eea

The explicit form for the
matrix in this particular representation
is given by (\ref{spin2mixedthree}). 
Then, one obtains
$\frac{21}{4(k+5)}$ in the second line elements.
One can apply for other $N$ values  
and it will turn out that the numerator of the above quantity  does 
depend on $N$ linearly as well as the constant term
while the denominator
is generalized to $4(k+N+2)$.
One can also use the formula with the correct $\hat{u}$ charge
\bea
\frac{3 (N^2+4 N+1)}{2 (N+2)(k+N+2)}-
\frac{2 (N-2) (N+1)}{2 N(k+N+2)}
-\frac{(2-\frac{N}{2})^2}{N (N+2) (k+N+2)}
=\frac{(2 N+15)}{4 (k+N+2)}
\nonu
\eea
where the quadratic Casimir 
$C^{(N+2)}(\tiny\yng(2,1))$ and $C^{(N)}(\tiny\yng(1,1))$ are used.
The quantum number for the
$l^{+}$ can be obtained by 
the above matrix calculation in (\ref{su2mixedthree}).
From the  eigenvalues $\frac{3}{4}$
appearing in the $24$, $26$, $28$, $30$  matrix elements
(and some linear combination between $35$ and $36$ matrix elements
and the linear combination between $37$ and $38$ matrix elements)
in (\ref{su2mixedthree}), one can see the above
$l^{+}$ quantum number $l^{+} =\frac{1}{2}$, a doublet under the $SU(2)_k$.
This also can be seen from the previous expression
$ ({\tiny\yng(1,1)},{\bf 2})_{2-\frac{N}{2}}$.
The trivial $l^-=0$ quantum number holds in this representation.
For the last eigenvalue corresponding to $\hat{u}$ charge,
the previous matrix calculation given in (\ref{u1chargemixedthree})
can be used.
The eigenvalues  
appearing in the $24$, $26$, $28$, $30$ matrix elements
(and other two from the linear combinations in the $35$, $36$, $37$
and $38$ matrix elements)
in (\ref{u1chargemixedthree}) imply that the $\hat{u}$ charge is given by
$\frac{1}{2}$.
Varying the $N$ values,
one finds that the $\hat{u}$ charge is linear in $N$
as well as the constant term.
This is also consistent with the representation 
$ ({\tiny\yng(1,1)},{\bf 2})_{2-\frac{N}{2}}$ where the subscript denotes
the $\hat{u}$ charge.
One observes that the above conformal dimension does not satisfy
the BPS bound.

%%%%%%%%%%%%%%%%%%%%%%%%%%%
\subsubsection{The $(\rm{mixed};\rm{symm})$ representation
with three and two boxes}
%%%%%%%%%%%%%%%%%%%%%%%%%%%%

Let us consider the
 higher representation 
 where the symmetric representation in
 $SU(N)$ $ ({\tiny\yng(2)},{\bf 2})_{2-\frac{N}{2}}$
survives
in the branching of (\ref{transtranstrans}).
The four eigenvalues can be summarized by
\bea
h (\tiny\yng(2,1);\tiny\yng(2)) & = & \frac{(2 N+7)}{4 (k+N+2)},
\nonu \\
%l^{+} (\tiny\yng(3);\tiny\yng(1,1,1)) & = &  ,\qquad
l^{+} (l^{+}+1)(\tiny\yng(2,1);\tiny\yng(2))  & = & \frac{3}{4},  
\nonu \\
%l^{-} (\tiny\yng(3);\tiny\yng(1,1,1)) & = & , \qquad
l^{-} (l^{-}+1)  (\tiny\yng(2,1);\tiny\yng(2)) & = & 0, 
\nonu \\
\hat{u}( \tiny\yng(2,1);\tiny\yng(2) ) & = & 2-\frac{N}{2}.
\nonu
\eea

The explicit form for the
matrix in this particular representation
is given by (\ref{spin2mixedthree}). 
Then, one obtains
$\frac{13}{4(k+5)}$ in the $3$, $4$, $10$, $11$, $15$, $16$,
$23$, $25$, $27$, $29$ (and two from $35$, $36$, $37$ and $38$)
elements.
One can apply for other $N$ values  
and it will turn out that the numerator of the above quantity  does 
depend on $N$ linearly as well as the constant term
while the denominator
is generalized to $4(k+N+2)$.
One can also use the formula with the correct $\hat{u}$ charge
\bea
\frac{3 (N^2+4 N+1)}{2 (N+2)(k+N+2)}-
\frac{2 (N-1) (N+2)}{2 N (k+N+2)}-
\frac{(2-\frac{N}{2})^2}{N (N+2) (k+N+2)}
=\frac{(2 N+7)}{4 (k+N+2)},
\nonu
\eea
where the quadratic Casimir 
$C^{(N+2)}(\tiny\yng(2,1))$ and $C^{(N)}(\tiny\yng(2))$ are used.
The quantum number for the
$l^{+}$ can be obtained by 
the above matrix calculation in (\ref{su2mixedthree}).
From the  eigenvalues $\frac{3}{4}$
appearing in the above  elements 
in (\ref{su2mixedthree}), one can see the above
$l^{+}$ quantum number $l^{+} =\frac{1}{2}$, a doublet under the $SU(2)_k$.
This also can be seen from the previous expression
$ ({\tiny\yng(2)},{\bf 2})_{2-\frac{N}{2}}$.
The trivial $l^-=0$ quantum number holds in this representation.
For the last eigenvalue corresponding to $\hat{u}$ charge,
the previous matrix calculation given in (\ref{u1chargemixedthree})
can be used.
The eigenvalues  
appearing in the previous matrix elements
in (\ref{u1chargemixedthree}) imply that the $\hat{u}$ charge is given by
$\frac{1}{2}$.
Varying the $N$ values,
one finds that the $\hat{u}$ charge is linear in $N$
as well as the constant term.
This is also consistent with the representation 
$ ({\tiny\yng(2)},{\bf 2})_{2-\frac{N}{2}}$ where the subscript denotes
the $\hat{u}$ charge.
One observes that the above conformal dimension does not satisfy
the BPS bound.

%%%%%%%%%%%%%%%%%%%%%%%%%%%
\subsubsection{The $({\rm mixed};f)$ representation
with three boxes}
%%%%%%%%%%%%%%%%%%%%%%%%%%%%

Let us consider the higher representation 
where the fundamental representation in $SU(N)$
$ ({\tiny\yng(1)},{\bf 1})_{1-N}$ or $ ({\tiny\yng(1)},{\bf 3})_{1-N}$
survives
in the branching of (\ref{transtranstrans}).
The four eigenvalues are given by
\bea
h (\tiny\yng(2,1);\tiny\yng(1)) & = & \frac{(N+2)}{(k+N+2)},
\nonu \\
%l^{+} (\tiny\yng(3);\tiny\yng(1,1,1)) & = &  ,\qquad
l^{+} (l^{+}+1)(\tiny\yng(2,1);\tiny\yng(1))  & = & 2, \qquad \mbox{or}
\qquad 0,  
\nonu \\
%l^{-} (\tiny\yng(3);\tiny\yng(1,1,1)) & = & , \qquad
l^{-} (l^{-}+1)  (\tiny\yng(2,1);\tiny\yng(1)) & = & 0, 
\nonu \\
\hat{u}( \tiny\yng(2,1);\tiny\yng(1) ) & = & 1-N.
\nonu
\eea

The explicit form for the
matrix in this particular representation
is given by (\ref{spin2mixedthree}). 
Then, one obtains
$\frac{5}{(k+5)}$ in the  $7$, $8$, $13$, $14$,
$17$, $18$, $31$, $32$, $33$, $34$, $39$ and $40$  elements.
One can apply for other $N$ values  
and it will  turn out that the numerator of the above quantity  does 
depend on $N$ linearly as well as the constant term
while the denominator
is generalized to $(k+N+2)$.
One can also use the formula with the correct $\hat{u}$ charge
\bea
\frac{3 (N^2+4 N+1)}{2 (N+2)(k+N+2)}-
\frac{(\frac{N}{2}-\frac{1}{2 N})}{(k+N+2)}
-\frac{(1-N)^2}{N (N+2) (k+N+2)}=
\frac{(N+2)}{(k+N+2)},
\nonu
\eea
where the quadratic Casimir 
$C^{(N+2)}(\tiny\yng(2,1))$ and $C^{(N)}(\tiny\yng(1))$ are used.
The quantum number for the
$l^{+}$ can be obtained by 
the above matrix calculation in (\ref{su2mixedthree}).
From the  eigenvalue $2$
appearing in the  $7$, $8$, $13$, $14$,
$17$, $18$, two linear combinations (between $31$, $32$, $33$, and $34$),
and $39$ elements 
in (\ref{su2mixedthree}), one can see the above
$l^{+}$ quantum number $l^{+} =1$, a triplet under the $SU(2)_k$.
This  also can be seen from the previous expression
$ ({\tiny\yng(1)},{\bf 3})_{1-N}$.
Furthermore,
from the eigenvalue $0$
appearing in the  two linear combinations (between $31$, $32$, $33$,
and $34$) and $40$ elements
in (\ref{su2mixedthree}), one can see the above
$l^{+}$ quantum number $l^{+} =0$, a singlet under the $SU(2)_k$.
This also can be seen from the previous expression
$ ({\tiny\yng(1)},{\bf 1})_{1-N}$.
The trivial $l^-=0$ quantum number holds in this representation.
For the last eigenvalue corresponding to $\hat{u}$ charge,
the previous matrix calculation given in (\ref{u1chargemixedthree})
can be used.
The eigenvalues  
appearing in the previous elements
in (\ref{u1chargemixedthree}) imply that the $\hat{u}$ charge is given by
$-2$.
Varying the $N$ values,
one finds that the $\hat{u}$ charge is linear in $N$
as well as the constant term.
This is also consistent with the representation 
$ ({\tiny\yng(1)},{\bf 1})_{1-N}$ or $ ({\tiny\yng(1)},{\bf 3})_{1-N}$
where the subscript denotes
the $\hat{u}$ charge.
One observes that the above conformal dimension does satisfy
the BPS bound with $l^+=1$ and $l^-=0$. There is no BPS bound
for $l^{\pm} =0$.

%%%%%%%%%%%%%%%%%%%%%%%%%%%
\subsubsection{The $(\rm{mixed};0)$ representation
  with three boxes
\label{mixedzero}}
%%%%%%%%%%%%%%%%%%%%%%%%%%%%

Let us consider the higher representation 
where the singlet representation in $SU(N)$
$ ({\bf 1},{\bf 2})_{-\frac{3N}{2}}$
survives
in the branching of (\ref{transtranstrans}).
The four eigenvalues are given by
\bea
h (\tiny\yng(2,1);0) & = & \frac{3 (2 N+1)}{4 (k+N+2)},
\nonu \\
%l^{+} (\tiny\yng(3);\tiny\yng(1,1,1)) & = &  ,\qquad
l^{+} (l^{+}+1)(\tiny\yng(2,1);0)  & = & \frac{3}{4},  
\nonu \\
%l^{-} (\tiny\yng(3);\tiny\yng(1,1,1)) & = & , \qquad
l^{-} (l^{-}+1)  (\tiny\yng(2,1);0) & = & 0, 
\nonu \\
\hat{u}( \tiny\yng(2,1);0 ) & = & -\frac{3N}{2}.
\label{eigenvaluesthreemixed}
\eea

The explicit form for the
matrix in this particular representation
is given by (\ref{spin2mixedthree}). 
Then, one obtains
$\frac{21}{4(k+5)}$ in the $19$ and $20$ elements.
One can apply for other $N$ values  
and it will turn out that the numerator of the above quantity  does 
depend on $N$ linearly as well as the constant term
while the denominator
is generalized to $4(k+N+2)$.
One can also use the formula with the correct $\hat{u}$ charge
\bea
\frac{3 (N^2+4 N+1)}{2 (N+2)(k+N+2)}-
\frac{(-\frac{3 N}{2})^2}{N (N+2) (k+N+2)}=
\frac{3 (2 N+1)}{4 (k+N+2)},
\nonu
\eea
where the quadratic Casimir 
$C^{(N+2)}(\tiny\yng(2,1))$ is used.
The quantum number for the
$l^{+}$ can be obtained by 
the above matrix calculation in (\ref{su2mixedthree}).
From the  eigenvalues $\frac{3}{4}$
appearing in the $19$, $20$  elements
in (\ref{su2mixedthree}), one can see the above
$l^{+}$ quantum number $l^{+} =\frac{1}{2}$, a doublet under the $SU(2)_k$.
This also can be seen from the previous expression
$ ({\bf 1},{\bf 2})_{-\frac{3N}{2}}$.
The trivial $l^-=0$ quantum number holds in this representation.
For the last eigenvalue corresponding to $\hat{u}$ charge,
the previous matrix calculation given in (\ref{u1chargemixedthree})
can be used.
The eigenvalues  
appearing in the $19$, $20$ elements
in (\ref{u1chargemixedthree}) imply that the $\hat{u}$ charge is given by
$-\frac{9}{2}$.
Varying the $N$ values,
one finds that the $\hat{u}$ charge is linear in $N$.
This is also consistent with the representation 
$ ({\bf 1},{\bf 2})_{-\frac{3N}{2}}$ where the subscript denotes
the $\hat{u}$ charge.
One observes that the above conformal dimension does not satisfy
the BPS bound.

%%%%%%%%%%%%%%%%%%%%%%%%%%%
\subsubsection{The $({\rm mixed};\overline{f})$ representation
with three boxes}
%%%%%%%%%%%%%%%%%%%%%%%%%%%%

Let us consider the higher representation
which arises from the product of $(\tiny\yng(2,1);0)$
and $(0;\overline{\tiny\yng(1)})$.
The former occurs in the subsection 
\ref{mixedzero} and the latter occurs in the subsection
\ref{0frep} together with the complex conjugation.

In this case, the corresponding
four eigenvalues are described by
\bea
h (\tiny\yng(2,1);\overline{\tiny\yng(1)}) & = & \frac{(k+3 N)}{2 (k+N+2)},
\nonu \\
%l^{+} (\tiny\yng(3);\tiny\yng(1,1,1)) & = &  ,\qquad
l^{+} (l^{+}+1)(\tiny\yng(2,1);\overline{\tiny\yng(1)})  & = & \frac{3}{4},  
\nonu \\
%l^{-} (\tiny\yng(3);\tiny\yng(1,1,1)) & = & , \qquad
l^{-} (l^{-}+1)  (\tiny\yng(2,1);\overline{\tiny\yng(1)}) & = & \frac{3}{4}, 
\nonu \\
\hat{u}( \tiny\yng(2,1);\overline{\tiny\yng(1)} ) & = & -2N-1.
\nonu
\eea

First of all, one can obtain the following $40 \times 40$ matrix
by calculating the commutator
$[T_0, Q_{-\frac{1}{2}}^{\bar{A}}]$ as in the subsection
\ref{ffbarrep}
\bea
&& \frac{1}{(5+k)} \mbox{diag} (1,1,\frac{1}{6},\frac{1}{6},1,
1,-\frac{2}{3},-\frac{2}{3},1,\frac{1}{6},
\frac{1}{6},1,-\frac{2}{3},-\frac{2}{3},\frac{1}{6},\frac{1}{6},
-\frac{2}{3},-\frac{2}{3};-\frac{3}{2},-\frac{3}{2};
\nonu \\
&& 1,1,\frac{1}{6},\frac{1}{6},\frac{1}{6},\frac{1}{6},\frac{1}{6},\frac{1}{6},\frac{1}{6},\frac{1}{6},-\frac{2}{3},-\frac{2}{3},-\frac{2}{3},-\frac{2}{3},\frac{1}{6},\frac{1}{6},\frac{1}{6},\frac{1}{6},-\frac{2}{3},-\frac{2}{3}).
\label{EExtramatrix}
\eea
The eigenvalues (the $N$ generalization is
straightforward to obtain) appearing in the $19$, $20$ elements
in (\ref{EExtramatrix})
provide the extra contribution as well as the sum of conformal dimensions
of  $(\tiny\yng(2,1);0)$
and $(0;\overline{\tiny\yng(1)})$.
They are given in (\ref{eigenvaluesthreemixed}) and (\ref{0ffoureigen})
respectively.
Then one obtains the final conformal dimension by adding the above
contribution appearing in (\ref{EExtramatrix}) as follows
\bea
\frac{(2 k+3)}{4 (k+N+2)}+\frac{3 (2 N+1)}{4 (k+N+2)}
-\frac{3}{2 (k+N+2)} = \frac{(k+3 N)}{2 (k+N+2)},
\nonu
\eea
as in (\ref{eigenforsymmf}).

It is also useful to interpret the above result
from the conformal dimension formula.
One determines the following result
\bea
\frac{3 (N^2+4 N+1)}{2 (N+2)(k+N+2)}
-\frac{(\frac{N}{2}-\frac{1}{2 N})}{(k+N+2)}
-\frac{(-\frac{1}{2} (N+2)-\frac{3 N}{2})^2}{N (N+2) (k+N+2)}
+\frac{1}{2} =\frac{(k+3 N)}{2 (k+N+2)}.
\nonu
\eea
Here we used the quadratic Casimirs for $C^{(N+2)}(\tiny\yng(2,1))$
and $C^{(N)}(\overline{\tiny\yng(1)})$. The correct $\hat{u}$
charge is inserted. The excitation number is given by $\frac{1}{2}$.

For the $l^+$ quantum number, due to the vanishing of $l^+$ in
$(0;\overline{\tiny\yng(1)})$, it turns out that
the $l^+$ is the same as the one($l^+=\frac{1}{2}$) in $(\tiny\yng(2,1);0)$. 
For the $l^-$ quantum number, due to the vanishing of $l^-$ in
 $(\tiny\yng(2,1);0)$, it turns out that
the $l^-$ is the same as the one($l^-=\frac{1}{2}$) in
$(0;\overline{\tiny\yng(1)})$. 
It is easy to see that
the above conformal dimension does not satisfy the BPS bound by substituting
$l^{\pm}=\frac{1}{2}$.
One can add each $\hat{u}$ charge and it is obvious that
the total $\hat{u}$ charge is given by $-\frac{3N}{2} -\frac{(N+2)}{2}$
which leads to the above result. Note that
the $\hat{u}$ charge for $(0;\overline{\tiny\yng(1)})$ is opposite to
the one for $(0;\tiny\yng(1))$.

%%%%%%%%%%%%%%%%%%%%%%%%%%%
\subsubsection{The $({\rm mixed};\overline{\rm symm})$ representation
with three and two boxes}
%%%%%%%%%%%%%%%%%%%%%%%%%%%%

Let us consider the higher representation
which arises from the product of $(\tiny\yng(2,1);0)$
and $(0;\overline{\tiny\yng(2)})$.
The former occurs in the subsection 
\ref{mixedzero} while the latter occurs in the subsection
\ref{0symmtwo} with complex conjugation.

The four eigenvalues are given by 
\bea
h (\tiny\yng(2,1);\overline{\tiny\yng(2)}) & = & \frac{(4 k+6 N-9)}{4 (k+N+2)},
\nonu \\
%l^{+} (\tiny\yng(3);\tiny\yng(1,1,1)) & = &  ,\qquad
l^{+} (l^{+}+1)(\tiny\yng(2,1);\overline{\tiny\yng(2)})  & = & \frac{3}{4},  
\nonu \\
%l^{-} (\tiny\yng(3);\tiny\yng(1,1,1)) & = & , \qquad
l^{-} (l^{-}+1)  (\tiny\yng(2,1);\overline{\tiny\yng(2)}) & = & 0, 
\nonu \\
\hat{u}( \tiny\yng(2,1);\overline{\tiny\yng(2)} ) & = &  -\frac{5N}{2}-2.
\nonu
\eea

One should calculate the commutator
$[T_0,  Q_{-\frac{1}{2}}^{1} Q_{-\frac{1}{2}}^{4}]$
and it turns out that
\bea
&& \frac{1}{(5+k)} \mbox{diag} (
2,2,\frac{1}{3},\frac{1}{3},2,2,-\frac{4}{3},-\frac{4}{3},2,\frac{1}{3},
\frac{1}{3},2,-\frac{4}{3},-\frac{4}{3},\frac{1}{3},\frac{1}{3},
-\frac{4}{3},-\frac{4}{3};-3,-3;
\nonu \\
&& 2,2,\frac{1}{3},\frac{1}{3},\frac{1}{3},\frac{1}{3},\frac{1}{3},\frac{1}{3},\frac{1}{3},\frac{1}{3},-\frac{4}{3},-\frac{4}{3},-\frac{4}{3},-\frac{4}{3},\frac{1}{3},\frac{1}{3},\frac{1}{3},\frac{1}{3},-\frac{4}{3},-\frac{4}{3}).
\label{EExtramatrix2}
\eea
The two eigenvalues (the $N$ generalization is
simply $ -\frac{3}{(N+k+2)}$) appearing in the $19$, $20$ elements
 in (\ref{EExtramatrix2})
give the extra contribution as well as the sum of conformal dimensions
of  $(\tiny\yng(2,1);0)$
and $(0;\overline{\tiny\yng(2)})$.
They are given in (\ref{eigenvaluesthreemixed}) and
(\ref{eigenvaluessymmtwo})
respectively.
Then one obtains the final conformal dimension by adding the above
contribution appearing in (\ref{EExtramatrix2}) as follows
\bea
\frac{3 (2 N+1)}{4 (k+N+2)}+\frac{k}{(k+N+2)}
-\frac{3}{(k+N+2)}= \frac{(4 k+6 N-9)}{4 (k+N+2)}.
\nonu
\eea
The conformal dimension formula implies that
{\small
  \bea
\frac{3 (N^2+4 N+1)}{2 (N+2)(k+N+2)}
-\frac{2 (N-1) (N+2)}{(2 N) (k+N+2)}
-\frac{(-(N+2)-\frac{3 N}{2})^2}{N (N+2) (k+N+2)}
+1 = \frac{(4 k+6 N-9)}{4 (k+N+2)}.
\nonu
\eea}
The quadratic Casimir
$C^{(N)}(\overline{\tiny\yng(2)})$ is the same as
$C^{(N)}(\tiny\yng(2))$ and the excitation number
is equal to $1$. The correct $\hat{u}$ charge is inserted.
For the $l^+$ quantum number, due to the vanishing of $l^+$ in
$(0;\overline{\tiny\yng(2)})$, 
the $l^+$ is the same as the one($l^+=\frac{1}{2}$) in $(\tiny\yng(2,1);0)$. 
For the $l^-$ quantum number, due to the vanishing of $l^-$ in
 $(\tiny\yng(2,1);0)$ and $(0;\overline{\tiny\yng(2)})$, 
the total $l^-$ is given by $l^-=0$, a singlet. 
The total $\hat{u}$ charge is given by $-\frac{3N}{2}-(N+2)$
which leads to the above result. Again,
the $\hat{u}$ charge for $(0;\overline{\tiny\yng(2)})$ is opposite to
the one for $(0;\tiny\yng(2))$.
One can easily see that the above conformal dimension
does not lead to the BPS bound with $l^{+} =\frac{1}{2}$ and $l^-=0$.

%%%%%%%%%%%%%%%%%%%%%%%%%%%
\subsubsection{The $({\rm mixed};\overline{\rm antisymm})$ representation
with three and two boxes}
%%%%%%%%%%%%%%%%%%%%%%%%%%%%

Let us consider the higher representation
which arises from the product of $(\tiny\yng(2,1);0)$
and $(0;\overline{\tiny\yng(1,1)})$.
The former occurs in the subsection 
\ref{mixedzero} while the latter occurs in the subsection
\ref{0antisymm} with complex conjugation.

The four eigenvalues are given by
\bea
h (\tiny\yng(2,1);\overline{\tiny\yng(1,1)}) & = & \frac{(4 k+6 N-1)}{4 (k+N+2)},
\nonu \\
%l^{+} (\tiny\yng(3);\tiny\yng(1,1,1)) & = &  ,\qquad
l^{+} (l^{+}+1)(\tiny\yng(2,1);\overline{\tiny\yng(1,1)})  & = & \frac{3}{4},  
\nonu \\
%l^{-} (\tiny\yng(3);\tiny\yng(1,1,1)) & = & , \qquad
l^{-} (l^{-}+1)  (\tiny\yng(2,1);\overline{\tiny\yng(1,1)}) & = & 2, 
\nonu \\
\hat{u}( \tiny\yng(2,1);\overline{\tiny\yng(1,1)} ) & = & -\frac{5N}{2}-2.
\nonu
\eea

One should calculate the commutator
$[T_0,  Q_{-\frac{1}{2}}^{1} Q_{-\frac{1}{2}}^{2}]$
and it turns out that
\bea
&& \frac{1}{(5+k)} \mbox{diag} (
2,2,\frac{1}{3},\frac{1}{3},2,2,-\frac{4}{3},-\frac{4}{3},2,\frac{1}{3},
\frac{1}{3},2,-\frac{4}{3},-\frac{4}{3},\frac{1}{3},\frac{1}{3},
-\frac{4}{3},-\frac{4}{3};-3,-3;
\nonu \\
&& 2,2,\frac{1}{3},\frac{1}{3},\frac{1}{3},\frac{1}{3},\frac{1}{3},\frac{1}{3},\frac{1}{3},\frac{1}{3},-\frac{4}{3},-\frac{4}{3},-\frac{4}{3},-\frac{4}{3},\frac{1}{3},\frac{1}{3},\frac{1}{3},\frac{1}{3},-\frac{4}{3},-\frac{4}{3}).
\label{EExtramatrix3}
\eea
The two eigenvalues (the $N$ generalization is
simply $ -\frac{3}{(N+k+2)}$) appearing in the $19$, $20$ elements
 in (\ref{EExtramatrix3})
give the extra contribution as well as the sum of conformal dimensions
of  $(\tiny\yng(2,1);0)$
and $(0;\overline{\tiny\yng(1,1)})$.
They are given in (\ref{eigenvaluesthreemixed}) and
(\ref{eigenvaluestwoantisymm})
respectively.
Then one obtains the final conformal dimension by adding the above
contribution appearing in (\ref{EExtramatrix3}) as follows
\bea
\frac{3 (2 N+1)}{4 (k+N+2)}+\frac{(k+2)}{(k+N+2)}-\frac{3}{(k+N+2)}=
\frac{(4 k+6 N-1)}{4 (k+N+2)}.
\nonu
\eea

The conformal dimension formula implies that
{\small
  \bea
\frac{3 (N^2+4 N+1)}{2 (N+2)(k+N+2)}
-\frac{2 (N-2) (N+1)}{(2 N) (k+N+2)}
-\frac{(-(N+2)-\frac{3 N}{2})^2}{N (N+2) (k+N+2)}
+ 1 =\frac{(4 k+6 N-1)}{4 (k+N+2)}.
\nonu
\eea}
The quadratic Casimir
$C^{(N)}(\overline{\tiny\yng(1,1)})$ is the same as
$C^{(N)}(\tiny\yng(1,1))$ and the excitation number
is equal to $1$. The correct $\hat{u}$ charge is inserted.
For the $l^+$ quantum number, due to the vanishing of $l^+$ in
$(0;\overline{\tiny\yng(1,1)})$, 
the $l^+$ is the same as the one($l^+=\frac{1}{2}$) in $(\tiny\yng(2,1);0)$. 
For the $l^-$ quantum number, due to the vanishing of $l^-$ in
 $(\tiny\yng(2,1);0)$, 
the total $l^-$ is given by $l^-=1$ in
$(\overline{\tiny\yng(1,1)})$, a triplet. 
The total $\hat{u}$ charge is given by $-\frac{3N}{2}-(N+2)$
which leads to the above result. Again,
the $\hat{u}$ charge for $(0;\overline{\tiny\yng(1,1)})$ is opposite to
the one for $(0;\tiny\yng(1,1))$.
One can easily see that the above conformal dimension
does not lead to the BPS bound with $l^{+} =\frac{1}{2}$ and $l^-=1$.

%%%%%%%%%%%%%%%%%%%%%%%%%%%
\subsubsection{The $({\rm mixed};\overline{\rm antisymm})$ representation
with three boxes}
%%%%%%%%%%%%%%%%%%%%%%%%%%%%

Let us consider the higher representation
which arises from the product of $(\tiny\yng(2,1);0)$
and $(0;\overline{\tiny\yng(1,1,1)})$.
The former occurs in the subsection 
\ref{mixedzero} while the latter occurs in the subsection
\ref{0antisymmthree} with complex conjugation.

The four eigenvalues are given by
\bea
h (\tiny\yng(2,1);\overline{\tiny\yng(1,1,1)}) & = & \frac{3 (k+N)}{2 (k+N+2)},
\nonu \\
%l^{+} (\tiny\yng(3);\tiny\yng(1,1,1)) & = &  ,\qquad
l^{+} (l^{+}+1)(\tiny\yng(2,1);\overline{\tiny\yng(1,1,1)})  & = & \frac{3}{4},  
\nonu \\
%l^{-} (\tiny\yng(3);\tiny\yng(1,1,1)) & = & , \qquad
l^{-} (l^{-}+1)  (\tiny\yng(2,1);\overline{\tiny\yng(1,1,1)}) & = & \frac{15}{4}, 
\nonu \\
\hat{u}( \tiny\yng(2,1);\overline{\tiny\yng(1,1,1)} ) & = & -3N-3.
\nonu
\eea

One should calculate the commutator
$[T_0,  Q_{-\frac{1}{2}}^{1} Q_{-\frac{1}{2}}^{2}  Q_{-\frac{1}{2}}^{3}]$
and it turns out that
\bea
&& \frac{1}{(5+k)} \mbox{diag} (
3,3,\frac{1}{2},\frac{1}{2},3,3,-2,-2,3,\frac{1}{2},
\frac{1}{2},3,-2,-2,\frac{1}{2},\frac{1}{2},-2,-2;-\frac{9}{2},-\frac{9}{2};
\nonu \\
&& 3,3,\frac{1}{2},\frac{1}{2},\frac{1}{2},\frac{1}{2},\frac{1}{2},\frac{1}{2},\frac{1}{2},\frac{1}{2},-2,-2,-2,-2,\frac{1}{2},\frac{1}{2},\frac{1}{2},\frac{1}{2},-2,-2).
\label{EEExtramatrix3}
\eea
The two eigenvalues (the $N$ generalization is
simply $ -\frac{9}{2(N+k+2)}$) appearing in the $19$, $20$ elements
 in (\ref{EEExtramatrix3})
give the extra contribution as well as the sum of conformal dimensions
of  $(\tiny\yng(2,1);0)$
and $(0;\overline{\tiny\yng(1,1,1)})$.
They are given in (\ref{eigenvaluesthreemixed}) and
(\ref{eigenvaluesthreeantisymm})
respectively.
Then one obtains the final conformal dimension by adding the above
contribution appearing in (\ref{EEExtramatrix3}) as follows
\bea
\frac{3 (2 N+1)}{4 (k+N+2)} +\frac{3 (2 k+5)}{4 (k+N+2)}
-\frac{9}{2 (k+N+2)} = \frac{3 (k+N)}{2 (k+N+2)}.
\nonu
\eea

The conformal dimension formula implies that
\bea
\frac{3 (N^2+4 N+1)}{2 (N+2)(k+N+2)}
-\frac{3 (N-3) (\frac{1}{N}+1)}{2 (k+N+2)}
-\frac{(-\frac{1}{2} 3 (N+2)-\frac{3 N}{2})^2}{N (N+2) (k+N+2)}
+ \frac{3}{2} = \frac{3 (k+N)}{2 (k+N+2)}.
\nonu
\eea
The quadratic Casimir
$C^{(N)}(\overline{\tiny\yng(1,1,1)})$ is the same as
$C^{(N)}(\tiny\yng(1,1,1))$ and the excitation number
is equal to $\frac{3}{2}$. The correct $\hat{u}$ charge is inserted.
For the $l^+$ quantum number, due to the vanishing of $l^+$ in
$(0;\overline{\tiny\yng(1,1,1)})$, 
the $l^+$ is the same as the one($l^+=\frac{1}{2}$) in $(\tiny\yng(2,1);0)$. 
For the $l^-$ quantum number, due to the vanishing of $l^-$ in
 $(\tiny\yng(2,1);0)$, 
the total $l^-$ is given by $l^-=\frac{3}{2}$ in
$(\overline{\tiny\yng(1,1,1)})$, a quartet. 
The total $\hat{u}$ charge is given by $-\frac{3N}{2}-\frac{3}{2}(N+2)$
which leads to the above result. Again,
the $\hat{u}$ charge for $(0;\overline{\tiny\yng(1,1,1)})$ is opposite to
the one for $(0;\tiny\yng(1,1,1))$.
One can easily see that the above conformal dimension
does not lead to the BPS bound with $l^{+} =\frac{1}{2}$ and
$l^-=\frac{3}{2}$
\footnote{For the higher representation
  $(\tiny\yng(2,1);\tiny\yng(1,1,1))$,
  the similar analysis can be done by calculating the
  $\hat{u}$ charge correctly.}.

%%%%%%%%%%%%%%%%%%%%%%%%%%%
\subsubsection{The $({\rm mixed};\overline{\rm mixed})$ representation
with three boxes}
%%%%%%%%%%%%%%%%%%%%%%%%%%%%

Let us consider the higher representation
which arises from the product of $(\tiny\yng(2,1);0)$
and $(0;\overline{\tiny\yng(2,1)})$.
The former occurs in the subsection 
\ref{mixedzero} while the latter occurs in the subsection
\ref{0mixed}.

The four eigenvalues are given by
\bea
h (\tiny\yng(2,1);\overline{\tiny\yng(2,1)}) & = & \frac{3 (k+N-2)}{2 (k+N+2)},
\nonu \\
%l^{+} (\tiny\yng(3);\tiny\yng(1,1,1)) & = &  ,\qquad
l^{+} (l^{+}+1)(\tiny\yng(2,1);\overline{\tiny\yng(2,1)})  & = & \frac{3}{4},  
\nonu \\
%l^{-} (\tiny\yng(3);\tiny\yng(1,1,1)) & = & , \qquad
l^{-} (l^{-}+1)  (\tiny\yng(2,1);\overline{\tiny\yng(2,1)}) & = & \frac{3}{4}, 
\nonu \\
\hat{u}( \tiny\yng(2,1);\overline{\tiny\yng(2,1)} ) & = & -3N-3.
\nonu
\eea

One should calculate the commutator
$[T_0,  Q_{-\frac{1}{2}}^{2} Q_{-\frac{1}{2}}^{4}  Q_{-\frac{1}{2}}^{1} ]$
and it turns out that
\bea
&& \frac{1}{(5+k)} \mbox{diag} (
3,3,\frac{1}{2},\frac{1}{2},3,3,-2,-2,3,\frac{1}{2},
\frac{1}{2},3,-2,-2,\frac{1}{2},\frac{1}{2},-2,-2;-\frac{9}{2},-\frac{9}{2};
\nonu \\
&& 3,3,\frac{1}{2},\frac{1}{2},\frac{1}{2},\frac{1}{2},\frac{1}{2},\frac{1}{2},\frac{1}{2},\frac{1}{2},-2,-2,-2,-2,\frac{1}{2},\frac{1}{2},\frac{1}{2},\frac{1}{2},-2,-2).
\label{EExtramatrix4}
\eea
The two eigenvalues (the $N$ generalization is
simply $ -\frac{9}{2(N+k+2)}$) appearing in the $19$, $20$ 
matrix elements in (\ref{EExtramatrix4})
give the extra contribution as well as the sum of conformal dimensions
of  $(\tiny\yng(2,1);0)$
and $(0;\overline{\tiny\yng(2,1)})$.
They are given in (\ref{eigenvaluesthreemixed}) and
(\ref{eigenvaluesmixed})
respectively.
Then one obtains the final conformal dimension by adding the above
contribution appearing in (\ref{EExtramatrix4}) as follows
\bea
\frac{3 (2 N+1)}{4 (k+N+2)} +
\frac{3 (2 k+1)}{4 (k+N+2)}
-\frac{9}{2 (k+N+2)}=
\frac{3 (k+N-2)}{2 (k+N+2)}.
\nonu
\eea

The conformal dimension formula implies that
\bea
\frac{3 (N^2+4 N+1)}{2 (N+2)(k+N+2)}-
\frac{3 (N^2-3)}{2 N}
-\frac{(-\frac{1}{2} 3 (N+2)-\frac{3 N}{2})^2}{N (N+2) (k+N+2)}+
\frac{3}{2} =
\frac{3 (k+N-2)}{2 (k+N+2)}.
\nonu
\eea
The quadratic Casimir
$C^{(N)}(\overline{\tiny\yng(2,1)})$ is used and the excitation number
is equal to $\frac{3}{2}$.
The correct $\hat{u}$ charge is inserted.
For the $l^+$ quantum number, due to the vanishing of $l^+$ in
$(0;\overline{\tiny\yng(2,1)})$, 
the $l^+$ is the same as the one($l^+=\frac{1}{2}$) in $(\tiny\yng(2,1);0)$. 
For the $l^-$ quantum number, due to the vanishing of $l^-$ in
 $(\tiny\yng(2,1);0)$, 
the total $l^-$ is given by $l^-=\frac{1}{2}$ in
$(\overline{\tiny\yng(2,1)})$, a doublet. 
The total $\hat{u}$ charge is given by $-\frac{3N}{2}-\frac{3}{2}(N+2)$
which leads to the above result.
%Again,
%the $\hat{u}$ charge for $(0;\overline{\tiny\yng(1,1)})$ is opposite to
%the one for $(0;\tiny\yng(1,1))$.
One can easily see that the above conformal dimension
does not lead to the BPS bound with $l^{\pm} =\frac{1}{2}$.

%%%%%%%%%%%%%%%%%%%%%%%%%%%
\subsubsection{Summary of this section}
%%%%%%%%%%%%%%%%%%%%%%%%%%%%

Let us describe the conformal dimension for the higher representation
arising 
from the product of $(\mbox{symm};0)$ and $(0;\overline{\mbox{antisymm}})$
where the number of box for the $\mbox{symm}$ representation
is given by $p$
and the number of box for the $\mbox{antisymm}$ representation is given by
$q$.
Then by substituting the quadratic Casimirs into the formula
and the excitation number is given by $\frac{q}{2}$, one obtains
the following expression
\bea
&& \frac{(N+1) p (N+p+2)}{2 (N+2) (k+N+2)} -\frac{(N+1) q (N-q)}{2 N (k+N+2)}
-\frac{(\frac{1}{2} (N+2) (-q)-\frac{N p}{2})^2}{N (N+2) (k+N+2)}
+\frac{q}{2}
\nonu \\
&& =\frac{2 k q+2 N p+p^2-2 p q+2 p+q^2+2 q}{4 (k+N+2)}.
\label{result}
\eea
According to the conditions
\bea
 p= 2l^+, \qquad q = 2l^-,
\nonu
\eea
the above expression (\ref{result}) reduces to the
BPS bound in (\ref{BPS}).

Now one can classify the possible combinations as follows:
\bea
&& 
p=0, q= 1,2,3; \qquad (0; \overline{\tiny\yng(1)}), \qquad
(0; \overline{\tiny\yng(1,1)}), \qquad
(0; \overline{\tiny\yng(1,1,1)}),
\nonu \\
&&
p=1, q= 0,1,2,3; \qquad
(\tiny\yng(1); 0), \qquad
(\tiny\yng(1); \overline{\tiny\yng(1)}), \qquad
(\tiny\yng(1); \overline{\tiny\yng(1,1)}), \qquad
(\tiny\yng(1); \overline{\tiny\yng(1,1,1)}),
\nonu \\
&&
p=2, q= 0,1,2,3; \qquad
(\tiny\yng(2); 0), \qquad
(\tiny\yng(2); \overline{\tiny\yng(1)}), \qquad
(\tiny\yng(2); \overline{\tiny\yng(1,1)}), \qquad
(\tiny\yng(2); \overline{\tiny\yng(1,1,1)}), \nonu \\
&&
p=3, q= 0,1,2,3; \qquad
(\tiny\yng(3); 0), \qquad
(\tiny\yng(3); \overline{\tiny\yng(1)}), \qquad
(\tiny\yng(3); \overline{\tiny\yng(1,1)}), \qquad
(\tiny\yng(3); \overline{\tiny\yng(1,1,1)}). 
\label{variousboxes}
\eea
Furthermore, there are also complex conjugated representations
for (\ref{variousboxes}).
 The quadratic Casimirs do not change and the $\hat{u}^2$
does not change.

Let us describe the conformal dimension for the higher representation
where the representation $\La_-$ appears in the branching rule of
$\La_+$.
The representation arises from $(\mbox{antisymm};\mbox{antisymm})$
where the number of box for the first $\mbox{antisymm}$ representation
is given by $p$
and the number of box for the second $\mbox{antisymm}$
representation is given by
$q$.
Then the formula implies 
\bea
&& \frac{(N+2+1) p (N-p+2)}{2 (N+2) (k+N+2)}-
\frac{(N+1) q (N-q)}{2 N (k+N+2)} -\frac{(q-\frac{N}{2})^2}{N (N+2) (k+N+2)}
\nonu \\
&& = \frac{(2 N^2 p-2 N^2 q-2 N p^2+10 N p+2 N q^2-6 N q-N-6 p^2+12 p+6 q^2)}
{4 (N+2) (k+N+2)}.
\label{compresult}
\eea
Under the further condition
\bea
q=p-1,
\nonu
\eea
the above result
(\ref{compresult}) reduces to
\bea
\frac{(2 N+3)}{4 (k+N+2)},
\nonu
\eea
which is equal to the BPS bound with $l^+=\frac{1}{2}$ and $l^-=0$:
\bea
p=1,2,3; \qquad (\tiny\yng(1),0), \qquad
(\tiny\yng(1,1); \tiny\yng(1)), \qquad
(\tiny\yng(1,1,1); \tiny\yng(1,1)).
\nonu
\eea
In this case also, the complex conjugated representations
are possible. The quadratic Casimirs do not change and the $\hat{u}^2$
does not change.

In summary of this section, 
the conformal dimensions for the higher representations
up to three boxes are described explicitly.
In next two Tables, its large $(N,k)$ 't Hooft like limit
is written and the particular ones with ``BPS'' notation are specified. 
The large $(N,k)$ 't Hooft-like limit
is defined by
\bea
N,k \rightarrow \infty, \qquad \la \equiv \frac{(N+1)}{(N+k+2)}
\qquad \mbox{fixed},
\label{largenk}
\eea
which will be used in Tables.

%%%%%%%%%%%%%%%%%%%
%\subsection{
%The higher  representations
%$(0;\yng(3))$
%}
%%%%%%%%%%%%%%%%%%%

%\bea
%\overline{\tiny\yng(3)}
%\nonu
%\eea

%%%%%%%%%%%%%%%%%%%
%\subsection{ The higher representation}
%%%%%%%%%%%%%%%%%%%

%\bea
%\overline{
%  \tiny
%\yng(2,1)}
%\nonu
%\eea

%%%%%%%%%%%%%%%%%%%
%\subsection{ The higher representation}
%%%%%%%%%%%%%%%%%%%

%\bea
%\overline{
%  \tiny
%\yng(1,1,1)}
%\nonu
%\eea

%%%%%%%%%%%%%%%%%%%
%\subsection{ The higher representation}
%%%%%%%%%%%%%%%%%%%

%\bea
%\tiny
%\yng(2,1,1,1)
%\nonu
%\eea

%%%%%%%%%%%%%%%%%%%
%\subsection{ The higher representation}
%%%%%%%%%%%%%%%%%%%

%\bea
%\overline{
%\tiny
%  \yng(2,1,1,1)}
%\nonu
%\eea

{\small
%%%%%%%%%%%%%%%%%%%%%%%%%%%%%%%%%%%%%%%%%%%%%%%%%%%%%%%%%%%%%%%%%%%
\begin{table}[ht]
\centering % used for centering table
\begin{tabular}{|c||c|c|c|c|c|c|c| } % centered columns (4 columns)
\hline %inserts double horizontal lines
$(\Lambda_+; \Lambda_-)$ & 0 &$\tiny\yng(1)$ &$\overline{\tiny\yng(1)}$ &
$ \tiny\yng(2)$ &$\tiny\yng(1,1)$ & $\overline{ \tiny\yng(2)}$
& $\overline{\tiny\yng(1,1)}$
%& $ \tiny\yng(3)$
%& $ \tiny\yng(2,1)$
%& $ \tiny\yng(1,1,1)$
%& $ \overline{\tiny\yng(3)}$
%& $ \overline{\tiny\yng(2,1)}$
%& $ \overline{\tiny\yng(1,1,1)}$ 
\\ [0.5ex] % inserts table
%heading
\hline \hline % inserts single horizontal line
$0$  & 0 & $(\frac{1-\lambda }{2})_{bps}$  & ${\bf (\frac{1-\la}{2})_{bps}}$ & $1-\la$ &  $(1-\la)_{bps}$ & ${\bf 1-\la}$ &  ${\bf (1-\la)_{bps}}$
%&&
%$\frac{3(1-\la)}{2}$ &
%$\frac{3(1-\la)}{2}$ & &${\bf \frac{3(1-\la)}{2}}$ &$
%{\bf \frac{3(1-\la)}{2}}$
\\ % inserting body of the table
\hline
$\tiny\yng(1)$ & $(\frac{\la}{2})_{bps}$ & $\frac{\la}{N}$ & $(\frac{1}{2})_{bps}$ &
$ {\bf \frac{2-\la}{2}}$& $ {\bf \frac{2-\la}{2}}$ &  $
{\bf \frac{2-\la}{2}}$ &
$ {\bf (\frac{2-\la}{2})_{bps}}$
%& & ${\bf \frac{3-2\la}{2}}$ & ${\bf \frac{3-2\la}{2}}$&
%& ${\bf \frac{3-2\la}{2}}$&  ${\bf \frac{3-2\la}{2}}$
\\
\hline
$\overline{\tiny\yng(1)}$ & ${\bf (\frac{\la}{2})_{bps}}$ &
${\bf (\frac{1}{2})_{bps}}$ & ${\bf \frac{\la}{N}}$ &${\bf
  \frac{2-\la}{2}}$ & ${\bf (\frac{2-\la}{2})_{bps}}$& ${\bf \frac{2-\la}{2}}$&
${\bf \frac{2-\la}{2}}$
%&&  ${\bf \frac{3-2\la}{2}}$&  ${\bf \frac{3-2\la}{2}}$&
%& ${\bf \frac{3-2\la}{2}}$ &  ${\bf \frac{3-2\la}{2}}$
\\
\hline
$ \tiny\yng(2)$ & $(\la)_{bps}$  &
$\frac{\la}{2}$ & $(\frac{\lambda +1}{2})_{bps} $
& $\frac{2\la}{N}$ & $1$ & $1$ & ${
  \bf (1)_{bps} }$
%&&  ${\bf \frac{3-\la}{2}}$&${\bf \frac{3-\la}{2}}$ & &${\bf \frac{3-\la}{2}}$ &% ${\bf \frac{3-\la}{2}}$
\\
\hline
$\tiny\yng(1,1)$ & $\la$ & $(\frac{\la}{2})_{bps}$ &$\frac{\lambda +1}{2} $ & $1$ & $\frac{2\la}{N}$ &${
  \bf 1}$ & $1$
%& & ${\bf \frac{3-\la}{2}}$& ${\bf \frac{3-\la}{2}}$&&
% ${\bf \frac{3-\la}{2}}$&  ${\bf \frac{3-\la}{2}}$
\\ 
\hline
$\overline{\tiny\yng(2)}$ & ${\bf (\la)_{bps}}$ &
${\bf (\frac{\la+1}{2})_{bps}}$ & ${\bf \frac{\la}{2}}$  & ${
\bf 1}$ & ${\bf (1)_{bps}}$ &
${\bf \frac{2\la}{N}} $& ${
  \bf 1}$
%&&${\bf \frac{3-\la}{2}}$ &${\bf \frac{3-\la}{2}}$ & & ${\bf \frac{3-\la}{2}}$& %${\bf \frac{3-\la}{2}}$
\\ 
\hline
$\overline{\tiny\yng(1,1)}$ &  ${\bf \la}$   & ${\bf \frac{\la+1}{2}}$  & ${\bf (\frac{\la}{2})_{bps}}$ & ${
\bf 1}$ & ${\bf 1}$ & ${\bf 1}$ &
${\bf \frac{2\la}{N}} $
%& & ${\bf \frac{3-\la}{2}}$& ${\bf \frac{3-\la}{2}}$&&
%${\bf \frac{3-\la}{2}}$& ${\bf \frac{3-\la}{2}}$
\\
\hline
$\tiny\yng(3)$ & $(\frac{3 \lambda }{2})_{bps} $ &$\la$ & $
(\frac{2\la +1}{2})_{bps}
$ & $\frac{\la}{2}$ & $\frac{\lambda +2}{2}$ &$\frac{\lambda +2}{2}$
&$(\frac{\lambda +2}{2})_{bps}$
%& $\frac{3 \la}{N}$ & $\frac{3}{2}$ & $\frac{3}{2}$ &&
%${\bf \frac{3}{2}}$ & ${\bf \frac{3}{2}}$
\\
\hline
$\tiny\yng(2,1)$ &$\frac{3 \lambda }{2} $ & $\la_{bps}$ & $\frac{2\la +1}{2}
$ & $\frac{\la}{2}$ & $\frac{\la}{2}$
& $\frac{\lambda +2}{2}$ & $\frac{\lambda +2}{2}$
%&& $\frac{3 \la}{N}$ &${\bf \frac{3}{2}}$ &
%& $\frac{3}{2}$ & $\frac{3}{2}$
\\ 
\hline
$\tiny\yng(1,1,1)$ & & $\la$ & & &$(\frac{\la}{2})_{bps}$ & &
%&&&$\frac{3 \la}{N}$ & & &
\\
\hline
$\overline{\tiny\yng(3)}$ & ${
  \bf (\frac{3\la}{2})_{bps} }$ & ${\bf (\frac{2\la+1}{2})_{bps}}$
& ${\bf \la}$ &
${\bf \frac{\la+2}{2}}$&${\bf (\frac{\la+2}{2})_{bps}}$
& ${\bf \frac{\la}{2}}$ &
${\bf \frac{\la+2}{2}}$
%&
%&${\bf \frac{3}{2}}$ &${\bf \frac{3}{2}}$ & ${
%\bf \frac{3\la}{N}}$ & ${\bf \frac{3}{2}}$& ${\bf \frac{3}{2}}$
\\
\hline
$\overline{\tiny\yng(2,1)}$ & ${
  \bf \frac{3\la}{2} }$  &${\bf \frac{2\la+1}{2}}$ & ${\bf \la_{bps}}$ &
${\bf \frac{\la+2}{2}}$&${\bf \frac{\la+2}{2}}$ & ${\bf \frac{\la}{2}}$ &
${\bf \frac{\la}{2}}$
%& & ${\bf \frac{3}{2}}$& ${\bf \frac{3}{2}}$& &  ${\bf \frac{3\la}{N}}$ &${\bf \%frac{3}{2}}$
\\ 
\hline
$\overline{\tiny\yng(1,1,1)}$ & & & ${
  \bf \la}$ & & & & ${
  \bf (\frac{\la}{2})_{bps} }$
%&&& & & & ${
%\bf \frac{3\la}{N}}$
\\
[1ex] % [1ex] adds vertical space
\hline %inserts single line
\end{tabular}
%\label{tableone} % is used to refer this table in the text
\caption{The eigenvalue $h$ under the large $(N,k)$ 't Hooft-like
  limit (\ref{largenk}).
  All the eigenvalues described in the section $3$ are presented in this
  Table and next one.
  Those in the footnotes in the section $3$ and some eigenvalues
  appearing in next sections are denoted by the
  boldface notation. The subscript ``bps'' stands for
  the conformal dimension satisfying the BPS bound
  (\ref{conformaldimension}) at finite $(N,k)$.
  According to the last one of
  the branching rule (\ref{transtranstrans}) (and its complex conjugated
  one), there is no singlet under the $SU(N)$. Therefore, there
  are no higher representations corresponding to the blanks that can be
  obtained from the product of zero and $(0;\La_{-})$ in this
  Table and next one. Of course, there are higher representations
  where the representation $\La_{-}$ appears in the branching of $\La_{+}=
  \mbox{antisymm}$ or $\overline{\mbox{antisymm}}$ with three boxes
  in these Tables. Note that there is a BPS bound for mixed representation
  described in the subsection $3.5.4$ (and its complex conjugated one).
} % title of Table
\end{table}}

{\small
%%%%%%%%%%%%%%%%%%%%%%%%%%%%%%%%%%%%%%%%%%%%%%%%%%%%%%%%%%%%%%%%%%%
\begin{table}[ht]
\centering % used for centering table
\begin{tabular}{|c||c|c|c|c|c|c| } % centered columns (4 columns)
\hline %inserts double horizontal lines
$(\Lambda_+; \Lambda_-)$  & $ \tiny\yng(3)$ & $ \tiny\yng(2,1)$
& $ \tiny\yng(1,1,1)$ & $ \overline{\tiny\yng(3)}$
& $ \overline{\tiny\yng(2,1)}$ & $ \overline{\tiny\yng(1,1,1)}$ 
\\ [0.5ex] % inserts table
%heading
\hline \hline % inserts single horizontal line
$0$  & &
$\frac{3(1-\la)}{2}$ &
$(\frac{3(1-\la)}{2})_{bps}$ & &${\bf \frac{3(1-\la)}{2}}$ &$
{\bf (\frac{3(1-\la)}{2})_{bps}}$
\\ % inserting body of the table
\hline
$\tiny\yng(1)$ &  & ${\bf \frac{3-2\la}{2}}$ & ${\bf \frac{3-2\la}{2}}$&
& ${\bf \frac{3-2\la}{2}}$&  ${\bf (\frac{3-2\la}{2})_{bps}}$  \\
\hline
$\overline{\tiny\yng(1)}$ &  &
${\bf \frac{3-2\la}{2}}$&  ${\bf (\frac{3-2\la}{2})_{bps}}$&
& ${\bf \frac{3-2\la}{2}}$  &  ${\bf \frac{3-2\la}{2}}$   \\
\hline
$ \tiny\yng(2)$ &  & ${\bf \frac{3-\la}{2}}$&${\bf \frac{3-\la}{2}}$ & &${\bf \frac{3-\la}{2}}$ & ${\bf (\frac{3-\la}{2})_{bps}}$ \\
\hline
$\tiny\yng(1,1)$ & & ${\bf \frac{3-\la}{2}}$& ${\bf \frac{3-\la}{2}}$&&
 ${\bf \frac{3-\la}{2}}$&  ${\bf \frac{3-\la}{2}}$ \\ 
\hline
$\overline{\tiny\yng(2)}$ & &${\bf \frac{3-\la}{2}}$
&${\bf (\frac{3-\la}{2})_{bps}}$ & & ${\bf \frac{3-\la}{2}}$& ${\bf \frac{3-\la}{2}}$  \\ 
\hline
$\overline{\tiny\yng(1,1)}$ &  & ${\bf \frac{3-\la}{2}}$& ${\bf \frac{3-\la}{2}}$&&
${\bf \frac{3-\la}{2}}$& ${\bf \frac{3-\la}{2}}$ \\
\hline
$\tiny\yng(3)$ & $ \frac{3\la}{N}$ &$\frac{3}{2}$ & $\frac{3}{2}$ & &  ${\bf \frac{3}{2}}$
& ${\bf (\frac{3}{2})_{bps}}$ \\
\hline
$\tiny\yng(2,1)$ & & $\frac{3 \la}{N}$ &${\bf \frac{3}{2}}$ &
& $\frac{3}{2}$ & $\frac{3}{2}$ \\ 
\hline
$\tiny\yng(1,1,1)$  &&&$\frac{3 \la}{N}$ & & & \\
\hline
$\overline{\tiny\yng(3)}$  &
&${\bf \frac{3}{2}}$ &${\bf (\frac{3}{2})_{bps}}$ & ${
\bf \frac{3\la}{N}}$ & ${\bf \frac{3}{2}}$& ${\bf \frac{3}{2}}$  \\
\hline
$\overline{\tiny\yng(2,1)}$  & & ${\bf \frac{3}{2}}$& ${\bf \frac{3}{2}}$& &  ${\bf \frac{3\la}{N}}$ &${\bf \frac{3}{2}}$ \\ 
\hline
$\overline{\tiny\yng(1,1,1)}$  && & & & & ${
\bf \frac{3\la}{N}}$ \\
[1ex] % [1ex] adds vertical space
\hline %inserts single line
\end{tabular}
%\label{tableone} % is used to refer this table in the text
\caption{The (continued) eigenvalue $h$   
 under the large $(N,k)$ 't Hooft-like limit (\ref{largenk}).
All the eigenvalues described in the section $3$ are presented in this
  Table and previous one.
  Those in the footnotes in the section $3$
and some eigenvalues
  appearing in next sections
  are denoted by the
  boldface notation. The subscript ``bps'' stands for
  the conformal dimension satisfying the BPS bound
  (\ref{conformaldimension}) at finite $(N,k)$.
  Because there are no $(0; \mbox{symm})$ or
$(0; \overline{\mbox{symm}})$
  representations
  with three boxes  from the description of the subsection $2.2.9$,
there
are no higher representations corresponding to the blanks
that can be obtained from the product of $(\La_{+};0)$ and zero in this
  Table.
Of course, there are higher representations
  where the representation $\La_{-}$ appears in the branching of $\La_{+}=
  \mbox{symm}$ or $\overline{\mbox{symm}}$ with three boxes.
 Note that there is a BPS bound for mixed representation
  described in the subsection $3.5.4$ (and its complex conjugated one).}
% title of Table
\end{table}}
%%%%%%%%%%%%%%%%%%%%%%%%%%%%%%%%%%%%%%%%%%%%%%%%%%%%%%%%%%%%%%%%%%%%%%

%%%%%%%%%%%%%%%%%%%%%%%%%%%%%%%%%%%%%%%%%%%%%%%%%%%%%%%%%%%%%%%%%%%%%
%%%%%%%%%%%%%%%%%%%%%%%%%%%%%%%%%%%%%%%%%%%%%%%%%%%%%%%%%%%%%%%%%%%%%%
%\section{ The higher representations in the $\frac{SU(N+2)}{SU(N) \times SU(%2)
%  \times U(1)}$ Wolf space coset}
%2%%%%%%%%%%%%%%%%%%%%%%%%%%%%%%%%%%%%%%%%%%%%%%%%%%%%%%%%%%%%%%%%%%%%%
%%%%%%%%%%%%%%%%%%%%%%%%%%%%%%%%%%%%%%%%%%%%%%%%%%%%%%%%%%%%%%%%%%%%%

%%%%%%%%%%%%%%%%%%%%%%%%%%%%%%%%%%%%%%%%%%%%%%%%%%%%%%%%
%%%%%%%%%%%%%%%%%%%
\section{ Review of eigenvalues in the minimal representations  with
  the higher spin-$1,2,3$ currents 
  %in the
  %higher representations
  in the $\frac{SU(N+2)}{SU(N) \times SU(2) \times U(1)}$ Wolf space coset}
%section4%%%%%%%%%%%%%%%%%%
%%%%%%%%%%%%%%%%%%%%%%%%%%%%%%%%%%%%%%%%%%%%%%%%%%%%%%%%

Let us describe the eigenvalues of
\bea
&& 1) \,\, \mbox{the zero mode of
the higher spin-$1$ current}: \, \, (\Phi_0^{(1)})_0,
\nonu \\
&& 2) \,\, \mbox{the zero mode of sum of the square of higher spin-$2$ current
}: \,\,  (V^{+})_0,
\nonu \\
&& 3) \,\,
\mbox{the zero mode of sum of the square of  other higher spin-$2$ current
}: \,\,
(V^{-})_0,
%\,\, \mbox{ and}
\nonu \\
&& 4) \,\, \mbox{the zero mode of the higher spin-$3$ current}: \,\,
(\Phi_2^{(1)})_0,
\nonu
\eea
where the higher spin-$2$ currents $V^{\pm i}(z)$
in the $SU(2)_k \times SU(2)_N$
basis are related to the ones  $\Phi_1^{(1), \mu\nu}(z)$ in the $SO(4)$ basis
\bea
V^{\pm 1}(z) & \equiv & i (\Phi_1^{(1),14} \mp \Phi_1^{(1),23})(z),
\nonu \\
V^{\pm 2}(z) & \equiv & -i (\Phi_1^{(1),24} \pm \Phi_1^{(1),13})(z),
\nonu \\
V^{\pm 3}(z) & \equiv & i (-\Phi_1^{(1),34} \pm \Phi_1^{(1),12})(z).
\nonu
\eea
Then one can construct the following two quantities
by summing over each $SU(2)$ adjoint indices
\bea
V^+(z) \equiv \sum_{i=1}^3 (V^{+i})^2(z), \qquad
V^{-}(z) \equiv \sum_{i=1}^3 (V^{-i})^2(z),
\nonu
\eea
which have the conformal dimension (or spin) of $4$.
The corresponding eigenvalues are described by
$\phi_0^{(1)}$, $v^+$, $v^-$ and $\phi_2^{(1)}$ respectively.
See also the relevant works in \cite{GH,GGHR}.

The higher spin $1$ current  
is described as \cite{AK1411} (see also \cite{Ahn1311,Ahn1408,Ahn1504}
for fixed $N$)
\bea
\Phi_0^{(1)} (z) &=&
-\frac{1}{2(k+N+2)} \, d^0_{\bar{a} \bar{b}} \,
f^{\bar{a} \bar{b}}_{\,\,\,\,\,\, c}  V^c  (z)
+ \frac{k}{2(k+N+2)^2} \, d^0_{\bar{a} \bar{b}} \, Q^{\bar{a}} \, Q^{\bar{b}} (z),
\label{finalspinone}
\eea
where the antisymmetric  $d$ tensor of rank $2$ is given by
$4N \times 4N$ matrix as follows:
\bea
d^0_{\bar{a} \bar{b}}  = 
\left(
\begin{array}{cccc}
0 & 0 & -1 & 0 \\
0 & 0 & 0 & -1 \\
1 & 0 & 0 & 0  \\
0 & 1 & 0 & 0 \\
\end{array}
\right).
%\label{dzero}
\nonu
\eea
Each element is $N \times N$ matrix. The locations of the nonzero
elements of this matrix are the same as the previous almost complex
structure $h^3_{\bar{a} \bar{b}}$ but numerical values are different from each
other. Note that the summation over
$c$ index in (\ref{finalspinone}) runs over the whole range of $SU(N+2)$
adjoint indices.

This higher spin $1$ current plays the role of the `generator'
of the next higher spin currents because one can construct them
using the OPEs between the spin $\frac{3}{2}$ currents of the
large ${\cal N}=4$ nonlinear superconformal algebra and  the
higher spin $1$ current.
That is, from the first order pole of the
OPE between $G^{\mu}(z)$ and the higher spin $1$
current $\Phi_0^{(1)} (w)$, one obtains $\Phi_{\frac{1}{2}}^{(1),\mu} (w)$
with minus sign \cite{AKK1703}. After that, one can calculate
the OPE between $G^{\mu}(z)$ and the higher spin $\frac{3}{2}$
current $\Phi_{\frac{1}{2}}^{(1),\nu} (w)$.
Then the first order pole will provide the next higher spin $2$ current
$\Phi_1^{(1),\mu \nu} (w)$. One can go further.
The OPE between  $G^{\mu}(z)$ and the higher spin $2$
current $\Phi_1^{(1),\nu \rho} (w)$ contains the first order pole
where the next higher spin $\frac{5}{2}$ current
$\delta^{\mu \nu} \Phi_{\frac{3}{2}}^{(1),\rho} (w)$ occurs.
Finally, one can calculate the OPE between
 $G^{\mu}(z)$ and the higher spin $\frac{5}{2}$
current $\Phi_{\frac{3}{2}}^{(1),\nu} (w)$ and then
the first order pole gives us to
the last higher spin $3$ current term
$\delta^{\mu \nu} \Phi_2^{(1)} (w)$. Once the normalization
of the higher spin $1$ current is fixed, then the normalization
for the higher spin $3$ current can be fixed in this way.

So far, although the complete closed form for the higher spin currents
in terms of the adjoint spin $1$ and $\frac{1}{2}$ currents
is not known, but their expressions for several $N$ values are known
explicitly.  They (which are
written explicitly for $N=3,5,7,9,11$ and maybe for $N=13$ for some other
cases) are enough to obtain all the results of this paper.

%%%%%%%%%%%%%%%%%%%%%%%%%%%%%%%%%
\subsection{ The eigenvalues in
  the $(0;f)$ and $(0;\overline{f})$ representations
\label{41subsection}}
%%%%%%%%%%%%%%%%%%%%%%%%%%%%%%%%%

The relevant subsection is given by the subsection \ref{0frep}.
The above four eigenvalues associated with
one of the minimal representations
can be summarized by
\bea
\phi_0^{(1)} (0;
  \tiny\yng(1)
) & = & -\frac{k}{(N+k+2)},
\nonu \\
 v^+(0;{ \tiny\yng(1)})  & = & \frac{24 k}{(k+N+2)^2},
\nonu \\
 v^-(0;{ \tiny\yng(1)})  & = & \frac{12 k (5 k+4 N+2)}{(k+N+2)^2},
\nonu \\
\phi_2^{(1)} (0;
  \tiny\yng(1)
) & = &
\frac{4k(12+28N+5N^{2}+14k+39kN {\bf +6 k N^2}+4 k^{2} {\bf +12k^2 N})}
  {3(2+k+N)^2(4+5k+5N+6kN)}.
\label{0feigenhigher}
\eea
For the first eigenvalue,
one should calculate the OPE
between the ``reduced'' $\Phi_0^{(1)}(z)$ and
$Q^{13}(w)$ (in $SU(5)$) and read off
the first order pole. See also (\ref{state0f}) and (\ref{5matrix}).
The coefficient of $Q^{13}(w)$ in the right hand side of
this OPE is the corresponding eigenvalue.
In other words, the zero mode of
$ d^0_{\bar{a} \bar{b}} \, Q^{\bar{a}} \, Q^{\bar{b}}$
in (\ref{finalspinone}) acting on this state gives
$-2(N+k+2)$.
For the second and third eigenvalues,
one calculates  the OPEs
between the ``reduced'' $V^{(\pm)}(z)$ and
$Q^{13}(w)$ (in $SU(5)$) and read off
the fourth order pole respectively.
The coefficients of $Q^{13}(w)$ in the right hand side of
these OPEs are the corresponding eigenvalues respectively.
For the last eigenvalue,
one computes 
 the OPE
between the ``reduced'' $\Phi_2^{(1)}(z)$ and
$Q^{13}(w)$ (in $SU(5)$) and read off
the third order pole.
Of course, all the higher spin currents do not
contain the spin-$1$ currents $V^a(z)$.

By counting the highest powers of $k$ or $N$ (the sum of powers
in $k$ and $N$ for the expressions containing both dependences)
in the numerators and
the denominators appearing in the above eigenvalues,
one can observe the behaviors under the
large $(N,k)$ 't Hooft like limit. 
Except the $v^+$ eigenvalue having $\frac{1}{N}$ dependence,
the remaining three eigenvalues
approach to the finite $\la$  dependent values \footnote{
Note that we make some boldface notation for the highest power of
$(N,k)$ in the numerator of the higher spin $3$ current in
(\ref{0feigenhigher}). We will observe that they will play the role of
the fundamental quantity in the sense that the eigenvalues of the higher
spin $3$ current for
any representation $(0;\La_-)$ will be a multiple of this quantity,
under the large $(N,k)$ 't Hooft like limit.}.

Similarly, the other four eigenvalues can be also obtained
from
\bea
\phi_0^{(1)} (0;
  \overline{\tiny\yng(1)}
) & = & \frac{k}{(N+k+2)},
\nonu \\
v^+(0;
\overline{\tiny\yng(1)}) & = & \frac{24 k}{(k+N+2)^2}, 
\nonu \\
v^-(0;
\overline{\tiny\yng(1)}) & = & \frac{12 k (5 k+4 N+2)}{(k+N+2)^2},
\nonu \\
\phi_2^{(1)} (0;
  \overline{\tiny\yng(1)}
) & = &
-\frac{4k(12+28N+5N^{2}+14k+39kN {\bf +6 k N^2}+4 k^{2} {\bf +12k^2 N})}
  {3(2+k+N)^2(4+5k+5N+6kN)}.
\label{zerofbareigenhigher}
\eea
Because the generators for the complex conjugated (antifundamental)
representation $\overline{\tiny\yng(1)}$
have an extra minus sign compared to the fundamental representation
$\tiny\yng(1)$, the eigenvalue for the odd higher spin currents
(corresponding to the first and the last ones)
have an extra minus sign and the ones for the even higher spin currents
(corresponding to the two middle ones)
remain the same compared to the results of the previous section
in (\ref{0feigenhigher}).

More explicitly, one can obtain the OPEs between the ``reduced''
higher spin currents and the $\frac{1}{2}$ current $Q^1(w)$.
By reading off the corresponding coefficients in the appropriate
poles, the above eigenvalues can be determined.
One can also analyze the large $(N,k)$ 't Hooft like limit for these
eigenvalues.

%%%%%%%%%%%%%%%%%%%%%%%%%%%%
\subsection{ The eigenvalues in the
  $(f;0)$ and $(\overline{f};0)$ representations}
%%%%%%%%%%%%%%%%%%%%%%%%%%%

The relevant subsection is given by the subsection \ref{f0subsectionname}.
The four eigenvalues can be described as
\bea
\phi_0^{(1)} (
  \tiny\yng(1);0
) & = & -\frac{N}{(N+k+2)},
\nonu \\
v^+ ({ \tiny\yng(1)};0)  & = & \frac{12 N (4 k+5 N+2)}{(k+N+2)^2},
\nonu \\
 v^- ({ \tiny\yng(1)};0)  & = & \frac{24 N}{(k+N+2)^2},
\nonu \\
\phi_2^{(1)} (
  \tiny\yng(1);0
) & = & -
\frac{4 N(12+28k+5k^{2}+14N+39kN {\bf +6k^{2}N}+4N{}^{2} {\bf +12kN{}^{2}})}
  {3(2+k+N)^2(4+5k+5N+6kN)}.
 \label{f0eigenhigher}
 \eea
 One obtains
 these eigenvalues
 by substituting the $SU(N+2)$ generators
 $T_{a^{\ast}}$ into the zero mode of the spin $1$ current
 $V_0^a$ in the corresponding ``reduced'' higher spin currents
 where all the $Q^a(z)$ dependent terms are ignored.
 Then one has the unitary matrix acting on the corresponding state and the
 diagonal elements of the last $2\times 2$ subdiagonal matrix
 provide the above eigenvalues.
 From the explicit form for the higher spin $1$ current in
 (\ref{finalspinone}), the corresponding eigenvalue implies that
 the zero mode of $d^0_{\bar{a} \bar{b}} \,
 f^{\bar{a} \bar{b}}_{\,\,\,\,\,\, c}  V^c$ acting on this state is equal to
 $2N$.
 The large $(N,k)$ 't Hooft like limit can be analyzed similarly
 \footnote{ In this case, the highest power of $(N,k)$ in the numerator
   of the higher spin $3$ current, denoted by the boldface
   notation will play the role of
the fundamental quantity in the sense that the eigenvalues of the higher
spin $3$ current for
any representation $(\La_+;0)$ will be a multiple of this quantity,
under the large $(N,k)$ 't Hooft like limit.}.
 
 As observed in \cite{AK1506}, under the
 symmetry $N \leftrightarrow k$ and $ 0 \leftrightarrow \tiny\yng(1)$,
 the eigenvalues become
 $\phi_0^{(1)} (
  \tiny\yng(1);0
) \rightarrow \phi_0^{(1)} (0;
  \tiny\yng(1))$, $v^+ ({ \tiny\yng(1)};0) \rightarrow
  v^- (0;{ \tiny\yng(1)})$, $v^- ({ \tiny\yng(1)};0) \rightarrow
  v^+ (0;{ \tiny\yng(1)})$ and
  $\phi_2^{(1)} (
  \tiny\yng(1);0
) \rightarrow - \phi_2^{(1)} (
  0;\tiny\yng(1)
)$.

When we consider the complex conjugated representation,    
the following results hold
\bea
\phi_0^{(1)} (
 \overline{ \tiny\yng(1)};0
) & = & \frac{N}{(N+k+2)},
\nonu \\
v^+ (\overline{ \tiny\yng(1)};0)  & = & \frac{12 N (4 k+5 N+2)}{(k+N+2)^2},
\nonu \\
 v^- (\overline{ \tiny\yng(1)};0)  & = & \frac{24 N}{(k+N+2)^2},
\nonu \\
\phi_2^{(1)} (
  \overline{\tiny\yng(1)};0
) & = & 
\frac{4 N(12+28k+5k^{2}+14N+39kN {\bf +6k^{2}N}+4N{}^{2} {\bf +12kN{}^{2}})}
  {3(2+k+N)^2(4+5k+5N+6kN)}.
 \label{fbarzeroeigenhigher}
\eea
According to the previous analysis, the
first eigenvalue (corresponding to the higher spin $1$ current)
and the last eigenvalue (corresponding to the higher spin $3$ current)
have the extra minus signs compared to the ones in (\ref{f0eigenhigher}).
  
%%%%%%%%%%%%%%%%%%%%%%%%%%%%%%%%%%%%%%%%%%%%%%%%%%%%%%%%
%%%%%%%%%%%%%%%%%%%
\section{ Eigenvalues for the higher representations  with
  the higher spin-$1,2,3$ currents
  %in the
  %higher representations
  in the $\frac{SU(N+2)}{SU(N) \times SU(2) \times U(1)}$ Wolf space coset}
%section5%%%%%%%%%%%%%%%%%%
%%%%%%%%%%%%%%%%%%%%%%%%%%%%%%%%%%%%%%%%%%%%%%%%%%%%%%%%

In this section, there are $22$ subsections where
we consider the explicit $22$ higher representations
and the same number of subsections with coincident higher representation
will appear in section $8$ for the linear case.

%%%%%%%%%%%%%%%%%%%%%%%%%%%%%%%%%%%%%%%%%%%%%
\subsection{The $(f;f)$ representation}
%%%%%%%%%%%%%%%%%%%%%%%%%%%%%%%%%%%%%%%%%%%%%

The relevant subsection on this higher representation
is given by $2.2.1$.
In this case,
when one takes the $N \times N $ subdiagonal unitary matrix
inside of $(N+2) \times (N+2)$ unitary matrix,
the corresponding diagonal elements for the higher spin currents
provide the following four
eigenvalues 
\bea
\phi_0^{(1)} (\tiny\yng(1);\tiny\yng(1)) & = & \frac{2}{(k+N+2)},
\nonu \\
v^+ (\tiny\yng(1);\tiny\yng(1)) & = & \frac{96 k}{(k+N+2)^2},
\nonu \\
v^- (\tiny\yng(1);\tiny\yng(1)) & = & \frac{96 N}{(k+N+2)^2},
\nonu \\
\phi_2^{(1)} (\tiny\yng(1);\tiny\yng(1)) & = &
\frac{8 (k-N) ({\bf 6 k N}+5 k+5 N+16)}{3 (k+N+2)^2 (6 k N+5 k+5 N+4)}.
\label{ffeigenvalues1}
\eea
Under the symmetry $ N \leftrightarrow k$ (with $\tiny\yng(1)
\leftrightarrow \tiny\yng(1)$),
the first eigenvalue in (\ref{ffeigenvalues1}) remains the same,
the second eigenvalue becomes the third one,
the third eigenvalue becomes the second one and
the last eigevalue remains the same with an extra sign change.
By power counting of $N$ and $k$, one sees that the above eigenvalues
behave as $\frac{1}{N}$ dependence under the large $(N,k)$ 't Hooft like
limit. However, this will play the role of next leading order and moreover
this term will be the fundamental quantity 
in the sense that the eigenvalues of the higher
spin $3$ current for
any representation $(\La_+;\La_+)$ will be a multiple of this quantity.
Here $\La_+$ is the symmetric or antisymmetric representation
and the number of boxes is arbitrary. Note the presence of
factor $(k-N)$ in the above
\footnote{
  \label{barbarfoot}
  One obtains
  \bea
  \phi_0^{(1)} (\overline{\tiny\yng(1)};\overline{\tiny\yng(1)}) & = &
  -\phi_0^{(1)} (\tiny\yng(1);\tiny\yng(1)),
\qquad
v^{\pm} (\overline{\tiny\yng(1)}; \overline{\tiny\yng(1)})  =  v^{\pm}
(\tiny\yng(1);\tiny\yng(1)),
\qquad
\phi_2^{(1)} (\overline{\tiny\yng(1)};\overline{\tiny\yng(1)})  = 
-\phi_2^{(1)} (\tiny\yng(1);\tiny\yng(1)).
\nonu
\eea
Note that there are sign changes in the higher spin currents
with odd spin.}.

%%%%%%%%%%%%%%%%%%%%%%%%%%%%%%%%%%%%%%%%%%%%%%%%%%%
\subsection{The $(f;\overline{f})$ representation}
%%%%%%%%%%%%%%%%%%%%%%%%%%%%%%%%%%%%%%%%%%%%%%%%%%%

The relevant subsection on this higher representation
is given by $2.2.2$.
The four eigenvalues corresponding to the zero modes
of the higher spin currents of spins $1, 4, 4$ and $3$
which act on
the representation $(\tiny\yng(1); \overline{\tiny\yng(1)})$
can be summarized by
{\small
\bea
\phi_0^{(1)} (\tiny\yng(1);\overline{\tiny\yng(1)}) & = &
\phi_0^{(1)} (\tiny\yng(1);0) +
\phi_0^{(1)} (0;\overline{\tiny\yng(1)})=
\frac{(k-N)}{(k+N+2)},
\nonu \\
v^+ (\tiny\yng(1);\overline{\tiny\yng(1)}) & = &
\frac{12 (4 k N+2 k+5 N^2-16 N+5)}{(k+N+2)^2},
\nonu \\
v^- (\tiny\yng(1);\overline{\tiny\yng(1)}) & = &
\frac{12 (5 k^2+4 k N-28 k+2 N+5)}{(k+N+2)^2},
\nonu \\
\phi_2^{(1)} (\tiny\yng(1);\overline{\tiny\yng(1)}) & = &
-\frac{8}{
  3 (k+N+2)^2 (6 k N+5 k+5 N+4)}
\nonu \\
& \times & ({\bf 6 k^3 N}+2 k^3 {\bf +6 k^2 N^2}+k^2 N-11 k^2
{\bf +6 k N^3}+
7 k N^2-32 k N-42 k \nonu \\
& + & 2 N^3-11 N^2-30 N-24).
\label{nonlinearffbar}
\eea}
The previous relations in (\ref{zerofbareigenhigher}) and
(\ref{f0eigenhigher}) are used.
Note that the eigenvalue in (\ref{nonlinearffbar})
for the higher spin $1$ current
does not have any contribution from the commutator
$[(\Phi_0^{(1)})_0, Q^{\bar{A}}_{-\frac{1}{2}}]$ because
the OPE between the corresponding higher spin $1$ current
and the spin $\frac{1}{2}$ current has only the first order pole.
See also (\ref{state0f}).
This provides only the eigenvalue for the representation $(0;\overline{
  \tiny\yng(1)})$. Also the term 
$ Q^{\bar{A}}_{-\frac{1}{2}} (\Phi_0^{(1)})_0$
acting on the representation $(\tiny\yng(1);0)$
gives the eigenvalue $\phi_0^{(1)} (\tiny\yng(1);0)$ with
$ Q^{\bar{A}}_{-\frac{1}{2}}$ acting on the state  $|(\tiny\yng(1);0)>$.
By inserting the overall factor into this state, one has
the final state associated with 
the representation $(\tiny\yng(1);\overline{\tiny\yng(1)})$.
Therefore, one arrives at the above eigenvalue for the higher spin $1$
current.

For the eigenvalues corresponding to the remaining higher spin currents,
there are the contributions from the lower order poles
appearing in the commutators,
$[(V^+)_0, Q^{\bar{A}}_{-\frac{1}{2}}]$,
$[(V^-)_0, Q^{\bar{A}}_{-\frac{1}{2}}]$ and $[(\Phi_2^{(1)})_0, 
  Q^{\bar{A}}_{-\frac{1}{2}}]$ \footnote{In other words, for example,
one has $[(\Phi_2^{(1)})_0, 
  Q^{\bar{A}}_{-\frac{1}{2}}] =\oint_{c_0} \frac{d w}{2 \pi i} w^{-\frac{1}{2}+
  (\frac{1}{2}-1)} \oint_{C_w} \frac{d z}{2 \pi i} z^{0+(3-1)}
\Phi_2^{(1)}(z)\,  Q^{\bar{A}}(w)$. One sees that the OPE
between $\Phi_2^{(1)}(z)$ and $ Q^{\bar{A}}(w)$ contains the first and second
order poles as well as the third order pole. They all contribute to the
eigenvalue.}. 
They can be summarized by
\bea
%& = & \frac{k}{(k+N+2)},
%\nonu \\
\delta v^+ (\tiny\yng(1);\overline{\tiny\yng(1)})
& = & -\frac{12 (18 N-5)}{(k+N+2)^2},
\nonu \\
\delta v^- (\tiny\yng(1);\overline{\tiny\yng(1)})
& = &  -\frac{60 (6 k-1)}{(k+N+2)^2},
\nonu \\
\delta \phi_2^{(1)} (\tiny\yng(1);\overline{\tiny\yng(1)})& = & \frac{8
  ({\bf 7 k^2 N}+6 k^2 {\bf +5 k N^2}+20 k N+16 k+6 N^2+12 N+8)}{
  (k+N+2)^2 (6 k N+5 k+5 N+4)}.
\label{ffbarcorrection1}
\eea
Once again, the
large $(N,k)$ 't Hooft like limits
for these extra contributions
lead to the $\frac{1}{N}$ behavior.
Then the above eigenvalues
are obtained from the relations,
\bea
v^{\pm} (\tiny\yng(1);\overline{\tiny\yng(1)})  & = &
v^{\pm} (\tiny\yng(1);0)+
v^{\pm} (0;\overline{\tiny\yng(1)}) +\delta v^{\pm}
(\tiny\yng(1);\overline{\tiny\yng(1)}),\nonu \\
\phi_2^{(1)} (\tiny\yng(1);\overline{\tiny\yng(1)}) & = &
\phi_2^{(1)} (\tiny\yng(1);0)+
\phi_2^{(1)} (0;\overline{\tiny\yng(1)})
+ \delta \phi_2^{(1)} (\tiny\yng(1);\overline{\tiny\yng(1)})
\label{ffbarcorrection}
\eea
respectively \footnote{
\label{footfbarfresult}
  One obtains
  \bea
%& = & \frac{k}{(k+N+2)},
%\nonu \\
\delta v^+ (\overline{\tiny\yng(1)};\tiny\yng(1))
& = & \frac{12 (18 N+5)}{(k+N+2)^2},
\qquad
\delta v^- (\overline{\tiny\yng(1)};\tiny\yng(1))
 =   \frac{60 (6 k+1)}{(k+N+2)^2},
\nonu \\
\delta \phi_2^{(1)} (\overline{\tiny\yng(1)};\tiny\yng(1))& = &
-\frac{8 ({\bf 5 k^2 N}+3 k^2 {\bf +k N^2}
  +10 k N+11 k+3 N^2+3 N+4)}{(k+N+2)^2 (6 k N+5 k+5 N+4)}.
\label{fbarfeigehhighercorrection}
\eea
Here
 the commutators,
$[(V^+)_0, Q^{\bar{A}^{\ast}}_{-\frac{1}{2}}]$,
$[(V^-)_0, Q^{\bar{A}^{\ast}}_{-\frac{1}{2}}]$ and $[(\Phi_2^{(1)})_0, 
  Q^{\bar{A}^{\ast}}_{-\frac{1}{2}}]$ are used.
Then one also has
\bea
\phi_0^{(1)} (\overline{\tiny\yng(1)};\tiny\yng(1)) & = &
\phi_0^{(1)} (\overline{\tiny\yng(1)};0) +
\phi_0^{(1)} (0;\tiny\yng(1)), \nonu \\
v^{\pm} (\overline{\tiny\yng(1)};\tiny\yng(1))  & = &
v^{\pm} (\overline{\tiny\yng(1)};0)+
v^{\pm} (0;\tiny\yng(1)) +\delta v^{\pm}
(\overline{\tiny\yng(1)};\tiny\yng(1)),\nonu \\
\phi_2^{(1)} (\overline{\tiny\yng(1)};\tiny\yng(1)) & = &
\phi_2^{(1)} (\overline{\tiny\yng(1)};0)+
\phi_2^{(1)} (0;\tiny\yng(1))
+ \delta \phi_2^{(1)} (\overline{\tiny\yng(1)};\tiny\yng(1)),
\nonu
\eea
together with (\ref{0feigenhigher}), (\ref{fbarzeroeigenhigher}),
and (\ref{fbarfeigehhighercorrection}).
Note that 
$\phi_2^{(1)} (\overline{\tiny\yng(1)};\tiny\yng(1))$ is not equal to
$\phi_2^{(1)} (\tiny\yng(1);\overline{\tiny\yng(1)})$ even in the
large $(N,k)$ 't Hooft like limit.}.
The previous relations (\ref{f0eigenhigher}) and
(\ref{zerofbareigenhigher}) can be used.

%%%%%%%%%%%%%%%%%%%%%%%%%%%%%%%%%%%%%%%%%%%%%%%%%%
\subsection{The $(f;\mbox{symm})$ representation}
%%%%%%%%%%%%%%%%%%%%%%%%%%%%%%%%%%%%%%%%%%%%%%%%%%

This higher representation can be obtained from the
product of
$(\tiny\yng(1);0)$ and $(0;\tiny\yng(2))$ \footnote{The subsections
  $5.3$-$5.6$ are not considered in previous sections $2$ and $3$.
  It is easy to obtain the corresponding eigenvalues on
  these higher representations. Some of them will appear in section $8$. }.
It turns out that the four eigenvalues are given by 
{\small
\bea
\phi_0^{(1)} (\tiny\yng(1);\tiny\yng(2)) & = &
\phi_0^{(1)} (\tiny\yng(1);0) +
\phi_0^{(1)} (0;\tiny\yng(2))= -\frac{(N+2k)}{(N+k+2)},
\nonu \\
v^+ (\tiny\yng(1);\tiny\yng(2)) & = & \frac{12
  (4 k N+8 k+5 N^2+38 N+20)}{(k+N+2)^2},
\nonu \\
v^- (\tiny\yng(1);\tiny\yng(2)) & = & \frac{24
  (4 k^2+4 k N+12 k+N)}{(k+N+2)^2},
\nonu \\
\phi_2^{(1)} (\tiny\yng(1);\tiny\yng(2)) & = &
\frac{4}{3 (k+N+2)^2 (6 k N+5 k+5 N+4)}
\nonu \\
& \times &
({\bf 24 k^3 N}-4 k^3 {\bf +6 k^2 N^2}+73 k^2 N-14 k^2
{\bf -12 k N^3}-17 k N^2\nonu \\
& - & 56 k N-72 k-4 N^3-32 N^2-156 N-96).
\label{fsymmeigenhigher}
\eea}
The relations in (\ref{f0eigenhigher}) and (\ref{zerosymmhigher})
which will appear later
are used.
Note that the eigenvalue (\ref{fsymmeigenhigher})
for the higher spin $1$ current
does not have any contribution from the commutator
$[(\Phi_0^{(1)})_0, Q^{13}_{-\frac{1}{2}}  Q^{16}_{-\frac{1}{2}} ]$ (see also
the subsection \ref{0symmtwo}) because
the OPE between the corresponding higher spin $1$ current
and the product of spin $\frac{1}{2}$ currents has only the first order pole.
This provides only the eigenvalue for the representation $(0;
  \tiny\yng(2))$. Also the term 
$ Q^{13}_{-\frac{1}{2}}  Q^{16}_{-\frac{1}{2}} (\Phi_0^{(1)})_0$
acting on the representation $(\tiny\yng(1);0)$
gives the eigenvalue $\phi_0^{(1)} (\tiny\yng(1);0)$ with
$ Q^{13}_{-\frac{1}{2}}  Q^{16}_{-\frac{1}{2}}$
acting on the state  $|(\tiny\yng(1);0)>$.
By inserting the overall factor into this state, one has
the final state associated with 
the representation $(\tiny\yng(1);\tiny\yng(2))$.
Therefore, one arrives at the above eigenvalue for the higher spin $1$
current.

For the eigenvalues corresponding to the remaining higher spin currents,
there are the contributions from the lower order poles
appearing in the commutators,
$[(V^+)_0, Q^{13}_{-\frac{1}{2}}  Q^{16}_{-\frac{1}{2}}]$,
$[(V^-)_0, Q^{13}_{-\frac{1}{2}}  Q^{16}_{-\frac{1}{2}}]$ and
$[(\Phi_2^{(1)})_0, 
  Q^{13}_{-\frac{1}{2}}  Q^{16}_{-\frac{1}{2}}]$. 
They can be summarized by
\bea
\delta
v^+ (\tiny\yng(1);\tiny\yng(2))
& = & \frac{48 (9 N+5)}{(k+N+2)^2},
\nonu \\
\delta
v^- (\tiny\yng(1);\tiny\yng(2))
& = & \frac{384 k}{(k+N+2)^2},
\nonu \\
\delta
\phi_2^{(1)} (\tiny\yng(1);\tiny\yng(2))
& = & \frac{8 ({\bf 10 k^2 N+2 k N^2}-k N-8 k-3 N^2-24 N-16)}{
  (k+N+2)^2 (6 k N+5 k+5 N+4)}.
\label{fsymmeigenhighercorrection}
\eea
It is easy to see that 
 the
large $(N,k)$ 't Hooft like limit
for these extra contributions
lead to the $\frac{1}{N}$ behavior
\footnote{
From the relations
\label{fbarsymmresultresult}
\bea
\delta
v^{\pm} (\overline{\tiny\yng(1)};\tiny\yng(2))
& = & \delta v^{\pm} (\tiny\yng(1);\tiny\yng(2)),
\qquad
\delta
\phi_2^{(1)} (\overline{\tiny\yng(1)};\tiny\yng(2))
 =  - \delta \phi_2^{(1)} (\tiny\yng(1);\tiny\yng(2)),
\label{fbarsymmeigenhighercorrection}
\eea
together with (\ref{fsymmeigenhighercorrection}),
one has
\bea
\phi_0^{(1)} (\overline{\tiny\yng(1)};\tiny\yng(2)) & = &
\phi_0^{(1)} (\overline{\tiny\yng(1)};0) +
\phi_0^{(1)} (0;\tiny\yng(2)), \nonu \\
v^{\pm} (\overline{\tiny\yng(1)};\tiny\yng(2))  & = &
v^{\pm} (\overline{\tiny\yng(1)};0)+
v^{\pm} (0;\tiny\yng(2)) +\delta v^{\pm}
(\overline{\tiny\yng(1)};\tiny\yng(2)),\nonu \\
\phi_2^{(1)} (\overline{\tiny\yng(1)};\tiny\yng(2)) & = &
\phi_2^{(1)} (\overline{\tiny\yng(1)};0)+
\phi_2^{(1)} (0;\tiny\yng(2))
+ \delta \phi_2^{(1)} (\overline{\tiny\yng(1)};\tiny\yng(2))
\nonu
\eea
with (\ref{fbarzeroeigenhigher}), (\ref{zerosymmhigher}) and
(\ref{fbarsymmeigenhighercorrection}).}.
Then the relation (\ref{fsymmeigenhigher}) can be obtained
from the previous ones in 
 (\ref{f0eigenhigher}) and (\ref{zerosymmhigher})
together with
(\ref{fsymmeigenhighercorrection}).
The highest power terms of $(N,k)$ in the higher spin $3$ current
can be described by the corresponding  terms in 
(\ref{f0eigenhigher}) and (\ref{zerosymmhigher}) respectively via
simple addition of them.

%%%%%%%%%%%%%%%%%%%%%%%%%%%%%%%%%%%%%%%%%%%%%%%%%%
\subsection{The $(f;\overline{\mbox{symm}})$ representation}
%%%%%%%%%%%%%%%%%%%%%%%%%%%%%%%%%%%%%%%%%%%%%%%%%%

This higher representation can be obtained from the
product of
$(\tiny\yng(1);0)$ and $(0;\overline{\tiny\yng(2)})$.
It turns out that the four eigenvalues are given by 
{\small
\bea
\phi_0^{(1)} (\tiny\yng(1);\overline{\tiny\yng(2)}) & = &
\phi_0^{(1)} (\tiny\yng(1);0) +
\phi_0^{(1)} (0;\overline{\tiny\yng(2)})= -\frac{(N-2k)}{(N+k+2)},
\nonu \\
v^+ (\tiny\yng(1);\overline{\tiny\yng(2)}) & = & \frac{12
  (4 k N+8 k+5 N^2-34 N-20)}{(k+N+2)^2},
\nonu \\
v^- (\tiny\yng(1);\overline{\tiny\yng(2)}) & = & \frac{24
  (4 k^2+4 k N-20 k+N)}{(k+N+2)^2},
\nonu \\
\phi_2^{(1)} (\tiny\yng(1);\overline{\tiny\yng(2)}) & = &
-\frac{4}{3 (k+N+2)^2 (6 k N+5 k+5 N+4)}
\nonu \\
& \times  &
({\bf 24 k^3 N}-4 k^3 {\bf +18 k^2 N^2}-61 k^2 N-50 k^2
{\bf +12 k N^3}-11 k N^2
\nonu \\
& - & 108 k N-36 k+4 N^3-4 N^2+48 N+48).
\label{fsymmbarnonlinear}
\eea}
The relations (\ref{f0eigenhigher}) and the footnote
\ref{zerosymmbareigenhigher} can be used.
Note that the eigenvalue for the higher spin $1$ current in
(\ref{fsymmbarnonlinear})
does not have any contribution from the commutator
$[(\Phi_0^{(1)})_0, Q^{1}_{-\frac{1}{2}}  Q^{4}_{-\frac{1}{2}} ]$
(see also
the subsection \ref{0symmtwo})
because
the OPE between the corresponding higher spin $1$ current
and the spin $\frac{1}{2}$ current has only the first order pole.
This provides only the eigenvalue for the representation $(0;
  \overline{\tiny\yng(2)})$. Also the term 
$ Q^{1}_{-\frac{1}{2}}  Q^{4}_{-\frac{1}{2}} (\Phi_0^{(1)})_0$
acting on the representation $(\tiny\yng(1);0)$
gives the eigenvalue $\phi_0^{(1)} (\tiny\yng(1);0)$ with
$ Q^{1}_{-\frac{1}{2}}  Q^{4}_{-\frac{1}{2}}$
acting on the state  $|(\tiny\yng(1);0)>$.
By inserting the overall factor into this state, one has
the final state associated with 
the representation $(\tiny\yng(1);\overline{\tiny\yng(2)})$.
Therefore, one arrives at the above eigenvalue for the higher spin $1$
current.

For the eigenvalues corresponding to the remaining higher spin currents,
there are the contributions from the lower order poles
appearing in the commutators,
$[(V^+)_0, Q^{1}_{-\frac{1}{2}}  Q^{4}_{-\frac{1}{2}}]$,
$[(V^-)_0, Q^{1}_{-\frac{1}{2}}  Q^{4}_{-\frac{1}{2}}]$ and
$[(\Phi_2^{(1)})_0, 
  Q^{1}_{-\frac{1}{2}}  Q^{4}_{-\frac{1}{2}}]$. 
They can be summarized by
\footnote{
\label{fbarsymmbarfoot}
  From the relations
  \bea
\delta
v^{\pm} (\overline{\tiny\yng(1)};\overline{\tiny\yng(2)})
& = & \delta v^{\pm} (\tiny\yng(1);\overline{\tiny\yng(2)}),
\qquad
\delta
\phi_2^{(1)} (\overline{\tiny\yng(1)};\overline{\tiny\yng(2)})
 =  - \delta \phi_2^{(1)} (\tiny\yng(1);\overline{\tiny\yng(2)}),
\label{fbarsymmbarcorrection}
\eea
one has
\bea
\phi_0^{(1)} (\overline{\tiny\yng(1)};\overline{\tiny\yng(2)}) & = &
\phi_0^{(1)} (\overline{\tiny\yng(1)};0) +
\phi_0^{(1)} (0;\overline{\tiny\yng(2)}), \nonu \\
v^{\pm} (\overline{\tiny\yng(1)};\overline{\tiny\yng(2)})  & = &
v^{\pm} (\overline{\tiny\yng(1)};0)+
v^{\pm} (0;\overline{\tiny\yng(2)}) +\delta v^{\pm}
(\overline{\tiny\yng(1)};\overline{\tiny\yng(2)}),\nonu \\
\phi_2^{(1)} (\overline{\tiny\yng(1)};\overline{\tiny\yng(2)}) & = &
\phi_2^{(1)} (\overline{\tiny\yng(1)};0)+
\phi_2^{(1)} (0;\overline{\tiny\yng(2)})
+ \delta \phi_2^{(1)} (\overline{\tiny\yng(1)};\overline{\tiny\yng(2)}),
\nonu
\eea
together with (\ref{fbarzeroeigenhigher}), the footnote
\ref{zerosymmbareigenhigher} and
(\ref{fbarsymmbarcorrection}).}
\bea
\delta
v^+ (\tiny\yng(1);\overline{\tiny\yng(2)})
& = & -\frac{48 (9 N-5)}{(k+N+2)^2},
\nonu \\
\delta
v^- (\tiny\yng(1);\overline{\tiny\yng(2)})
& = & -\frac{384 k}{(k+N+2)^2},
\nonu \\
\delta
\phi_2^{(1)} (\tiny\yng(1);\overline{\tiny\yng(2)})
& = & \frac{8 ({\bf 14 k^2 N}+6 k^2{\bf +10 k N^2}+19 k N+2 k+3
  N^2-6 N-8)}{(k+N+2)^2 (6 k N+5 k+5 N+4)}.
\label{fsymmbareigenhighercorrection}
\eea
Then the relation (\ref{fsymmbarnonlinear}) can be obtained
from the previous ones in 
 (\ref{f0eigenhigher}) and the footnote \ref{zerosymmbareigenhigher}
together with
(\ref{fsymmbareigenhighercorrection}).
The highest power terms of $(N,k)$ in the last eigenvalue of
(\ref{fsymmbareigenhighercorrection}) is twice of the ones in
(\ref{ffbarcorrection1}). 
It is easy to see that
the highest power terms of $(N,k)$ in the higher spin $3$ current
can be described by the corresponding  terms in 
(\ref{f0eigenhigher}) and the footnote
(\ref{zerosymmbareigenhigher}) respectively and it turns out that
it is given by simple addition of two contributions.

%%%%%%%%%%%%%%%%%%%%%%%%%%%%%%%%%%%%%%%%%%%%%%%%%%%
{\subsection{The $(f;\mbox{antisymm})$ representation}
%%%%%%%%%%%%%%%%%%%%%%%%%%%%%%%%%%%%%%%%%%%%%%%%%%%

  It turns out that the four eigenvalues are given by 
{\small
\bea
\phi_0^{(1)} (\tiny\yng(1);\tiny\yng(1,1)) & = &
\phi_0^{(1)} (\tiny\yng(1);0) +
\phi_0^{(1)} (0;\tiny\yng(1,1))= -\frac{(N+2k)}{(N+k+2)},
\nonu \\
v^+ (\tiny\yng(1);\tiny\yng(1,1)) & = & \frac{12
  (4 k N+8 k+5 N^2+38 N+20)}{(k+N+2)^2},
\nonu \\
v^- (\tiny\yng(1);\tiny\yng(1,1)) & = & \frac{8
  (16 k^2+12 k N+108 k+3 N+20)}{(k+N+2)^2},
\nonu \\
\phi_2^{(1)} (\tiny\yng(1);\tiny\yng(1,1)) & = &
\frac{4}{3 (k+N+2)^2 (6 k N+5 k+5 N+4)} \nonu \\
& \times &
({\bf 24 k^3 N}-4 k^3 {\bf +6 k^2 N^2}+145 k^2 N-2 k^2 {\bf -12 k N^3}-17 k N^2
\nonu \\
& + & 196 k N+156 k-4 N^3+40 N^2+24 N+48).
\label{fantisymmeigenhigher}
\eea}  
See also the subsection \ref{0antisymm}.
The first and the second eigenvalues in (\ref{fantisymmeigenhigher})
are the same as the ones
in (\ref{fsymmeigenhigher}) respectively.
This implies that there is no difference between
$\tiny\yng(2)$ or $\tiny\yng(1,1)$ as long as these eigenvalues are
concerned.
%The second eigenvalue is the same as the one in (\ref{fsymmeigenhigher})
%up to the highest power of $N$ and $k$.
Furthermore, the last eigenvalue has common behavior with
the one in (\ref{fsymmeigenhigher}) because they contain
$(24k^3 N + 6k^2 N^2-12k N^3)$ in the numerators.

Note that the eigenvalue for the higher spin $1$ current
does not have any contribution from the commutator
$[(\Phi_0^{(1)})_0, Q^{13}_{-\frac{1}{2}}  Q^{14}_{-\frac{1}{2}} ]$ because
the OPE between the corresponding higher spin $1$ current
and the product of spin $\frac{1}{2}$ currents has only the first order pole.
This provides only the eigenvalue for the representation $(0;
  \tiny\yng(1,1))$. Also the term 
$ Q^{13}_{-\frac{1}{2}}  Q^{14}_{-\frac{1}{2}} (\Phi_0^{(1)})_0$
acting on the representation $(\tiny\yng(1);0)$
gives the eigenvalue $\phi_0^{(1)} (\tiny\yng(1);0)$ with
$ Q^{13}_{-\frac{1}{2}}  Q^{14}_{-\frac{1}{2}}$
acting on the state  $|(\tiny\yng(1);0)>$.
By inserting the overall factor into this state, one has
the final state associated with 
the representation $(\tiny\yng(1);\tiny\yng(1,1))$.
Therefore, one arrives at the above eigenvalue for the higher spin $1$
current.

For the eigenvalues corresponding to the remaining higher spin currents,
there are the contributions from the lower order poles
appearing in the commutators,
$[(V^+)_0, Q^{13}_{-\frac{1}{2}}  Q^{14}_{-\frac{1}{2}}]$,
$[(V^-)_0, Q^{13}_{-\frac{1}{2}}  Q^{14}_{-\frac{1}{2}}]$ and
$[(\Phi_2^{(1)})_0, 
  Q^{13}_{-\frac{1}{2}}  Q^{14}_{-\frac{1}{2}}]$. 
They can be summarized by
\bea
\delta
v^+ (\tiny\yng(1);\tiny\yng(1,1))
& = & \frac{48 (9 N+5)}{(k+N+2)^2},
\nonu \\
\delta
v^- (\tiny\yng(1);\tiny\yng(1,1))
& = & \frac{32 (26 k+5)}{(k+N+2)^2},
\nonu \\
\delta
\phi_2^{(1)} (\tiny\yng(1);\tiny\yng(1,1))
& = & \frac{8 ({\bf 10 k^2 N+2 k N^2}+23 k N+22 k+9 N^2+6 N+8)}{(k+N+2)^2 (6 k N+5 k+5 N+4)}.
\label{fantisymmeigenhighercorrection}
\eea 
The first eigenvalue in (\ref{fantisymmeigenhighercorrection}) is the same as the one in
(\ref{fsymmeigenhighercorrection}).
For the last eigenvalue, the behavior of highest power of
$(N,k)$ is the same as the one in (\ref{fsymmeigenhighercorrection})
\footnote{
From the relations
\label{fbarantisymmresultresult}
\bea
\delta
v^{\pm} (\overline{\tiny\yng(1)};\tiny\yng(1,1))
& = & \delta v^{\pm} (\tiny\yng(1);\tiny\yng(1,1)),
\qquad
\delta
\phi_2^{(1)} (\overline{\tiny\yng(1)};\tiny\yng(1,1))
 =  - \delta \phi_2^{(1)} (\tiny\yng(1);\tiny\yng(1,1)),
\label{correction}
\eea
one has
\bea
\phi_0^{(1)} (\overline{\tiny\yng(1)};\tiny\yng(1,1)) & = &
\phi_0^{(1)} (\overline{\tiny\yng(1)};0) +
\phi_0^{(1)} (0;\tiny\yng(1,1)), \nonu \\
v^{\pm} (\overline{\tiny\yng(1)};\tiny\yng(1,1))  & = &
v^{\pm} (\overline{\tiny\yng(1)};0)+
v^{\pm} (0;\tiny\yng(1,1)) +\delta v^{\pm}
(\overline{\tiny\yng(1)};\tiny\yng(1,1)),\nonu \\
\phi_2^{(1)} (\overline{\tiny\yng(1)};\tiny\yng(1,1)) & = &
\phi_2^{(1)} (\overline{\tiny\yng(1)};0)+
\phi_2^{(1)} (0;\tiny\yng(1,1))
+ \delta \phi_2^{(1)} (\overline{\tiny\yng(1)};\tiny\yng(1,1)),
\nonu
\eea
together with (\ref{fbarzeroeigenhigher}),
(\ref{zeroantisymmhigher}), and (\ref{correction}).}.
Then the relations (\ref{fantisymmeigenhigher})
can be obtained from (\ref{f0eigenhigher}) and (\ref{zeroantisymmhigher})
together with (\ref{fantisymmeigenhighercorrection}).

%%%%%%%%%%%%%%%%%%%%%%%%%%%%%%%%%%%%%%%%%%%%%%%%%%%%%%%%%
\subsection{The $(f;\overline{\mbox{antisymm}})$ representation}
%%%%%%%%%%%%%%%%%%%%%%%%%%%%%%%%%%%%%%%%%%%%%%%%%%%%%%%%%

Similarly, the four eigenvalues are given by 
{\small
\bea
\phi_0^{(1)} (\tiny\yng(1);\overline{\tiny\yng(1,1)}) & = &
\phi_0^{(1)} (\tiny\yng(1);0) +
\phi_0^{(1)} (0;\overline{\tiny\yng(1,1)})= -\frac{(N-2k)}{(N+k+2)},
\nonu \\
v^+ (\tiny\yng(1);\overline{\tiny\yng(1,1)}) & = & \frac{12 (4 k N+8 k+5 N^2-34 N-20)}{(k+N+2)^2},
\nonu \\
v^- (\tiny\yng(1);\overline{\tiny\yng(1,1)}) & = & \frac{8 (16 k^2+12 k N-100 k+3 N+20)}{(k+N+2)^2},
\nonu \\
\phi_2^{(1)} (\tiny\yng(1);\overline{\tiny\yng(1,1)}) & = &
-\frac{4}{3 (k+N+2)^2 (6 k N+5 k+5 N+4)}
\nonu \\
& \times &
({\bf 24 k^3 N}-4 k^3 {\bf +18 k^2 N^2}+11 k^2 N-38 k^2
{\bf +12 k N^3}-11 k N^2
\nonu \\
& - & 144 k N-168 k+4 N^3-76 N^2-132 N-96).
\label{fantisymmbareigenhigher}
\eea}
The relations in (\ref{f0eigenhigher}) and the
footnote \ref{zeroantisymmbarexpression} are needed.
The first and the second eigenvalues in (\ref{fantisymmbareigenhigher}) are
the same as the ones in (\ref{fsymmbarnonlinear}).
Furthermore, the last eigenvalue has common behavior with
the one in (\ref{fsymmbarnonlinear}) because they contain
$(24k^3 N + 18 k^2 N^2 +12 k N^3)$ in the numerators.
See also the subsection \ref{0antisymm}.

Note that the eigenvalue for the higher spin $1$ current
does not have any contribution from the commutator
$[(\Phi_0^{(1)})_0, Q^{1}_{-\frac{1}{2}}  Q^{2}_{-\frac{1}{2}} ]$ because
the OPE between the corresponding higher spin $1$ current
and the spin $\frac{1}{2}$ current has only the first order pole.
This provides only the eigenvalue for the representation $(0;
  \overline{\tiny\yng(1,1)})$. Also the term 
$ Q^{1}_{-\frac{1}{2}}  Q^{2}_{-\frac{1}{2}} (\Phi_0^{(1)})_0$
acting on the representation $(\tiny\yng(1);0)$
gives the eigenvalue $\phi_0^{(1)} (\tiny\yng(1);0)$ with
$ Q^{1}_{-\frac{1}{2}}  Q^{2}_{-\frac{1}{2}}$
acting on the state  $|(\tiny\yng(1);0)>$.
By inserting the overall factor into this state, one has
the final state associated with 
the representation $(\tiny\yng(1);\overline{\tiny\yng(1,1)})$.
Therefore, one arrives at the above eigenvalue for the higher spin $1$
current.

For the eigenvalues corresponding to the remaining higher spin currents,
there are the contributions from the lower order poles
appearing in the commutators,
$[(V^+)_0, Q^{1}_{-\frac{1}{2}}  Q^{2}_{-\frac{1}{2}}]$,
$[(V^-)_0, Q^{1}_{-\frac{1}{2}}  Q^{2}_{-\frac{1}{2}}]$ and
$[(\Phi_2^{(1)})_0, 
  Q^{1}_{-\frac{1}{2}}  Q^{2}_{-\frac{1}{2}}]$. 
They can be summarized by
\bea
\delta
v^+ (\tiny\yng(1);\overline{\tiny\yng(1,1)})
& = & -\frac{48 (9 N-5)}{(k+N+2)^2},
\nonu \\
\delta
v^- (\tiny\yng(1);\overline{\tiny\yng(1,1)})
& = & -\frac{32 (26 k-5)}{(k+N+2)^2},
\nonu \\
\delta
\phi_2^{(1)} (\tiny\yng(1);\overline{\tiny\yng(1,1)})
& = & \frac{8 ({\bf 14 k^2 N}+6 k^2 {\bf +10 k N^2}
  +43 k N+32 k+15 N^2+24 N+16)}{(k+N+2)^2 (6 k N+5 k+5 N+4)}.
\label{fantisymmbareigenhighercorrection}
\eea
The first eigenvalue in (\ref{fantisymmbareigenhighercorrection}) is the same as the one in
(\ref{fsymmbareigenhighercorrection}).
For the last eigenvalue, the behavior of highest power of
$(N,k)$ is the same as the one in (\ref{fsymmbareigenhighercorrection})
\footnote{
\label{footcor1}
  From the relations
  \bea
\delta
v^{\pm} (\overline{\tiny\yng(1)};\overline{\tiny\yng(1,1)})
& = & \delta v^{\pm} (\tiny\yng(1);\overline{\tiny\yng(1,1)}),
\qquad
\delta
\phi_2^{(1)} (\overline{\tiny\yng(1)};\overline{\tiny\yng(1,1)})
 =  - \delta \phi_2^{(1)} (\tiny\yng(1);\overline{\tiny\yng(1,1)}),
\label{cor1}
\eea
one has
\bea
\phi_0^{(1)} (\overline{\tiny\yng(1)};\overline{\tiny\yng(1,1)}) & = &
\phi_0^{(1)} (\overline{\tiny\yng(1)};0) +
\phi_0^{(1)} (0;\overline{\tiny\yng(1,1)}), \nonu \\
v^{\pm} (\overline{\tiny\yng(1)};\overline{\tiny\yng(1,1)})  & = &
v^{\pm} (\overline{\tiny\yng(1)};0)+
v^{\pm} (0;\overline{\tiny\yng(1,1)}) +\delta v^{\pm}
(\overline{\tiny\yng(1)};\overline{\tiny\yng(1,1)}),\nonu \\
\phi_2^{(1)} (\overline{\tiny\yng(1)};\overline{\tiny\yng(1,1)}) & = &
\phi_2^{(1)} (\overline{\tiny\yng(1)};0)+
\phi_2^{(1)} (0;\overline{\tiny\yng(1,1)})
+ \delta \phi_2^{(1)} (\overline{\tiny\yng(1)};\overline{\tiny\yng(1,1)}),
\nonu
\eea
together with (\ref{fbarzeroeigenhigher}), (\ref{cor1}) and 
the footnote \ref{zeroantisymmbarexpression}.}.
Then one obtains the relations in (\ref{fantisymmbareigenhigher})
from the relations in (\ref{f0eigenhigher}) and the
footnote \ref{zeroantisymmbarexpression} with
(\ref{fantisymmbareigenhighercorrection}).

%%%%%%%%%%%%%%%%%%%%%%%%%%%%%%%%%%%%%%%%%%%%%%%%%%%%
\subsection{The $(\mbox{symm}; 0)$ representation}
%%%%%%%%%%%%%%%%%%%%%%%%%%%%%%%%%%%%%%%%%%%%%%%%%%%%

The relevant subsection on this higher representation
is given by $2.2.3$.
This higher representation can be obtained from the
product of the minimal representation $(f;0)$ and itself.
The four eigenvalues with this representation can be
described as 
\bea
\phi_0^{(1)} (\tiny\yng(2);0) & = & -\frac{2 N}{(k+N+2)},
\nonu \\
 v^+ (\tiny\yng(2);0) & = & \frac{32 N (3 k+4 N+1)}{(k+N+2)^2},
\nonu \\
 v^- (\tiny\yng(2);0) & = &  \frac{96 N}{(k+N+2)^2},
\nonu \\
\phi_2^{(1)} (\tiny\yng(2);0) & = & -\frac{8 N
  ({\bf 6 k^2 N}+5 k^2 {\bf +12 k N^2}+45 k N+43 k-2 N^2-N+12)}
    {3 (k+N+2)^2 (6 k N+5 k+5 N+4)}.
\label{symmzerohigher}
\eea
It is easy to see that the first eigenvalue is the twice of
the one in (\ref{f0eigenhigher}) at finite $(N,k)$.
Except the third eigenvalue, the remaining ones in (\ref{symmzerohigher})
behave as $\la$ dependent constant values
under the large $(N,k)$ 't Hooft like limit.
Furthermore, if one sees the last eigenvalue closely,  
one observes that the highest power terms in the numerator
are given by $-8N(6 k^2 N +12 k N^2)$ which is the {\bf twice} of
the ones in (\ref{f0eigenhigher}). Note that
the denominators in both expressions are the same
at finite $(N,k)$. This implies that
the eigenvalue for the higher spin $3$  current
in this higher representation can be interpreted as the additive
quantum number and it is given by the sum of each eigenvalue
for the higher representation in the minimal representation
$(\tiny\yng(1);0)$ in (\ref{f0eigenhigher})
under the large $(N,k)$ 't Hooft like limit
\footnote{
\label{symmbareigenhigher}
  For the similar higher representation $(
  \overline{\mbox{symm}};0)$, one has the following
  results
  \bea
  \phi_0^{(1)} (\overline{\tiny\yng(2)};0) & = &
  -\phi_0^{(1)} (\tiny\yng(2);0),
\qquad
 v^{\pm} (\overline{\tiny\yng(2)};0)  =   v^{\pm} (\tiny\yng(2);0),
\qquad
% v^- (\overline{\tiny\yng(2)};0) & = &   v^- (\tiny\yng(2);0),
%\nonu \\
\phi_2^{(1)} (\overline{\tiny\yng(2)};0)  = 
-\phi_2^{(1)} (\tiny\yng(2);0)
\nonu
\eea
together with (\ref{symmzerohigher}).}.

%%%%%%%%%%%%%%%%%%%%%%%%%%%%%%%%%%%%%%%%%%%%%%%%%%%%%%
\subsection{The $(\mbox{antisymm}; 0)$ representation}
%%%%%%%%%%%%%%%%%%%%%%%%%%%%%%%%%%%%%%%%%%%%%%%%%%%%%%

The relevant subsection on this higher representation
is given by $2.2.4$.
The remaining higher representation obtained from the
product of the minimal representation $(f;0)$ and itself
is given by this higher representation
and the four eigenvalues can be summarized by
\bea
\phi_0^{(1)} (\tiny\yng(1,1);0) & = & -\frac{2 N}{(k+N+2)},
\nonu \\
 v^+ (\tiny\yng(1,1);0) & = & \frac{96 N (k+N-1)}{(k+N+2)^2},
\nonu \\
 v^- (\tiny\yng(1,1);0) & = &  \frac{96 N}{(k+N+2)^2},
\nonu \\
\phi_2^{(1)} (\tiny\yng(1,1);0) & = & -\frac{8 N
  ({\bf 6 k^2 N}+5 k^2 {\bf +12 k N^2}+9 k N-11 k-2 N^2-7 N-12)}{
  3 (k+N+2)^2 (6 k N+5 k+5 N+4)}.
\label{antisymmzerohigher}
\eea
One observes that the first eigenvalue and the third eigenvalue
in (\ref{antisymmzerohigher}) are
the same as the ones in
(\ref{symmzerohigher}) respectively.  
This implies that there is no difference between $(\tiny\yng(2))$
or $(\tiny\yng(1,1))$ for the representation $\La_{+}$.
The third eigenvalue behaves as $\frac{1}{N}$ under the
large $(N,k)$ 't Hooft like limit.
The last eigenvalue shares the common behavior with the one in
(\ref{symmzerohigher}). Although the exact expressions in both cases
are different from each other at finite
(N,k), 
the highest power terms in the numerator
given by $-8N(6 k^2 N +12 k N^2)$ are the same as the ones in
(\ref{symmzerohigher}).
In other words,  they are {\bf twice} of
the ones in (\ref{f0eigenhigher}).
Therefore, this eigenvalue is 
 the sum of each eigenvalue
in the minimal representation
$(\tiny\yng(1);0)$  in (\ref{f0eigenhigher})
under the large $(N,k)$ 't Hooft like limit
\footnote{
\label{antisymmbareigenhigher}
  For the similar higher representation $(
  \overline{\mbox{antisymm}};0)$, the following
  results are satisfied
  \bea
  \phi_0^{(1)} (\overline{\tiny\yng(1,1)};0) & = &
  -\phi_0^{(1)} (\tiny\yng(1,1);0),
\qquad
 v^{\pm} (\overline{\tiny\yng(1,1)};0)  =   v^{\pm} (\tiny\yng(1,1);0),
\qquad
% v^- (\overline{\tiny\yng(2)};0) & = &   v^- (\tiny\yng(2);0),
%\nonu \\
\phi_2^{(1)} (\overline{\tiny\yng(1,1)};0)  = 
-\phi_2^{(1)} (\tiny\yng(1,1);0)
\nonu
\eea
together with (\ref{antisymmzerohigher}).}.

%%%%%%%%%%%%%%%%%%%%%%%%%%%%%%%%%%%%%%%%%%%%%%%%%%
\subsection{The $(0; \mbox{symm})$ representation}
%%%%%%%%%%%%%%%%%%%%%%%%%%%%%%%%%%%%%%%%%%%%%%%%%%

The relevant subsection on this higher representation
is given by $2.2.5$.
This higher representation can be obtained from the
product of the minimal representation $(0;f)$ and itself.
The four eigenvalues with this representation can be
described as 
\bea
\phi_0^{(1)} (0;\tiny\yng(2)) & = & -\frac{2k}{(N+k+2)},
\nonu \\
 v^+ (0;\tiny\yng(2)) & = & \frac{96 k}{(k+N+2)^2},
\nonu \\
 v^- (0;\tiny\yng(2)) & = &  \frac{96 k (k+N-1)}{(k+N+2)^2},
\nonu \\
\phi_2^{(1)} (0;\tiny\yng(2)) & = &
\frac{8 k ( {\bf 12 k^2 N}-2 k^2 {\bf +6 k N^2}+9 k N-7 k+5 N^2-11 N-12)}
     {3 (k+N+2)^2 (6 k N+5 k+5 N+4)}.
\label{zerosymmhigher}
\eea
It is easy to see that the first eigenvalue is the twice of
the one in (\ref{0feigenhigher}) at finite $(N,k)$.
Except the second eigenvalue, the remaining ones in (\ref{zerosymmhigher})
behave as $\la$ dependent constant values
under the large $(N,k)$ 't Hooft like limit.
The first two eigenvalues are related to
the first and the third eigenvalues in (\ref{symmzerohigher})
by considering the exchange of $N$ and $k$.
Furthermore, 
one observes that the highest power terms in the numerator
are given by $8k(12 k^2 N +6 k N^2)$ which is the {\bf twice} of
the ones in (\ref{0feigenhigher}).
Compared to the expression in (\ref{symmzerohigher}),
this quantity can be obtained from the corresponding
one by using the symmetry under the $N \leftrightarrow k$
with extra minus sign.
Note that
the denominators in both expressions are the same
at finite $(N,k)$. This implies that
the eigenvalue for the higher spin $3$  current
in this higher representation can be interpreted as the additive
quantum number and it is given by the sum of each eigenvalue
for the higher representation in the minimal representation
$(0;\tiny\yng(1))$ in (\ref{0feigenhigher})
under the large $(N,k)$ 't Hooft like limit
\footnote{
\label{zerosymmbareigenhigher}
  For the similar complex conjugated higher representation,
one obtains the following results
  \bea
  \phi_0^{(1)} (0;\overline{\tiny\yng(2)}) & = &
  -\phi_0^{(1)} (0;\tiny\yng(2)),
\qquad
 v^{\pm} (0;\overline{\tiny\yng(2)})  =   v^{\pm} (0;\tiny\yng(2)),
\qquad
% v^- (\overline{\tiny\yng(2)};0) & = &   v^- (\tiny\yng(2);0),
%\nonu \\
\phi_2^{(1)} (0;\overline{\tiny\yng(2)})  = 
-\phi_2^{(1)} (0;\tiny\yng(2))
\nonu
\eea
together with (\ref{zerosymmhigher}).  
%\bea
%\phi_0^{(1)} (0;\overline{\tiny\yng(2)}) & = & \frac{2k}{(N+k+2)},
%\nonu \\
%v^+ (0;\overline{\tiny\yng(2)}) & = & \frac{96 k}{(k+N+2)^2},
%\nonu \\
%v^- (0;\overline{\tiny\yng(2)}) & = &  \frac{96 k (k+N-1)}{(k+N+2)^2},
%\nonu \\
%\phi_2^{(1)} (0;\overline{\tiny\yng(2)}) & = &
%-\frac{8 k (12 k^2 N-2 k^2+6 k N^2+9 k N-7 k+5 N^2-11 N-12)}
%{3 (k+N+2)^2 (6 k N+5 k+5 N+4)}
%\nonu
%\eea
}.

%%%%%%%%%%%%%%%%%%%%%%%%%%%%%%%%%%%%%%%%%%%%%%%%%%%%%%%
\subsection{The $(0; \mbox{antisymm})$ representation}
%%%%%%%%%%%%%%%%%%%%%%%%%%%%%%%%%%%%%%%%%%%%%%%%%%%%%%%

The relevant subsection on this higher representation
is given by $2.2.6$.
The remaining higher representation obtained from the
product of the minimal representation $(0;f)$ and itself
is given by this higher representation
and the four eigenvalues can be summarized by
\bea
\phi_0^{(1)} (0;\tiny\yng(1,1)) & = & -\frac{2k}{(N+k+2)},
\nonu \\
v^+ (0;\tiny\yng(1,1)) & = & \frac{96 k}{(k+N+2)^2},
\nonu \\
v^- (0;\tiny\yng(1,1)) & = & \frac{32 k (4 k+3 N+1)}{(k+N+2)^2},
\nonu \\
\phi_2^{(1)} (0;\tiny\yng(1,1)) & = &
\frac{8 k ({\bf 12 k^2 N}-2 k^2 {\bf +6 k N^2}+45 k N-k+5 N^2+43 N+12)}{
  3 (k+N+2)^2 (6 k N+5 k+5 N+4)}.
\label{zeroantisymmhigher}
\eea
One observes that the first eigenvalue and the second eigenvalue are
the same as the ones in
(\ref{zerosymmhigher}) respectively.  
The second eigenvalue behaves as $\frac{1}{N}$ under the
large $(N,k)$ 't Hooft like limit.
The last eigenvalue in (\ref{zeroantisymmhigher})
shares the common behavior with the one in
(\ref{zerosymmhigher}). Although the exact expressions in both cases
are different from each other at finite
$(N,k)$, 
the highest power terms in the numerator
given by $8k(12 k^2 N +6 k N^2)$ are the same as the ones in
(\ref{zerosymmhigher}).
They are {\bf twice} of
the ones in (\ref{0feigenhigher}).
Therefore, this eigenvalue is 
 the sum of each eigenvalue
in the minimal representation
$(0;\tiny\yng(1))$
 in (\ref{0feigenhigher})
under the large $(N,k)$ 't Hooft like limit
\footnote{
\label{zeroantisymmbarexpression}
  For the similar higher representation,
the following results can be obtained
  \bea
  \phi_0^{(1)} (0;\overline{\tiny\yng(1,1)}) & = &
  -\phi_0^{(1)} (0;\tiny\yng(1,1)),
\qquad
 v^{\pm} (0;\overline{\tiny\yng(1,1)})  =   v^{\pm} (0;\tiny\yng(1,1)),
\qquad
% v^- (\overline{\tiny\yng(2)};0) & = &   v^- (\tiny\yng(2);0),
%\nonu \\
\phi_2^{(1)} (0;\overline{\tiny\yng(1,1)})  = 
-\phi_2^{(1)} (0;\tiny\yng(1,1))
\nonu
\eea
together with (\ref{zeroantisymmhigher}).}.
%\bea
%\phi_0^{(1)} (0;\overline{\tiny\yng(1,1)}) & = & \frac{2k}{(N+k+2)},
%\nonu \\
%v^+ (0;\overline{\tiny\yng(1,1)}) & = & \frac{96 k}{(k+N+2)^2},
%\nonu \\
%v^- (0;\overline{\tiny\yng(1,1)}) & = & \frac{32 k (4 k+3 N+1)}{(k+N+2)%^2},
%\nonu \\
%\phi_2^{(1)} (0;\overline{\tiny\yng(1,1)}) & = &
%-
%\frac{8 k (12 k^2 N-2 k^2+6 k N^2+45 k N-k+5 N^2+43 N+12)}{
%  3 (k+N+2)^2 (6 k N+5 k+5 N+4)}
%\nonu
%\eea

%%%%%%%%%%%%%%%%%%%%%%%%%%%%%%%%%%%%%%%%%%%%%%%%%%%%%%%%%%%%%
\subsection{The $(\mbox{symm}; \mbox{symm})$ representation}
%%%%%%%%%%%%%%%%%%%%%%%%%%%%%%%%%%%%%%%%%%%%%%%%%%%%%%%%%%%%%

The relevant subsection on this higher representation
is given by $3.1.1$.
One can describe the following eigenvalues  
\bea
\phi_0^{(1)} (\tiny\yng(2);\tiny\yng(2)) & = & \frac{4}{(N+k+2)},
\nonu \\
 v^+ (\tiny\yng(2);\tiny\yng(2)) & = & \frac{192 (k-1)}{(k+N+2)^2},
\nonu \\
 v^- (\tiny\yng(2);\tiny\yng(2)) & = &  \frac{192 (N+1)}{(k+N+2)^2},
\nonu \\
\phi_2^{(1)} (\tiny\yng(2);\tiny\yng(2)) & = &
\frac{16 ( {\bf 6 k^2 N}+5 k^2 {\bf -6 k N^2}-18 k N+13 k-5 N^2-43 N-12)}{
  3 (k+N+2)^2 (6 k N+5 k+5 N+4)}.
\label{symmsymmnonlinear}
\eea
One can construct the matrices
(\ref{spin2symmetric}),
(\ref{su2symmetric}) and (\ref{u1chargesymm})
for the corresponding (higher spin) currents.
Under the symmetry $ N \leftrightarrow k$ (with $\tiny\yng(2)
\leftrightarrow \tiny\yng(2)$),
the first eigenvalue remains the same, and 
the second eigenvalue becomes the third one and  
the third eigenvalue becomes the second one
by ignoring
the constant term in the numerator.
The last eigenvalue in (\ref{symmsymmnonlinear}) remains the same with an extra sign change
if we consider only the case where the total power of $N$ and $k$
is given by $3$: $(6k^2N-6k N^2)=6(k-N)k N$.
Then one can see that the eigenvalue of the higher spin $3$ current
is the {\bf twice} of the one in (\ref{ffeigenvalues1}) under the large
$(N,k)$ 't Hooft like limit.
By power counting of $N$ and $k$, one sees that the above eigenvalues
behave as $\frac{1}{N}$ dependence under the large $(N,k)$ 't Hooft like
limit \footnote{
\label{symmbarsymmbarfoot}
  The following relations hold
  \bea
  \phi_0^{(1)} (\overline{\tiny\yng(2)};\overline{\tiny\yng(2)}) & = &
  -\phi_0^{(1)} (\tiny\yng(2);\tiny\yng(2)),
\qquad
v^{\pm} (\overline{\tiny\yng(2)};\overline{\tiny\yng(2)})  = 
v^{\pm} (\tiny\yng(2);\tiny\yng(2)),
\qquad
\phi_2^{(1)} (\overline{\tiny\yng(2)};\overline{\tiny\yng(2)})  = 
-\phi_2^{(1)} (\tiny\yng(2);\tiny\yng(2))
  \nonu
  \eea
with (\ref{symmsymmnonlinear}).}.

%%%%%%%%%%%%%%%%%%%%%%%%%%%%%%%%%%%%%%%%%%%%%%%%%%%%%
\subsection{The $(\mbox{symm}; f)$ representation}
%%%%%%%%%%%%%%%%%%%%%%%%%%%%%%%%%%%%%%%%%%%%%%%%%%%%%

The relevant subsection on this higher representation
is given by $3.1.2$.
The following eigenvalues in this higher representation
can be determined by
\bea
\phi_0^{(1)} (\tiny\yng(2);\tiny\yng(1)) & = & -\frac{(N-2)}{(N+k+2)},
\nonu \\
v^+ (\tiny\yng(2);\tiny\yng(1)) & = & \frac{12
  (4 k N+16 k+5 N^2+14 N+12)}{(k+N+2)^2},
\nonu \\
 v^- (\tiny\yng(2);\tiny\yng(1)) & = &  \frac{24 (5 N-4)}{(k+N+2)^2},
\nonu \\
\phi_2^{(1)} (\tiny\yng(2);\tiny\yng(1)) & = & -\frac{4
  }{3 (k+N+2)^2 (6 k N+5 k+5 N+4)}
\nonu \\
& \times & ({\bf 6 k^2 N^2}-7 k^2 N-10 k^2 {\bf +12 k N^3}+87 k N^2+106 k N-44 k
\nonu \\
& + & 4 N^3+
  42 N^2+140 N+24).
\label{symmfnonlinear}
\eea
One observes that the first, the second and the last eigenvalues
in (\ref{symmfnonlinear})
coincide with the ones in (\ref{f0eigenhigher}) if
one takes the higher order terms in the numerators
respectively.
The third eigenvalue behaves as $\frac{1}{N}$
under the large $(N,k)$ 't Hooft like limit
\footnote{
\label{symmfrelated}
  The following relations can be obtained
  \bea
  \phi_0^{(1)} (\overline{\tiny\yng(2)};\overline{\tiny\yng(1)}) & = &
  -\phi_0^{(1)} (\tiny\yng(2);\tiny\yng(1)),
\qquad
v^{\pm} (\overline{\tiny\yng(2)};\overline{\tiny\yng(1)})  = 
v^{\pm} (\tiny\yng(2);\tiny\yng(1)),
\qquad
\phi_2^{(1)} (\overline{\tiny\yng(2)};\overline{\tiny\yng(1)})  = 
-\phi_2^{(1)} (\tiny\yng(2);\tiny\yng(1))
  \nonu
  \eea
with (\ref{symmfnonlinear}).}.

%%%%%%%%%%%%%%%%%%%%%%%%%%%%%%%%%%%%%%%%%%%%%%%%%%%%%%%%%%%%%%
\subsection{The $(\mbox{symm}; \overline{f})$ representation}
%%%%%%%%%%%%%%%%%%%%%%%%%%%%%%%%%%%%%%%%%%%%%%%%%%%%%%%%%%%%%%

The relevant subsection on this higher representation
is given by $3.1.4$.
The four eigenvalues corresponding to the zero modes
of the higher spin currents of spins $1, 4, 4$ and $3$
which act on
the representation $(\tiny\yng(2); \overline{\tiny\yng(1)})$
can be summarized by
{\small  
\bea
\phi_0^{(1)} (\tiny\yng(2);\overline{\tiny\yng(1)}) & = &
\phi_0^{(1)} (\tiny\yng(2);0)+
\phi_0^{(1)} (0;\overline{\tiny\yng(1)})=
\frac{(k-2N)}
    {(k+N+2)},
\nonu \\
 v^+ (\tiny\yng(2);\overline{\tiny\yng(1)}) & = & \frac{8 (12 k N+3 k+16 N^2-52 N+20)}{(k+N+2)^2},
\nonu \\
v^- (\tiny\yng(2);\overline{\tiny\yng(1)}) & = &  \frac{12
  (5 k^2+4 k N-58 k+8 N+20)}{(k+N+2)^2},
\nonu \\
\phi_2^{(1)} (\tiny\yng(2);\overline{\tiny\yng(1)}) & = &
-\frac{4 }{3 (k+N+2)^2 (6 k N+5 k+5 N+4)}
\nonu \\
& \times &
({\bf 12 k^3 N}+4 k^3 {\bf +18 k^2 N^2}-35 k^2 N-64 k^2
{\bf +24 k N^3}+35 k N^2\nonu \\
& - & 132 k N
-180 k-4 N^3-62 N^2-120 N-96).
\label{symmfbarhigher}
\eea}
The relations (\ref{zerofbareigenhigher}) and (\ref{symmzerohigher})
are used.
Note that the eigenvalue for the higher spin $1$ current in
(\ref{symmfbarhigher})
does not have any contribution from the commutator
$[(\Phi_0^{(1)})_0, Q^{\bar{A}}_{-\frac{1}{2}}]$ because
the OPE between the corresponding higher spin $1$ current
and the spin $\frac{1}{2}$ current has only the first order pole.
This provides only the eigenvalue for the representation $(0;\overline{
  \tiny\yng(1)})$. Also the term 
$ Q^{\bar{A}}_{-\frac{1}{2}} (\Phi_0^{(1)})_0$
acting on the representation $(\tiny\yng(2);0)$
gives the eigenvalue $\phi_0^{(1)} (\tiny\yng(2);0)$ with
$ Q^{\bar{A}}_{-\frac{1}{2}}$ acting on the state  $|(\tiny\yng(2);0)>$.
By inserting the overall factor into this state, one has
the final state associated with 
the representation $(\tiny\yng(2);\overline{\tiny\yng(1)})$.
Therefore, one arrives at the above eigenvalue for the higher spin $1$
current.
All the eigenvalues survive even under the
large $(N,k)$ 't Hooft like limit.
As expected, one can rewrite the highest power terms in terms of
$(24k N^3 + 12 k^2 N^2)$ and $k(12k^2 N +6 k N^2)$.
The former comes from (\ref{symmzerohigher}) and the latter comes from
(\ref{zerofbareigenhigher}).

For the eigenvalues corresponding to the remaining higher spin currents,
there are the contributions from the lower order poles
appearing in the commutators,
$[(V^+)_0, Q^{\bar{A}}_{-\frac{1}{2}}]$,
$[(V^-)_0, Q^{\bar{A}}_{-\frac{1}{2}}]$ and $[(\Phi_2^{(1)})_0,
  Q^{\bar{A}}_{-\frac{1}{2}}]$. 
They can be summarized by
\bea
%& = & \frac{k}{(k+N+2)},
%\nonu \\
\delta v^+ (\tiny\yng(2);\overline{\tiny\yng(1)})
& = & -\frac{32 (14 N-5)}{(k+N+2)^2},
\nonu \\
\delta
v^- (\tiny\yng(2);\overline{\tiny\yng(1)})
& = &  -\frac{240 (3 k-1)}{(k+N+2)^2},
\nonu \\
\delta
\phi_2^{(1)} (\tiny\yng(2);\overline{\tiny\yng(1)})
& = & \frac{8
  ({\bf 14 k^2 N}+13 k^2 {\bf +10 k N^2}+41 k N+32 k+10 N^2+24 N+16)}{
  (k+N+2)^2 (6 k N+5 k+5 N+4)}.
\label{symmfbarcorrection}
\eea
In particular, the second and last eigenvalues
in (\ref{symmfbarcorrection})
are twice of the ones in (\ref{ffbarcorrection1}) as long as the
highest power of $(N,k)$ is concerned. 
Then as done in (\ref{ffbarcorrection}), the final
eigenvalues are determined in (\ref{symmfbarhigher}). 
Of course, the eigenvalues in (\ref{symmfbarcorrection})
vanish under the large $(N,k)$ 't Hooft like limit
\footnote{
\label{symmbarfrelation}
  One obtains
  \bea
%& = & \frac{k}{(k+N+2)},
%\nonu \\
\delta v^+ (\overline{\tiny\yng(2)};\tiny\yng(1))
& = & \frac{32 (14 N+5)}{(k+N+2)^2},
\qquad
\delta v^- (\overline{\tiny\yng(2)};\tiny\yng(1))
 =  \frac{240 (3 k+1)}{(k+N+2)^2},
\nonu \\
\delta \phi_2^{(1)} (\overline{\tiny\yng(2)};\tiny\yng(1))& = &
-\frac{8 ({\bf 10 k^2 N}+5 k^2 {\bf +2 k N^2}+19 k N+22 k+8 N^2+6 N+8)}
{(k+N+2)^2 (6 k N+5 k+5 N+4)}.
\label{symmbarfeigehhighercorrection1}
\eea
Here
 the commutators,
 $[(V^{\pm})_0, Q^{\bar{A}^{\ast}}_{-\frac{1}{2}}]$ and
 $[(\Phi_2^{(1)})_0,
  Q^{\bar{A}^{\ast}}_{-\frac{1}{2}}]$ are used. 
Then one also has
\bea
\phi_0^{(1)} (\overline{\tiny\yng(2)};\tiny\yng(1)) & = &
\phi_0^{(1)} (\overline{\tiny\yng(2)};0) +
\phi_0^{(1)} (0;\tiny\yng(1)), \nonu \\
v^{\pm} (\overline{\tiny\yng(2)};\tiny\yng(1))  & = &
v^{\pm} (\overline{\tiny\yng(2)};0)+
v^{\pm} (0;\tiny\yng(1)) +\delta v^{\pm}
(\overline{\tiny\yng(2)};\tiny\yng(1)),\nonu \\
\phi_2^{(1)} (\overline{\tiny\yng(2)};\tiny\yng(1)) & = &
\phi_2^{(1)} (\overline{\tiny\yng(2)};0)+
\phi_2^{(1)} (0;\tiny\yng(1))
+ \delta \phi_2^{(1)} (\overline{\tiny\yng(2)};\tiny\yng(1)),
\nonu
\eea
together with the footnote
\ref{symmbareigenhigher}, (\ref{0feigenhigher}),
and (\ref{symmbarfeigehhighercorrection1}).
}.
As before, the
relations (\ref{symmfbarhigher})
can be determined 
by (\ref{zerofbareigenhigher}) and (\ref{symmzerohigher})
together with (\ref{symmfbarcorrection}).
%\footnote{
%From the relations
%  \bea
%\delta
%v^{\pm} (\overline{\tiny\yng(2)};\overline{\tiny\yng(1)})
%& = & \delta v^{\pm} (\tiny\yng(2);\overline{\tiny\yng(1)}),
%\qquad
%\delta
%\phi_2^{(1)} (\overline{\tiny\yng(2)};\overline{\tiny\yng(1)})
% =  - \delta \phi_2^{(1)} (\tiny\yng(2);\overline{\tiny\yng(1)}),
%\label{cor2}
%\eea
%one has
%\bea
%\phi_0^{(1)} (\overline{\tiny\yng(2)};\overline{\tiny\yng(1)}) & = &
%\phi_0^{(1)} (\overline{\tiny\yng(2)};0) +
%\phi_0^{(1)} (0;\overline{\tiny\yng(1)}), \nonu \\
%v^{\pm} (\overline{\tiny\yng(2)};\overline{\tiny\yng(1)})  & = &
%v^{\pm} (\overline{\tiny\yng(2)};0)+
%v^{\pm} (0;\overline{\tiny\yng(1)}) +\delta v^{\pm}
%(\overline{\tiny\yng(2)};\overline{\tiny\yng(1)}),\nonu \\
%\phi_2^{(1)} (\overline{\tiny\yng(2)};\overline{\tiny\yng(1)}) & = &
%\phi_2^{(1)} (\overline{\tiny\yng(2)};0)+
%\phi_2^{(1)} (0;\overline{\tiny\yng(1)})
%+ \delta \phi_2^{(1)} (\overline{\tiny\yng(2)};\overline{\tiny\yng(1)}).
%\nonu
%\eea
%together with the footnote \ref{symmbareigenhigher}, (\ref{cor2}) and 
%(\ref{zerofbareigenhigher}).}. 

%%%%%%%%%%%%%%%%%%%%%%%%%%%%%%%%%%%%%%%%%%%%%%%%%%%%%%%%%%%%%%%%%%
\subsection{The $(\mbox{symm}; \mbox{antisymm})$ representation}
%%%%%%%%%%%%%%%%%%%%%%%%%%%%%%%%%%%%%%%%%%%%%%%%%%%%%%%%%%%%%%%%%%

The relevant subsection on this higher representation
is given by $3.1.5$.
The four eigenvalues are characterized by
\bea
\phi_0^{(1)} (\tiny\yng(2);\tiny\yng(1,1)) & = &
\phi_0^{(1)} (\tiny\yng(2);0)+
\phi_0^{(1)} (0;\tiny\yng(1,1))=
-\frac{2 (k+N)}
    {(k+N+2)},
\nonu \\
v^+ (\tiny\yng(2);\tiny\yng(1,1)) & = & \frac{32
  (3 k N+3 k+4 N^2+29 N+20)}{(k+N+2)^2},
\nonu \\
v^- (\tiny\yng(2);\tiny\yng(1,1)) & = &  \frac{32
  (4 k^2+3 k N+53 k+3 N+20)}{(k+N+2)^2},
\nonu \\
\phi_2^{(1)} (\tiny\yng(2);\tiny\yng(1,1)) & = &
\frac{8 }{3 (k+N+2)^2 (6 k N+5 k+5 N+4)}
\nonu \\
& \times & ({\bf 12 k^3 N}-2 k^3+100 k^2 N-7 k^2
{\bf -12 k N^3}-28 k N^2+132 k N
\nonu \\
& + & 144 k+2 N^3+67 N^2+24 N+48).
\label{symmantisymmnonlinear}
\eea
The relations in (\ref{symmzerohigher})
and (\ref{zeroantisymmhigher}) are used.
In the third eigenvalue of (\ref{symmantisymmnonlinear}),
the quantities in the highest power of
$(N, k)$ are the same as the ones in (\ref{fantisymmeigenhigher}).
For the highest power of $(N,k)$ in the last eigenvalue,
the $12 k^3 N$ is the same as the one in (\ref{fantisymmeigenhigher})
while the $-12 k N^3$ is the twice of the one in
(\ref{fantisymmeigenhigher}).

Here the relevant extra contributions from the various commutators
are used in these calculations
\bea
\delta v^+ (\tiny\yng(2);\tiny\yng(1,1)) & = &
\frac{128 (7 N+5)}{(k+N+2)^2},
\nonu \\
\delta
v^- (\tiny\yng(2);\tiny\yng(1,1))
& = &
\frac{128 (13 k+5)}{(k+N+2)^2},
\nonu \\
\delta
\phi_2^{(1)} (\tiny\yng(2);\tiny\yng(1,1))
& = & \frac{16 ({\bf 10 k^2 N}-k^2 {\bf +2 k N^2}+22 k N+22 k+11 N^2+6 N+8)}
     {(k+N+2)^2 (6 k N+5 k+5 N+4)}.
\label{corrcorr}
\eea
The last eigenvalue in (\ref{corrcorr}) for the highest power of
$(N,k)$ is twice of the one in (\ref{fantisymmeigenhighercorrection})
\footnote{
\label{cor3foot}
  From the relations
  \bea
\delta
v^{\pm} (\overline{\tiny\yng(2)};\tiny\yng(1,1))
& = & \delta v^{\pm} (\tiny\yng(2);\tiny\yng(1,1)),
\qquad
\delta
\phi_2^{(1)} (\overline{\tiny\yng(2)};\tiny\yng(1,1))
 =  - \delta \phi_2^{(1)} (\tiny\yng(2);\tiny\yng(1,1)),
\label{cor3}
\eea
one has
\bea
\phi_0^{(1)} (\overline{\tiny\yng(2)};\tiny\yng(1,1)) & = &
\phi_0^{(1)} (\overline{\tiny\yng(2)};0) +
\phi_0^{(1)} (0;\tiny\yng(1,1)), \nonu \\
v^{\pm} (\overline{\tiny\yng(2)};\tiny\yng(1,1))  & = &
v^{\pm} (\overline{\tiny\yng(2)};0)+
v^{\pm} (0;\tiny\yng(1,1)) +\delta v^{\pm}
(\overline{\tiny\yng(2)};\tiny\yng(1,1)),\nonu \\
\phi_2^{(1)} (\overline{\tiny\yng(2)};\tiny\yng(1,1)) & = &
\phi_2^{(1)} (\overline{\tiny\yng(2)};0)+
\phi_2^{(1)} (0;\tiny\yng(1,1))
+ \delta \phi_2^{(1)} (\overline{\tiny\yng(2)};\tiny\yng(1,1)),
\nonu
\eea
together with the footnote \ref{symmbareigenhigher}, (\ref{cor3}) and 
(\ref{zeroantisymmhigher}).}.
In (\ref{symmantisymmnonlinear}),
the relations (\ref{symmzerohigher}),
(\ref{zeroantisymmhigher})
and (\ref{corrcorr}) are combined together. 

%%%%%%%%%%%%%%%%%%%%%%%%%%%%%%%%%%%%%%%%%%%%%%%%%%%%%%%%%%%%%%
\subsection{The $(\mbox{symm}; \overline{\mbox{antisymm}})$ representation}
%%%%%%%%%%%%%%%%%%%%%%%%%%%%%%%%%%%%%%%%%%%%%%%%%%%%%%%%%%%%%%%

The foue eigenvalues are given by \footnote{One can analyze the
  eigenvalues on this particular higher representation in the context of
sections $2$ and $3$.}
\bea
\phi_0^{(1)} (\tiny\yng(2);\overline{\tiny\yng(1,1)}) & = &
\phi_0^{(1)} (\tiny\yng(2);0)+
\phi_0^{(1)} (0;\overline{\tiny\yng(1,1)})=
-\frac{2 (-k+N)}
    {(k+N+2)},
\nonu \\
v^+ (\tiny\yng(2);\overline{\tiny\yng(1,1)}) & = & \frac{32 (3 k N+3 k+4 N^2-27 N+20)}{(k+N+2)^2},
\nonu \\
v^- (\tiny\yng(2);\overline{\tiny\yng(1,1)}) & = &  \frac{32 (4 k^2+3 k N-51 k+3 N+20)}{(k+N+2)^2},
\nonu \\
\phi_2^{(1)} (\tiny\yng(2);\overline{\tiny\yng(1,1)}) & = &
-\frac{8}{3 (k+N+2)^2 (6 k N+5 k+5 N+4)}
\nonu \\
& \times &
({\bf 12 k^3 N}-2 k^3 {\bf +12 k^2 N^2}-34 k^2 N-43 k^2 {\bf +12 k N^3}
-10 k N^2
\nonu \\
& - & 178 k N-180 k-2 N^3-79 N^2-132 N-96).
\label{symmantisymmbarnonlinear}
\eea
The third eigenvalue of (\ref{symmantisymmbarnonlinear})
in the highest power of $(N,k)$
is the same as the one in (\ref{fantisymmbareigenhigher}).
The highest power terms in the last eigenvalue of
(\ref{symmantisymmbarnonlinear})
are twice of the ones in (\ref{nonlinearffbar}).

We have
\bea
\delta v^+ (\tiny\yng(2);\overline{\tiny\yng(1,1)})
& = & -\frac{128 (7 N-5)}{(k+N+2)^2},
\nonu \\
\delta
v^- (\tiny\yng(2);\overline{\tiny\yng(1,1)})
& = & -\frac{128 (13 k-5)}{(k+N+2)^2},
\nonu \\
\delta
\phi_2^{(1)} (\tiny\yng(2);\overline{\tiny\yng(1,1)})
& = &
\frac{16 ({\bf 14 k^2 N}+7 k^2 {\bf +10 k N^2}+44 k N+32 k+13 N^2+24 N+16)}{
  (k+N+2)^2 (6 k N+5 k+5 N+4)}.
\label{deldel}
\eea
The highest powers of $(N,k)$ in the second and third eigenvalues
in (\ref{deldel}) are twice of the ones in
(\ref{fantisymmbareigenhighercorrection})
\footnote{
\label{cor4foot}
  From the relations
  \bea
\delta
v^{\pm} (\overline{\tiny\yng(2)};\overline{\tiny\yng(1,1)})
& = & \delta v^{\pm} (\tiny\yng(2);\overline{\tiny\yng(1,1)}),
\qquad
\delta
\phi_2^{(1)} (\overline{\tiny\yng(2)};\overline{\tiny\yng(1,1)})
 =  - \delta \phi_2^{(1)} (\tiny\yng(2);\overline{\tiny\yng(1,1)}),
\label{cor4}
\eea
one has
\bea
\phi_0^{(1)} (\overline{\tiny\yng(2)};\overline{\tiny\yng(1,1)}) & = &
\phi_0^{(1)} (\overline{\tiny\yng(2)};0) +
\phi_0^{(1)} (0;\overline{\tiny\yng(1,1)}), \nonu \\
v^{\pm} (\overline{\tiny\yng(2)};\overline{\tiny\yng(1,1)})  & = &
v^{\pm} (\overline{\tiny\yng(2)};0)+
v^{\pm} (0;\overline{\tiny\yng(1,1)}) +\delta v^{\pm}
(\overline{\tiny\yng(2)};\overline{\tiny\yng(1,1)}),\nonu \\
\phi_2^{(1)} (\overline{\tiny\yng(2)};\overline{\tiny\yng(1,1)}) & = &
\phi_2^{(1)} (\overline{\tiny\yng(2)};0)+
\phi_2^{(1)} (0;\overline{\tiny\yng(1,1)})
+ \delta \phi_2^{(1)} (\overline{\tiny\yng(2)};\overline{\tiny\yng(1,1)}),
\nonu
\eea
together with the footnote \ref{symmbareigenhigher}, (\ref{cor4}) and 
the footnote \ref{zeroantisymmbarexpression}.}. 
In (\ref{symmantisymmbarnonlinear}),
the relations (\ref{symmzerohigher}),
the footnote \ref{zeroantisymmbarexpression}
and (\ref{deldel}) are combined together. 

%%%%%%%%%%%%%%%%%%%%%%%%%%%%%%%%%%%%%%%%%%%%%%%%%%%%%%%%%%%%%%%%%%%%%%%%
\subsection{The $(\mbox{symm}; \overline{\mbox{symm}})$ representation}
%%%%%%%%%%%%%%%%%%%%%%%%%%%%%%%%%%%%%%%%%%%%%%%%%%%%%%%%%%%%%%%%%%%%%%%%

The relevant subsection on this higher representation
is given by $3.1.6$.
The four eigenvalues are
\bea
\phi_0^{(1)} (\tiny\yng(2);\overline{\tiny\yng(2)}) & = &
\phi_0^{(1)} (\tiny\yng(2);0)+
\phi_0^{(1)} (0;\overline{\tiny\yng(2)})=
\frac{2(-N+k)}{(k+N+2)},
\nonu \\
v^+ (\tiny\yng(2);\overline{\tiny\yng(2)}) & = &
\frac{32 (3 k N+3 k+4 N^2-27 N+20)}{(k+N+2)^2},
\nonu \\
v^- (\tiny\yng(2);\overline{\tiny\yng(2)}) & = &  \frac{96
  (k^2+k N-9 k+N)}{(k+N+2)^2},
\nonu \\
\phi_2^{(1)} (\tiny\yng(2);\overline{\tiny\yng(2)}) & = &
-\frac{8}{3 (k+N+2)^2 (6 k N+5 k+5 N+4)} \nonu \\
& \times & ({\bf 12 k^3 N}-2 k^3 {\bf +12 k^2 N^2}-70 k^2 N-49 k^2
{\bf +12 k N^3}
\nonu \\
& - & 10 k N^2-88 k N-24 k-2 N^3-7 N^2+48 N+48).
\label{symmsymmbarnonlinear}
\eea
The last eigenvalue with boldface notation in
(\ref{symmsymmbarnonlinear})
is the same as the one in
(\ref{symmantisymmbarnonlinear}). 
The third eigenvalue has the common behavior
with the one in (\ref{fsymmbarnonlinear}) under the large $(N,k)$
't Hooft like limit.
Note that the eigenvalue for the higher spin $1$ current
does not have any contribution from the commutator
$[(\Phi_0^{(1)})_0, Q^{1}_{-\frac{1}{2}}  Q^{4}_{-\frac{1}{2}}]$ because
the OPE between the corresponding higher spin $1$ current
and the spin $\frac{1}{2}$ current has only the first order pole.
This provides only the eigenvalue for the representation $(0;\overline{
  \tiny\yng(2)})$. Also the term 
$ Q^{1}_{-\frac{1}{2}}  Q^{4}_{-\frac{1}{2}} (\Phi_0^{(1)})_0$
acting on the representation $(\tiny\yng(2);0)$
gives the eigenvalue $\phi_0^{(1)} (\tiny\yng(2);0)$ with
$ Q^{1}_{-\frac{1}{2}}  Q^{4}_{-\frac{1}{2}}$
acting on the state  $|(\tiny\yng(2);0)>$.
By inserting the overall factor into this state, one has
the final state associated with 
the representation $(\tiny\yng(2);\overline{\tiny\yng(2)})$.
Therefore, one arrives at the above eigenvalue for the higher spin $1$
current.

For the eigenvalues corresponding to the remaining higher spin currents,
there are the contributions from the lower order poles
appearing in the commutators,
$[(V^+)_0, Q^{1}_{-\frac{1}{2}} Q^{4}_{-\frac{1}{2}}]$,
$[(V^-)_0,  Q^{1}_{-\frac{1}{2}} Q^{4}_{-\frac{1}{2}}]$
and $[(\Phi_2^{(1)})_0,  Q^{1}_{-\frac{1}{2}} Q^{4}_{-\frac{1}{2}}]$. 
They can be summarized by
\bea
\delta
v^+ (\tiny\yng(2);\overline{\tiny\yng(2)})
& = & -\frac{128 (7 N-5)}{(k+N+2)^2}, \nonu \\
\delta
v^- (\tiny\yng(2);\overline{\tiny\yng(2)})
& = & -\frac{768 k}{(k+N+2)^2}, \nonu \\
\delta
\phi_2^{(1)} (\tiny\yng(2);\overline{\tiny\yng(2)})
& = & \frac{16 ({\bf 14 k^2 N}+7 k^2 {\bf +10 k N^2}
  +20 k N+2 k+N^2-6 N-8)}{
  (k+N+2)^2 (6 k N+5 k+5 N+4)}.
\label{symmsymmbarhigher}
\eea
In particular, the last eigenvalue in (\ref{symmsymmbarhigher}) for the highest power
of $(N,k)$ in the numerator is exactly four times
the ones in (\ref{ffbarcorrection1}) or twice of
the ones in (\ref{symmfbarcorrection})
\footnote{
\label{footnewcorrection}
  One obtains
  \bea
%& = & \frac{k}{(k+N+2)},
%\nonu \\
\delta v^+ (\overline{\tiny\yng(2)};\tiny\yng(2))
& = & \frac{128 (7 N+5)}{(k+N+2)^2},
\qquad
\delta v^- (\overline{\tiny\yng(2)};\tiny\yng(2))
 =  \frac{768 k}{(k+N+2)^2},
\nonu \\
\delta \phi_2^{(1)} (\overline{\tiny\yng(2)};\tiny\yng(2))& = &
-\frac{16 ({\bf 10 k^2 N}-k^2{\bf +2 k N^2}-2 k N-8 k-N^2-24 N-16)}
{(k+N+2)^2 (6 k N+5 k+5 N+4)}.
\label{newcorrection}
\eea
The commutators
$[(V^{\pm})_0,  Q^{13}_{-\frac{1}{2}} Q^{16}_{-\frac{1}{2}}]$
and $[(\Phi_2^{(1)})_0,  Q^{13}_{-\frac{1}{2}} Q^{16}_{-\frac{1}{2}}]$
are used.
Then one also has
\bea
\phi_0^{(1)} (\overline{\tiny\yng(2)};\tiny\yng(2)) & = &
\phi_0^{(1)} (\overline{\tiny\yng(2)};0) +
\phi_0^{(1)} (0;\tiny\yng(2)), \nonu \\
v^{\pm} (\overline{\tiny\yng(2)};\tiny\yng(2))  & = &
v^{\pm} (\overline{\tiny\yng(2)};0)+
v^{\pm} (0;\tiny\yng(2)) +\delta v^{\pm}
(\overline{\tiny\yng(2)};\tiny\yng(2)),\nonu \\
\phi_2^{(1)} (\overline{\tiny\yng(2)};\tiny\yng(2)) & = &
\phi_2^{(1)} (\overline{\tiny\yng(2)};0)+
\phi_2^{(1)} (0;\tiny\yng(2))
+ \delta \phi_2^{(1)} (\overline{\tiny\yng(2)};\tiny\yng(2)),
\nonu
\eea
together with the footnote
\ref{symmbareigenhigher}, (\ref{zerosymmhigher}),
and (\ref{newcorrection}).
}.
Note that the second eigenvalue of (\ref{symmsymmbarhigher})
is the twice of the one in (\ref{fsymmbareigenhighercorrection}).
The relations (\ref{symmzerohigher}), the footnote
\ref{zerosymmbareigenhigher} and
(\ref{symmsymmbarhigher}) are used
in (\ref{symmsymmbarnonlinear}).

%%%%%%%%%%%%%%%%%%%%%%%%%%%%%%%%%%%%%%%%%%%%%%%%%%%%%%%%%%%%%%%%%%%%%
\subsection{The $(\mbox{antisymm}; \mbox{antisymm})$ representation}
%%%%%%%%%%%%%%%%%%%%%%%%%%%%%%%%%%%%%%%%%%%%%%%%%%%%%%%%%%%%%%%%%%%%%%

The relevant subsection on this higher representation
is given by $3.2.1$.
The four eigenvalues are given by
\bea
\phi_0^{(1)} (\tiny\yng(1,1);\tiny\yng(1,1)) & = & \frac{4}{(k+N+2)},
\nonu \\
v^+ (\tiny\yng(1,1);\tiny\yng(1,1)) & = & \frac{192 (k+1)}{(k+N+2)^2},
\nonu \\
 v^- (\tiny\yng(1,1);\tiny\yng(1,1)) & = &   \frac{192 (N-1)}{(k+N+2)^2},
\nonu \\
\phi_2^{(1)} (\tiny\yng(1,1);\tiny\yng(1,1)) & = & \frac{16
  ( {\bf 6 k^2 N}+5 k^2 {\bf -6 k N^2}+18 k N+43 k-5 N^2-13 N+12)}
    {3 (k+N+2)^2 (6 k N+5 k+5 N+4)}.
\label{antiantinonlinear}
\eea
Under the symmetry $ N \leftrightarrow k$ (with $\tiny\yng(1,1)
\leftrightarrow \tiny\yng(1,1)$),
the first eigenvalue in (\ref{antiantinonlinear}) remains the same, and 
the second eigenvalue becomes the third one, and  
the third eigenvalue becomes the second one
by ignoring
the constant term in the numerator.
The last eigenvalue in (\ref{antiantinonlinear}) remains the same with an extra sign change
if we consider only the case where the total power of $N$ and $k$
is given by $3$: $(6k^2N-6k N^2)$. 
Compared to the ones in (\ref{symmsymmnonlinear}), the eigenvalues
looks similar to each other.
The large $(N,k)$ behavior is the same.
By power counting of $N$ and $k$, one sees that the above eigenvalues
behave as $\frac{1}{N}$ dependence under the large $(N,k)$ 't Hooft like
limit
\footnote{
\label{antibarantibarfoot}
  It turns out that the following relations hold
  \bea
  \phi_0^{(1)} (\overline{\tiny\yng(1,1)};\overline{\tiny\yng(1,1)}) & = &
  -\phi_0^{(1)} (\tiny\yng(1,1);\tiny\yng(1,1)),
\qquad
v^{\pm} (\overline{\tiny\yng(1,1)};\overline{\tiny\yng(1,1)})  = 
v^{\pm} (\tiny\yng(1,1);\tiny\yng(1,1)),
\qquad
\phi_2^{(1)} (\overline{\tiny\yng(1,1)};\overline{\tiny\yng(1,1)})  = 
-\phi_2^{(1)} (\tiny\yng(1,1);\tiny\yng(1,1))
  \nonu
  \eea
with (\ref{antiantinonlinear}).}.

%%%%%%%%%%%%%%%%%%%%%%%%%%%%%%%%%%%%%%%%%%%%%%%%%%%%%%%%%
\subsection{The $(\mbox{antisymm}; f)$ representation}
%%%%%%%%%%%%%%%%%%%%%%%%%%%%%%%%%%%%%%%%%%%%%%%%%%%%%%%%

The relevant subsection on this higher representation
is given by $3.2.2$. The four eigenvalues are
\bea
\phi_0^{(1)} (\tiny\yng(1,1);\tiny\yng(1)) & = & -\frac{(N-2)}{(k+N+2)},
\nonu \\
v^+ (\tiny\yng(1,1);\tiny\yng(1)) & = & \frac{12
  (4 k N+5 N^2-10 N+12)}{(k+N+2)^2},
\nonu \\
 v^- (\tiny\yng(1,1);\tiny\yng(1)) & = &   \frac{24 (5 N-4)}{(k+N+2)^2},
\label{antifnonlinear}
 \\
\phi_2^{(1)} (\tiny\yng(1,1);\tiny\yng(1)) & = & -\frac{4 (N-2)
  ({\bf 6 k^2 N}+5 k^2 {\bf +12 k N^2}+39 k N+28 k+4 N^2+14 N+12)}{
  3 (k+N+2)^2 (6 k N+5 k+5 N+4)}.
\nonu
\eea
One observes that the first, the second and the last eigenvalues
in (\ref{antifnonlinear}) coincide with the ones in (\ref{f0eigenhigher}) if
one takes the higher order terms in the numerators
respectively.
The third eigenvalue behaves as $\frac{1}{N}$
under the large $(N,k)$ 't Hooft like limit
\footnote{
  \label{antisymmantisymmfoot}
  It turns out that the following relations hold
  \bea
  \phi_0^{(1)} (\overline{\tiny\yng(1,1)};\overline{\tiny\yng(1)}) & = &
  -\phi_0^{(1)} (\tiny\yng(1,1);\tiny\yng(1)),
\qquad
v^{\pm} (\overline{\tiny\yng(1,1)};\overline{\tiny\yng(1)})  = 
v^{\pm} (\tiny\yng(1,1);\tiny\yng(1)),
\qquad
\phi_2^{(1)} (\overline{\tiny\yng(1,1)};\overline{\tiny\yng(1)})  = 
-\phi_2^{(1)} (\tiny\yng(1,1);\tiny\yng(1))
  \nonu
  \eea
with (\ref{antifnonlinear}).}.
The large $(N,k)$ behavior of the eigenvalue for the higher spin $3$
current is the same as the one in (\ref{symmfnonlinear}). 

%%%%%%%%%%%%%%%%%%%%%%%%%%%%%%%%%%%%%%%%%%%%%%%%%%%%%%%%%%%%%%%%%%
\subsection{The $(\mbox{antisymm}; \overline{f})$ representation}
%%%%%%%%%%%%%%%%%%%%%%%%%%%%%%%%%%%%%%%%%%%%%%%%%%%%%%%%%%%%%%%%%%

The relevant subsection on this higher representation
is given by $3.2.4$. The four eigenvalues are given by
{\small
\bea
\phi_0^{(1)} (\tiny\yng(1,1);\overline{\tiny\yng(1)}) & = &
\phi_0^{(1)} (\tiny\yng(1,1);0)+
\phi_0^{(1)} (0;\overline{\tiny\yng(1)})=
\frac{ (k-2N)}{k+N+2},
\nonu \\
v^+ (\tiny\yng(1,1);\overline{\tiny\yng(1)}) & = & \frac{24
  (4 k N+k+4 N^2-20 N)}{(k+N+2)^2},
\nonu \\
v^- (\tiny\yng(1,1);\overline{\tiny\yng(1)}) & = &
\frac{12 (5 k^2+4 k N-58 k+8 N+20)}{(k+N+2)^2},
\nonu \\
\phi_2^{(1)} (\tiny\yng(1,1);\overline{\tiny\yng(1)}) & = &
-\frac{4}{3 (k+N+2)^2 (6 k N+5 k+5 N+4)}
\nonu \\
& \times & ({\bf 12 k^3 N}+4 k^3 {\bf +18 k^2 N^2}-35 k^2 N-40 k^2
{\bf +24 k N^3}
\nonu \\
& - & 37 k N^2-192 k N-120 k-4 N^3-74 N^2-108 N-48).
\label{antifbarnonlinear}
\eea}
The relations (\ref{antisymmzerohigher})
and (\ref{zerofbareigenhigher}) are used.
The large $(N,k)$ behavior of the eigenvalue in (\ref{antifbarnonlinear})
for the higher
spin $3$ current is the same as the one in (\ref{symmfbarhigher}).
Note that the eigenvalue for the higher spin $1$ current
does not have any contribution from the commutator
$[(\Phi_0^{(1)})_0, Q^{\bar{A}}_{-\frac{1}{2}}]$ because
the OPE between the corresponding higher spin $1$ current
and the spin $\frac{1}{2}$ current has only the first order pole.
This provides only the eigenvalue for the representation $(0;\overline{
  \tiny\yng(1)})$. Also the term 
$ Q^{\bar{A}}_{-\frac{1}{2}} (\Phi_0^{(1)})_0$
acting on the representation $(\tiny\yng(1,1);0)$
gives the eigenvalue $\phi_0^{(1)} (\tiny\yng(1,1);0)$ with
$ Q^{\bar{A}}_{-\frac{1}{2}}$ acting on the state  $|(\tiny\yng(1,1);0)>$.
By inserting the overall factor into this state, one has
the final state associated with 
the representation $(\tiny\yng(1,1);\overline{\tiny\yng(1)})$.
Therefore, one arrives at the above eigenvalue for the higher spin $1$
current.
All the eigenvalues survive even under the
large $(N,k)$ 't Hooft like limit.

For the eigenvalues corresponding to the remaining higher spin currents,
there are the contributions from the lower order poles
appearing in the commutators,
$[(V^+)_0, Q^{\bar{A}}_{-\frac{1}{2}}]$,
$[(V^-)_0, Q^{\bar{A}}_{-\frac{1}{2}}]$ and $[(\Phi_2^{(1)})_0,
Q^{\bar{A}}_{-\frac{1}{2}}]$. 
They can be summarized by
\bea
%& = & \frac{k}{(k+N+2)},
%\nonu \\
\delta
v^+ (\tiny\yng(1,1);\overline{\tiny\yng(1)})
& = & -\frac{384 N}{(k+N+2)^2},
\nonu \\
\delta
v^- (\tiny\yng(1,1);\overline{\tiny\yng(1)})
& = &  -\frac{240 (3 k-1)}{(k+N+2)^2},
\nonu \\
\delta
\phi_2^{(1)} (\tiny\yng(1,1);\overline{\tiny\yng(1)})
& = & \frac{8
  ({\bf 14 k^2 N}+9 k^2 {\bf +10 k N^2}
  +33 k N+22 k+10 N^2+14 N+8)}{(k+N+2)^2 (6 k N+5 k+5 N+4)}.
\label{deldeldel}
\eea
In particular, the second and last eigenvalues
in (\ref{deldeldel})
are twice of the ones in (\ref{ffbarcorrection1}),
and the ones in (\ref{symmfbarcorrection}) as long as the
highest power of $(N,k)$ is concerned. 
Then as done in (\ref{ffbarcorrection}), the final
eigenvalues are determined in (\ref{antifbarnonlinear}). 
Of course, the eigenvalues in (\ref{deldeldel})
vanish under the large $(N,k)$ 't Hooft like limit
\footnote{
\label{antibarffootnote}
  One obtains
  \bea
%& = & \frac{k}{(k+N+2)},
%\nonu \\
\delta v^+ (\overline{\tiny\yng(1,1)};\tiny\yng(1))
& = & \frac{384 N}{(k+N+2)^2},
\qquad
\delta v^- (\overline{\tiny\yng(1,1)};\tiny\yng(1))
 =  \frac{240 (3 k+1)}{(k+N+2)^2},
\nonu \\
\delta \phi_2^{(1)} (\overline{\tiny\yng(1,1)};\tiny\yng(1))& = &
-\frac{8 ({\bf 10 k^2 N}+9 k^2 {\bf +2 k N^2}+27 k N+32 k+8 N^2+16 N+16)}
{(k+N+2)^2 (6 k N+5 k+5 N+4)},
\label{newcorrection2}
\eea
where the commutators
$[(V^{\pm})_0, Q^{\bar{A}^{\ast}}_{-\frac{1}{2}}]$ and
$[(\Phi_2^{(1)})_0,
Q^{\bar{A}^{\ast}}_{-\frac{1}{2}}]$ are used. 
Then one also has
\bea
\phi_0^{(1)} (\overline{\tiny\yng(1,1)};\tiny\yng(1)) & = &
\phi_0^{(1)} (\overline{\tiny\yng(1,1)};0) +
\phi_0^{(1)} (0;\tiny\yng(1)), \nonu \\
v^{\pm} (\overline{\tiny\yng(1,1)};\tiny\yng(1))  & = &
v^{\pm} (\overline{\tiny\yng(1,1)};0)+
v^{\pm} (0;\tiny\yng(1)) +\delta v^{\pm}
(\overline{\tiny\yng(1,1)};\tiny\yng(1)),\nonu \\
\phi_2^{(1)} (\overline{\tiny\yng(1,1)};\tiny\yng(1)) & = &
\phi_2^{(1)} (\overline{\tiny\yng(1,1)};0)+
\phi_2^{(1)} (0;\tiny\yng(1))
+ \delta \phi_2^{(1)} (\overline{\tiny\yng(1,1)};\tiny\yng(1)),
\nonu
\eea
together with the footnote
\ref{antisymmbareigenhigher}, (\ref{0feigenhigher}),
and (\ref{newcorrection2}).
}.

%%%%%%%%%%%%%%%%%%%%%%%%%%%%%%%%%%%%%%%%%%%%%%%%%%%%%%%%%%%%%%%%%
\subsection{The $(\mbox{antisymm}; \mbox{symm})$ representation}
%%%%%%%%%%%%%%%%%%%%%%%%%%%%%%%%%%%%%%%%%%%%%%%%%%%%%%%%%%%%%%%%%

The relevant subsection on this higher representation
is given by $3.2.5$. The four eigenvalues are summarized by
\bea
\phi_0^{(1)} (\tiny\yng(1,1);\tiny\yng(2)) & = &
\phi_0^{(1)} (\tiny\yng(1,1);0)+
\phi_0^{(1)} (0;\tiny\yng(2))=
-\frac{2 (k+N)}{k+N+2},
\nonu \\
v^+ (\tiny\yng(1,1);\tiny\yng(2)) & = & \frac{96 (k N+k+N^2+7 N)}{(k+N+2)^2},
\nonu \\
v^- (\tiny\yng(1,1);\tiny\yng(2)) & = & \frac{96 (k^2+k N+7 k+N)}{(k+N+2)^2},
\nonu \\
\phi_2^{(1)} (\tiny\yng(1,1); \tiny\yng(2)) & = &
\frac{8}
     {3 (k+N+2)^2 (6 k N+5 k+5 N+4)}
     \nonu \\
     & \times & ({\bf 12 k^3 N}-2 k^3+64 k^2 N+11 k^2 {\bf -12 k N^3}+8 k N^2
     +36 k N\nonu \\
     & + & 2 N^3+N^2-72 N-48).
\label{antisymmnonlinear}
\eea
Again the relations (\ref{antisymmzerohigher})
and (\ref{zerosymmhigher}) are used.
The third eigenvalue (\ref{antisymmnonlinear}) in the highest power of $(N,k)$
is the {\bf twice} of the one in (\ref{fsymmeigenhigher}).
The large $(N,k)$ behavior of the eigenvalue for the
higher spin $3$ current is the same as the one in
(\ref{symmantisymmnonlinear}). 

The following results can be obtained similarly 
\bea
\delta
v^+ (\tiny\yng(1,1);\tiny\yng(2))
& = & \frac{768 N}{(k+N+2)^2}, \nonu \\
\delta
v^- (\tiny\yng(1,1);\tiny\yng(2))
& = & \frac{768 k}{(k+N+2)^2}, \nonu \\
\delta
\phi_2^{(1)} (\tiny\yng(1,1); \tiny\yng(2))
& = & \frac{16 ({\bf 10 k^2 N}+3 k^2{\bf +2 k N^2}+6 k N+2 k-N^2-14 N-8)}{
  (k+N+2)^2 (6 k N+5 k+5 N+4)}.
\label{otherdel}
\eea
The second eigenvalue in (\ref{otherdel}) is the twice as the one in
(\ref{fsymmeigenhighercorrection})
while the last eigenvalue shares the common value in the
highest power of $(N,k)$ with the twice of the one in
(\ref{fsymmeigenhighercorrection})
\footnote{
\label{cor5foot}
  From the relations
  \bea
\delta
v^{\pm} (\overline{\tiny\yng(1,1)};\tiny\yng(2))
& = & \delta v^{\pm} (\tiny\yng(2);\tiny\yng(2)),
\qquad
\delta
\phi_2^{(1)} (\overline{\tiny\yng(1,1)};\tiny\yng(2))
 =  - \delta \phi_2^{(1)} (\tiny\yng(1,1);\tiny\yng(2)),
\label{cor5}
\eea
one has
\bea
\phi_0^{(1)} (\overline{\tiny\yng(1,1)};\tiny\yng(2)) & = &
\phi_0^{(1)} (\overline{\tiny\yng(1,1)};0) +
\phi_0^{(1)} (0;\tiny\yng(2)), \nonu \\
v^{\pm} (\overline{\tiny\yng(1,1)};\tiny\yng(2))  & = &
v^{\pm} (\overline{\tiny\yng(1,1)};0)+
v^{\pm} (0;\tiny\yng(2)) +\delta v^{\pm}
(\overline{\tiny\yng(1,1)};\tiny\yng(2)),\nonu \\
\phi_2^{(1)} (\overline{\tiny\yng(1,1)};\tiny\yng(2)) & = &
\phi_2^{(1)} (\overline{\tiny\yng(1,1)};0)+
\phi_2^{(1)} (0;\tiny\yng(2))
+ \delta \phi_2^{(1)} (\overline{\tiny\yng(1,1)};\tiny\yng(2)),
\nonu
\eea
together with the footnote \ref{antisymmbareigenhigher}, (\ref{cor5}) and 
 (\ref{zerosymmhigher}).}.
Then the above expressions (\ref{antisymmnonlinear}) can be obtained explicitly
by using 
the relations (\ref{antisymmzerohigher}),
 (\ref{zerosymmhigher}) and (\ref{otherdel}).

%%%%%%%%%%%%%%%%%%%%%%%%%%%%%%%%%%%%%%%%%%%%%%%%%%%%%%%%%%%%%%%%
\subsection{The $(\mbox{antisymm}; \overline{\mbox{symm}})$
  representation}
%%%%%%%%%%%%%%%%%%%%%%%%%%%%%%%%%%%%%%%%%%%%%%%%%%%%%%%%%%%%%%%%

The four eigenvalues are
\footnote{One can also analyze the
  eigenvalues on this particular higher representation in the context of
sections $2$ and $3$.}
\bea
\phi_0^{(1)} (\tiny\yng(1,1);\overline{\tiny\yng(2)}) & = &
\phi_0^{(1)} (\tiny\yng(1,1);0)+
\phi_0^{(1)} (0;\overline{\tiny\yng(2)})=
-\frac{2 (-k+N)}{k+N+2},
\nonu \\
v^+ (\tiny\yng(1,1);\overline{\tiny\yng(2)}) & = &
\frac{96 (k N+k+N^2-9 N)}{(k+N+2)^2},
\nonu \\
v^- (\tiny\yng(1,1);\overline{\tiny\yng(2)}) & = & \frac{96
  (k^2+k N-9 k+N)}{(k+N+2)^2},
\nonu \\
\phi_2^{(1)} (\tiny\yng(1,1); \overline{\tiny\yng(2)}) & = &
-\frac{8}{3 (k+N+2)^2 (6 k N+5 k+5 N+4)}
\nonu \\
& \times &
({\bf 12 k^3 N}-2 k^3 {\bf+12 k^2 N^2}-70 k^2 N-25 k^2
{\bf +12 k N^3}-46 k N^2
\nonu \\
& - & 94 k N+36 k-2 N^3-13 N^2+84 N+96).
\label{antisymmsymmbarnonlinear}
\eea
The last eigenvalue with boldface notation
in (\ref{antisymmsymmbarnonlinear}) is the same as the one in
(\ref{symmantisymmbarnonlinear}) or (\ref{symmsymmbarnonlinear}).
Here the previous relations (\ref{antisymmzerohigher})
and the footnote \ref{zerosymmbareigenhigher} are used.
In particular, the large $(N,k)$ behavior in the eigenvalue
for the higher spin $3$ current is the same as the one in
(\ref{symmantisymmbarnonlinear}).

The following relations hold
\bea
\delta
v^+ (\tiny\yng(1,1);\overline{\tiny\yng(2)})
& = & -\frac{768 N}{(k+N+2)^2},
\nonu \\
\delta
v^- (\tiny\yng(1,1);\overline{\tiny\yng(2)})
& = & -\frac{768 k}{(k+N+2)^2},
\nonu \\
\delta
\phi_2^{(1)} (\tiny\yng(1,1); \overline{\tiny\yng(2)})
& = & \frac{16 ({\bf 14 k^2 N}+3 k^2{\bf +10 k N^2}
  +12 k N-8 k+N^2-16 N-16)}{(k+N+2)^2 (6 k N+5 k+5 N+4)}.
\label{varvar}
\eea
The highest power of $(N,k)$ in the last eigenvalue of (\ref{varvar})
is the twice of the ones in (\ref{fantisymmbareigenhighercorrection})
\footnote{
\label{cor6foot}
  From the relations
  \bea
\delta
v^{\pm} (\overline{\tiny\yng(1,1)};\overline{\tiny\yng(2)})
& = & \delta v^{\pm} (\tiny\yng(2);\overline{\tiny\yng(2)}),
\qquad
\delta
\phi_2^{(1)} (\overline{\tiny\yng(1,1)};\overline{\tiny\yng(2)})
 =  - \delta \phi_2^{(1)} (\tiny\yng(1,1);\overline{\tiny\yng(2)}),
\label{cor6}
\eea
one has
\bea
\phi_0^{(1)} (\overline{\tiny\yng(1,1)};\overline{\tiny\yng(2)}) & = &
\phi_0^{(1)} (\overline{\tiny\yng(1,1)};0) +
\phi_0^{(1)} (0;\overline{\tiny\yng(2)}), \nonu \\
v^{\pm} (\overline{\tiny\yng(1,1)};\overline{\tiny\yng(2)})  & = &
v^{\pm} (\overline{\tiny\yng(1,1)};0)+
v^{\pm} (0;\overline{\tiny\yng(2)}) +\delta v^{\pm}
(\overline{\tiny\yng(1,1)};\overline{\tiny\yng(2)}),\nonu \\
\phi_2^{(1)} (\overline{\tiny\yng(1,1)};\overline{\tiny\yng(2)}) & = &
\phi_2^{(1)} (\overline{\tiny\yng(1,1)};0)+
\phi_2^{(1)} (0;\overline{\tiny\yng(2)})
+ \delta \phi_2^{(1)} (\overline{\tiny\yng(1,1)};\overline{\tiny\yng(2)}),
\nonu
\eea
together with the footnote \ref{antisymmbareigenhigher}, (\ref{cor6}) and 
 the footnote \ref{zerosymmbareigenhigher}.}.

%%%%%%%%%%%%%%%%%%%%%%%%%%%%%%%%%%%%%%%%%%%%%%%%%%%%%%%%%%%%%%%%%%%%%%%
\subsection{The $(\mbox{antisymm}; \overline{\mbox{antisymm}})$ representation}
%%%%%%%%%%%%%%%%%%%%%%%%%%%%%%%%%%%%%%%%%%%%%%%%%%%%%%%%%%%%%%%%%%%%%%%

The relevant subsection on this higher representation
is given by $3.2.6$. The four eigenvalues are
\bea
\phi_0^{(1)} (\tiny\yng(1,1);\overline{\tiny\yng(1,1)}) & = &
\phi_0^{(1)} (\tiny\yng(1,1);0)+
\phi_0^{(1)} (0;\overline{\tiny\yng(1,1)})=
\frac{2(k-N)}{(k+N+2)},
\nonu \\
v^+ (\tiny\yng(1,1);\overline{\tiny\yng(1,1)}) & = &
\frac{96 (k N+k+N^2-9 N)}{(k+N+2)^2},
\nonu \\
v^- (\tiny\yng(1,1);\overline{\tiny\yng(1,1)}) & = &
\frac{32 (4 k^2+3 k N-51 k+3 N+20)}{(k+N+2)^2},
\nonu \\
\phi_2^{(1)} (\tiny\yng(1,1); \overline{\tiny\yng(1,1)}) & = &
-\frac{8}{3 (k+N+2)^2 (6 k N+5 k+5 N+4)}
\nonu \\
& \times & ({\bf 12 k^3 N}-2 k^3 {\bf +12 k^2 N^2}-34 k^2 N-19 k^2
{\bf +12 k N^3}
-46 k N^2\nonu \\
& - & 184 k N-120 k-2 N^3-85 N^2-96 N-48).
\label{antiantibarnonlinear}
\eea
The last eigenvalue with boldface notation in
(\ref{antiantibarnonlinear})
is the same as the one in
(\ref{symmantisymmbarnonlinear}), (\ref{symmsymmbarnonlinear})
or (\ref{antisymmsymmbarnonlinear}).
See also (\ref{antisymmantisymmbarresult}).
Note that the eigenvalue for the higher spin $1$ current
does not have any contribution from the commutator
$[(\Phi_0^{(1)})_0, Q^{1}_{-\frac{1}{2}}  Q^{2}_{-\frac{1}{2}}]$ because
the OPE between the corresponding higher spin $1$ current
and the spin $\frac{1}{2}$ current has only the first order pole.
This provides only the eigenvalue for the representation $(0;\overline{
  \tiny\yng(1,1)})$. Also the term 
$ Q^{1}_{-\frac{1}{2}}  Q^{2}_{-\frac{1}{2}} (\Phi_0^{(1)})_0$
acting on the representation $(\tiny\yng(1,1);0)$
gives the eigenvalue $\phi_0^{(1)} (\tiny\yng(1,1);0)$ with
$ Q^{1}_{-\frac{1}{2}}  Q^{2}_{-\frac{1}{2}}$
acting on the state  $|(\tiny\yng(1,1);0)>$.
By inserting the overall factor into this state, one has
the final state associated with 
the representation $(\tiny\yng(1,1);\overline{\tiny\yng(1,1)})$.
Therefore, one arrives at the above eigenvalue for the higher spin $1$
current.

For the eigenvalues corresponding to the remaining higher spin currents,
there are the contributions from the lower order poles
appearing in the commutators,
$[(V^+)_0, Q^{1}_{-\frac{1}{2}}  Q^{2}_{-\frac{1}{2}}]$,
$[(V^-)_0, Q^{1}_{-\frac{1}{2}}  Q^{2}_{-\frac{1}{2}}]$ and
$[(\Phi_2^{(1)})_0,
  Q^{1}_{-\frac{1}{2}}  Q^{2}_{-\frac{1}{2}}]$. 
They can be summarized by
\bea
\delta
v^+ (\tiny\yng(1,1);\overline{\tiny\yng(1,1)})
&= & -\frac{768 N}{(k+N+2)^2}, \nonu \\
\delta
v^- (\tiny\yng(1,1);\overline{\tiny\yng(1,1)})
& = & -\frac{128 (13 k-5)}{(k+N+2)^2}, \nonu \\
\delta
\phi_2^{(1)} (\tiny\yng(1,1); \overline{\tiny\yng(1,1)})
& = & \frac{16 ({\bf 14 k^2 N}+3 k^2 {\bf +10 k N^2}
  +36 k N+22 k+13 N^2+14 N+8)}{
  (k+N+2)^2 (6 k N+5 k+5 N+4)}.
\label{finaldel}
\eea
It is obvious that the last eigenvalue for the highest power
of $(N,k)$ in the numerator in (\ref{finaldel}) is exactly four times
the ones in (\ref{ffbarcorrection1}), twice of
the ones in (\ref{symmfbarcorrection}), or
the ones in (\ref{symmsymmbarhigher})
\footnote{
\label{footlast}
  One obtains
  \bea
%& = & \frac{k}{(k+N+2)},
%\nonu \\
\delta v^+ (\overline{\tiny\yng(1,1)};\tiny\yng(1,1))
& = & \frac{768 N}{(k+N+2)^2},
\qquad
\delta v^- (\overline{\tiny\yng(1,1)};\tiny\yng(1,1))
 =  \frac{128 (13 k+5)}{(k+N+2)^2},
\nonu \\
\delta \phi_2^{(1)} (\overline{\tiny\yng(1,1)};\tiny\yng(1,1))& = &
-\frac{16 ({\bf 10 k^2 N}+3 k^2{\bf +2 k N^2}
  +30 k N+32 k+11 N^2+16 N+16)}{(k+N+2)^2 (6 k N+5 k+5 N+4)}.
\label{newcorrection1}
\eea
The commutators
$[(V^{\pm})_0, Q^{13}_{-\frac{1}{2}}  Q^{14}_{-\frac{1}{2}}]$ and
$[(\Phi_2^{(1)})_0,
  Q^{13}_{-\frac{1}{2}}  Q^{14}_{-\frac{1}{2}}]$ are used. 
Then one also has
\bea
\phi_0^{(1)} (\overline{\tiny\yng(1,1)};\tiny\yng(1,1)) & = &
\phi_0^{(1)} (\overline{\tiny\yng(1,1)};0) +
\phi_0^{(1)} (0;\tiny\yng(1,1)), \nonu \\
v^{\pm} (\overline{\tiny\yng(1,1)};\tiny\yng(1,1))  & = &
v^{\pm} (\overline{\tiny\yng(1,1)};0)+
v^{\pm} (0;\tiny\yng(1,1)) +\delta v^{\pm}
(\overline{\tiny\yng(1,1)};\tiny\yng(1,1)),\nonu \\
\phi_2^{(1)} (\overline{\tiny\yng(1,1)};\tiny\yng(1,1)) & = &
\phi_2^{(1)} (\overline{\tiny\yng(1,1)};0)+
\phi_2^{(1)} (0;\tiny\yng(1,1))
+ \delta \phi_2^{(1)} (\overline{\tiny\yng(1,1)};\tiny\yng(1,1)),
\nonu
\eea
together with the footnote
\ref{antisymmbareigenhigher}, (\ref{zeroantisymmhigher}),
and (\ref{newcorrection1}).
}.
Then the above result (\ref{antiantibarnonlinear})
can be obtained from (\ref{antisymmzerohigher}),
the footnote \ref{zeroantisymmbarexpression} and (\ref{finaldel}).
The large $(N,k)$ behavior in the eigenvalue for the
higher spin $3$ current is the same as the one in
(\ref{symmantisymmbarnonlinear}).

In summary of this section,
we present the relevant Tables where the large $(N,k)$ 't Hooft limit
for the eigenvalues is taken.
We will observe that for the linear case, the corresponding
eigenvalues behave as exactly same as these Tables under the large
$(N,k)$ 't Hooft like limit.
At finite $(N,k)$, the eigenvalues, $\phi_0^{(1)}$ and $v^{\pm}$
in linear case coincide with the ones in the nonlinear case.
Only the eigenvalues $\phi_2^{(1)}$ are different from each other
at finite $(N,k)$.

For the eigenvalue for the higher spin $1$ current
on the representation $(\La_+;\La_-)$
which can be obtained the product of
$(\La_+;0)$ and $(0;\La_-)$,
one obtains
\bea
\phi_0^{(1)} (\La_{+}; \La_{-}) =
\phi_0^{(1)} (\La_{+}; 0) +
\phi_0^{(1)} (0; \La_{-})=
\Bigg[\frac{ \mp |\La_+| N \mp |\La_-| k}{(N+k+2)}
  \Bigg]
(\La_{+}; \La_{-}).
\label{form}
\eea
There are four cases depending on whether the representation
$\La_{\pm}$ is given by the multiple product of $\tiny\yng(1)$ or
$\overline{\tiny\yng(1)}$. We have minus (plus) sign for the former
(latter) in (\ref{form}). 
We also denote the number of boxes as $|\La_{\pm}|$.

%%%%%%%%%%%%%%%%%%%%%%%%%%%%%%%%%%%%%%%%%%%%%%%%%%%%%%%%%%%%%%%%%%%
\begin{table}[ht]
\centering % used for centering table
\begin{tabular}{|c||c|c|c|c|c|c|c| } % centered columns (4 columns)
\hline %inserts double horizontal lines
$(\Lambda_+; \Lambda_-)$ & 0 &$\tiny\yng(1)$ &$\overline{\tiny\yng(1)}$ &
$ \tiny\yng(2)$ &$\tiny\yng(1,1)$ & $\overline{ \tiny\yng(2)}$
& $\overline{\tiny\yng(1,1)}$
\\ [0.5ex] % inserts table
%heading
\hline \hline % inserts single horizontal line
$0$  &0 & $-(1-\la)$ & $(1-\la)$ & $-2(1-\la)$ &$-2(1-\la)$ & $2(1-\la)$  &   
$2(1-\la)$
\\ % inserting body of the table
\hline
$\tiny\yng(1)$ & $-\la$ & $\frac{2\la}{N}$  & $(1-2\la)$ & $-(2-\la)$ &
$-(2-\la)$& $(2-3\la)$ & $(2-3\la)$  \\
\hline
$\overline{\tiny\yng(1)}$ & $\la$ & $-(1-2\la)$ & $-\frac{2\la}{N}$  &
$-(2-3\la)$ & $-(2-3\la)$ & $(2-\la)$ & $(2-\la)$  \\
\hline
$ \tiny\yng(2)$ &  $-2\la$ & $-\la$ & $(1-3\la)$ & $\frac{4\la}{N}$ &
$-2$ & $2(1-2\la)$ & $2(1-2\la)$\\
\hline
$\tiny\yng(1,1)$ & $-2\la$ &$-\la$  &$(1-3\la)$ & $-2$ &$\frac{4\la}{N}$ &
$2(1-2\la)$
& $2(1-2\la)$  \\ 
\hline
$\overline{\tiny\yng(2)}$ & $2\la$ & $-(1-3\la)$ & $\la$ &
$-2(1-2\la)$&$-2(1-2\la)$ & $-\frac{4\la}{N}$ & $2$ \\ 
\hline
$\overline{\tiny\yng(1,1)}$ & $2\la$ &$-(1-3\la)$  &$\la$ &
$-2(1-2\la)$& $-2(1-2\la)$ & $2$ &  $-\frac{4\la}{N}$ \\ 
[1ex] % [1ex] adds vertical space
\hline %inserts single line
\end{tabular}
%\label{tableone} % is used to refer this table in the text
\caption{The eigenvalue $\phi_0^{(1)}$   
  under the large $(N,k)$ 't Hooft-like limit (\ref{largenk}).
   The eigenvalue with $\La_{+} = \La_{-}$
  can be written in terms of the multiple of the eigenvalue of $(f;f)$
  or $(\overline{f};\overline{f})$. The $\frac{1}{N}$ behavior in this case
  is written explicitly in this Table and next ones.
   The general structure for the eigenvalue with $\La_{+} \neq \La_{-}$
   (in the product of $(\La_+;0)$ and $(0;\La_-)$)
   is given by the linear combinations
   of the one of $(0;f)$ (or $(0;\overline{f})$) and the one of
   $(f;0)$ (or $(\overline{f};0)$).
  Then each coefficient depends on the the number of boxes in $\La_{+}$
  and $\La_{-}$.
 When the representation $\La_{-}$ appears in the branching of $\La_{+}$,
  the eigenvalue leads to the representation $(|\La_+|-|\La_{-}|;0)$
  where $|\La_{\pm}|$ denotes the number of boxes.
  %Also one has plus sign for the fundamental representation
  %while minus sign for the complex conjugated (anti fundamental)
  %representation(******). 
 We also present the eigenvalues in terms of the 't Hooft coupling
  constant $\la$.
See also the first Table in section $9$.
} % title of Table
\end{table}
\begin{table}[ht]
\centering % used for centering table
\begin{tabular}{|c||c|c|c| } % centered columns (4 columns)
\hline %inserts double horizontal lines
$(\Lambda_+; \Lambda_-)$ & 0 &$\tiny\yng(1)$ &$\overline{\tiny\yng(1)}$
\\ [0.5ex] % inserts table
%heading
\hline \hline % inserts single horizontal line
$0$  &0 & $\frac{24}{N} (1-\lambda ) \lambda$  &  $\frac{24}{N} (1-\lambda) \lambda$
\\ % inserting body of the table
\hline
$\tiny\yng(1)$ &$12\la(4+\la)$ & $\frac{96}{N}(1-\la)\la$
& $12\la(4+\la)$   \\
\hline
$\overline{\tiny\yng(1)}$ &$12\la(4+\la)$  &$12\la(4+\la)$ &  $\frac{96}{N}(1-\la)\la $  \\
\hline
$ \tiny\yng(2)$ & $32 \lambda  (\lambda +3)$ & $12 \lambda  (\lambda +4)$
& $32 \lambda  (\lambda +3)$  \\
\hline
$\tiny\yng(1,1)$ & $96\la$ &$12 \lambda  (\lambda +4)$ &
$96 \la$    \\ 
\hline
$\overline{\tiny\yng(2)}$ &$32\la(\la+3)$ & $32\la(\la+3)$ & $12\la(\la+4)$ \\ 
\hline
$\overline{\tiny\yng(1,1)}$ & $96\la$  & $96\la$ &   $12\la(\la+4)$ \\ 
[1ex] % [1ex] adds vertical space
\hline %inserts single line
\end{tabular}
%\label{tableone} % is used to refer this table in the text
\caption{The eigenvalue $v^+$
  under the large $(N,k)$ 't Hooft-like limit (\ref{largenk}).
  Due to the behavior of $\frac{1}{N}$ in the representation
  $(0;\La_{-})$, the final eigenvalues for
  the representation $(\La_+;\La_-)$ (in the product of
  $(\La_+;0)$ and $(0;\La_-)$)
  come from the contribution from $(\La_+;0)$.
Furthermore, for the representation $(\La_+;\La_+)$,
the eigenvalues are given by the multiple of
the fundamental quantity which is given by the eigenvalue for the
representation $(f;f)$ (or $(\overline{f};\overline{f})$).
 When the representation $\La_{-}$ appears in the branching of $\La_{+}$,
  the eigenvalue leads to the representation $(|\La_+|-|\La_{-}|;0)$
  where $|\La_{\pm}|$ denotes the number of boxes.
} % title of Table
\end{table}
%%%%%%%%%%%%%%%%%%%%%%%%%%%%%%%%%%%%%%%%%%%%%%%%%%%%%%%%%%%%%%%%%%%%%%

%%%%%%%%%%%%%%%%%%%%%%%%%%%%%%%%%%%%%%%%%%%%%%%%%%%%%%%%%%%%%%%%%%%
\begin{table}[ht]
\centering % used for centering table
\begin{tabular}{|c||c|c|c|c| } % centered columns (4 columns)
\hline %inserts double horizontal lines
$(\Lambda_+; \Lambda_-)$ & 
$ \tiny\yng(2)$ &$\tiny\yng(1,1)$ & $\overline{ \tiny\yng(2)}$
& $\overline{\tiny\yng(1,1)}$
\\ [0.5ex] % inserts table
%heading
\hline \hline % inserts single horizontal line
$0$   &  $\frac{96}{N} (1-\lambda) \lambda$ &  $\frac{96}{N} (1-\lambda) \lambda$ & $\frac{96}{N} (1-\lambda) \lambda$ &
 $\frac{96}{N} (1-\lambda) \lambda$
\\ % inserting body of the table
\hline
$\tiny\yng(1)$  & $12\la(\la+4)$ & $12\la(\la+4)$ & $12\la(\la+4)$ &  $12\la(\la+4)$ \\
\hline
$\overline{\tiny\yng(1)}$ &  $12\la(\la+4)$ & $12\la(\la+4)$ & $12\la(\la+4)$ &   $12\la(\la+4)$  \\
\hline
$ \tiny\yng(2)$ &
$\frac{192}{N} (1-\lambda) \lambda $&
$32 \lambda  (\lambda +3)$ & $32 \lambda  (\lambda +3)$ & $32 \lambda  (\lambda +3)$ \\
\hline
$\tiny\yng(1,1)$  & $96\la$ & $\frac{192}{N} (1-\la)\la$ & $96\la$ & $96\la$  \\ 
\hline
$\overline{\tiny\yng(2)}$ & $32 \lambda  (\lambda +3)$ & $32 \lambda  (\lambda +3)$ &$\frac{192}{N} (1-\lambda) \lambda $ &  $32 \lambda  (\lambda +3)$  \\ 
\hline
$\overline{\tiny\yng(1,1)}$ &$96\la$  &$96\la$ &$96\la$ & $\frac{192}{N} (1-\lambda) \lambda $  \\ 
[1ex] % [1ex] adds vertical space
\hline %inserts single line
\end{tabular}
%\label{tableone} % is used to refer this table in the text
\caption{The (continued) eigenvalue $v^+$
  under the large $(N,k)$ 't Hooft-like limit (\ref{largenk}).
   Due to the behavior of $\frac{1}{N}$ in the representation
  $(0;\La_{-})$, the final eigenvalues for
  the representation $(\La_+;\La_-)$ (in the product of
  $(\La_+;0)$ and $(0;\La_-)$)
  come from the contribution from $(\La_+;0)$.
Furthermore, for the representation $(\La_+;\La_+)$,
the eigenvalues are given by the multiple of
the fundamental quantity which is given by the eigenvalue for the
representation $(f;f)$.
On the other hand, for the  representation $(\La_+;\La_-)$
where the representation $\La_-$ appears in the branching of
$\La_+$, the eigenvalues can be written in terms of the eigenvalue for
$(f;0)$.
} % title of Table
\end{table}
\begin{table}[ht]
\centering % used for centering table
\begin{tabular}{|c||c|c|c| } % centered columns (4 columns)
\hline %inserts double horizontal lines
$(\Lambda_+; \Lambda_-)$ & 0 &$\tiny\yng(1)$ &$\overline{\tiny\yng(1)}$
\\ [0.5ex] % inserts table
%heading
\hline \hline % inserts single horizontal line
$0$  &0 & $12 (\lambda -5) (\lambda -1)$  &
 $12 (\lambda -5) (\lambda -1)$
\\ % inserting body of the table
\hline
$\tiny\yng(1)$ & $\frac{24}{N} \la^2$ & $ \frac{96}{N} \la^2
$ & $12 (\lambda -5) (\lambda -1)$   \\
\hline
$\overline{\tiny\yng(1)}$ & $\frac{24}{N} \la^2$  & $12 (\lambda -5) (\lambda -1)$ &   $ \frac{96}{N} \la^2
$ \\
\hline
$ \tiny\yng(2)$ & $\frac{96}{N} \lambda^2$  &  $\frac{120}{N} \la^2$ &
$12(\la-5)(\la-1)$
 \\
\hline
$\tiny\yng(1,1)$ & $\frac{96}{N} \lambda^2$ & $\frac{120}{N} \la^2$  &
$12 (\lambda -5) (\lambda -1)$
 \\ 
\hline
$\overline{\tiny\yng(2)}$ & $\frac{96}{N} \lambda^2$   &$12 (\lambda -5) (\lambda -1)$ &  $\frac{120}{N} \la^2$  \\ 
\hline
$\overline{\tiny\yng(1,1)}$ & $\frac{96}{N} \lambda^2$   &$12 (\lambda -5) (\lambda -1)$ &  $\frac{120}{N} \la^2$  \\ 
[1ex] % [1ex] adds vertical space
\hline %inserts single line
\end{tabular}
%\label{tableone} % is used to refer this table in the text
\caption{The eigenvalue $v^-$ 
  under the large $(N,k)$ 't Hooft-like limit (\ref{largenk}).
   Due to the behavior of $\frac{1}{N}$ in the representation
  $(\La_{+};0)$, the final eigenvalues for
  the representation $(\La_+;\La_-)$ (in the product of
  $(\La_+;0)$ and $(0;\La_-)$)
  come from the contribution from $(0;\La_-)$.
  One can see the $\frac{1}{N}$ behavior in the representation
  $(\La_+;\La_+)$.
} % title of Table
\end{table}
%%%%%%%%%%%%%%%%%%%%%%%%%%%%%%%%%%%%%%%%%%%%%%%%%%%%%%%%%%%%%%%%%%%%%%

%%%%%%%%%%%%%%%%%%%%%%%%%%%%%%%%%%%%%%%%%%%%%%%%%%%%%%%%%%%%%%%%%%%
\begin{table}[ht]
\centering % used for centering table
\begin{tabular}{|c||c|c|c|c| } % centered columns (4 columns)
\hline %inserts double horizontal lines
$(\Lambda_+; \Lambda_-)$ & 
$ \tiny\yng(2)$ &$\tiny\yng(1,1)$ & $\overline{ \tiny\yng(2)}$
& $\overline{\tiny\yng(1,1)}$
\\ [0.5ex] % inserts table
%heading
\hline \hline % inserts single horizontal line
$0$ 
&$96 (1-\lambda )$ & $32 (\lambda -4) (\lambda -1)$ & $96 (1-\lambda)$ &
$32 (\lambda -4) (\lambda -1)$
\\ % inserting body of the table
\hline
$\tiny\yng(1)$  &$96 (1-\lambda )$ &$32 (\lambda -4) (\lambda -1)$ &$96 (1-\lambda )$ & $32 (\lambda -4) (\lambda -1)$ \\
\hline
$\overline{\tiny\yng(1)}$  &$96 (1-\lambda )$ &$32 (\lambda -4) (\lambda -1)$ &$96 (1-\lambda )$ & $32 (\lambda -4) (\lambda -1)$ \\
\hline
$ \tiny\yng(2)$ &
$\frac{192}{N} \la^2$&$32 (\lambda -4) (\lambda -1)$ & $96 (1-\lambda)$ & $32 (\lambda -4) (\lambda -1)$ \\
\hline
$\tiny\yng(1,1)$ & $96 (1-\lambda)$ & $\frac{192}{N} \la^2$ &$96 (1-\lambda)$  & $32 (\lambda -4) (\lambda -1)$ \\ 
\hline
$\overline{\tiny\yng(2)}$ &$96 (1-\lambda)$  & $32 (\lambda -4) (\lambda -1)$ &$\frac{192}{N} \la^2$ &  $32 (\lambda -4) (\lambda -1)$  \\ 
\hline
$\overline{\tiny\yng(1,1)}$  &$96 (1-\lambda)$  & $32 (\lambda -4) (\lambda -1)$ &$96 (1-\lambda)$  & $\frac{192}{N} \la^2$ \\ 
[1ex] % [1ex] adds vertical space
\hline %inserts single line
\end{tabular}
%\label{tableone} % is used to refer this table in the text
\caption{
The (continued) eigenvalue $v^-$ 
  under the large $(N,k)$ 't Hooft-like limit (\ref{largenk}).
   Due to the behavior of $\frac{1}{N}$ in the representation
  $(\La_{+};0)$, the final eigenvalues for
  the representation $(\La_+;\La_-)$ (in the product of
  $(\La_+;0)$ and $(0;\La_-)$)
  come from the contribution from $(0;\La_-)$.
  One can see the $\frac{1}{N}$ behavior in the representation
  $(\La_+;\La_+)$.
} % title of Table
\end{table}
\begin{table}[ht]
\centering % used for centering table
{\small
\begin{tabular}{|c||c|c|c| } % centered columns (4 columns)
\hline %inserts double horizontal lines
$(\La_+;\La_-)$ & 0 &$\tiny\yng(1)$ &$\overline{\tiny\yng(1)}$
\\ [0.5ex] % inserts table
%heading
\hline \hline % inserts single horizontal line
$0$  &0 & $\phi_2^{(1)}(0;\tiny\yng(1))$  &  $-\phi_2^{(1)}(0;\tiny\yng(1))$
\\ % inserting body of the table
 & & $=\frac{4}{3}(1-\la)(2-\la)$ & $=-\frac{4}{3}(1-\la)(2-\la)$  \\
\hline
$\tiny\yng(1)$ &  $\phi_2^{(1)}(\tiny\yng(1);0)$ &
 $\phi_2^{(1)}(\tiny\yng(1);\tiny\yng(1))
$& $\phi_2^{(1)}[(\tiny\yng(1);0) -(0;\tiny\yng(1))]
$   \\
& $=-\frac{4}{3} \la(\la+1)$ & $=-\frac{8}{3N } \la (2\lambda -1)$ & $=
-\frac{8}{3}(\lambda ^2-\lambda +1)$ \\
\hline
$\overline{\tiny\yng(1)}$ & $-\phi_2^{(1)}(\tiny\yng(1);0)$ &
$-\phi_2^{(1)}[(\tiny\yng(1);0) -(0;\tiny\yng(1))]
$ 
&  $-\phi_2^{(1)}(\tiny\yng(1);\tiny\yng(1))
$  \\
&$=\frac{4}{3} \la(\la+1)$ & $=
\frac{8}{3}(\lambda ^2-\lambda +1)$ & $=\frac{8}{3N } \la (2\lambda -1)$ \\
\hline
$ \tiny\yng(2)$ & $2\phi_2^{(1)}(\tiny\yng(1);0)$  & $\phi_2^{(1)}(\tiny\yng(1);0) $  & $  \phi_2^{(1)}[2 (\tiny\yng(1);0) -
(0;\tiny\yng(1))]$ \\
& $=-\frac{8}{3} \lambda  (\lambda +1)$ & $=-\frac{4}{3} \la(\la+1)$ &
$=-\frac{4}{3} (3 \lambda ^2-\lambda +2)$ \\
\hline
$\tiny\yng(1,1)$ & $2\phi_2^{(1)}(\tiny\yng(1);0)$ &
 $\phi_2^{(1)}(\tiny\yng(1);0) $
&$ \phi_2^{(1)}[2(\tiny\yng(1);0) -
(0;\tiny\yng(1))]$
\\
& $=-\frac{8}{3} \lambda  (\lambda +1)$ & $=-\frac{4}{3} \la(\la+1)$ &
$=-\frac{4}{3} (3 \lambda ^2-\lambda +2)$ \\
\hline
$\overline{\tiny\yng(2)}$ &$-2\phi_2^{(1)}(\tiny\yng(1);0)$ &
$ -\phi_2^{(1)}[2(\tiny\yng(1);0) -
(0;\tiny\yng(1))]$&  $-\phi_2^{(1)}(\tiny\yng(1);0) $  \\ 
&$=\frac{8}{3} \lambda  (\lambda +1)$ &$=\frac{4}{3} (3 \lambda ^2-\lambda +2)$ & $=\frac{4}{3} \la(\la+1)$ \\
\hline
$\overline{\tiny\yng(1,1)}$ & $-2\phi_2^{(1)}(\tiny\yng(1);0)$  &
$ -\phi_2^{(1)}[2(\tiny\yng(1);0) -
(0;\tiny\yng(1))]$
&  $-\phi_2^{(1)}(\tiny\yng(1);0) $  \\
&$=\frac{8}{3} \lambda  (\lambda +1)$ &$=\frac{4}{3} (3 \lambda ^2-\lambda +2)$ &  $=\frac{4}{3} \la(\la+1)$\\
[1ex] % [1ex] adds vertical space
\hline %inserts single line
\end{tabular}}
%\label{tableone} % is used to refer this table in the text
\caption{The eigenvalue $\phi_2^{(1)}$
  under the large $(N,k)$ 't Hooft-like limit (\ref{largenk}).
  The general structure for the eigenvalue with $\La_{+} \neq \La_{-}$
  in the product of $(\La_{+};0)$ and $(0;\La_{-})$
  is given by the linear combinations
  of the one of $(0;f)$ (or $(0;\overline{f})$) and the one of $(f;0)$
  (or $(\overline{f};0)$).
  Then each coefficient depends on the the number of boxes in $\La_{+}$
  and $\La_{-}$.
  %Also one has plus sign for the fundamental representation
  %while minus sign for the complex conjugated (anti fundamental)
  %representation.
  When the representation $\La_{-}$ appears in the branching of $\La_{+}$,
  the eigenvalue leads to the representation $(|\La_+|-|\La_{-}|;0)$
  where $|\La_{\pm}|$ denotes the number of boxes.
  The eigenvalue with $\La_{+} = \La_{-}$
  can be written in terms of the multiple of the eigenvalue of $(f;f)$
  or $(\overline{f};\overline{f})$. The $\frac{1}{N}$ behavior in this case
  is written explicitly in this Table and next one.
 We also present the eigenvalues in terms of the 't Hooft coupling
  constant $\la$.
} % title of Table
\end{table}
%%%%%%%%%%%%%%%%%%%%%%%%%%%%%%%%%%%%%%%%%%%%%%%%%%%%%%%%%%%%%%%%%%%%%%

%%%%%%%%%%%%%%%%%%%%%%%%%%%%%%%%%%%%%%%%%%%%%%%%%%%%%%%%%%%%%%%%%%%
\begin{table}[ht]
\centering % used for centering table
{\small
\begin{tabular}{|c||c|c|c|c| } % centered columns (4 columns)
  \hline %inserts double horizontal lines
  %$(\La_+;\La_-)$
  &
$ \tiny\yng(2)$ &$\tiny\yng(1,1)$ & $\overline{ \tiny\yng(2)}$
& $\overline{\tiny\yng(1,1)}$
\\ [0.5ex] % inserts table
%heading
\hline \hline % inserts single horizontal line
$0$   & $2 \phi_2^{(1)}(0;\tiny\yng(1))$ &
$2 \phi_2^{(1)}(0;\tiny\yng(1))$ & $-2 \phi_2^{(1)}(0;\tiny\yng(1))$  &
 $-2 \phi_2^{(1)}(0;\tiny\yng(1))$
\\ % inserting body of the table
& $=\frac{8}{3} (1-\lambda ) (2-\lambda )$ &$=\frac{8}{3} (1-\lambda ) (2-\lambda )$ &$=-\frac{8}{3} (1-\lambda ) (2-\lambda )$ & $=-\frac{8}{3} (1-\lambda ) (2-\lambda )$ \\
\hline
$\tiny\yng(1)$  &$ \phi_2^{(1)}[(\tiny\yng(1);0) +
2 (0;\tiny\yng(1))]$ & $ \phi_2^{(1)}[(\tiny\yng(1);0) +
2 (0;\tiny\yng(1))]$ &$ \phi_2^{(1)}[(\tiny\yng(1);0) -
2 (0;\tiny\yng(1))]$ & $ \phi_2^{(1)}[(\tiny\yng(1);0) -
2 (0;\tiny\yng(1))]$  \\
& $=\frac{4}{3} (\la^2-7\la +4)$ & $=\frac{4}{3} (\la^2-7\la +4)$ &
 $=-\frac{4}{3} (3\la^2-5\la +4)$ &  $=-\frac{4}{3} (3\la^2-5\la +4)$ \\
\hline
$\overline{\tiny\yng(1)}$    &$ -\phi_2^{(1)}[(\tiny\yng(1);0) -
2 (0;\tiny\yng(1))]$ & $ -\phi_2^{(1)}[(\tiny\yng(1);0) -
  2 (0;\tiny\yng(1))]$ &$ -\phi_2^{(1)}[(\tiny\yng(1);0) +
2 (0;\tiny\yng(1))]$ & $ -\phi_2^{(1)}[(\tiny\yng(1);0) +
2 (0;\tiny\yng(1))]$ \\
& $=\frac{4}{3} (3\la^2-5\la +4)$ &  $=\frac{4}{3} (3\la^2-5\la +4)$  &
$=-\frac{4}{3} (\la^2-7\la +4)$& $=-\frac{4}{3} (\la^2-7\la +4)$ \\
\hline
$ \tiny\yng(2)$
& $2\phi_2^{(1)}(\tiny\yng(1);\tiny\yng(1))$ &
$2 \phi_2^{(1)}[(\tiny\yng(1);0) +
(0;\tiny\yng(1))]$ &
$2 \phi_2^{(1)}[(\tiny\yng(1);0) -
  (0;\tiny\yng(1) )]$ & $2 \phi_2^{(1)}[(\tiny\yng(1);0) -
  (0;\tiny\yng(1) )]$ \\
 &$=-\frac{16}{3 N} \lambda  (2 \lambda -1)$ & $=-\frac{16}{3}  (2 \lambda -1)$ & $=-\frac{16}{3}(\lambda ^2-\lambda +1)$ &  $=-\frac{16}{3}(\lambda ^2-\lambda +1)$\\
\hline
$\tiny\yng(1,1)$ & $2 \phi_2^{(1)}[(\tiny\yng(1);0) +
(0;\tiny\yng(1))]$ &
$ 2\phi_2^{(1)}(\tiny\yng(1);\tiny\yng(1))$ &
 $
2 \phi_2^{(1)}[(\tiny\yng(1);0) -(0;\tiny\yng(1) )] $
&   $
2 \phi_2^{(1)}[(\tiny\yng(1);0) -(0;\tiny\yng(1) )]
$  \\
& $=-\frac{16}{3}  (2 \lambda -1)$ & $=-\frac{16}{3 N} \lambda  (2 \lambda -1)$ &$=-\frac{16}{3}(\lambda ^2-\lambda +1)$ & $=-\frac{16}{3}(\lambda ^2-\lambda +1)$ \\
\hline
$\overline{\tiny\yng(2)}$  &
$-2 \phi_2^{(1)}[(\tiny\yng(1);0) -
  (0;\tiny\yng(1) )]$ &$-2 \phi_2^{(1)}[(\tiny\yng(1);0) -
  (0;\tiny\yng(1) )]$ &$-2\phi_2^{(1)}(\tiny\yng(1);\tiny\yng(1))$ &
$-2 \phi_2^{(1)}[(\tiny\yng(1);0) +
(0;\tiny\yng(1))]$ \\ 
&$=\frac{16}{3}(\lambda ^2-\lambda +1)$ &$=\frac{16}{3}(\lambda ^2-\lambda +1)$ &$=\frac{16}{3 N} \lambda  (2 \lambda -1)$ &
$=\frac{16}{3}  (2 \lambda -1)$
\\
\hline
$\overline{\tiny\yng(1,1)}$  &$-2 \phi_2^{(1)}[(\tiny\yng(1);0) -
  (0;\tiny\yng(1) )]$ &$-2 \phi_2^{(1)}[(\tiny\yng(1);0) -
  (0;\tiny\yng(1) )]$ &$-2 \phi_2^{(1)}[(\tiny\yng(1);0) +
(0;\tiny\yng(1))]$ & $-2\phi_2^{(1)}(\tiny\yng(1);\tiny\yng(1))$  \\ 
&$=\frac{16}{3}(\lambda ^2-\lambda +1)$ &$=\frac{16}{3}(\lambda ^2-\lambda +1)$ &$=\frac{16}{3}  (2 \lambda -1)$ &$=\frac{16}{3 N} \lambda  (2 \lambda -1)$ \\
[1ex] % [1ex] adds vertical space
\hline %inserts single line
\end{tabular}}
%\label{tableone} % is used to refer this table in the text
\caption{
The (continued) eigenvalue $\phi_2^{(1)}$
  under the large $(N,k)$ 't Hooft-like limit (\ref{largenk}).
  The general structure for the eigenvalue with $\La_{+} \neq \La_{-}$
  in the product of $(\La_{+};0)$ and $(0;\La_{-})$
  is given by the linear combinations
  of the one of $(0;f)$ (or $(0;\overline{f})$) and the one of $(f;0)$
  (or $(\overline{f};0)$).
  Then each coefficient depends on the the number of boxes in $\La_{+}$
  and $\La_{-}$.
  %Also one has plus sign for the fundamental representation
  %while minus sign for the complex conjugated (anti fundamental)
  %representation.
  When the representation $\La_{-}$ appears in the branching of $\La_{+}$,
  the eigenvalue leads to the representation $(|\La_+|-|\La_{-}|;0)$
  where $|\La_{\pm}|$ denotes the number of boxes.
  The eigenvalue with $\La_{+} = \La_{-}$
  can be written in terms of the multiple of the eigenvalue of $(f;f)$
  or $(\overline{f};\overline{f})$. The $\frac{1}{N}$ behavior in this case
  is written explicitly in this Table and previous one.
 We also present the eigenvalues in terms of the 't Hooft coupling
  constant $\la$.
} % title of Table
\end{table}
%%%%%%%%%%%%%%%%%%%%%%%%%%%%%%%%%%%%%%%%%%%%%%%%%%%%%%%%%%%%%%%%%%%%%%

%\bea
%\phi_2^{(1)}
%(\tiny\yng(1)\tiny\yng(1) \cdots \tiny\yng(1);
%\tiny\yng(1)\tiny\yng(1) \cdots \tiny\yng(1))
% & = &  n \, 
%\phi_2^{(1)}(\tiny\yng(1);\tiny\yng(1))
%\nonu \\
%\phi_2^{(1)}
%(\overline{\tiny\yng(1)}\overline{\tiny\yng(1)} \cdots
%\overline{\tiny\yng(1)};
%\overline{\tiny\yng(1)}\overline{\tiny\yng(1)} \cdots \overline{\tiny\yng(1)}%)
% & = &  -n \, 
%\phi_2^{(1)}(\tiny\yng(1);\tiny\yng(1))
%\nonu
%\eea

%%%%%%%%%%%%%%%%%%%%%%%%%%%%%%%%%%%%%%%%%%%%%%%%%%%%%%%%%%%%%%%%%%%%%
%%%%%%%%%%%%%%%%%%%%%%%%%%%%%%%%%%%%%%%%%%%%%%%%%%%%%%%%%%%%%%%%%%%%%%
%\section{
%  The minimal and
%  higher representations in the $\frac{SU(5)}{SU(3)}$ coset}
%2%%%%%%%%%%%%%%%%%%%%%%%%%%%%%%%%%%%%%%%%%%%%%%%%%%%%%%%%%%%%%%%%%%%%%
%%%%%%%%%%%%%%%%%%%%%%%%%%%%%%%%%%%%%%%%%%%%%%%%%%%%%%%%%%%%%%%%%%%%%

%%%%%%%%%%%%%%%%%%%%%%%%%%%%%%%%%%%%%%%%%%%%%%%%%%%%%%%%%%%%%%%%%%%%%
%%%%%%%%%%%%%%%%%%%%%%%%%%%%%%%%%%%%%%%%%%%%%%%%%%%%%%%%%%%%%%%%%%%%%%
\section{ Review of the minimal and
  higher representations in the $\frac{SU(N+2)}{SU(N)}$
  coset}
%section6%%%%%%%%%%%%%%%%%%%%%%%%%%%%%%%%%%%%%%%%%%%%%%%%%%%%%%%%%%%%%%%%%%%%%
%%%%%%%%%%%%%%%%%%%%%%%%%%%%%%%%%%%%%%%%%%%%%%%%%%%%%%%%%%%%%%%%%%%%%

%The stress energy tensor spin-$2$ current has different eigenvalues.

The explicit relation between the
$11$ currents of the large ${\cal N}=4$ ``nonlinear'' superconformal
algebra and the $16$ currents (with boldface notation) of the
large ${\cal N}=4$ ``linear'' superconformal algebra
is described by \cite{GS}
\bea
T^{\mu \nu}(z) & = &
{\bf T^{\mu \nu}}(z)-\frac{2i}{(2+k+N)}\,{\bf \Gamma^{\mu} \Gamma^{\nu}}(z),
\nonu\\
G^{\mu}(z) & = & {\bf G^{\mu}}(z)-\frac{2i}{(2+k+N)}\,{\bf U\Gamma^{\mu}}(z)
\nonu \\
& - & \varepsilon^{\mu\nu\rho\si} \Bigg[  
\frac{4i}{3(2+k+N)^2}{\bf \Gamma^{\nu}\Gamma^{\rho}\Gamma^{\si}}
-\frac{1}{(2+k+N)}{\bf T^{\nu\rho} \Gamma^{\si}}
\Bigg](z),
\nonu\\
T(z) & = & 
{\bf T}(z)+\frac{1}{(2+k+N)}\,\Bigg[\,{\bf UU}-{\bf
  \partial \Gamma^{\mu} \Gamma^{\mu}}\,\Bigg](z).
\label{gsformula}
\eea
Here 
the four fermionic spin $\frac{1}{2}$ currents are given by
\cite{Saulina,AK1506}
\bea
{\bf \Gamma}^0 (z) =-\frac{i }{4(N+1)}
h^j_{\tilde{a} \tilde{b} }
f^{\tilde{a} \tilde{b}}_{\,\,\,\,\,\, \tilde{c}} h^{j \tilde{c}}_{ \,\,\,\, \tilde{d} } Q^{\tilde{d}}(z),
\qquad
{\bf \Gamma}^j (z) =-\frac{i }{4(N+1)} h^j_{\tilde{a} \tilde{b} }
f^{\tilde{a} \tilde{b}}_{\,\,\,\,\,\, \tilde{c}} Q^{\tilde{c}} (z),
\label{Gamma}
\eea
where $j=1,2,3$ and there is no sum over $j$ in the first equation of
(\ref{Gamma}) \footnote{One should change the index structures
  as follows: ${\bf \Gamma^0} \rightarrow -i {\bf \Gamma^2}$,
  ${\bf \Gamma^1} \rightarrow
  -i {\bf \Gamma^3}$,
  ${\bf \Gamma^2} \rightarrow i {\bf \Gamma^4}$ and
  ${\bf \Gamma^3} \rightarrow -i {\bf \Gamma^1}$ in order to use
  (\ref{gsformula}) from (\ref{Gamma}).
  We introduce the coset $\frac{G}{H}=\frac{SU(N+2)}{SU(N)}$ notation 
  $\tilde{a}=(\bar{a}, \hat{a})$ 
where the $\bar{a}$ index runs over $4N$ values
as before
and the index $\hat{a}$ associates with the  
$2 \times 2$ matrix corresponding to $SU(2)\times U(1)$
and runs over $4$ values. 
The $4 \times 4$ matrices $h^{i}_{\hat{a}  \hat{b}}$
appearing in (\ref{Gamma})
are given in  \cite{AK1506}.
}.
The bosonic spin $1$ current is given by
\bea
{\bf U} (z) =-\frac{1}{4(N+1)} h^j_{\tilde{a} \tilde{b} }
f^{\tilde{a} \tilde{b}}_{\,\,\,\,\,\, \tilde{c}} h^{j \tilde{c}}_{ \,\,\,\, \tilde{d} } \left[
 V^{\tilde{d}}
-\frac{1}{2(k+N+2)}  
f^{\tilde{d} }_{\,\,\,\, \tilde{e} \tilde{f}} Q^{\tilde{e}} Q^{\tilde{f}}
\right](z),
%\label{ulinear}
\nonu
\eea
where there is no sum over the index $j$.
Of course, the $11$ currents in (\ref{11currents})
or (\ref{gsformula}) are regular in the OPEs between them
and the spin $\frac{1}{2}$ currents
${\bf \Gamma}^{\mu}(z)$ and the spin $1$ current ${\bf U}(z)$.

The spin $2$ current in the ${\cal N}=4$ linear superconformal
algebra has extra two terms in (\ref{gsformula}),
compared to the one in the
${\cal N}=4$ nonlinear superconformal algebra.
Moreover, the OPE between the current ${\bf \Gamma^{\mu}}(z)$ 
and the adjoint spin $\frac{1}{2}$ current $Q^{\bar{A}^{\ast}}(w)$
is regular.
Therefore, the ${\bf \Gamma^{\mu}}$ term does not contribute to the eigenvalue
for the
representation $(0;\La_-)$.
Furthermore,  the current ${\bf \Gamma^{\mu}}(z)$ contains
only  the adjoint spin $\frac{1}{2}$ current $Q^{\bar{A}^{\ast}}(w)$
from (\ref{gsformula}).
This implies that
the ${\bf \Gamma^{\mu}}$ term does not contribute to the eigenvalue
for the
representation $(\La_+;0)$ also.
Then one obtains the following identity
\bea
    {\bf h} (\La_+;\La_-) = h (\La_+;\La_-) + \frac{1}{(N+k+2)}
    \frac{1}{N(N+2)}  \hat{u}^2 (\La_+;\La_-).
\nonu
\eea
Among four eigenvalues studied in sections $2$ and $3$, the eigenvalues
for the spin $2$ current are different from the ones in sections $6$, $7$
and $8$. The quantum numbers $l^{\pm}$ is the same because
the ${\bf \Gamma^{\mu}}$ dependent terms  in the
adjoint spin $1$ current (\ref{gsformula}) do not contribute to the
final eigenvalue equations.
Therefore, we present only the eigenvalue ${\bf h}(\La_+;\La_-)$
with the eigenvalues for the higher spin $3$ current
in next sections.

%%%%%%%%%%%%%%%%%%%%%%%%%%%%%%%%%%%%%%%%%%%%%%%%%%%%%%%%
%%%%%%%%%%%%%%%%%%%
\section{ Review of eigenvalues for the minimal representations  with
  the higher spin-$1,2,3$ currents 
  %in the
  %higher representations
  in the $\frac{SU(N+2)}{SU(N)}$  coset}
%section7%%%%%%%%%%%%%%%%%%
%%%%%%%%%%%%%%%%%%%%%%%%%%%%%%%%%%%%%%%%%%%%%%%%%%%%%%%%

The higher spin $1$ current
in the linear version is the same as the previous higher spin
$1$ current in (\ref{finalspinone}).
The higher spin $2$ current
in the linear version behaves as the one in the nonlinear
version exactly as long as the eigenvalues are concerned
(note that the higher spin $\frac{3}{2}$ currents contain
the product of adjoint spin $1$ current and adjoint spin $\frac{1}{2}$
current. See also the equation $(4.7)$ of \cite{AKK1703}).
Therefore, one needs to have the eigenvalues for
the higher spin $3$ current only.
In next two sections, the previous eigenvalues
$\phi_0^{(1)}$ and $v^{\pm}$ acting on $(\La_+;\La_-)$ described in sections
$4$ and $5$
are still valid.

Recall that the higher spin $3$ current in the linear version
\cite{BCG}
is 
\bea
{\bf \Phi_{2}^{(1)}}(z) & = &  \Phi_{2}^{(1)}(z)-c_{7}\,
    {\bf L\Phi_{0}^{(1)}}(z)-c_{8}\, \partial^{2} {\bf \Phi_{0}^{(1)}}(z)-
    c_{9}\, \partial {\bf U\Phi_{0}^{(1)}}(z)-c_{10}\,
    {\bf U } \partial {\bf \Phi_{0}^{(1)}}(z)
    \label{Phinonandlin}
    \\
&-& c_{11}\,{\bf U U \Phi_{0}^{(1)}}(z)
-c_{12}\, \partial {\bf \Gamma^{\mu}\Phi_{\frac{1}{2}}^{(1),\mu}}(z)-
c_{13}\,{\bf \Gamma^{\mu}} \partial {\bf \Phi_{\frac{1}{2}}^{(1),\mu}}(z)-
c_{14}\,  \partial {\bf \Gamma^{\mu}\Gamma^{\mu}\Phi_{0}^{(1)}}(z)
\nonu \\
& \rightarrow &
 \Phi_{2}^{(1)}(z)-c_{7}\,
    {\bf L\Phi_{0}^{(1)}}(z)-c_{8}\, \partial^{2} {\bf \Phi_{0}^{(1)}}(z)-
    c_{9}\, \partial {\bf U\Phi_{0}^{(1)}}(z)-c_{10}\,
    {\bf U } \partial {\bf \Phi_{0}^{(1)}}(z)
    \nonu 
    \\
&-& c_{11}\,{\bf U U \Phi_{0}^{(1)}}(z),
\nonu 
    \eea
    where
    the $c_{12}$, $c_{13}$ and $c_{14}$ terms
    do not contribute to
    the eigenvalues for the representation $(\La_+;\La_-)$.
    Recall that the higher spin $\frac{3}{2}$ currents
    $ {\bf \Phi_{\frac{1}{2}}^{(1),\mu}}(z)$ contain
the product of adjoint spin $1$ current and adjoint spin $\frac{1}{2}$
current. 
    Here the $(N,k)$ dependent coefficients are given by
\bea
c_ {7} & = & \frac{16 (k-N)}{(6 k N+5 k+5 N+4)}, \qquad
c_ {8}  = -\frac{(k-N) (6 k N+29 k+29 N+52)}{3 (k+N+2) (6 k N+5 k+5 N+4)},
\nonu\\
c_ {9}& = & -\frac{4}{(k+N+2)}, \qquad
c_ {10} = \frac{4}{(k+N+2)}, \nonu\\
c_ {11} & = & \frac{16 (k-N)}{(k+N+2) (6 k N+5 k+5 N+4)}, \nonu\\
c_ {12}& = & \frac{6 i}{(2 + k + N)}, \qquad
c_ {13} = -\frac{2 i}{(2 + k + N)},  \nonu\\
c_ {14} & = &
-\frac{16 (k-N)}{(k+N+2) (6 k N+5 k+5 N+4)}.
\label{ccoeff}
\eea

In the primary basis \cite{AK1509}, the following quantity can be
constructed 
\bea
    {\bf \widetilde\Phi_{2}^{(1)}}(z) & \equiv &
    {\bf \Phi_{2}^{(1)}}(z)+
p_{1}\:\partial^{2}{\bf \Phi_{0}^{(1)}}(z)+p_{2}\:{\bf L}{\bf \Phi_{0}^{(1)}}(z)
\nonu \\
& \rightarrow &
 \Phi_{2}^{(1)}(z) + (p_2 -c_{7})\,
    {\bf L\Phi_{0}^{(1)}}(z) +(p_1-c_{8})\, \partial^{2} {\bf \Phi_{0}^{(1)}}(z)-
    c_{9}\, \partial {\bf U\Phi_{0}^{(1)}}(z) \nonu \\
    & - & c_{10}\,
    {\bf U } \partial {\bf \Phi_{0}^{(1)}}(z)
- c_{11}\,{\bf U U \Phi_{0}^{(1)}}(z)
\label{widetildenot}
\eea
where the two quantities are introduced
{\small
\bea
p_{1} & \equiv &  -\frac{(k-N) (3 k N+16 k+16 N+29)}{3 (k+N+2)
  (3 k N+4 k+4 N+5)}, \qquad
p_{2}  \equiv  \frac{8 (k-N)}{(3 k N+4 k+4 N+5)}.
\label{p1p2}
\eea}
The relation (\ref{Phinonandlin}) is used in (\ref{widetildenot}).
In the remaining parts of this paper, the higher spin $3$ current
is given by (\ref{widetildenot}). We will examine the
corresponding eigenvalues by considering both
the higher spin $3$ current in the nonlinear version and
the other remaining parts separately in order to see how they
($\Phi_2^{(1)}(z)$ and $ {\bf \widetilde\Phi_{2}^{(1)}}(z)$) behave under the
large $(N,k)$ 't Hooft like limit. 

%%%%%%%%%%%%%%%%%%%%%%%%%%%%%%%
\subsection{ The eigenvalues in the $(0;f)$
and $(0;\overline{f})$ representations}
%%%%%%%%%%%%%%%%%%%%%%%%%%%%%%%

The relevant subsections are given by \ref{0frep} and \ref{41subsection}.
The two eigenvalues associated with
one of the minimal representations
can be summarized by
{\small
  \bea
{\bf h}  (0;
  \tiny\yng(1)
) & = & \frac{(N k+ 2N+1)}{2N(N+k+2)},
\nonu \\
 {\bf \phi_2^{(1)}} (0;
  \tiny\yng(1)
) & = &
   \frac{4 k
     ({\bf 6 k^2 N^2}+5 k^2 N {\bf +3 k N^3}
     +24 k N^2+16 k N-3 k+4 N^3+23 N^2+18 N)}
    {3 N (k+N+2)^2 (3 k N+4 k+4 N+5)}.
    \label{0feigenlinear}   
\eea}
For the first eigenvalue,
one should calculate the OPE
between ${\bf T}(z)$ and
$Q^{13}(w)$ (in $SU(5)$) and read off
the second order pole.
The coefficient of $Q^{13}(w)$ in the right hand side of
this OPE is the corresponding eigenvalue.
For the second eigenvalue,
one computes 
 the OPE
between ${\bf \widetilde\Phi_2^{(1)}}(z)$ and
$Q^{13}(w)$ (in $SU(5)$) and read off
the third order pole.
Of course, all the higher spin currents do not
contain the spin-$1$ currents $V^a(z)$.
By counting the highest powers of $k$ or $N$ (the sum of powers
in $k$ and $N$ for the expressions containing both dependences)
in the numerators and
the denominators appearing in the above eigenvalues,
one can observe the behaviors under the
large $(N,k)$ 't Hooft like limit and  
these eigenvalues
approach to the finite values \footnote{The highest power terms
  of the eigenvalue
  for the higher spin $3$ current are exactly the same as the ones in
  (\ref{0feigenhigher}) although the denominators are little different
  from each other at finite $(N,k)$.}.

Similarly, the other two eigenvalues can be also obtained
from
{\small
  \bea
{\bf h} (0;
  \overline{\tiny\yng(1)}
) & = &   \frac{(N k+ 2N+1)}{2N(N+k+2)},
\nonu \\
 {\bf \phi_2^{(1)}} (0;
  \overline{\tiny\yng(1)}
) & = & - \frac{4 k
    ({\bf 6 k^2 N^2}+5 k^2 N {\bf +3 k N^3}
    +24 k N^2+16 k N-3 k+4 N^3+23 N^2+18 N)}
    {3 N (k+N+2)^2 (3 k N+4 k+4 N+5)}.
\label{0fbareigenhigherlinear}
\eea}
Because the generators for the complex conjugated (antifundamental)
representation $\overline{\tiny\yng(1)}$
have an extra minus sign compared to the fundamental representation
$\tiny\yng(1)$, the eigenvalue for the odd higher spin currents
(corresponding to the second)
have an extra minus sign and the ones for the even higher spin currents
(corresponding to the first)
remain the same compared to the results of the previous paragraph.
More explicitly, one can obtain the OPEs between the ``reduced''
(higher) spin currents and the $\frac{1}{2}$ current $Q^1(w)$.
By reading off the corresponding coefficients in the appropriate
poles, the above eigenvalues can be determined.

%%%%%%%%%%%%%%%%%%%%%
\subsection{The eigenvalues in the $(f;0)$
and $(\overline{f};0)$ representations}
%%%%%%%%%%%%%%%%%%%%%

The relevant subsections are given by \ref{f0subsectionname}
and $4.2$.
The two eigenvalues can be described as
\bea
 {\bf h} (
  \tiny\yng(1);0
) & = & \frac{(N+1)(N+3)}{2(N+2)(N+k+2)},
\nonu \\
 {\bf \phi_2^{(1)}} (
  \tiny\yng(1);0
) & = & 
  -\frac{4}{3 (N+2) (k+N+2)^2 (3 k N+4 k+4 N+5)}
  \nonu \\
  & \times &
  N ({\bf 3 k^2 N^2}+10 k^2 N+8 k^2{\bf +6 k N^3}+36 k N^2+71 k N+43 k
  \nonu \\
  & + & 5 N^3+26 N^2+50 N+30).
 \label{f0eigenhigherlinear}
 \eea
 One obtains
 these eigenvalues
 by substituting the $SU(N+2)$ generators
 $T_{a^{\ast}}$ into the zero mode of the spin $1$ current
 $V_0^a$ in the corresponding ``reduced'' (higher) spin currents
 where all the $Q^a(z)$ dependent terms are ignored.
 Then one has the unitary matrix acting on the corresponding state and the
 diagonal elements of the last $2\times 2$ subdiagonal matrix
 provide the above eigenvalues.
As observed in \cite{AK1506}, under the
 symmetry $N \leftrightarrow k$ and $ 0 \leftrightarrow \tiny\yng(1)$,
 the eigenvalues do not remain the same, compared to (\ref{0feigenlinear}).
 However, the large $(N,k)$ 't Hooft limit provide the coincident values
 with minus sign
 \footnote{One can  use $h (
   \tiny\yng(1);0)
   + \frac{1}{(N+k+2)} \frac{1}{N(N+2)} \hat{u}^2
   (\tiny\yng(1);0)=\frac{(2N+3)}{4(N+k+2)} +
   \frac{(-\frac{N}{2})^2}{(N+k+2)} \frac{1}{N(N+2)} $  to obtain the
 eigenvalue in (\ref{f0eigenhigherlinear}).}. 
 
When we consider the complex conjugated representation,    
the following results hold 
 \bea
 {\bf h} (
\overline{  \tiny\yng(1)};0
) & = & \frac{(N+1)(N+3)}{2(N+2)(N+k+2)},
\nonu \\
 {\bf \phi_2^{(1)}} (
  \overline{\tiny\yng(1)};0
) & = & \frac{4}{3 (N+2) (k+N+2)^2 (3 k N+4 k+4 N+5)}
  \nonu \\
  & \times &
  N ({\bf 3 k^2 N^2}+10 k^2 N+8 k^2 {\bf +6 k N^3}+36 k N^2+71 k N+43 k
  \nonu \\
  & + & 5 N^3+26 N^2+50 N+30).
 \label{fbarzerolinearexpression}
\eea
According to the previous analysis, the
first eigenvalue remains the same
and the second eigenvalue
has the extra minus signs,
compared to the ones in (\ref{f0eigenhigherlinear}).

%%%%%%%%%%%%%%%%%%%%%%%%%%%%%%%%%%%%%%%%%%%%%%%%%%%%%%%%
%%%%%%%%%%%%%%%%%%%
\section{ Eigenvalues for the higher representations  with
  the higher spin-$1,2,3$ currents
  %in the
  %higher representations
  in the $\frac{SU(N+2)}{SU(N)}$  coset}
%section8%%%%%%%%%%%%%%%%%%
%%%%%%%%%%%%%%%%%%%%%%%%%%%%%%%%%%%%%%%%%%%%%%%%%%%%%%%%

In this section, there are $22$ subsections where
we consider the explicit $22$ higher representations
as done in section $5$. The same analysis done in section $5$
is applied to this section. Note
that  each subsection in this section corresponds to
each section of section $5$.
For the representations where $\La_-$ appears in the branching of
$\La_+$, we write down the eigenvalues for the higher spin $3$
current completely. On the other hand, for the
representations where the product of $(\La_+;0)$ and $(0;\La_-)$
occurs, we describe some relations
for them rather than presenting the complete expressions
because they are rather involved due to the presence of imaginary $i$.

%%%%%%%%%%%%%%%%%%%%%%%%%%%%%%%%%%%%%%%%%%%%%%
\subsection{The $(f;f)$ representation}
%%%%%%%%%%%%%%%%%%%%%%%%%%%%%%%%%%%%%%%%%%%%%%

In this case,
when one takes the $N \times N $ subdiagonal unitary matrix
inside of $(N+2) \times (N+2)$ unitary matrix,
the corresponding diagonal elements for the higher spin currents
provide the following two
eigenvalues 
\bea
{\bf h} (
  \tiny\yng(1);\tiny\yng(1)) 
  & = &  \frac{(N+1)^2}{N (N+2) (k+N+2)},
\nonu \\
 {\bf \phi_2^{(1)}} (
\tiny\yng(1);\tiny\yng(1))
& = & \phi_2^{(1)} (
\tiny\yng(1);\tiny\yng(1)) + \Delta (
\tiny\yng(1);\tiny\yng(1))
\nonu \\
  & = & -\frac{8 (N-k) ({\bf 3 k N^3}+10 k N^2+8 k N+4 N^3+19 N^2+22 N+6)}{
    3 N (N+2) (k+N+2)^2 (3 k N+4 k+4 N+5)}.
\label{boxboxlinear}
\eea
The previous relation in (\ref{ffeigenvalues1}) is used. 
The first eigenvalue can be interpreted as follows.
By recalling that the stress energy tensor in the linear version
has the extra contribution from the spin $1$ current,
one can write down the precise relation from (\ref{ffeigenvalues})
\bea
&&   h (
   \tiny\yng(1);\tiny\yng(1))
   + \frac{1}{(N+k+2)} \frac{1}{N(N+2)} \hat{u}^2
   (\tiny\yng(1);\tiny\yng(1))
 =  \frac{1}{(N+k+2)}+
 \frac{1}{N (N+2) (k+N+2)},
 \nonu
 %\\
 %&& =
 % {\bf h} (
 %  0;\tiny\yng(1))
%+  {\bf h} (
%   \tiny\yng(1);0)
%    =  \frac{(k N+2 N+1)}{2 N (k+N+2)} +
%\frac{(N+1) (N+3)}{2 (N+2) (k+N+2)}
%   \nonu
 \eea
 which is exactly the same as the first eigenvalue
 of (\ref{boxboxlinear}).
By power counting of $N$ and $k$, one sees that the above eigenvalues
behave as $\frac{1}{N}$ dependence under the large $(N,k)$ 't Hooft like
limit.

Let us introduce the difference between the higher spin $3$ currents
in the linear and nonlinear version from (\ref{widetildenot})
\bea
{\bf \Phi}(z) \equiv \Bigg[
 (p_2 -c_{7})\,
    {\bf L\Phi_{0}^{(1)}} +(p_1-c_{8})\, \partial^{2} {\bf \Phi_{0}^{(1)}}-
    c_{9}\, \partial {\bf U\Phi_{0}^{(1)}} 
     -  c_{10}\,
    {\bf U } \partial {\bf \Phi_{0}^{(1)}}
- c_{11}\,{\bf U U \Phi_{0}^{(1)}} \Bigg](z).
\label{BigPhi}
\eea
The coefficients are given by (\ref{ccoeff}) and (\ref{p1p2}).
Let us denote the eigenvalue of the zero mode of this field 
as $\Delta$ and it turns out that the eigenvalue
in the above higher representation
is described as 
\bea
\Delta (
\tiny\yng(1);\tiny\yng(1)) & = &
-\frac{16 (k-N) ({\bf 3 k N^2}-5 k {\bf +3 N^3}+12 N^2+7 N-4)}{
  N (N+2) (k+N+2)^2 (3 k N+4 k+4 N+5) (6 k N+5 k+5 N+4)}.
\label{corrff}
\eea
This can be obtained as follows.
Let us denote the zero mode eigenvalues for 
${\bf L\Phi_{0}^{(1)}}$, $\partial^{2} {\bf \Phi_{0}^{(1)}}$,
 $\partial {\bf U\Phi_{0}^{(1)}}$,
   $ {\bf U } \partial {\bf \Phi_{0}^{(1)}}$,
and ${\bf U U \Phi_{0}^{(1)}}$
as
${\bf l \phi_{0}^{(1)}}$, $\partial^{2} {\bf \phi_{0}^{(1)}}$,
 $\partial {\bf u \phi_{0}^{(1)}}$,
   $ {\bf u } \partial {\bf \phi_{0}^{(1)}}$,
and ${\bf u u \phi_{0}^{(1)}}$.
Then these eigenvalues for the representation
$(\tiny\yng(1),\tiny\yng(1))$ are given by 
\bea
&& \Bigg[\frac{2 (k N^2+2 k N+N^3+5 N^2+6 N+1)}{N (N+2) (k+N+2)^2},
  \frac{4}{(k+N+2)},
  -\frac{2 i}{\sqrt{N (N+2)} (k+N+2)}, \nonu \\
&& -\frac{2 i}{\sqrt{N (N+2)} (k+N+2)}, 
 -\frac{2}{N (N+2) (k+N+2)} \Bigg].
\label{eigenvalues}
\eea
Then by substituting the coefficients (\ref{ccoeff}) and (\ref{p1p2})
into the relation (\ref{BigPhi}) together with
(\ref{eigenvalues}) one sees the final result in (\ref{corrff}).
Note that although the third and fourth eigenvalues in (\ref{eigenvalues})
contain the imaginary $i$ with same $(N,k)$ coefficients,
due to the vanishing of $(c_9 + c_{10})$ in (\ref{ccoeff}),
there is no contribution
for these eigenvalues.
It is easy to see that the simple power counting
of $(N,k)$ implies that (\ref{corrff}) behaves as
$\frac{1}{N}$ under the large $(N,k)$ 't Hooft limit
\footnote{One has the following eigenvalue in the complex conjugated
  representation $(\overline{\tiny\yng(1)};\overline{\tiny\yng(1)}))$
  \bea
{\bf \phi_2^{(1)}} (
\overline{\tiny\yng(1)};\overline{\tiny\yng(1)})
& = & \phi_2^{(1)} (
\overline{\tiny\yng(1)};\overline{\tiny\yng(1)}) + \Delta (
\overline{\tiny\yng(1)};\overline{\tiny\yng(1)})
= -\phi_2^{(1)} (
\tiny\yng(1);\tiny\yng(1)) - \Delta (
\tiny\yng(1);\tiny\yng(1))
\nonu
\eea
together with (\ref{ffeigenvalues1}) and (\ref{corrff}).
See also the footnote \ref{barbarfoot}.
The extra minus sign comes from the spin $3$ associated with
the cubic generators of $SU(N+2)$.}.
In particular, the highest power of $(N,k)$ in the eigenvalue
for the higher spin $3$ current in (\ref{boxboxlinear})
plays the role of next leading order $\frac{1}{N}$ and this will be
the basic quantity for the arbitrary representation $(\La_{+};\La_{+})$
because the eigenvalue will be the multiple of the eigenvalue for
$(\tiny\yng(1),\tiny\yng(1))$.

%%%%%%%%%%%%%%%%%%%%%%%%%%%%%%%%%%%%%%%%%%%%%%%%%%%%
\subsection{The $(f;\overline{f})$ representation}
%%%%%%%%%%%%%%%%%%%%%%%%%%%%%%%%%%%%%%%%%%%%%%%%%%%

The two eigenvalues corresponding to the zero modes
of the (higher spin) currents of spins $2$ and $3$
which act on
the representation $(\tiny\yng(1); \overline{\tiny\yng(1)})$
can be summarized by
{\small
  \bea
  {\bf h} (
  \tiny\yng(1);\overline{\tiny\yng(1)}) 
  &= & \frac{(k N^2+2 k N+N^3+6 N^2+8 N+2)}{2 N (N+2) (k+N+2)},
  \nonu \\ 
 {\bf \phi_2^{(1)}} (
  \tiny\yng(1);\overline{\tiny\yng(1)}) 
&= &  \phi_2^{(1)} (
  \tiny\yng(1);\overline{\tiny\yng(1)}) +
\Delta(\tiny\yng(1);0) + \Delta(0;\overline{\tiny\yng(1)}) +
  \frac{4}{9}
\Bigg[ \frac{18 i \sqrt{\frac{N}{N+2}}}{(k+N+2)}
 \nonu \\
& 
 + & \frac{9}{(N+2) (k+N+2)^2 (3 k N+4 k+4 N+5) (6 k N+5 k+5 N+4)}
 \nonu \\
 & \times &
 (N-k) ({\bf -27 k^2 N^2}-70 k^2 N-39 k^2 {\bf +9 k N^3}+4 k N^2-31 k N-12 k
 \nonu \\
 & + & 2 N^3+8 N^2+20 N+24) \Bigg].
  \label{ffbareigenhigherlinear}
\eea}
The eigenvalue for the higher spin $3$ current
in the nonlinear version can be found in (\ref{nonlinearffbar}).
Note that the eigenvalue for the  spin $2$ current
does not have any contribution from the commutator
$[({\bf T})_0, Q^{\bar{A}}_{-\frac{1}{2}}]$.
This provides only the eigenvalue for the representation $(0;\overline{
  \tiny\yng(1)})$. Also the term 
$ Q^{\bar{A}}_{-\frac{1}{2}} ({\bf T})_0$
acting on the representation $(\tiny\yng(1);0)$
gives the eigenvalue ${\bf h} (\tiny\yng(1);0)$ with
$ Q^{\bar{A}}_{-\frac{1}{2}}$ acting on the state  $|(\tiny\yng(1);0)>$.
By inserting the overall factor into this state, one has
the final state associated with 
the representation $(\tiny\yng(1);\overline{\tiny\yng(1)})$.
Therefore, one arrives at the above eigenvalue for the spin $2$
current
\footnote{
  One can calculate the eigenvalue equations \cite{AKK1703} for each term
  in (\ref{BigPhi})
  \bea
      {\bf l \phi_{0}^{(1)}} (\tiny\yng(1);0)
& = &
 -\frac{N \left(2 k N+4 k+3 N^2+12 N+11\right)}{2 (N+2) (k+N+2)^2}
   (\tiny\yng(1);0) = -  {\bf l \phi_{0}^{(1)}} (\overline{\tiny\yng(1)};0),
%\label{c7c11cal-1}
\nonu
 \\
\partial^{2} {\bf \phi_{0}^{(1)}} (\tiny\yng(1);0)
& = & 
 -\frac{2 N}{(k+N+2)}
 (\tiny\yng(1);0) = - \partial^{2} {\bf \phi_{0}^{(1)}}
 (\overline{\tiny\yng(1)};0),
\nonu \\
 \partial {\bf u \phi_{0}^{(1)}}  (\tiny\yng(1);0)
& = &  
-\frac{i N \sqrt{\frac{N}
      {N+2}}}{2 (k+N+2)}
(\tiny\yng(1);0) =  -\partial {\bf u \phi_{0}^{(1)}}
(\overline{\tiny\yng(1)};0),
\nonu \\
 {\bf u } \partial {\bf \phi_{0}^{(1)}} (\tiny\yng(1);0)
& = &
-\frac{i N \sqrt{\frac{N}{N+2}}}
  {2 (k+N+2)}
  (\tiny\yng(1);0) =- 
  {\bf u } \partial {\bf \phi_{0}^{(1)}} (\overline{\tiny\yng(1)};0),
\nonu \\
{\bf u u \phi_{0}^{(1)}} (\tiny\yng(1);0)
 & = & 
\frac{N^2}{4 (N+2) (k+N+2)}
   (\tiny\yng(1);0)= - {\bf u u \phi_{0}^{(1)}} (\overline{\tiny\yng(1)};0).
\label{knownresult1}
\eea
Similarly, one has the following eigenvalue equations
\bea
{\bf l \phi_{0}^{(1)}} (0;\tiny\yng(1))
& = &
 -\frac{k (3 k N+2 N^2+6 N+1)}{2 N (k+N+2)^2}
   (0;\tiny\yng(1)) = - {\bf l \phi_{0}^{(1)}} (0;\overline{\tiny\yng(1)}),
%\label{c7c11cal-2}
\nonu \\
\partial^{2} {\bf \phi_{0}^{(1)}} (0;\tiny\yng(1))
& = & 
 -\frac{2 k}{(k+N+2)}
 (0;\tiny\yng(1)) = - \partial^{2} {\bf \phi_{0}^{(1)}}
 (0;\overline{\tiny\yng(1)}),
 \nonu \\
\partial {\bf u \phi_{0}^{(1)}} (0;\tiny\yng(1))
& = &  
-\frac{i k \sqrt{\frac{N+2}{N}}}{2 (k+N+2)}
(0;\tiny\yng(1)) =\partial {\bf u \phi_{0}^{(1)}}
(0;\overline{\tiny\yng(1)}),
\nonu \\
 {\bf u } \partial {\bf \phi_{0}^{(1)}} (0;\tiny\yng(1))
& = &
-\frac{i k \sqrt{\frac{N+2}{N}}}{2 (k+N+2)}
(0;\tiny\yng(1))=  {\bf u } \partial {\bf \phi_{0}^{(1)}}
(0;\overline{\tiny\yng(1)}),
\nonu \\
 {\bf u u \phi_{0}^{(1)}}  (0;\tiny\yng(1))
& = & 
\frac{k (N+2)}{4 N (k+N+2)}
(0;\tiny\yng(1))=- {\bf u u \phi_{0}^{(1)}}
(0;\overline{\tiny\yng(1)}).
\label{knownresult2}
\eea
In the third and fourth eigenvalues,
there are no sign changes under the $\tiny\yng(1) \leftrightarrow
\overline{\tiny\yng(1)}$.
}.
As before, one can describe this eigenvalue in the context of
the corresponding eigenvalue nonlinear version. That is,
\bea
&&   h (
   \tiny\yng(1);\overline{\tiny\yng(1)})
   + \frac{1}{(N+k+2)} \frac{1}{N(N+2)} \hat{u}^2
   (\tiny\yng(1);\overline{\tiny\yng(1)})
 =  \frac{1}{2}+
 \frac{(-N-1)^2}{N (N+2) (k+N+2)}
 \nonu \\
 && =
  {\bf h} (
   0;\overline{\tiny\yng(1)})
+  {\bf h} (
   \tiny\yng(1);0)
    =  \frac{(k N+2 N+1)}{2 N (k+N+2)} +
\frac{(N+1) (N+3)}{2 (N+2) (k+N+2)}.
 \nonu
\eea
In the first line, the relations in (\ref{ffbareigenvalues}) are used.
In the second line, the previous relations in (\ref{0fbareigenhigherlinear})
and (\ref{f0eigenhigherlinear}) are used.

For the eigenvalue corresponding to the higher spin $3$ current,
there exists the contribution from the lower order poles
appearing in the commutator $[({\bf \Phi_2^{(1)}})_0,
  Q^{\bar{A}}_{-\frac{1}{2}}]$. 
One describes the following eigenvalues for each term appearing in
(\ref{BigPhi})
{\small
\bea
&& \Bigg[-\frac{(2 k N^2+2 k N-3 k+6 N^2+15 N+6)}{2 (N+2) (k+N+2)^2},0,
\frac{i \sqrt{\frac{N}{N+2}} (k+3 N+6)}{2 (k+N+2)},
-\frac{i \sqrt{\frac{N}{N+2}} (3 k+N+2)}{2 (k+N+2)},
\nonu \\
&& \frac{(-7 k N-12 k+5 N^2+16 N+12)}{4 (N+2) (k+N+2)} \Bigg].
\label{commutator1}
\eea}
Note that the eigenvalues in the third and the fourth are different
from each other.
This is the reason why there exists an imaginary $i$ term in the eigenvalue
in (\ref{ffbareigenhigherlinear}).
Then by substituting the coefficients (\ref{ccoeff}) and (\ref{p1p2})
into the relation (\ref{BigPhi}) together with
(\ref{commutator1}) one sees the final result in
(\ref{ffbareigenhigherlinear}).

Furthermore, there are $\Delta(\tiny\yng(1);0)$ and $\Delta(0;\overline{
\tiny\yng(1)})$. They can be obtained explicitly.
Once again, the
large $(N,k)$ 't Hooft like limit
for $\Delta(\tiny\yng(1);0) + \Delta(0;\overline{\tiny\yng(1)})$   
lead to the $\frac{1}{N}$ behavior.
%\footnote{%In the third and fourth eigenvalues,
%there are no sign changes under the $\tiny\yng(1) \leftrightarrow
%\overline{\tiny\yng(1)}$.
Note that the eigenvalues appearing in (\ref{knownresult1})
and (\ref{knownresult2}) are finite but the coefficients
in (\ref{ccoeff}) and (\ref{p1p2}) behave as $\frac{1}{N}$.
Similarly, one can easily see that 
due to the finiteness in (\ref{commutator1}),
the final extra terms in (\ref{ffbareigenhigherlinear})
vanish under the large $(N,k)$ 't Hooft limit.
Therefore, the eigenvalues for the higher spin $3$ current
in this higher representation in both linear and nonlinear versions
are the same as each other
\footnote{
In the similar higher representation, one obtains
  \bea
  {\bf h} (
  \overline{\tiny\yng(1)} ;\tiny\yng(1)) 
  &= & \frac{(k N^2+2 k N+N^3+6 N^2+8 N+2)}{2 N (N+2) (k+N+2)},
  \nonu \\ 
 {\bf \phi_2^{(1)}} (
  \overline{\tiny\yng(1)};\tiny\yng(1)) 
&= &  \phi_2^{(1)} (
  \overline{\tiny\yng(1)};\tiny\yng(1)) +
\Delta(\overline{\tiny\yng(1)};0) + \Delta(0;\tiny\yng(1)) +
  \frac{4}{9}
\Bigg[ -\frac{18 i \sqrt{\frac{N}{N+2}}}{(k+N+2)}
 \nonu \\
& 
 + & \frac{9}{(N+2) (k+N+2)^2 (3 k N+4 k+4 N+5) (6 k N+5 k+5 N+4)}
 \nonu \\
 & \times &
 (k-N)
 (27 k^2 N^2+86 k^2 N+57 k^2+9 k N^3+64 k N^2+119 k N
 \nonu \\
 & + & 48 k+14 N^3+44 N^2+20 N-24)
 \Bigg],
\nonu
\eea
where the last two terms in the eigenvalue of higher spin $3$ current
can be determined from the following eigenvalues 
\bea
&& \Bigg[ -\frac{(4 k N^2+10 k N+3 k+6 N^2+15 N+6)}{2 (N+2) (k+N+2)^2},0,
-\frac{i \sqrt{\frac{N}{N+2}} (k+3 N+6)}{2 (k+N+2)},
\frac{i \sqrt{\frac{N}{N+2}} (3 k+N+2)}{2 (k+N+2)},
\nonu \\
&& -\frac{(5 k N+12 k-N^2-8 N-12)}{4 (N+2) (k+N+2)} \Bigg].
\nonu
\eea
See also (\ref{fbarzerolinearexpression}) and
(\ref{0feigenlinear}) for the eigenvalue of the spin $2$ current.
The eigenvalues
$\Delta(\overline{\tiny\yng(1)};0)$ and $\Delta(0;\tiny\yng(1)) $
can be determined from the previous relations (\ref{knownresult1})
and (\ref{knownresult2}).
Furthermore, the footnote \ref{footfbarfresult} is needed for
the explicit expression of $\phi_2^{(1)} (
  \overline{\tiny\yng(1)};\tiny\yng(1))$.
}.

%%%%%%%%%%%%%%%%%%%%%%%%%%%%%%%%%%%%%%%%%%%%%%
\subsection{The $(f;\mbox{symm})$ representation}
%%%%%%%%%%%%%%%%%%%%%%%%%%%%%%%%%%%%%%%%%%%%%%

This higher representation can be obtained from the
product of
$(\tiny\yng(1);0)$ and $(0;\tiny\yng(2))$.
It turns out that the two eigenvalues are given by 
{\small
  \bea
  {\bf h} (
  \tiny\yng(1);\tiny\yng(2)) 
  &= & \frac{(2 k N^2+4 k N+N^3+6 N^2+11 N+8)}{2 N (N+2) (k+N+2)},
  \nonu \\ 
 {\bf \phi_2^{(1)}} (
  \tiny\yng(1);\tiny\yng(2)) 
&= &  \phi_2^{(1)} (
  \tiny\yng(1);\tiny\yng(2)) +
\Delta(\tiny\yng(1);0) + \Delta(0;\tiny\yng(2)) +
  \frac{4}{9}
\Bigg[ -\frac{36 i \sqrt{\frac{N}{N+2}}}{(k+N+2)}
 \nonu \\
& 
 + & \frac{18}{(N+2) (k+N+2)^2 (3 k N+4 k+4 N+5) (6 k N+5 k+5 N+4)}
 \nonu \\
 & \times &
 (k-N) ({\bf 45 k^2 N^2}+146 k^2 N+105 k^2+31 k N^2+104 k N
 \nonu \\
  & + & 78 k-7 N^3-46 N^2-88 N-48)
 \Bigg].
\label{fsymmlinear}
\eea}
The previous relation (\ref{fsymmeigenhigher})
can be inserted.
Note that the eigenvalue for the spin $2$ current
does not have any contribution from the commutator
$[({\bf T})_0, Q^{13}_{-\frac{1}{2}}  Q^{16}_{-\frac{1}{2}} ]$.
This provides only the eigenvalue for the representation $(0;
  \tiny\yng(2))$. Also the term 
$ Q^{13}_{-\frac{1}{2}}  Q^{16}_{-\frac{1}{2}} ({\bf T})_0$
acting on the representation $(\tiny\yng(1);0)$
gives the eigenvalue ${\bf h} (\tiny\yng(1);0)$ with
$ Q^{13}_{-\frac{1}{2}}  Q^{16}_{-\frac{1}{2}}$
acting on the state  $|(\tiny\yng(1);0)>$.
By inserting the overall factor into this state, one has
the final state associated with 
the representation $(\tiny\yng(1);\tiny\yng(2))$.
Therefore, one arrives at the above eigenvalue for the spin $2$
current.

Furthermore, one can interpret the above eigenvalue of
spin $2$ current from the one in the nonlinear version as follows:
\bea
&&   h (
   \tiny\yng(1);\tiny\yng(2))
   + \frac{1}{(N+k+2)} \frac{1}{N(N+2)} \hat{u}^2
   (\tiny\yng(1);\tiny\yng(2))
 =  \frac{(2N+4k+7)}{4(N+k+2)}+
 \frac{(-\frac{N}{2}+N+2)^2}{N (N+2) (k+N+2)}
 \nonu \\
 && =
  {\bf h} (
   0;\tiny\yng(2))
+  {\bf h} (
   \tiny\yng(1);0)
    =  \frac{(k N+ N+2)}{ N (k+N+2)} +
\frac{(N+1) (N+3)}{2 (N+2) (k+N+2)}.
 \nonu
\eea
In the first line, the first result
can be obtained from the conformal dimension formula in previous
sections and the $\hat{u}$ charge is the sum of
$\hat{u}$ charges of $(
   0;\tiny\yng(2))$ and $(
   \tiny\yng(1);0)$.
   In the second line the previous relation (\ref{f0eigenhigherlinear})
   is used and the relation (\ref{0symmeigenlinear})
   is used.

   Furthermore, there are $\Delta(\tiny\yng(1);0)$
   and $\Delta(0;
   \tiny\yng(2))$. The former can be obtained from the
   relation (\ref{knownresult1})
   while the latter is given by (\ref{Delta0symm}).
   
   For the second eigenvalue
   corresponding to the higher spin $3$ current,
there is the contribution from the lower order poles
appearing in the commutator, $[({\bf \Phi_2^{(1)}})_0, 
  Q^{13}_{-\frac{1}{2}}  Q^{16}_{-\frac{1}{2}}]$. 
By using the following eigenvalues with the appropriate coefficients
{\small
  \bea
&& \Bigg[-\frac{(4 k N^2+10 k N+3 k+3 N^2+12 N+12)}{(N+2) (k+N+2)^2} ,0,
-\frac{i \sqrt{\frac{N}{N+2}} (k+3 N+6)}{k+N+2},
\frac{i \sqrt{\frac{N}{N+2}} (3 k+N+2)}{k+N+2},
\nonu \\
&&  -\frac{(11 k N+24 k-4 N^2-20 N-24)}{2 (N+2) (k+N+2)}\Bigg],
\nonu
\eea}
one determines the last two terms in the eigenvalue of the higher
spin $3$ current in (\ref{fsymmlinear})
\footnote{For the similar higher representation, one has
  \bea
{\bf \phi_2^{(1)}} (
\overline{\tiny\yng(1)};\tiny\yng(2))
& = & \phi_2^{(1)} (
\overline{\tiny\yng(1)};\tiny\yng(2)) + \Delta (
\overline{\tiny\yng(1)};0) +
 \Delta ( 0;\tiny\yng(2))
\nonu
\eea
plus the last terms in (\ref{fsymmlinear}) with an extra minus sign
because the cubic generators can have an extra minus sign under the
complex conjugated representation.
Note that the quantity $\phi_2^{(1)} (
\overline{\tiny\yng(1)};\tiny\yng(2))$
can be obtained in the footnote \ref{fbarsymmresultresult}.
Of course, 
$\Delta(\overline{\tiny\yng(1)};0)$ 
can be determined from the previous relation (\ref{knownresult1}).
%\bea
%&& \Bigg[
%  \frac{(4 k N^2+10 k N+3 k+3 N^2+12 N+12)}{(N+2) (k+N+2)^2},
%  0,
%  \frac{i \sqrt{\frac{N}{N+2}} (k+3 N+6)}{k+N+2},
%  -\frac{i \sqrt{\frac{N}{N+2}} (3 k+N+2)}{k+N+2},
%  \nonu \\
%  &&
%\frac{(11 k N+24 k-4 N^2-20 N-24)}{2 (N+2) (k+N+2)}
%  \Bigg].
%\nonu
%\eea
}.

%%%%%%%%%%%%%%%%%%%%%%%%%%%%%%%%%%%%%%%%%%%%%%
\subsection{The $(f;\overline{\mbox{symm}})$ representation}
%%%%%%%%%%%%%%%%%%%%%%%%%%%%%%%%%%%%%%%%%%%%%%

This higher representation can be obtained from the
product of
$(\tiny\yng(1);0)$ and $(0;\overline{\tiny\yng(2)})$.
It turns out that the two eigenvalues are given by 
{\small
  \bea
  {\bf h} (
  \tiny\yng(1);\overline{\tiny\yng(2)}) 
  &= & \frac{(2 k N^2+4 k N+N^3+6 N^2+11 N+8)}{2 N (N+2) (k+N+2)},
  \nonu \\ 
 {\bf \phi_2^{(1)}} (
  \tiny\yng(1);\overline{\tiny\yng(2)}) 
&= &  \phi_2^{(1)} (
  \tiny\yng(1);\overline{\tiny\yng(2)}) +
\Delta(\tiny\yng(1);0) + \Delta(0;\overline{\tiny\yng(2)}) +
  \frac{4}{9}
\Bigg[ \frac{36 i \sqrt{\frac{N}{N+2}}}{(k+N+2)}
 \nonu \\
& 
 + & \frac{18}{(N+2) (k+N+2)^2 (3 k N+4 k+4 N+5) (6 k N+5 k+5 N+4)}
 \nonu \\
 & \times &
 (k-N)
 ({\bf 45 k^2 N^2}+130 k^2 N+87 k^2 {\bf -18 k N^3}-37 k N^2+16 k N
 \nonu \\
 & + & 42 k-23 N^3-98 N^2-128 N-48)
 \Bigg].
\label{fsymmbarlinear}
\eea}
The previous relation in (\ref{fsymmbarnonlinear})
can be inserted.
The eigenvalue for the spin $2$ current
does not have any contribution from the commutator
$[({\bf T})_0, Q^{1}_{-\frac{1}{2}}  Q^{4}_{-\frac{1}{2}} ]$.
This provides only the eigenvalue for the representation $(0;
  \overline{\tiny\yng(2)})$. Also the term 
$ Q^{1}_{-\frac{1}{2}}  Q^{4}_{-\frac{1}{2}} ({\bf T})_0$
acting on the representation $(\tiny\yng(1);0)$
gives the eigenvalue ${\bf h} (\tiny\yng(1);0)$ with
$ Q^{1}_{-\frac{1}{2}}  Q^{4}_{-\frac{1}{2}}$
acting on the state  $|(\tiny\yng(1);0)>$.
By inserting the overall factor into this state, one has
the final state associated with 
the representation $(\tiny\yng(1);\overline{\tiny\yng(2)})$.
Therefore, one arrives at the above eigenvalue for the spin $2$
current.

Furthermore, one can interpret the above eigenvalue of
spin $2$ current from the one in the nonlinear version as follows:
\bea
&&   h (
   \tiny\yng(1);\overline{\tiny\yng(2)})
   + \frac{1}{(N+k+2)} \frac{1}{N(N+2)} \hat{u}^2
   (\tiny\yng(1);\overline{\tiny\yng(2)})
 =  \frac{(2N+4k-1)}{4(N+k+2)}+
 \frac{(-\frac{N}{2}-N-2)^2}{N (N+2) (k+N+2)}
 \nonu \\
 && =
  {\bf h} (
   0;\overline{\tiny\yng(2)})
+  {\bf h} (
   \tiny\yng(1);0)
    =  \frac{(k N+ N+2)}{ N (k+N+2)} +
\frac{(N+1) (N+3)}{2 (N+2) (k+N+2)}.
 \nonu
\eea
In the first line, the first result
can be obtained from the conformal dimension formula in previous
sections and the $\hat{u}$ charge is the sum of
$\hat{u}$ charges of $(
   0;\overline{\tiny\yng(2)})$ and $(
   \tiny\yng(1);0)$.
   In the second line the previous relation (\ref{f0eigenhigherlinear})
   is used and the relation (\ref{0symmeigenlinear})
   is used.

Furthermore, there are $\Delta(\tiny\yng(1);0)$
   and $\Delta(0;
   \overline{\tiny\yng(2)})$. The former can be obtained from the
   relation (\ref{knownresult1})
   while the latter is given by (\ref{Delta0symm})
   together with the footnote \ref{0symmbarres}.   
   
   For the eigenvalue corresponding to the higher spin $3$ current
   (the corresponding eigenvalue of the
   higher spin $3$ current in the nonlinear version appears in
   (\ref{fsymmbarnonlinear})),
there is the contribution from the lower order poles
appearing in the commutator,
$[({\bf \Phi_2^{(1)}})_0, 
  Q^{1}_{-\frac{1}{2}}  Q^{4}_{-\frac{1}{2}}]$. 
From the following eigenvalues for each term in (\ref{BigPhi})
{\small
  \bea
&& \Bigg[-\frac{(2 k N^2+2 k N-3 k+3 N^2+12 N+12)}{(N+2) (k+N+2)^2}, 0,
\frac{i \sqrt{\frac{N}{N+2}} (k+3 N+6)}{(k+N+2)},
-\frac{i \sqrt{\frac{N}{N+2}} (3 k+N+2)}{(k+N+2)},
\nonu \\
&&  -\frac{(13 k N+24 k-8 N^2-28 N-24)}{2 (N+2) (k+N+2)}\Bigg],
\nonu
\eea}
and the known results for the coefficients in (\ref{ccoeff}) and (\ref{p1p2}),
the above eigenvalue of (\ref{fsymmbarlinear}) can be observed
\footnote{One has the following eigenvalues for the similar higher
  representation
  \bea
{\bf \phi_2^{(1)}} (
\overline{\tiny\yng(1)};\overline{\tiny\yng(2)})
& = & \phi_2^{(1)} (
\overline{\tiny\yng(1)};\overline{\tiny\yng(2)}) + \Delta (
\overline{\tiny\yng(1)};0) +
 \Delta ( 0;\overline{\tiny\yng(2)})
\nonu
\eea
plus the last terms in (\ref{fsymmbarlinear}) with an extra minus sign.
Note that the quantity $\phi_2^{(1)} (
\overline{\tiny\yng(1)};\overline{\tiny\yng(2)})$
can be obtained in the footnote \ref{fbarsymmbarfoot}.
Of course, 
$\Delta(\overline{\tiny\yng(1)};0)$ 
can be determined from the previous relation (\ref{knownresult1}).
Moreover $\Delta(0;\overline{\tiny\yng(2)})$ 
can be determined from the footnote \ref{0symmbarres}.
%\bea
%&& \Bigg[
%  \frac{(2 k N^2+2 k N-3 k+3 N^2+12 N+12)}{(N+2) (k+N+2)^2},
%  0,
%  -\frac{i \sqrt{\frac{N}{N+2}} (k+3 N+6)}{k+N+2},
%  \frac{i \sqrt{\frac{N}{N+2}} (3 k+N+2)}{k+N+2},
%  \nonu \\
%  &&
%\frac{(13 k N+24 k-8 N^2-28 N-24)}{2 (N+2) (k+N+2)}
%  \Bigg].
%\nonu
%\eea
}.

%%%%%%%%%%%%%%%%%%%%%%%%%%%%%%%%%%%%%%%%%%%%%%
\subsection{The $(f;\mbox{antisymm})$ representation}
%%%%%%%%%%%%%%%%%%%%%%%%%%%%%%%%%%%%%%%%%%%%%%

The two eigenvalues are given by
{\small
  \bea
  {\bf h} (
  \tiny\yng(1);\tiny\yng(1,1)) 
  &= & \frac{(2 k N^2+4 k N+N^3+10 N^2+19 N+8)}{2 N (N+2) (k+N+2)},
  \nonu \\ 
 {\bf \phi_2^{(1)}} (
  \tiny\yng(1);\tiny\yng(1,1)) 
&= &  \phi_2^{(1)} (
  \tiny\yng(1);\tiny\yng(1,1)) +
\Delta(\tiny\yng(1);0) + \Delta(0;\tiny\yng(1,1)) +
  \frac{4}{9}
\Bigg[ -\frac{36 i \sqrt{\frac{N}{N+2}}}{(k+N+2)}
 \nonu \\
& 
 + & \frac{18}{(N+2) (k+N+2)^2 (3 k N+4 k+4 N+5) (6 k N+5 k+5 N+4)}
 \nonu \\
 & \times &
 (k-N)
 ({\bf 45 k^2 N^2}+146 k^2 N+105 k^2 {\bf +49 k N^2}+140 k N\nonu \\
 & + & 78 k+11 N^3+26 N^2-16 N-48)
 \Bigg].
\label{fantisymmlinear}
\eea}
The relation in (\ref{fantisymmeigenhigher})
can be inserted.
Note that the eigenvalue for the spin $2$ current
does not have any contribution from the commutator
$[({\bf T})_0, Q^{13}_{-\frac{1}{2}}  Q^{14}_{-\frac{1}{2}} ]$. 
This provides only the eigenvalue for the representation $(0;
  \tiny\yng(1,1))$. Also the term 
$ Q^{13}_{-\frac{1}{2}}  Q^{14}_{-\frac{1}{2}} ({\bf T})_0$
acting on the representation $(\tiny\yng(1);0)$
gives the eigenvalue ${\bf h} (\tiny\yng(1);0)$ with
$ Q^{13}_{-\frac{1}{2}}  Q^{14}_{-\frac{1}{2}}$
acting on the state  $|(\tiny\yng(1);0)>$.
By inserting the overall factor into this state, one has
the final state associated with 
the representation $(\tiny\yng(1);\tiny\yng(1,1))$.
Therefore, one arrives at the above eigenvalue for the spin $2$
current.

Furthermore, one can interpret the above eigenvalue of
spin $2$ current from the one in the nonlinear version as follows:
\bea
&&   h (
   \tiny\yng(1);\tiny\yng(1,1))
   + \frac{1}{(N+k+2)} \frac{1}{N(N+2)} \hat{u}^2
   (\tiny\yng(1);\tiny\yng(1,1))
 =  \frac{(2N+4k+15)}{4(N+k+2)}+
 \frac{(-\frac{N}{2}+N+2)^2}{N (N+2) (k+N+2)}
 \nonu \\
 && =
  {\bf h} (
   0;\tiny\yng(1,1))
+  {\bf h} (
   \tiny\yng(1);0)
    =  \frac{(k N+ 3N+2)}{ N (k+N+2)} +
\frac{(N+1) (N+3)}{2 (N+2) (k+N+2)}.
 \nonu
\eea
In the first line, the first result
can be obtained from the conformal dimension formula in previous
sections and the $\hat{u}$ charge is the sum of
$\hat{u}$ charges of $(
   0;\tiny\yng(1,1))$ and $(
   \tiny\yng(1);0)$.
   In the second line the previous relation (\ref{f0eigenhigherlinear})
   is used and the relation (\ref{0antisymmeigenlinear})
   is used.

Furthermore, there are $\Delta(\tiny\yng(1);0)$
   and $\Delta(0;
   \tiny\yng(1,1))$. The former can be obtained from the
   relation (\ref{knownresult1})
   while the latter is given by (\ref{Delta0antisymm}).   

For the eigenvalues corresponding to the remaining higher spin currents,
there are the contributions from the lower order poles
appearing in the commutator $[({\bf \Phi_2^{(1)}})_0, 
  Q^{13}_{-\frac{1}{2}}  Q^{14}_{-\frac{1}{2}}]$. 
By combining the following eigenvalues with the various
coefficients
{\small
  \bea
&& \Bigg[-\frac{(4 k N^2+10 k N+3 k+9 N^2+24 N+12)}{(N+2) (k+N+2)^2},
  0,
-\frac{i \sqrt{\frac{N}{N+2}} (k+3 N+6)}{(k+N+2)},
\frac{i \sqrt{\frac{N}{N+2}} (3 k+N+2)}{(k+N+2)},
\nonu \\
&&
-\frac{(11 k N+24 k-4 N^2-20 N-24)}{2 (N+2) (k+N+2)}
\Bigg],
\nonu
\eea}
the last two terms in eigenvalue of the higher spin $3$ current 
of (\ref{fantisymmlinear})
can be determined explicitly
\footnote{One has
  \bea
{\bf \phi_2^{(1)}} (
\overline{\tiny\yng(1)};\tiny\yng(1,1))
& = & \phi_2^{(1)} (
\overline{\tiny\yng(1)};\tiny\yng(1,1)) + \Delta (
\overline{\tiny\yng(1)};0) +
 \Delta ( 0;\tiny\yng(1,1))
\nonu
\eea
plus the last terms
 in (\ref{fantisymmlinear}) with an extra minus sign.
Note that the quantity $\phi_2^{(1)} (
\overline{\tiny\yng(1)};\tiny\yng(1,1))$
can be obtained in the footnote \ref{fbarantisymmresultresult}.
Of course, 
$\Delta(\overline{\tiny\yng(1)};0)$ 
can be determined from the previous relation (\ref{knownresult1}).
Moreover $\Delta(0;\tiny\yng(1,1))$ 
can be determined from  (\ref{Delta0antisymm}).
%\bea
%&& \Bigg[
%  \frac{(4 k N^2+10 k N+3 k+9 N^2+24 N+12)}{(N+2) (k+N+2)^2},
%  0,
%  \frac{i \sqrt{\frac{N}{N+2}} (k+3 N+6)}{k+N+2},
%  -\frac{i \sqrt{\frac{N}{N+2}} (3 k+N+2)}{k+N+2},
%  \nonu \\
%  &&
%\frac{(11 k N+24 k-4 N^2-20 N-24)}{2 (N+2) (k+N+2)}
%  \Bigg].
%\nonu
%\eea
}.

%%%%%%%%%%%%%%%%%%%%%%%%%%%%%%%%%%%%%%%%%%%%%%
\subsection{The $(f;\overline{\mbox{antisymm}})$ representation}
%%%%%%%%%%%%%%%%%%%%%%%%%%%%%%%%%%%%%%%%%%%%%%

The two eigenvalues are described by
{\small
  \bea
  {\bf h} (
  \tiny\yng(1);\overline{\tiny\yng(1,1)}) 
  &= & \frac{(2 k N^2+4 k N+N^3+10 N^2+19 N+8)}{2 N (N+2) (k+N+2)},
  \nonu \\ 
 {\bf \phi_2^{(1)}} (
  \tiny\yng(1);\overline{\tiny\yng(1,1)}) 
&= &  \phi_2^{(1)} (
  \tiny\yng(1);\overline{\tiny\yng(1,1)}) +
\Delta(\tiny\yng(1);0) + \Delta(0;\overline{\tiny\yng(1,1)}) +
  \frac{4}{9}
\Bigg[ \frac{36 i \sqrt{\frac{N}{N+2}}}{(k+N+2)}
 \nonu \\
& 
 + & \frac{18}{(N+2) (k+N+2)^2 (3 k N+4 k+4 N+5) (6 k N+5 k+5 N+4)}
 \nonu \\
 & \times &
 (k-N)
 ({\bf 45 k^2 N^2}+130 k^2 N+87 k^2 {\bf -18 k N^3}-19 k N^2+52 k N
 \nonu \\
 & + & 42 k-5 N^3-26 N^2-56 N-48)
 \Bigg].
\label{fantisymmbarlinear}
\eea}
One can insert the relation
(\ref{fantisymmbareigenhigher}).
Note that the eigenvalue for the spin $2$ current
does not have any contribution from the commutator
$[({\bf T})_0, Q^{1}_{-\frac{1}{2}}  Q^{2}_{-\frac{1}{2}} ]$. 
This provides only the eigenvalue for the representation $(0;
  \overline{\tiny\yng(1,1)})$. Also the term 
$ Q^{1}_{-\frac{1}{2}}  Q^{2}_{-\frac{1}{2}} ({\bf T})_0$
acting on the representation $(\tiny\yng(1);0)$
gives the eigenvalue ${\bf h} (\tiny\yng(1);0)$ with
$ Q^{1}_{-\frac{1}{2}}  Q^{2}_{-\frac{1}{2}}$
acting on the state  $|(\tiny\yng(1);0)>$.
By inserting the overall factor into this state, one has
the final state associated with 
the representation $(\tiny\yng(1);\overline{\tiny\yng(1,1)})$.
Therefore, one arrives at the above eigenvalue for the spin $2$
current.

Furthermore, one can interpret the above eigenvalue of
spin $2$ current from the one in the nonlinear version as follows:
\bea
&&   h (
   \tiny\yng(1);\overline{\tiny\yng(1,1)})
   + \frac{1}{(N+k+2)} \frac{1}{N(N+2)} \hat{u}^2
   (\tiny\yng(1);\overline{\tiny\yng(1,1)})
 =  \frac{(2N+4k+7)}{4(N+k+2)}+
 \frac{(-\frac{N}{2}-N-2)^2}{N (N+2) (k+N+2)}
 \nonu \\
 && =
  {\bf h} (
   0;\overline{\tiny\yng(1,1)})
+  {\bf h} (
   \tiny\yng(1);0)
    =  \frac{(k N+ 3N+2)}{ N (k+N+2)} +
\frac{(N+1) (N+3)}{2 (N+2) (k+N+2)}.
 \nonu
\eea
In the first line, the first result
can be obtained from the conformal dimension formula in previous
sections and the $\hat{u}$ charge is the sum of
$\hat{u}$ charges of $(
   0; \overline{\tiny\yng(1,1)})$ and $(
   \tiny\yng(1);0)$.
   In the second line the previous relation (\ref{f0eigenhigherlinear})
   is used and the relation (\ref{0antisymmeigenlinear})
   is used.

Furthermore, there are $\Delta(\tiny\yng(1);0)$
   and $\Delta(0;
  \overline{ \tiny\yng(1,1)})$. The former can be obtained from the
   relation (\ref{knownresult1})
   while the latter is given by (\ref{Delta0antisymm}) with the footnote
   \ref{0antisymmbarresult}.   

For the eigenvalues corresponding to the remaining higher spin currents,
there are the contributions from the lower order poles
appearing in the commutator $[({\bf \Phi_2^{(1)}})_0, 
  Q^{1}_{-\frac{1}{2}}  Q^{2}_{-\frac{1}{2}}]$. 
By combining the following eigenvalues with the various
coefficients
{\small
  \bea
&& \Bigg[-\frac{(2 k N^2+2 k N-3 k+9 N^2+24 N+12)}{(N+2) (k+N+2)^2},
  0,
\frac{i \sqrt{\frac{N}{N+2}} (k+3 N+6)}{(k+N+2)},
-\frac{i \sqrt{\frac{N}{N+2}} (3 k+N+2)}{(k+N+2)},
\nonu \\
&&
-\frac{(13 k N+24 k-8 N^2-28 N-24)}{2 (N+2) (k+N+2)}
\Bigg],
\nonu
\eea}
the last two terms in eigenvalue of the higher spin $3$ current 
of (\ref{fantisymmbarlinear})
can be determined explicitly
\footnote{One has
  \bea
{\bf \phi_2^{(1)}} (
\overline{\tiny\yng(1)};\overline{\tiny\yng(1,1)})
& = & \phi_2^{(1)} (
\overline{\tiny\yng(1)};\overline{\tiny\yng(1,1)}) + \Delta (
\overline{\tiny\yng(1)};0) +
 \Delta ( 0;\overline{\tiny\yng(1,1)})
\nonu
\eea
plus the last terms
 in (\ref{fantisymmbarlinear}) with an extra minus sign.
Note that the quantity $\phi_2^{(1)} (
\overline{\tiny\yng(1)};\overline{\tiny\yng(1,1)})$
can be obtained in the footnote \ref{footcor1}.
Of course, 
$\Delta(\overline{\tiny\yng(1)};0)$ 
can be determined from the previous relation (\ref{knownresult1}).
Moreover $\Delta(0; \overline{\tiny\yng(1,1)})$ 
can be determined from  (\ref{Delta0antisymm}) together with the
footnote \ref{0antisymmbarresult}.
%\bea
%&& \Bigg[
%  \frac{(2 k N^2+2 k N-3 k+9 N^2+24 N+12)}{(N+2) (k+N+2)^2},
%  0,
%  -\frac{i \sqrt{\frac{N}{N+2}} (k+3 N+6)}{k+N+2},
%  \frac{i \sqrt{\frac{N}{N+2}} (3 k+N+2)}{k+N+2},
%  \nonu \\
%  &&
%\frac{(13 k N+24 k-8 N^2-28 N-24)}{2 (N+2) (k+N+2)}
%  \Bigg].
%\nonu
%\eea
}

%%%%%%%%%%%%%%%%%%%%%%%%%%%%%%%%%%%%%%%%%%%%%%%%%%%%
\subsection{The $(\mbox{symm}; 0)$ representation}
%%%%%%%%%%%%%%%%%%%%%%%%%%%%%%%%%%%%%%%%%%%%%%%%%%%%

This higher representation can be obtained from the
product of the minimal representation $(f;0)$ and itself.
The two eigenvalues with this representation can be
described as 
\bea
{\bf h} (
  \tiny\yng(2);0) 
  & = & \frac{(N+1) (N+4)}{(N+2) (k+N+2)},
\nonu \\
{\bf \phi_2^{(1)}} (
  \tiny\yng(2);0) 
  & = &
  \phi_2^{(1)} (
  \tiny\yng(2);0) +  \Delta (
  \tiny\yng(2);0) 
  \nonu \\
  & = & -
  \frac{8}{3 (N+2) (k+N+2)^2 (3 k N+4 k+4 N+5)}
  \nonu \\
  & \times &
  N ({\bf 3 k^2 N^2}+10 k^2 N+8 k^2{\bf +6 k N^3}+39 k N^2+89 k N+58 k\nonu \\
  & + & 2 N^3+8 N^2+35 N+30).
  \label{symmzerolinearres}
\eea
The relation (\ref{symmzerohigher}) is used.
It is easy to see that the first eigenvalue is the twice of
the one in (\ref{f0eigenhigherlinear}) under the $(N,k)$ 't Hooft limit.
One also realizes that 
the description from the nonlinear analysis for the spin $2$ current
implies the following expression (with the help of (\ref{eigenvaluestwosymm}))
\bea
&&   h (
   \tiny\yng(2);0)
   + \frac{1}{(N+k+2)} \frac{1}{N(N+2)} \hat{u}^2
   (\tiny\yng(2);0)
 =  \frac{(N+2)}{(N+k+2)}+
 \frac{N^2}{N (N+2) (k+N+2)},
 \nonu
 \eea
which coincides with the above eigenvalue in (\ref{symmzerolinearres}). 
Furthermore, if one sees the last eigenvalue closely,  
one observes that the highest power terms in the numerator
(\ref{symmzerolinearres})
are given by $-8N(3 k^2 N^2  +6 k N^3)$ which is the twice of
the ones in (\ref{f0eigenhigherlinear}). Note that
the denominators in both expressions are the same
at finite $(N,k)$. This implies that
the eigenvalue for the higher spin $3$  current
in this higher representation can be interpreted as the additive
quantum number and it is given by the sum of each eigenvalue
for the higher representation in the minimal representation
$(\tiny\yng(1);0)$ under the large $(N,k)$ 't Hooft like limit.

The difference between the nonlinear and linear cases
which vanishes in the large $N$ 't Hooft limit  
is given by 
\bea
 \Delta (
 \tiny\yng(2);0)  & = &
 -\frac{16 N (k-N) ({\bf 3 k N^2}-7 k N-12 k{\bf -3 N^3}-13 N^2-32 N-24)}{
  (N+2) (k+N+2)^2 (3 k N+4 k+4 N+5) (6 k N+5 k+5 N+4)}.
\label{Deltasymm0}
\eea
More explicitly, the eigenvalues for each term in (\ref{BigPhi})
have the following form
\bea
&& \Bigg[
  -\frac{2 N (k N+2 k+2 N^2+9 N+8)}{(N+2) (k+N+2)^2},
  -\frac{4 N}{(k+N+2)},
  -\frac{2 i N \sqrt{\frac{N}{N+2}}}{(k+N+2)},
  \nonu \\
  && -\frac{2 i N \sqrt{\frac{N}{N+2}}}{(k+N+2)},
  \frac{2 N^2}{(N+2) (k+N+2)}
  \Bigg].
\nonu
\eea
As described before, although the third and fourth eigenvalues 
contain the imaginary $i$, they are the same  
and their coefficients are opposite to each other.
This leads to the absence of imaginary $i$ in (\ref{Deltasymm0}).
One can easily see that under the large $(N,k)$ 't Hooft like limit
both eigenvalues for the higher spin $3$ currents in the nonlinear and linear
versions coincide with each other because the expression in
(\ref{Deltasymm0}) behaves as $\frac{1}{N}$
\footnote{One has
  \bea
{\bf \phi_2^{(1)}} (
\overline{\tiny\yng(2)};0)
& = & \phi_2^{(1)} (
\overline{\tiny\yng(2)};0) + \Delta (
\overline{\tiny\yng(2)};0). 
\nonu
\eea
Note that here
the quantity $\Delta (
\overline{\tiny\yng(2)};0)$
can be obtained from the corresponding quantity
$\Delta (
\tiny\yng(2);0)$
by changing the sign. We need to have
the footnote \ref{symmbareigenhigher} for the quantity
$\phi_2^{(1)} (
\overline{\tiny\yng(2)};0)$.}.

%%%%%%%%%%%%%%%%%%%%%%%%%%%%%%%%%%%%%%%%%%%%%%%%%%%%%%%
\subsection{The $(\mbox{antisymm}; 0)$ representation}
%%%%%%%%%%%%%%%%%%%%%%%%%%%%%%%%%%%%%%%%%%%%%%%%%%%%%%%

This higher representation can be obtained from the
product of the minimal representation $(f;0)$ and itself.
The two eigenvalues with this representation can be
described as 
\bea
{\bf h} (
  \tiny\yng(1,1);0) 
  & = & \frac{N (N+3)}{(N+2) (k+N+2)},
\nonu \\
{\bf \phi_2^{(1)}} (
\tiny\yng(1,1);0)
  & = &
   \phi_2^{(1)} (
  \tiny\yng(1,1);0) +  \Delta (
  \tiny\yng(1,1);0) 
  \nonu \\
  & = & -\frac{8}{3 (N+2) (k+N+2)^2 (3 k N+4 k+4 N+5)}
  \nonu \\
  & \times &
  N ({\bf 3 k^2 N^2}+10 k^2 N+8 k^2{\bf +6 k N^3}+21 k N^2+17 k N-14 k
  \nonu \\
  & + & 2 N^3-4 N^2-19 N-30).
\label{antisymmzerolinearres}
\eea
It is easy to see that the first eigenvalue is the twice of
the one in (\ref{f0eigenhigherlinear}) under the $(N,k)$ 't Hooft limit.
One also realizes that 
the description from the nonlinear analysis for the spin $2$ current
implies the following expression (with the help of the subsection
\ref{twoantisymmetric}: in particular (\ref{antidoubleboxzero}))
\bea
&&   h (
   \tiny\yng(1,1);0)
   + \frac{1}{(N+k+2)} \frac{1}{N(N+2)} \hat{u}^2
   (\tiny\yng(1,1);0)
 =  \frac{N}{(N+k+2)}+
 \frac{N^2}{N (N+2) (k+N+2)}.
 \nonu
\eea
which coincides with the above eigenvalue in (\ref{antisymmzerolinearres}). 
Furthermore, if one sees the last eigenvalue closely
(\ref{antisymmzerolinearres}),  
one observes that the highest power terms in the numerator
are given by $-8N(3 k^2 N^2  +6 k N^3)$ which is the twice of
the ones in (\ref{f0eigenhigherlinear}). Note that
the denominators in both expressions are the same
at finite $(N,k)$. This implies that
the eigenvalue for the higher spin $3$  current
in this higher representation can be interpreted as the additive
quantum number and it is given by the sum of each eigenvalue
for the higher representation in the minimal representation
$(\tiny\yng(1);0)$ under the large $(N,k)$ 't Hooft like limit.

The difference between the nonlinear and linear cases
which vanishes in the large $N$ 't Hooft limit  
is given by
\bea
\Delta (
 \tiny\yng(1,1);0)  & = &
-\frac{16 N^2 (k-N) ({\bf 3 k N}-k{\bf -3 N^2}-7 N-8)}{
  (N+2) (k+N+2)^2 (3 k N+4 k+4 N+5) (6 k N+5 k+5 N+4)}.
\label{Deltaantisymm0}
\eea
This can be determined by the following eigenvalues 
\bea
&& \Bigg[
-\frac{2 N (k N+2 k+2 N^2+7 N+4)}{(N+2) (k+N+2)^2},
-\frac{4 N}{(k+N+2)},
-\frac{2 i N \sqrt{\frac{N}{N+2}}}{(k+N+2)},
\nonu \\
&&
-\frac{2 i N \sqrt{\frac{N}{N+2}}}{(k+N+2)},
\frac{2 N^2}{(N+2) (k+N+2)}
\Bigg],
\nonu
\eea
together with the coefficients.
Note that the highest order terms in (\ref{Deltaantisymm0})
are the same as the ones in (\ref{Deltasymm0}).
One can easily see that the above quantity
(\ref{Deltaantisymm0}) vanishes under the large $(N,k)$ 't Hooft
limit
\footnote{Similarly, one has
  \bea
{\bf \phi_2^{(1)}} (
\overline{\tiny\yng(1,1)};0)
& = & \phi_2^{(1)} (
\overline{\tiny\yng(1,1)};0) + \Delta (
\overline{\tiny\yng(1,1)};0). 
\nonu
\eea
The quantity $\Delta (
\overline{\tiny\yng(1,1)};0)$
can be obtained from the corresponding quantity
$\Delta (
\tiny\yng(1,1);0)$
by changing the sign. 
The footnote \ref{antisymmbareigenhigher} for the quantity
$\phi_2^{(1)} (
\overline{\tiny\yng(1,1)};0)$ can be used.
}.

%%%%%%%%%%%%%%%%%%%%%%%%%%%%%%%%%%%%%%%%%%%%%%%%%%%%
\subsection{The $(0; \mbox{symm})$ representation}
%%%%%%%%%%%%%%%%%%%%%%%%%%%%%%%%%%%%%%%%%%%%%%%%%%%%

This higher representation can be obtained from the
product of the minimal representation $(0;f)$ and itself.
The two eigenvalues with this representation can be
described as 
\bea
{\bf h} (
  0;\tiny\yng(2)) 
  & = & \frac{(k N+N+2)}{N (k+N+2)},
\nonu \\
{\bf \phi_2^{(1)}} (
  0;\tiny\yng(2)) 
  & = &
  \phi_2^{(1)} (
  0;\tiny\yng(2)) +  \Delta (
  0;\tiny\yng(2)) 
\label{0symmeigenlinear}
  \\
  & = &
  \frac{8 k ({\bf 6 k^2 N^2}+2 k^2 N{\bf +3 k N^3}+9 k N^2-8 k N-12 k+4 N^3-N^2-3 N)}{
    3 N (k+N+2)^2 (3 k N+4 k+4 N+5)}.
\nonu
\eea
The relation (\ref{zerosymmhigher}) is inserted.
It is easy to see that the first eigenvalue is the twice of
the one in (\ref{0feigenlinear}) under the large $(N,k)$ 't Hooft limit.
One also realizes that 
the description from the nonlinear analysis for the spin $2$ current
implies the following expression (with the help of 
(\ref{eigenvaluessymmtwo}))
\bea
&&   h (
   0;\tiny\yng(2))
   + \frac{1}{(N+k+2)} \frac{1}{N(N+2)} \hat{u}^2
   (0;\tiny\yng(2))
 =  \frac{k}{(N+k+2)}+
 \frac{(N+2)^2}{N (N+2) (k+N+2)},
 \nonu
\eea
which is equal to the eigenvalue in (\ref{0symmeigenlinear}).
Furthermore, 
one observes that the highest power terms in the numerator
(\ref{0symmeigenlinear})
are given by $8k(6 k^2 N^2 +3 k N^3)$ which is the twice of
the ones in (\ref{0feigenlinear}).
This implies that
the eigenvalue for the higher spin $3$  current
in this higher representation can be interpreted as the additive
quantum number and it is given by the sum of each eigenvalue
for the higher representation in the minimal representation
$(0;\tiny\yng(1))$ under the large $(N,k)$ 't Hooft like limit.

The difference between the nonlinear and linear cases
which vanishes in the large $N$ 't Hooft limit  
is given by 
\bea
 \Delta (
 0;\tiny\yng(2))
 & = &
\frac{16 k (k-N) ({\bf 3 k^2 N-3 k N^2}-11 k N-10 k-5 N^2-14 N-8)}{
  N (k+N+2)^2 (3 k N+4 k+4 N+5) (6 k N+5 k+5 N+4)}.
\label{Delta0symm}
\eea
Here the eigenvalues for each term appearing in (\ref{BigPhi}) 
are used 
\bea
&& \Bigg[
  -\frac{2 k (2 k N+N^2+3 N+2)}{N (k+N+2)^2},
  -\frac{4 k}{(k+N+2)},
  -\frac{2 i k \sqrt{\frac{N+2}{N}}}{(k+N+2)},
  \nonu \\
  &&
  -\frac{2 i k \sqrt{\frac{N+2}{N}}}{(k+N+2)},
  \frac{2k (N+2)}{N (k+N+2)}
  \Bigg].
\label{eigenvaluesfive}
\eea
Again, the large $(N,k)$ 't Hooft limit
of the above eigenvalue (\ref{Delta0symm})
behaves as $\frac{1}{N}$
and therefore, both eigenvalues for the higher spin $3$ currents
in the nonlinear and linear versions have the same value
\footnote{
\label{0symmbarres}
  One has the following eigenvalues for similar higher representation
  \bea
{\bf \phi_2^{(1)}} (0;\overline{\tiny\yng(2)})
& = & \phi_2^{(1)} (0;
\overline{\tiny\yng(2)}) + \Delta (
0;\overline{\tiny\yng(2)}). 
\nonu
\eea
The quantity $\phi_2^{(1)} (0;
\overline{\tiny\yng(2)})$ can be read off from the footnote
\ref{zerosymmbareigenhigher} while the quantity
$\Delta (
0;\overline{\tiny\yng(2)})$ can be obtained from the eigenvalues
in (\ref{eigenvaluesfive}) by changing the signs in the first, second,
and last eigenvalues together with the coefficients
in (\ref{ccoeff}) and (\ref{p1p2}).}.

%%%%%%%%%%%%%%%%%%%%%%%%%%%%%%%%%%%%%%%%%%%%%%%%%%%%%%%
\subsection{The $(0; \mbox{antisymm})$ representation}
%%%%%%%%%%%%%%%%%%%%%%%%%%%%%%%%%%%%%%%%%%%%%%%%%%%%%%%%

This higher representation can be obtained from the
product of the minimal representation $(0;f)$ and itself.
The two eigenvalues with this representation can be
described as
\bea
{\bf h} (
  0;\tiny\yng(1,1)) 
  & = &  \frac{(k N+3 N+2)}{N (k+N+2)},
\nonu \\
{\bf \phi_2^{(1)}}  (
  0;\tiny\yng(1,1)) 
 & = &
  \phi_2^{(1)} (
  0;\tiny\yng(1,1)) +  \Delta (
  0;\tiny\yng(1,1)) 
 \label{0antisymmeigenlinear}
  \\
  & = & 
  \frac{8 k ({\bf 6 k^2 N^2}+2 k^2 N {\bf +3 k N^3}
    +27 k N^2+4 k N-12 k+4 N^3+35 N^2+27 N)}{3 N (k+N+2)^2 (3 k N+4 k+4 N+5)}.
 \nonu
\eea
The relation (\ref{zeroantisymmhigher}) is inserted.
It is easy to see that the first eigenvalue is the twice of
the one in (\ref{0feigenlinear}) under the large $(N,k)$ 't Hooft limit.
One also realizes that 
the description from the nonlinear analysis for the spin $2$ current
implies the following expression (with the help of 
(\ref{eigenvaluestwoantisymm}))
\bea
&&   h (
   0;\tiny\yng(1,1))
   + \frac{1}{(N+k+2)} \frac{1}{N(N+2)} \hat{u}^2
   (0;\tiny\yng(1,1))
 =  \frac{(k+2)}{(N+k+2)}+
 \frac{(N+2)^2}{N (N+2) (k+N+2)}.
 \nonu
 \eea
Furthermore, 
one observes that the highest power terms in the numerator
(\ref{0antisymmeigenlinear})
are given by $8k(6 k^2 N^2 +3 k N^3)$ which is the twice of
the ones in (\ref{0feigenlinear}).
This implies that
the eigenvalue for the higher spin $3$  current
in this higher representation can be interpreted as the additive
quantum number and it is given by the sum of each eigenvalue
for the higher representation in the minimal representation
$(0;\tiny\yng(1))$ under the large $(N,k)$ 't Hooft like limit 

The difference between the nonlinear and linear cases
which vanishes in the large $N$ 't Hooft limit  
is given by
\bea
 \Delta (
  0;\tiny\yng(1,1)) & = &
\frac{16 k (k-N) ({\bf 3 k^2 N-3 k N^2}-5 k N-10 k+N^2-2 N-8)}{
  N (k+N+2)^2 (3 k N+4 k+4 N+5) (6 k N+5 k+5 N+4)}.
\label{Delta0antisymm}
\eea
Here the corresponding eigenvalues are given by
\bea
&& \Bigg[
-\frac{2 k (2 k N+N^2+5 N+2)}{N (k+N+2)^2},
-\frac{4 k}{(k+N+2)},
-\frac{2 i k \sqrt{\frac{N+2}{N}}}{(k+N+2)},
\nonu \\
&& -\frac{2 i k \sqrt{\frac{N+2}{N}}}{(k+N+2)},
\frac{2k (N+2)}{N (k+N+2)}
\Bigg].
\label{fiveeigenvalues}
\eea
Note that the highest order terms in (\ref{Delta0antisymm})
are the same as the ones in (\ref{Delta0symm}).
Because the large $(N,k)$ 't Hooft limit
of the above eigenvalue (\ref{Delta0antisymm})
behaves as $\frac{1}{N}$,
both eigenvalues for the higher spin $3$ currents
in the nonlinear and linear versions have the same value
\footnote{
\label{0antisymmbarresult}
  One has
  \bea
{\bf \phi_2^{(1)}} (0;\overline{\tiny\yng(1,1)})
& = & \phi_2^{(1)} (0;
\overline{\tiny\yng(1,1)}) + \Delta (
0;\overline{\tiny\yng(1,1)}). 
\nonu
\eea
The quantity $\phi_2^{(1)} (0;
\overline{\tiny\yng(1,1)})$ can be read off from the footnote
\ref{zeroantisymmbarexpression} while the quantity
$\Delta (
0;\overline{\tiny\yng(1,1)})$ can be obtained from the eigenvalues
in (\ref{fiveeigenvalues}) by changing the signs in the first, second,
and last eigenvalues together with the coefficients
in (\ref{ccoeff}) and (\ref{p1p2}).
}.

%%%%%%%%%%%%%%%%%%%%%%%%%%%%%%%%%%%%%%%%%%%%%%%%%%%%%%%%%%%%
\subsection{The $(\mbox{symm}; \mbox{symm})$ representation}
%%%%%%%%%%%%%%%%%%%%%%%%%%%%%%%%%%%%%%%%%%%%%%%%%%%%%%%%%%%%%

The two eigenvalues are given by
\bea
{\bf h} (
  \tiny\yng(2);\tiny\yng(2)) 
  & = & \frac{2 (N^2+2 N+2)}{N (N+2) (k+N+2)},
\nonu \\
{\bf \phi_2^{(1)}} (
  \tiny\yng(2);\tiny\yng(2)) 
 & = &
   \phi_2^{(1)} (
  \tiny\yng(2);\tiny\yng(2)) +  \Delta (
  \tiny\yng(2);\tiny\yng(2)) 
  \nonu \\
  & = &
  \frac{16}{3 N (N+2) (k+N+2)^2 (3 k N+4 k+4 N+5)}
  \nonu \\
  & \times & 
  ({\bf 3 k^2 N^3}+10 k^2 N^2+8 k^2 N{\bf -3 k N^4}-15 k N^3-13 k N^2+10 k N
  \nonu \\
  & + & 24 k-4 N^4-37 N^3-73 N^2-54 N).
\label{symmsymmeigenlinear}
\eea
Note that the eigenvalue
$ \phi_2^{(1)} (
\tiny\yng(2);\tiny\yng(2))$ is given by
(\ref{symmsymmnonlinear}).
The highest power terms of $(N,k)$ in the eigenvalue of the higher
spin $3$ current
(\ref{symmsymmeigenlinear})
are twice of the terms in (\ref{boxboxlinear}).
The description from the nonlinear analysis for the spin $2$ current
implies the following expression (with the help of 
(\ref{doubleboxsymm}))
\bea
&&   h (
   \tiny\yng(2);\tiny\yng(2))
   + \frac{1}{(N+k+2)} \frac{1}{N(N+2)} \hat{u}^2
   (\tiny\yng(2);\tiny\yng(2))
 =  \frac{2}{(N+k+2)}+
 \frac{4}{N (N+2) (k+N+2)}.
 \nonu
\eea

By realizing that the following eigenvalues hold in
this higher representation
\bea
&& \Bigg[
  \frac{4(k N^2+2 k N+N^3+6 N^2+8 N+4)}{N (N+2) (k+N+2)^2},
  \frac{8}{(k+N+2)},-\frac{8 i}{\sqrt{N (N+2)} (k+N+2)},
  \nonu \\
  && -\frac{8 i}{\sqrt{N (N+2)} (k+N+2)},
  -\frac{16}{\left(N^2+2 N\right) (k+N+2)}\Bigg], 
  \label{symmsymmEigen}
\eea
and putting together with the various known coefficients,
the difference between the nonlinear and linear cases
in the eigenvalue for the higher spin $3$
current (which vanishes in the large $(N,k)$ 't Hooft limit)  
can be summarized by
{\small
  \bea
 \Delta (
 \tiny\yng(2);\tiny\yng(2))
 & = &
  - 
  \frac{64 (k-N) ({\bf 3 k N^2}-6 k N-10 k {\bf +3 N^3}
    +12 N^2+2 N-8)}{N (N+2) (k+N+2)^2 (3 k N+4 k+4 N+5) (6 k N+5 k+5 N+4)}.
\label{Deltasymmsymm}
\eea}
By inserting this value and the eigenvalue for the higher spin $3$ current
in the nonlinear version, the final eigenvalue in the linear version
can be written as the one in (\ref{symmsymmeigenlinear}).
Furthermore, the highest power terms of $(N,k)$ in (\ref{Deltasymmsymm})
are the four times of the ones in (\ref{corrff}). 
\footnote{One has
  \bea
{\bf \phi_2^{(1)}} (
\overline{\tiny\yng(2)};\overline{\tiny\yng(2)})
& = & \phi_2^{(1)} (
\overline{\tiny\yng(2)};\overline{\tiny\yng(2)}) + \Delta (
\overline{\tiny\yng(2)};\overline{\tiny\yng(2)}).
\nonu
\eea
The quantity $\phi_2^{(1)} (\overline{\tiny\yng(2)};
\overline{\tiny\yng(2)})$ can be read off from the footnote
\ref{symmbarsymmbarfoot} while the quantity
$\Delta (
\overline{\tiny\yng(2)};\overline{\tiny\yng(2)})$
can be obtained from the expression
in (\ref{Deltasymmsymm}) by changing the sign.
}

%%%%%%%%%%%%%%%%%%%%%%%%%%%%%%%%%%%%%%%%%%%%%%%%%%%%%%%
\subsection{The $(\mbox{symm}; f)$ representation}
%%%%%%%%%%%%%%%%%%%%%%%%%%%%%%%%%%%%%%%%%%%%%%%%%%%%%%%

The two eigenvalues are summarized by
{\small
  \bea
{\bf h} (
  \tiny\yng(2);\tiny\yng(1)) 
  & = &  \frac{(N+1) (N^2+7 N+2)}{2 N (N+2) (k+N+2)},
\nonu \\
{\bf \phi_2^{(1)}} (
  \tiny\yng(2);\tiny\yng(1)) 
 & = &
    \phi_2^{(1)} (
  \tiny\yng(2);\tiny\yng(1)) +  \Delta (
  \tiny\yng(2);\tiny\yng(1)) 
  \nonu \\
  & = &
  -\frac{4}{3 N (N+2) (k+N+2)^2 (3 k N+4 k+4 N+5) (6 k N+5 k+5 N+4)}
  \nonu \\
  & \times &
  ({\bf 18 k^3 N^5}+39 k^3 N^4-34 k^3 N^3-174 k^3 N^2-188 k^3 N+
  {\bf 36 k^2 N^6}+
  405 k^2 N^5 \nonu \\
  & + & 1334 k^2 N^4+1493 k^2 N^3+271 k^2 N^2-452 k^2 N-60 k^2+
  60 k N^6+643 k N^5 \nonu \\
  & + & 3823 k N^3+2172 k N^2+
  220 k N-48 k+25 N^6+280 N^5+1248 N^4+2012 N^3 \nonu \\
  & + & 1304 N^2+288 N+2565 k N^4).
\label{symmfeigenlinear}
\eea}
Note that the eigenvalue
$ \phi_2^{(1)} (
\tiny\yng(2);\tiny\yng(1))$ is given by
(\ref{symmfnonlinear}).
One observes that these eigenvalues
(\ref{symmfeigenlinear})
are twice of the ones in (\ref{f0eigenhigherlinear}) respectively if
one takes the highest order terms in the numerators.
It is easy to see this behavior in the spin $2$ current. 
For the higher spin $3$ current, more explicitly,
the corresponding terms in the
second eigenvalue in (\ref{symmfeigenlinear})
are given by $(18k^3 N^5 + 36 k^2  N^6)=
6 k N^3 (3 k^2 N^2 +6 k N^3)$
and the denominator contains the fator $N(6 k N+5 k+5 N+4) \rightarrow
6 k N^2$,
compared to the one in (\ref{f0eigenhigherlinear}). 
Therefore, one observes that the corresponding
quantity  $N (3 k^2 N^2 +6 k N^3)$ appears in the eigenvalue in
(\ref{f0eigenhigherlinear}).

The description from the nonlinear analysis for the spin $2$ current
implies the following expression (with the help of 
(\ref{symmfeigen}))
\bea
&&   h (
   \tiny\yng(2);\tiny\yng(1))
   + \frac{1}{(N+k+2)} \frac{1}{N(N+2)} \hat{u}^2
   (\tiny\yng(2);\tiny\yng(1))
 =  \frac{(2N+11)}{4(N+k+2)}+
 \frac{(-\frac{N}{2}+1)^2}{N (N+2) (k+N+2)}.
 \nonu
\eea
It is obvious that this is equal to the eigenvalue in (\ref{symmfeigenlinear}).

The difference between the nonlinear and linear cases
which vanishes in the large $(N,k)$ 't Hooft limit  
is given by
\bea
 \Delta (
 \tiny\yng(2);\tiny\yng(1)) & = &- 
 \frac{4}{N (N+2) (k+N+2)^2 (3 k N+4 k+4 N+5) (6 k N+5 k+5 N+4)}
 \nonu \\
 & \times & (k-N)
 ({\bf 6 k^2 N^3}-6 k^2 N^2-36 k^2 N{\bf +6 k N^4}-26 k N^3-15 k N^2-20 k
 {\bf -3 N^5} \nonu \\
 & - &
 20 N^4-34 N^3+108 N^2+136 N-16).
\label{Deltasymmf}
\eea
Here as before, the following eigenvalues, corresponding to
the five terms in (\ref{BigPhi}), are used in this calculation 
{\small
  \bea
&& \Bigg[
 -\frac{(2 k N^2+4 k N+3 N^4+10 N^3-15 N^2-32 N-4)}{2 N (N+2) (k+N+2)^2},
  -\frac{(2 N-4)}{(k+N+2)},
  \nonu \\
  && -  \frac{i (N-2)^2}{2 \sqrt{N (N+2)} (k+N+2)},
   -\frac{i (N-2)^2}{2 \sqrt{N (N+2)} (k+N+2)},
  \frac{(N-2)^3}{4 N (N+2) (k+N+2)}\Bigg]. 
  \nonu
\eea}
Then the final eigenvalue for the higher spin $3$ current
in the linear version
can be obtained by combining the one in the nonlinear version
and the (\ref{Deltasymmf})
as in (\ref{symmfeigenlinear})
\footnote{One can consider the following eigenvalues
  \bea
{\bf \phi_2^{(1)}} (
\overline{\tiny\yng(2)};\overline{\tiny\yng(1)})
& = & \phi_2^{(1)} (
\overline{\tiny\yng(2)};\overline{\tiny\yng(1)}) + \Delta (
\overline{\tiny\yng(2)};\overline{\tiny\yng(1)}),
\nonu
\eea
where
the quantity $\phi_2^{(1)} (\overline{\tiny\yng(2)};
\overline{\tiny\yng(1)})$ can be read off from the footnote
\ref{symmfrelated} while the quantity
$\Delta (
\overline{\tiny\yng(2)};\overline{\tiny\yng(1)})$
can be obtained from the expression
in (\ref{Deltasymmf}) by changing the sign.
}.

%%%%%%%%%%%%%%%%%%%%%%%%%%%%%%%%%%%%%%%%%%%%%%%%%%%%%%%%%%%%%%%
\subsection{The $(\mbox{symm}; \overline{f})$ representation}
%%%%%%%%%%%%%%%%%%%%%%%%%%%%%%%%%%%%%%%%%%%%%%%%%%%%%%%%%%%%%%%

The two eigenvalues are given by 
{\small
  \bea
{\bf h} (
  \tiny\yng(2);\overline{\tiny\yng(1)}) 
  &= &
\frac{(k N^2+2 k N+2 N^3+12 N^2+13 N+2)}{2 N (N+2) (k+N+2)},
  \nonu \\
{\bf \phi_2^{(1)}} (
  \tiny\yng(2);\overline{\tiny\yng(1)}) 
&= & \phi_2^{(1)} (
  \tiny\yng(2);\overline{\tiny\yng(1)}) +
  \Delta(\tiny\yng(2);0) + \Delta(0;\overline{\tiny\yng(1)}) +
  \frac{4}{9} \Bigg[ \frac{36 i \sqrt{\frac{N}{N+2}}}{(k+N+2)}
\nonu \\
&  + &
\frac{18}{(N+2) (k+N+2)^2 (3 k N+4 k+4 N+5) (6 k N+5 k+5 N+4)}
\nonu \\
& \times & (N-k) ({\bf -30 k^2 N^2}-71 k^2 N-36 k^2 {\bf +15 k N^3}
+23 k N^2-11 k N-6 k
\nonu \\
& + & 10 N^3+34 N^2+40 N+24) \Bigg].
\label{symmfbarlinearresult}
\eea}
The relation (\ref{symmfbarhigher}) is used.
Note that the eigenvalue for the  spin $2$ current
does not have any contribution from the commutator
$[({\bf T})_0, Q^{\bar{A}}_{-\frac{1}{2}}]$.
This provides only the eigenvalue for the representation $(0;\overline{
  \tiny\yng(1)})$. Also the term 
$ Q^{\bar{A}}_{-\frac{1}{2}} ({\bf T})_0$
acting on the representation $(\tiny\yng(2);0)$
gives the eigenvalue ${\bf h} (\tiny\yng(2);0)$ with
$ Q^{\bar{A}}_{-\frac{1}{2}}$ acting on the state  $|(\tiny\yng(2);0)>$.
By inserting the overall factor into this state, one has
the final state associated with 
the representation $(\tiny\yng(2);\overline{\tiny\yng(1)})$.
Therefore, one arrives at the above eigenvalue for the spin $2$
current.
As before, one can describe this eigenvalue in the context of
the corresponding eigenvalue nonlinear version. That is,
\bea
&&   h (
   \tiny\yng(2);\overline{\tiny\yng(1)})
   + \frac{1}{(N+k+2)} \frac{1}{N(N+2)} \hat{u}^2
   (\tiny\yng(2);\overline{\tiny\yng(1)})
 =  \frac{(4N+2k+7)}{4(N+k+2)}+
 \frac{(-\frac{3N}{2}-1)^2}{N (N+2) (k+N+2)}
 \nonu \\
 && =
  {\bf h} (
   0;\overline{\tiny\yng(1)})
+  {\bf h} (
   \tiny\yng(2);0)
    =  \frac{(k N+2 N+1)}{2 N (k+N+2)} +
\frac{(N+1) (N+4)}{ (N+2) (k+N+2)}.
   \nonu
\eea
In the first line the previous relations in (\ref{eigenforsymmf})
and in the second line,
the relations (\ref{0fbareigenhigherlinear})
and (\ref{symmzerolinearres}) are used.

For the eigenvalue corresponding to the higher spin $3$ current,
there exists the contribution from the lower order poles
appearing in the commutator $[({\bf \Phi_2^{(1)}})_0,
  Q^{\bar{A}}_{-\frac{1}{2}}]$. 
One describes the following eigenvalues for each term appearing in
(\ref{BigPhi})
{\small
\bea
&& \Bigg[
  -\frac{(2 k N^2+k N-4 k+6 N^2+15 N+6)}{(N+2) (k+N+2)^2},0,
  \frac{i \sqrt{\frac{N}{N+2}} (k+3 N+6)}{(k+N+2)},
  -\frac{i \sqrt{\frac{N}{N+2}} (3 k+N+2)}{(k+N+2)},
  \nonu \\
  && \frac{(-8 k N-12 k+7 N^2+20 N+12)}{2 (N+2) (k+N+2)} \Bigg].
\label{symmfbarEigen}
\eea}
Note that the eigenvalues in the third and the fourth are different
from each other.
This is the reason why there exists an imaginary $i$ term in the eigenvalue
in (\ref{symmfbarlinearresult}).
Then by substituting the coefficients (\ref{ccoeff}) and (\ref{p1p2})
into the relation (\ref{BigPhi}) together with
(\ref{symmfbarEigen}) one sees the final result in
(\ref{symmfbarlinearresult}).

Furthermore, there are $\Delta(\tiny\yng(2);0)$ and $\Delta(0;\overline{
\tiny\yng(1)})$. They can be obtained explicitly
from (\ref{Deltasymm0}) and (\ref{knownresult2}) respectively.
\footnote{For similar higher representation, one obtains
  \bea
{\bf \phi_2^{(1)}} (
\overline{\tiny\yng(2)};\tiny\yng(1))
& = & \phi_2^{(1)} (
\overline{\tiny\yng(2)};\tiny\yng(1)) + \Delta (
\overline{\tiny\yng(2)};0) +
 \Delta ( 0;\tiny\yng(1))
\nonu
\eea
plus the contributions from the following eigenvalues
together with the previous coefficients
\bea
&& \Bigg[
  \frac{(4 k N^2+11 k N+4 k+6 N^2+15 N+6)}{(N+2) (k+N+2)^2},
  0,
  \frac{i \sqrt{\frac{N}{N+2}} (k+3 N+6)}{(k+N+2)},
  -\frac{i \sqrt{\frac{N}{N+2}} (3 k+N+2)}{(k+N+2)},
  \nonu \\
  &&
\frac{(4 k N+12 k+N^2-4 N-12)}{2 (N+2) (k+N+2)}
  \Bigg].
\nonu
\eea
Note that the quantity
$\phi_2^{(1)} (
\overline{\tiny\yng(2)};\tiny\yng(1))$ can be obtained from
the footnote \ref{symmbarfrelation},
the quantity $\Delta (
\overline{\tiny\yng(2)};0)$ can be determined from
(\ref{Deltasymm0}) with an extra minus sign and the
quantity $
 \Delta ( 0;\tiny\yng(1))$ can be read off from (\ref{knownresult2}).
}.

%%%%%%%%%%%%%%%%%%%%%%%%%%%%%%%%%%%%%%%%%%%%%%%%%%%%%%%%%%%%%%%%%%%
\subsection{The $(\mbox{symm}; \mbox{antisymm})$ representation}
%%%%%%%%%%%%%%%%%%%%%%%%%%%%%%%%%%%%%%%%%%%%%%%%%%%%%%%%%%%%%%%%%%%

The two eigenvalues are given by
{\small
  \bea
 {\bf h} (
  \tiny\yng(2);\tiny\yng(1,1)) 
  &= &
\frac{(k N^2+2 k N+N^3+8 N^2+12 N+4)}{N (N+2) (k+N+2)},
  \nonu \\
{\bf \phi_2^{(1)}} (
  \tiny\yng(2);\tiny\yng(1,1)) 
&= & \phi_2^{(1)} (
  \tiny\yng(2);\tiny\yng(1,1)) +
  \Delta(\tiny\yng(2);0) + \Delta(0;\tiny\yng(1,1)) +
 \frac{16}{9} \Bigg[-\frac{18 i \sqrt{\frac{N}{N+2}}}{(k+N+2)}
\nonu \\
&-& 
\frac{9 }{(N+2) (k+N+2)^2 (3 k N+4 k+4 N+5) (6 k N+5 k+5 N+4)}
\nonu \\
& \times &
(N-k) ({\bf 42 k^2 N^2}+145 k^2 N+108 k^2 {\bf +6 k N^3}+68 k N^2+160 k N+84 k
\nonu \\
& + & 19 N^3+52 N^2+4 N-48)\Bigg].
\label{symmantisymmlinear}
\eea}
Note that the eigenvalue
$ \phi_2^{(1)} (
\tiny\yng(2);\tiny\yng(1,1))$ is given by
(\ref{symmantisymmnonlinear}).
As before, one can describe this eigenvalue in the context of
the corresponding eigenvalue in nonlinear version. That is,
\bea
&&   h (
   \tiny\yng(2);\tiny\yng(1,1))
   + \frac{1}{(N+k+2)} \frac{1}{N(N+2)} \hat{u}^2
   (\tiny\yng(2);\tiny\yng(1,1))
 =  \frac{(N+k+6)}{(N+k+2)}+
 \frac{4}{N (N+2) (k+N+2)}
 \nonu \\
 && =
  {\bf h} (
   0;\tiny\yng(1,1))
+  {\bf h} (
   \tiny\yng(2);0)
    =  \frac{(k N+3 N+2)}{ N (k+N+2)} +
\frac{(N+1) (N+4)}{ (N+2) (k+N+2)}.
   \nonu
\eea
In the first line the previous relations in (\ref{symmantisymmexp})
and in the second line,
the relations (\ref{0antisymmeigenlinear})
and (\ref{symmzerolinearres}) are used.

The contributions from the following eigenvalues
together with the previous coefficients
{\small
\bea
&& \Bigg[-\frac{2 (4 k N^2+11 k N+4 k+9 N^2+24 N+12)}
  {(N+2) (k+N+2)^2}, 0, -\frac{2 i \sqrt{\frac{N}{N+2}} (k+3 N+6)}{(k+N+2)},
  \nonu \\
  && \frac{2 i \sqrt{\frac{N}{N+2}} (3 k+N+2)}{(k+N+2)},
   \frac{2 (-5 k N-12 k+N^2+8 N+12)}{(N+2) (k+N+2)} \Bigg],
\label{symmantisymmEigen}
\eea}
provide the 
last two terms of (\ref{symmantisymmlinear})
\footnote{It is straightforward to see that one has
  \bea
{\bf \phi_2^{(1)}} (
\overline{\tiny\yng(2)};\tiny\yng(1,1))
& = & \phi_2^{(1)} (
\overline{\tiny\yng(2)};\tiny\yng(1,1)) + \Delta (
\overline{\tiny\yng(2)};0) +
 \Delta ( 0;\tiny\yng(1,1))
\nonu
\eea
plus other contributions from the following eigenvalues
\bea
&& \Bigg[
  \frac{2 (4 k N^2+11 k N+4 k+9 N^2+24 N+12)}{(N+2) (k+N+2)^2},
  0,
 \frac{2 i \sqrt{\frac{N}{N+2}} (k+3 N+6)}{(k+N+2)},
  -\frac{2 i \sqrt{\frac{N}{N+2}} (3 k+N+2)}{(k+N+2)},
  \nonu \\
  &&
-\frac{2 (-5 k N-12 k+N^2+8 N+12)}{(N+2) (k+N+2)}
  \Bigg],
\nonu
\eea
together with the previous coefficients. 
The quantity
$\phi_2^{(1)} (
\overline{\tiny\yng(2)};\tiny\yng(1,1))$ can be obtained from
the footnote \ref{cor3foot},
the quantity $\Delta (
\overline{\tiny\yng(2)};0)$ can be determined from
(\ref{Deltasymm0}) with an extra minus sign and the
quantity $
 \Delta ( 0;\tiny\yng(1,1))$ can be read off from (\ref{Delta0antisymm}).
}.

%%%%%%%%%%%%%%%%%%%%%%%%%%%%%%%%%%%%%%%%%%%%%%
\subsection{The $(\mbox{symm};\overline{\mbox{antisymm}})$ representation}
%%%%%%%%%%%%%%%%%%%%%%%%%%%%%%%%%%%%%%%%%%%%%%

The two eigenvalues are given by
{\small
  \bea
 {\bf h} (
  \tiny\yng(2);\overline{\tiny\yng(1,1)}) 
  &= & \frac{(k N^2+2 k N+N^3+8 N^2+12 N+4)}{N (N+2) (k+N+2)},
  \nonu \\
{\bf \phi_2^{(1)}} (
  \tiny\yng(2);\overline{\tiny\yng(1,1)}) 
&= & \phi_2^{(1)} (
  \tiny\yng(2);\overline{\tiny\yng(1,1)}) +
  \Delta(\tiny\yng(2);0) + \Delta(0;\overline{\tiny\yng(1,1)}) +
 \frac{16}{9} \Bigg[\frac{18 i \sqrt{\frac{N}{N+2}}}{(k+N+2)}
\nonu \\
&-& 
\frac{9 }{(N+2) (k+N+2)^2 (3 k N+4 k+4 N+5) (6 k N+5 k+5 N+4)}
\nonu \\
& \times &
(N-k)({\bf 36 k^2 N^2}+125 k^2 N+108 k^2 {\bf -36 k N^3}-122 k N^2-100 k N
\nonu \\
& + & 12 k-67 N^3-304 N^2-436 N-192) \Bigg].
\label{symmantisymmbarlinear}
\eea}
Note that the eigenvalue
$ \phi_2^{(1)} (
\tiny\yng(2);\overline{\tiny\yng(1,1)})$ is given by
(\ref{symmantisymmbarnonlinear}).
The contributions from the following eigenvalues
together with the previous coefficients
{\small
\bea
&& \Bigg[ \frac{2 (2 k N^2+k N-4 k+9 N^2+24 N+12)}{(N+2) (k+N+2)^2},
  0,\frac{2 i \sqrt{\frac{N}{N+2}} (k+3 N+6)}{(k+N+2)},
-\frac{2 i \sqrt{\frac{N}{N+2}} (3 k+N+2)}{(k+N+2)},
  \nonu \\
&& \frac{2 (-7 k N-12 k+5 N^2+16 N+12)}{(N+2) (k+N+2)}
  \Bigg],
\nonu
\eea}
provide the last two terms in (\ref{symmantisymmbarlinear})
\footnote{For similar higher representation, one has
  \bea
{\bf \phi_2^{(1)}} (
\overline{\tiny\yng(2)};\overline{\tiny\yng(1,1)})
& = & \phi_2^{(1)} (
\overline{\tiny\yng(2)};\overline{\tiny\yng(1,1)}) + \Delta (
\overline{\tiny\yng(2)};0) +
 \Delta ( 0;\overline{\tiny\yng(1,1)})
\nonu
\eea
plus the other contributions which can be
determined by the following eigenvalues
\bea
&& \Bigg[
  -\frac{2 (2 k N^2+k N-4 k+9 N^2+24 N+12)}{(N+2) (k+N+2)^2},
  0,
 -\frac{2 i \sqrt{\frac{N}{N+2}} (k+3 N+6)}{(k+N+2)},
  \frac{2 i \sqrt{\frac{N}{N+2}} (3 k+N+2)}{(k+N+2)},
  \nonu \\
  &&
-\frac{2 (-7 k N-12 k+5 N^2+16 N+12)}{(N+2) (k+N+2)}
  \Bigg].
\nonu
\eea
The quantity
$\phi_2^{(1)} (
\overline{\tiny\yng(2)};\overline{\tiny\yng(1,1)})$ can be obtained from
the footnote \ref{cor4foot},
the quantity $\Delta (
\overline{\tiny\yng(2)};0)$ can be determined from
(\ref{Deltasymm0}) with an extra minus sign and the
quantity $
\Delta ( 0;\overline{\tiny\yng(1,1)})$ can be read off from
(\ref{Delta0antisymm}) with an extra minus sign.
}.

%%%%%%%%%%%%%%%%%%%%%%%%%%%%%%%%%%%%%%%%%%%%%%%%%%%%%%%%%%%%%%%%%%%%%%%%
\subsection{The $(\mbox{symm}; \overline{\mbox{symm}})$ representation}
%%%%%%%%%%%%%%%%%%%%%%%%%%%%%%%%%%%%%%%%%%%%%%%%%%%%%%%%%%%%%%%%%%%%%%%%

The two eigenvalues are given by
{\small
  \bea
  {\bf h} (
  \tiny\yng(2);\overline{\tiny\yng(2)}) 
  &= &
\frac{(k N^2+2 k N+N^3+6 N^2+8 N+4)}{N (N+2) (k+N+2)},
  \nonu \\ 
{\bf \phi_2^{(1)}} (
  \tiny\yng(2);\overline{\tiny\yng(2)}) 
&= &  \phi_2^{(1)} (
  \tiny\yng(2);\overline{\tiny\yng(2)}) +
  \Delta(\tiny\yng(2);0) + \Delta(0;\overline{\tiny\yng(2)}) +
 \frac{16}{9} \Bigg[\frac{18 i \sqrt{\frac{N}{N+2}}}{(k+N+2)}
\nonu \\
&+ &  
\frac{9}{(N+2) (k+N+2)^2 (3 k N+4 k+4 N+5) (6 k N+5 k+5 N+4)}
\nonu \\
& \times & (k-N)
({\bf 48 k^2 N^2}+131 k^2 N+84 k^2 {\bf -24 k N^3}-56 k N^2-4 k N+36 k
-31 N^3\nonu \\
& - &  124 N^2-148 N-48)\Bigg].
\label{symmsymmbarlinearresult}
\eea}
Note that the eigenvalue
$ \phi_2^{(1)} (
\tiny\yng(2);\overline{\tiny\yng(2)})$ is given by
(\ref{symmsymmbarnonlinear}).
Note that the eigenvalue for the spin $2$ current
does not have any contribution from the commutator
$[({\bf T})_0, Q^{1}_{-\frac{1}{2}}  Q^{4}_{-\frac{1}{2}}]$.
This provides only the eigenvalue for the representation $(0;\overline{
  \tiny\yng(2)})$. Also the term 
$ Q^{1}_{-\frac{1}{2}}  Q^{4}_{-\frac{1}{2}} ({\bf T})_0$
acting on the representation $(\tiny\yng(2);0)$
gives the eigenvalue ${\bf h} (\tiny\yng(2);0)$ with
$ Q^{1}_{-\frac{1}{2}}  Q^{4}_{-\frac{1}{2}}$
acting on the state  $|(\tiny\yng(2);0)>$.
By inserting the overall factor into this state, one has
the final state associated with 
the representation $(\tiny\yng(2);\overline{\tiny\yng(2)})$.
Therefore, one arrives at the above eigenvalue for the spin $2$
current.

Furthermore, the above eigenvalue for the spin $2$ current
can be interpreted as the one in the nonlinear version 
\bea
&&   h (
   \tiny\yng(2);\overline{\tiny\yng(2)})
   + \frac{1}{(N+k+2)} \frac{1}{N(N+2)} \hat{u}^2
   (\tiny\yng(2);\overline{\tiny\yng(2)})
 =  \frac{(N+k)}{(N+k+2)}+
 \frac{(-2N-2)^2}{N (N+2) (k+N+2)}
 \nonu \\
 && =
  {\bf h} (
   0;\overline{\tiny\yng(2)})
+  {\bf h} (
   \tiny\yng(2);0)
    =  \frac{(k N+ N+2)}{ N (k+N+2)} +
\frac{(N+1) (N+4)}{ (N+2) (k+N+2)}.
   \nonu
\eea
In the first line, the relations in (\ref{symmsymmbarexpression}) 
are used and similarly
in the second line, the relations (\ref{0symmeigenlinear}) 
and (\ref{symmzerolinearres}) are used also.

The 
$ \Delta(0;\overline{\tiny\yng(2)}) $
can be obtained from
$ \Delta(0;\tiny\yng(2)) $
by changing the sign.
As before, the last two terms of (\ref{symmsymmbarlinearresult})
can be obtained from
the following eigenvalues 
{\small
  \bea
&& \Bigg[ -\frac{2 (2 k N^2+k N-4 k+3 N^2+12 N+12)}
  {(N+2) (k+N+2)^2}, 0,
  \frac{2 i \sqrt{\frac{N}{N+2}} (k+3 N+6)}{(k+N+2)},
  -\frac{2 i \sqrt{\frac{N}{N+2}} (3 k+N+2)}{(k+N+2)},
  \nonu \\
&&   \frac{2 (-7 k N-12 k+5 N^2+16 N+12)}{(N+2) (k+N+2)} \Bigg],
\label{symmsymmbarEigen}
\eea}
with the previous coefficients
\footnote{
For similar higher representation,
  one has
  \bea
{\bf \phi_2^{(1)}} (
\overline{\tiny\yng(2)};\tiny\yng(2))
& = & \phi_2^{(1)} (
\overline{\tiny\yng(2)};\tiny\yng(2)) + \Delta (
\overline{\tiny\yng(2)};0) +
 \Delta ( 0;\tiny\yng(2))
\nonu
\eea
plus other contributions from the eigenvalues
\bea
&& \Bigg[
  \frac{2 (4 k N^2+11 k N+4 k+3 N^2+12 N+12)}{(N+2) (k+N+2)^2},
  0,
 \frac{2 i \sqrt{\frac{N}{N+2}} (k+3 N+6)}{(k+N+2)},
  -\frac{2 i \sqrt{\frac{N}{N+2}} (3 k+N+2)}{(k+N+2)},
  \nonu \\
  &&
-\frac{2 (-5 k N-12 k+N^2+8 N+12)}{(N+2) (k+N+2)}
  \Bigg].
\nonu
\eea
The quantity
$\phi_2^{(1)} (
\overline{\tiny\yng(2)};\tiny\yng(2))$ can be obtained from
the footnote \ref{footnewcorrection},
the quantity $\Delta (
\overline{\tiny\yng(2)};0)$ can be determined from
(\ref{Deltasymm0}) with an extra minus sign and the
quantity $
\Delta ( 0;\tiny\yng(2))$ can be read off from
(\ref{Delta0symm}).
}.

%%%%%%%%%%%%%%%%%%%%%%%%%%%%%%%%%%%%%%%%%%%%%%%%%%%%%%%%%%%%%%%%%%%%%
\subsection{The $(\mbox{antisymm}; \mbox{antisymm})$ representation}
%%%%%%%%%%%%%%%%%%%%%%%%%%%%%%%%%%%%%%%%%%%%%%%%%%%%%%%%%%%%%%%%%%%%%

The two eigenvalues are 
\bea
{\bf h} (
  \tiny\yng(1,1);\tiny\yng(1,1)) 
  & = & \frac{2 (N^2+2 N+2)}{N (N+2) (k+N+2)},
\nonu \\
{\bf \phi_2^{(1)}} (
  \tiny\yng(1,1);\tiny\yng(1,1)) 
 & = &
  \phi_2^{(1)} (
  \tiny\yng(1,1);\tiny\yng(1,1)) +  \Delta (
  \tiny\yng(1,1);\tiny\yng(1,1)) 
  \nonu \\
  & = & \frac{16}{3 N (N+2) (k+N+2)^2 (3 k N+4 k+4 N+5)}
  \nonu \\
  & \times & ({\bf 3 k^2 N^3}+10 k^2 N^2+8 k^2 N {\bf -3 k N^4}
  +3 k N^3+47 k N^2+
  58 k N+24 k \nonu \\
  & - & 4 N^4-13 N^3+5 N^2+6 N).
\label{antiantilinear}
\eea
The highest power terms of the last term in the
eigenvalue of the higher spin $3$ current in (\ref{antiantilinear})
are twice of the ones in (\ref{boxboxlinear}).
Note that the eigenvalue
$ \phi_2^{(1)} (
\tiny\yng(1,1);\tiny\yng(1,1))$ is given by
(\ref{antiantinonlinear}).
The eigenvalue for the spin $2$ current can be interpreted
in terms of the one in the nonlinear version as follows:
\bea
&&   h (
   \tiny\yng(1,1);\tiny\yng(1,1))
   + \frac{1}{(N+k+2)} \frac{1}{N(N+2)} \hat{u}^2
   (\tiny\yng(1,1);\tiny\yng(1,1))
 =  \frac{2}{(N+k+2)}+
 \frac{4}{N (N+2) (k+N+2)}.
 \nonu
\eea
The previous relations in (\ref{antisymmantisymmresult})
are used in this calculation and this is equal to the one in
(\ref{antiantilinear}).

The difference between the nonlinear and linear cases
which vanishes in the large $N$ 't Hooft limit  
is obtained as follows:
{\small
  \bea
\Delta (
\tiny\yng(1,1);\tiny\yng(1,1))
& = &
-\frac{64 (k-N) ({\bf 3 k N^2}-6 k N-10 k {\bf +3 N^3}+12 N^2+2 N-8)}{
  N (N+2) (k+N+2)^2 (3 k N+4 k+4 N+5) (6 k N+5 k+5 N+4)},
\label{Deltaantianti}
\eea}
which is equal to the one in (\ref{Deltasymmsymm}).
The following eigenvalues are used in this expression
\bea
&& \Bigg[
  \frac{4 (k N^2+2 k N+N^3+6 N^2+8 N+4)}{N (N+2) (k+N+2)^2},
\frac{8}{(k+N+2)}, 
-\frac{8 i}{\sqrt{N^2+2 N} (k+N+2)},
\nonu \\
&&
-\frac{8 i}{\sqrt{N^2+2 N} (k+N+2)},
-\frac{16}{N (N+2) (k+N+2)}
\Bigg],
\nonu
\eea
together with the previous coefficients.
These eigenvalues are the same as the ones in (\ref{symmsymmEigen})
\footnote{One has, for similar higher representation,
  \bea
{\bf \phi_2^{(1)}} (
\overline{\tiny\yng(1,1)};\overline{\tiny\yng(1,1)})
& = & \phi_2^{(1)} (
\overline{\tiny\yng(1,1)};\overline{\tiny\yng(1,1)}) + \Delta (
\overline{\tiny\yng(1,1)};\overline{\tiny\yng(1,1)}).
\nonu
\eea
The quantity $
\phi_2^{(1)} (
\overline{\tiny\yng(1,1)};\overline{\tiny\yng(1,1)})
$ can be obtained from the footnote \ref{antibarantibarfoot}
and the quantity
$\Delta (
\overline{\tiny\yng(1,1)};\overline{\tiny\yng(1,1)})$
can be determined from (\ref{Deltaantianti}) with an extra minus sign.
}.

%%%%%%%%%%%%%%%%%%%%%%%%%%%%%%%%%%%%%%%%%%%%%%%%%%%%%%%%%%%
\subsection{The $(\mbox{antisymm}; f)$ representation}
%%%%%%%%%%%%%%%%%%%%%%%%%%%%%%%%%%%%%%%%%%%%%%%%%%%%%%%%%%%

The two eigenvalues are described by
\bea
{\bf h} (
  \tiny\yng(1,1);\tiny\yng(1)) 
  & = & \frac{(N^3+4 N^2+N+2)}{2 N (N+2) (k+N+2)},
\nonu \\
{\bf \phi_2^{(1)}}  (
  \tiny\yng(1,1);\tiny\yng(1)) 
 & = &
    \phi_2^{(1)} (
  \tiny\yng(1,1);\tiny\yng(1)) +  \Delta (
  \tiny\yng(1,1);\tiny\yng(1)) 
  \nonu \\
  & = &
  -\frac{4}{3 N (N+2) (k+N+2)^2 (3 k N+4 k+4 N+5) (6 k N+5 k+5 N+4)}
  \nonu \\
  &\times&
  ( {\bf 18 k^3 N^5}+39 k^3 N^4-34 k^3 N^3-174 k^3 N^2-188 k^3 N+
  {\bf 36 k^2 N^6} +189 k^2 N^5
  \nonu \\
  & + &
  146 k^2 N^4-643 k^2 N^3-1097 k^2 N^2-692 k^2 N-60 k^2+60 k N^6+247 k N^5
  \nonu \\
  & + & 177 k N^4-893 k N^3-1272 k N^2-572 k N-48 k+25 N^6+100 N^5+84 N^4
  \nonu \\
  & - & 424 N^3-592 N^2-192 N).
\label{antisymmflinear}
\eea
Note that the eigenvalue
$ \phi_2^{(1)} (
\tiny\yng(1,1);\tiny\yng(1))$ is given by
(\ref{antifnonlinear}).
Note that
the eigenvalue for the spin $2$ current
can be also obtained the one in the nonlinear version
as follows:
\bea
&&   h (
   \tiny\yng(1,1);\tiny\yng(1))
   + \frac{1}{(N+k+2)} \frac{1}{N(N+2)} \hat{u}^2
   (\tiny\yng(1,1);\tiny\yng(1))
 =  \frac{(2N+3)}{4(N+k+2)}+
 \frac{(-\frac{N}{2}+1)^2}{N (N+2) (k+N+2)},
 \nonu
\eea
where the relations in (\ref{antifexpression})
are used.

One observes that these eigenvalues
are twice of the ones in (\ref{f0eigenhigherlinear}) respectively if
one takes the highest order terms in the numerators.
It is easy to see this behavior in the spin $2$ current. 
For the higher spin $3$ current, more explicitly,
the corresponding terms in the
second eigenvalue in (\ref{antisymmflinear})
are given by $(18k^3 N^5 + 36 k^2  N^6)=
6 k N^3 (3 k^2 N^2 +6 k N^3)$
and the denominator contains the factor $N(6 k N+5 k+5 N+4) \rightarrow
6 k N^2$,
compared to the one in (\ref{f0eigenhigherlinear}). 
Therefore, one observes that the corresponding
quantity  $N (3 k^2 N^2 +6 k N^3)$ appears in the eigenvalue in
(\ref{f0eigenhigherlinear}).

The difference between the nonlinear and linear cases
which vanishes in the large $N$ 't Hooft limit  
is given by
\bea
  \Delta (
  \tiny\yng(1,1);\tiny\yng(1)) & = & - 
  \frac{4}{N (N+2) (k+N+2)^2 (3 k N+4 k+4 N+5) (6 k N+5 k+5 N+4)}
  \nonu \\
  & \times & (k-N)
  ({\bf 6 k^2 N^3}-6 k^2 N^2-36 k^2 N{\bf +6 k N^4}-14 k N^3-15 k N^2-
  48 k N-20 k
  \nonu \\
  & {\bf -} & {\bf 3 N^5}-8 N^4-10 N^3 +  60 N^2+40 N-16).
\label{Deltaantisymmf}
\eea
Here the following eigenvalues corresponding to the various
fields in (\ref{BigPhi}) are used
{\small
  \bea
&& \Bigg[
  -\frac{(2 k N^2+4 k N+3 N^4+6 N^3-15 N^2-16 N-4)}{2 N (N+2) (k+N+2)^2},
  -\frac{2 (N-2)}{(k+N+2)},
  \nonu \\
  && -\frac{i (N-2)^2}{2 \sqrt{N^2+2 N} (k+N+2)},
   -\frac{i (N-2)^2}{2 \sqrt{N^2+2 N} (k+N+2)},
\frac{(N-2)^3}{4 N (N+2) (k+N+2)}
  \Bigg].
\nonu
\eea}
One sees that the highest power terms in (\ref{Deltaantisymmf})
are the same as the ones in (\ref{Deltasymmf})
\footnote{For the similar higher representation, one has
  \bea
{\bf \phi_2^{(1)}} (
\overline{\tiny\yng(1,1)};\overline{\tiny\yng(1)})
& = & \phi_2^{(1)} (
\overline{\tiny\yng(1,1)};\overline{\tiny\yng(1)}) + \Delta (
\overline{\tiny\yng(1,1)};\overline{\tiny\yng(1)})
\nonu
\eea
The qunatity $
\phi_2^{(1)} (
\overline{\tiny\yng(1,1)};\overline{\tiny\yng(1)})$
can be obtained from the footnote \ref{antisymmantisymmfoot}
and the quantity $\Delta (
\overline{\tiny\yng(1,1)};\overline{\tiny\yng(1)})$
can be obtained from (\ref{Deltaantisymmf}) with an extra minus sign.
}.

%%%%%%%%%%%%%%%%%%%%%%%%%%%%%%%%%%%%%%%%%%%%%%%%%%%%%%%%%%%%%%%%%%
\subsection{The $(\mbox{antisymm}; \overline{f})$ representation}
%%%%%%%%%%%%%%%%%%%%%%%%%%%%%%%%%%%%%%%%%%%%%%%%%%%%%%%%%%%%%%%%%%

The two eigenvalues are given by
{\small
  \bea
  {\bf h} (
  \tiny\yng(1,1);\overline{\tiny\yng(1)}) 
  &= &
\frac{(k N^2+2 k N+2 N^3+8 N^2+5 N+2)}{2 N (N+2) (k+N+2)},
  \nonu \\
{\bf \phi_2^{(1)}} (
  \tiny\yng(1,1);\overline{\tiny\yng(1)}) 
&= &  \phi_2^{(1)} (
  \tiny\yng(1,1);\overline{\tiny\yng(1)}) +
  \Delta(\tiny\yng(1,1);0) + \Delta(0;\overline{\tiny\yng(1)}) +
\frac{4}{9} \Bigg[\frac{36 i \sqrt{\frac{N}{N+2}}}{(k+N+2)}
 \nonu \\
& + &  
\frac{18 }{(N+2) (k+N+2)^2 (3 k N+4 k+4 N+5)
  (6 k N+5 k+5 N+4)} \nonu \\
& \times &  (k-N)
({\bf 30 k^2 N^2}+77 k^2 N+48 k^2 {\bf -15 k N^3}-17 k N^2+35 k N+30 k-10 N^3
\nonu \\
& - &  34 N^2-40 N-24) \Bigg].
\label{antisymmflinear1}
\eea}
Note that the eigenvalue
$ \phi_2^{(1)} (
\tiny\yng(1,1);\overline{\tiny\yng(1)})$ is given by
(\ref{antifbarnonlinear}).
One sees that the highest power terms
of $(N,k)$ in the last term of the
eigenvalue (\ref{antisymmflinear1})
for the higher spin $3$ current also appear in
(\ref{symmfbarlinearresult}).

For the eigenvalue of spin $2$ current, one obtains the following
analysis from the nonlinear case
\bea
&&   h (
   \tiny\yng(1,1);\overline{\tiny\yng(1)})
   + \frac{1}{(N+k+2)} \frac{1}{N(N+2)} \hat{u}^2
   (\tiny\yng(1,1);\overline{\tiny\yng(1)})
 =  \frac{(4N+2k-1)}{4(N+k+2)}+
 \frac{(-\frac{3N}{2}-1)^2}{N (N+2) (k+N+2)}
 \nonu \\
 && =
  {\bf h} (
   0;\overline{\tiny\yng(1)})
+  {\bf h} (
   \tiny\yng(1,1);0)
   =  \frac{(k N+ 2N+1)}{ 2N (k+N+2)} +
\frac{N (N+3)}{ (N+2) (k+N+2)}.
   \nonu
   \eea
The relations in    
(\ref{antisymmfbarexpression}) are used in the first line and
the previous relations
(\ref{0fbareigenhigherlinear}) and (\ref{antisymmzerolinearres})
are also used in the second line.

One uses the following eigenvalues 
\bea
&& \Bigg[
-\frac{(2 k N^2+3 k N+6 N^2+15 N+6)}{(N+2) (k+N+2)^2}, 0, 
\frac{i \sqrt{\frac{N}{N+2}} (k+3 N+6)}{(k+N+2)},
-\frac{i \sqrt{\frac{N}{N+2}} (3 k+N+2)}{(k+N+2)},
\nonu \\
&& \frac{(-8 k N-12 k+7 N^2+20 N+12)}{2 (N+2) (k+N+2)} \Bigg],
\label{fivres}
\eea
in order to obtain the last two terms in the eigenvalue
of the higher spin $3$ current in (\ref{antisymmflinear1}).
The last eigenvalue of (\ref{fivres})
is the same as the one in (\ref{symmfbarEigen})
\footnote{One obtains 
  \bea
{\bf \phi_2^{(1)}} (
\overline{\tiny\yng(1,1)};\tiny\yng(1))
& = & \phi_2^{(1)} (
\overline{\tiny\yng(1,1)};\tiny\yng(1)) + \Delta (
\overline{\tiny\yng(1,1)};0) +
 \Delta ( 0;\tiny\yng(1))
\nonu
\eea
plus the contributions from the following
eigenvalues with the coefficients
\bea
&& \Bigg[
  \frac{(4 k N^2+9 k N+6 N^2+15 N+6)}{(N+2) (k+N+2)^2},
  0,
  \frac{i \sqrt{\frac{N}{N+2}} (k+3 N+6)}{(k+N+2)},
  -\frac{i \sqrt{\frac{N}{N+2}} (3 k+N+2)}{(k+N+2)},
  \nonu \\
  &&
\frac{(4 k N+12 k+N^2-4 N-12)}{2 (N+2) (k+N+2)}
  \Bigg].
\nonu
\eea
One can read off 
$\phi_2^{(1)} (
\overline{\tiny\yng(1,1)};\tiny\yng(1))$
from the footnote \ref{antibarffootnote},  $ \Delta (
\overline{\tiny\yng(1,1)};0)$ from (\ref{Deltaantisymm0})
with minus sign, and $ 
 \Delta ( 0;\tiny\yng(1))$ from (\ref{knownresult2}). 
}.

%%%%%%%%%%%%%%%%%%%%%%%%%%%%%%%%%%%%%%%%%%%%%%%%%%%%%%%%%%%%%%%%%
\subsection{The $(\mbox{antisymm}; \mbox{symm})$ representation}
%%%%%%%%%%%%%%%%%%%%%%%%%%%%%%%%%%%%%%%%%%%%%%%%%%%%%%%%%%%%%%%%%

The two eigenvalues are given by
{\small
  \bea
   {\bf h} (
  \tiny\yng(1,1);\tiny\yng(2)) 
&= & \frac{(k N^2+2 k N+N^3+4 N^2+4 N+4)}{N (N+2) (k+N+2)},
  \nonu \\
{\bf \phi_2^{(1)}} (
  \tiny\yng(1,1);\tiny\yng(2)) 
&= &  \phi_2^{(1)} (
  \tiny\yng(1,1);\tiny\yng(2)) +
\Delta(\tiny\yng(1,1);0) + \Delta(0;\tiny\yng(2)) +
  \frac{16}{9} \Bigg[
-\frac{18 i \sqrt{\frac{N}{N+2}}}{(k+N+2)}
\nonu \\
&+ & 
\frac{9 }{(N+2) (k+N+2)^2 (3 k N+4 k+4 N+5) (6 k N+5 k+5 N+4)} 
\nonu \\
& \times & (k-N) ({\bf 42 k^2 N^2}+139 k^2 N+96 k^2 {\bf +6 k N^3}+44 k N^2+100 k N
\nonu \\
& + &  60 k+N^3-20 N^2-68 N-48) \Bigg].
\label{antisymmsymmlinear}
  \eea
  }
Note that the eigenvalue
$ \phi_2^{(1)} (
\tiny\yng(1,1);\tiny\yng(2))$ is given by
(\ref{antisymmnonlinear}).
The eigenvalue for the spin $2$ current can be also obtained
from the  one in the nonlinear version
as follows:
\bea
&&   h (
   \tiny\yng(1,1);\tiny\yng(2))
   + \frac{1}{(N+k+2)} \frac{1}{N(N+2)} \hat{u}^2
   (\tiny\yng(1,1);\tiny\yng(2))
 =  1+
 \frac{4}{N (N+2) (k+N+2)}
 \nonu \\
 && =
  {\bf h} (
   0;\tiny\yng(2))
+  {\bf h} (
   \tiny\yng(1,1);0)
   =  \frac{(k N+ N+2)}{ N (k+N+2)} +
\frac{N (N+3)}{ (N+2) (k+N+2)}.
   \nonu
\eea
We use the relations in (\ref{antisymmsymmeigen})
in the first line and the relations in 
(\ref{0symmeigenlinear}) and (\ref{antisymmzerolinearres})
in the second line.

The highest power of $(N,k)$ in the last term of the eigenvalue
of higher spin $3$ current (\ref{antisymmsymmlinear}) is the same as the one in
(\ref{symmantisymmlinear}).
Furthermore, the following eigenvalues 
{\small
\bea
&& \Bigg[ -\frac{2 (4 k N^2+9 k N+3 N^2+12 N+12)}{(N+2) (k+N+2)^2},
  0, -\frac{2 i \sqrt{\frac{N}{N+2}} (k+3 N+6)}{(k+N+2)},
  \frac{2 i \sqrt{\frac{N}{N+2}} (3 k+N+2)}{(k+N+2)},
  \nonu \\
  && \frac{2 (-5 k N-12 k+N^2+8 N+12)}{(N+2) (k+N+2)} \Bigg]
\label{fivres1}
\eea}
are used for the last two terms in (\ref{antisymmsymmlinear}).
The last eigenvalue of (\ref{fivres1})
is the same as the one in (\ref{symmantisymmEigen})
\footnote{In this case, also for the similar higher
  representation, one has
  \bea
{\bf \phi_2^{(1)}} (
\overline{\tiny\yng(1,1)};\tiny\yng(2))
& = & \phi_2^{(1)} (
\overline{\tiny\yng(1,1)};\tiny\yng(2)) + \Delta (
\overline{\tiny\yng(1,1)};0) +
 \Delta ( 0;\tiny\yng(2))
\nonu
\eea
plus other contributions from the following eigenvalues
\bea
&& \Bigg[
  \frac{2 (4 k N^2+9 k N+3 N^2+12 N+12)}{(N+2) (k+N+2)^2},
  0,
  \frac{2 i \sqrt{\frac{N}{N+2}} (k+3 N+6)}{(k+N+2)},
  -\frac{2 i \sqrt{\frac{N}{N+2}} (3 k+N+2)}{(k+N+2)},
  \nonu \\
  &&
-\frac{2 (-5 k N-12 k+N^2+8 N+12)}{(N+2) (k+N+2)}
  \Bigg],
\nonu
\eea
together with the previous coefficients.
The eigenvalue $\phi_2^{(1)} (
\overline{\tiny\yng(1,1)};\tiny\yng(2))$
can be obtained from the footnote \ref{cor5foot},
the eigenvalue $\Delta (
\overline{\tiny\yng(1,1)};0)$
can be determined from (\ref{Deltaantisymm0}) with the extra minus sign, 
and the eigenvalue $ \Delta ( 0;\tiny\yng(2))$
can be read off from (\ref{Delta0symm}).
}.

%%%%%%%%%%%%%%%%%%%%%%%%%%%%%%%%%%%%%%%%%%%%%%
\subsection{The $(\mbox{antisymm};\overline{\mbox{symm}})$ representation}
%%%%%%%%%%%%%%%%%%%%%%%%%%%%%%%%%%%%%%%%%%%%%%

The two eigenvalues are given by
{\small
  \bea
   {\bf h} (
  \tiny\yng(1,1);\overline{\tiny\yng(2)}) 
&= & \frac{(k N^2+2 k N+N^3+4 N^2+4 N+4)}{N (N+2) (k+N+2)},
  \nonu \\
{\bf \phi_2^{(1)}} (
  \tiny\yng(1,1);\overline{\tiny\yng(2)}) 
&= &  \phi_2^{(1)} (
  \tiny\yng(1,1);\overline{\tiny\yng(2)}) +
\Delta(\tiny\yng(1,1);0) + \Delta(0;\overline{\tiny\yng(2)}) +
  \frac{16}{9} \Bigg[
\frac{18 i \sqrt{\frac{N}{N+2}}}{(k+N+2)}
\nonu \\
&+ & 
\frac{9 }{(N+2) (k+N+2)^2 (3 k N+4 k+4 N+5) (6 k N+5 k+5 N+4)} 
\nonu \\
& \times & (k-N)
({\bf 48 k^2 N^2}+137 k^2 N+96 k^2 {\bf -24 k N^3}-50 k N^2+20 k N
\nonu \\
& + & 60 k-31 N^3-124 N^2-148 N-48)
\Bigg].
\label{antisymmsymmbarlinear}
  \eea
  }
The highest power terms of $(N,k)$ in the last term of
the eigenvalue of
higher spin $3$  current (\ref{antisymmsymmbarlinear})
are exactly the same as the ones in (\ref{symmsymmbarlinearresult}).
Note that the eigenvalue
$ \phi_2^{(1)} (
\tiny\yng(1,1);\overline{\tiny\yng(2)})$ in (\ref{antisymmsymmbarlinear})
is given by
(\ref{antisymmsymmbarnonlinear}).
It is easy to check that the above
last two terms of (\ref{antisymmsymmbarlinear})
can be determined by using the following eigenvalues
{\small
\bea
&& \Bigg[-\frac{2 (2 k N^2+3 k N+3 N^2+12 N+12)}{(N+2) (k+N+2)^2} ,
  0,\frac{2 i \sqrt{\frac{N}{N+2}} (k+3 N+6)}{(k+N+2)} ,
  -\frac{2 i \sqrt{\frac{N}{N+2}} (3 k+N+2)}{(k+N+2)},
  \nonu \\
  && \frac{2 (-7 k N-12 k+5 N^2+16 N+12)}{(N+2) (k+N+2)} \Bigg],
\label{fivres2}
\eea}
together with the coefficients.
The last eigenvalue of (\ref{fivres2})
is the same as the one in (\ref{symmsymmbarEigen})
\footnote{One has following eigenvalues 
  \bea
{\bf \phi_2^{(1)}} (
\overline{\tiny\yng(1,1)};\overline{\tiny\yng(2)})
& = & \phi_2^{(1)} (
\overline{\tiny\yng(1,1)};\overline{\tiny\yng(2)}) + \Delta (
\overline{\tiny\yng(1,1)};0) +
 \Delta ( 0;\overline{\tiny\yng(2)})
\nonu
\eea
plus the contributions of the following eigenvalues 
\bea
&& \Bigg[
  \frac{2 (2 k N^2+3 k N+3 N^2+12 N+12)}{(N+2) (k+N+2)^2},
  0,
  -\frac{2 i \sqrt{\frac{N}{N+2}} (k+3 N+6)}{(k+N+2)},
  \frac{2 i \sqrt{\frac{N}{N+2}} (3 k+N+2)}{(k+N+2)},
  \nonu \\
  &&
-\frac{2 (-7 k N-12 k+5 N^2+16 N+12)}{(N+2) (k+N+2)}
  \Bigg].
\nonu
\eea
The eigenvalue $
\phi_2^{(1)} (
\overline{\tiny\yng(1,1)};\overline{\tiny\yng(2)})$
is given by the footnote \ref{cor6foot},
the eigenvalue $\Delta (
\overline{\tiny\yng(1,1)};0)$ is given by  (\ref{Deltaantisymm0})
with minus sign,
and the eigenvalue $
\Delta ( 0;\overline{\tiny\yng(2)})$ is given by
(\ref{Delta0symm}) with an extra minus sign.
}.

%%%%%%%%%%%%%%%%%%%%%%%%%%%%%%%%%%%%%%%%%%%%%%%%%%%%%%%%%%%%%%%%%%%%%%%%%%%%
\subsection{The $(\mbox{antisymm}; \overline{\mbox{antisymm}})$
  representation}
%%%%%%%%%%%%%%%%%%%%%%%%%%%%%%%%%%%%%%%%%%%%%%%%%%%%%%%%%%%%%%%%%%%%%%%%%%

The two eigenvalues are given by
{\small
  \bea
     {\bf h} (
  \tiny\yng(1,1);\overline{\tiny\yng(1,1)}) 
  &= & \frac{(k N^2+2 k N+N^3+6 N^2+8 N+4)}{N (N+2) (k+N+2)},
  \nonu \\
  {\bf \phi_2^{(1)}} (
  \tiny\yng(1,1);\overline{\tiny\yng(1,1)}) 
&= &  \phi_2^{(1)} (
  \tiny\yng(1,1);\overline{\tiny\yng(1,1)}) +
  \Delta(\tiny\yng(1,1);0) + \Delta(0;\overline{\tiny\yng(1,1)}) +
 \frac{16}{9}  \Bigg[ \frac{18 i \sqrt{\frac{N}{N+2}}}{(k+N+2)}
  \nonu \\
&+ & 
  \frac{9}{(N+2) (k+N+2)^2 (3 k N+4 k+4 N+5) (6 k N+5 k+5 N+4)}
  \nonu \\
  & \times & (k-N)
  ({\bf 48 k^2 N^2}+137 k^2 N+96 k^2 {\bf -24 k N^3}-32 k N^2+56 k N
  \nonu \\
  & + & 60 k-13 N^3-52 N^2-76 N-48)
  %(6 k^3 N^2+9 k^3 N+12 k^2 N^3+111 k^2 N^2+236 k^2 N+
  %132 k^2+6 k N^4+57 k N^3 \nonu \\
  %& + & 232 k N^2+362 k N+168 k+27 N^4+140 N^3+254 N^2+176 N+24)
  \Bigg].
\label{antisymmantisymmbarlinear}
  \eea
}
Note that the eigenvalue
$ \phi_2^{(1)} (
\tiny\yng(1,1);\overline{\tiny\yng(1,1)})$ is given by
(\ref{antiantibarnonlinear}).
One sees that the eigenvalue for the spin $2$ current
can be interpreted as the one in the nonlinear version
\bea
&&   h (
   \tiny\yng(1,1);\overline{\tiny\yng(1,1)})
   + \frac{1}{(N+k+2)} \frac{1}{N(N+2)} \hat{u}^2
   (\tiny\yng(1,1);\overline{\tiny\yng(1,1)})
 =  \frac{(k+N)}{(k+N+2)} +
 \frac{(-2N-2)^2}{N (N+2) (k+N+2)}
  \nonu \\
 && = {\bf h} (
   0;\overline{\tiny\yng(2)})
+  {\bf h} (
   \tiny\yng(1,1);0)
    =  \frac{(k N+ 3N+2)}{ N (k+N+2)} +
\frac{N (N+3)}{ (N+2) (k+N+2)}.
   \nonu
\eea
In the first line, the conformal dimension formula
and the additive property of $\hat{u}$ charge
can be used and in the second line, the relations
(\ref{0symmeigenlinear}) and
(\ref{antisymmzerolinearres})
can be used.

The eigenvalue $\Delta(\overline{\tiny\yng(1,1)};0)$ is given by
(\ref{Deltaantisymm0}) with minus sign and 
the eigenvalue
$ \Delta(0;\overline{\tiny\yng(1,1)}) $
can be obtained from
$ \Delta(0;\tiny\yng(1,1)) $ in (\ref{Delta0antisymm})
by changing the sign.  

Moreover, the last two terms in (\ref{antisymmantisymmbarlinear})
can be obtained from the following eigenvalues
{\small
\bea
&& \Bigg[
  -\frac{2 (2 k N^2+3 k N+9 N^2+24 N+12)}{(N+2) (k+N+2)^2},0,
  \frac{2 i \sqrt{\frac{N}{N+2}} (k+3 N+6)}{(k+N+2)},
  -\frac{2 i \sqrt{\frac{N}{N+2}} (3 k+N+2)}{(k+N+2)},
  \nonu \\
  &&
\frac{2 (-7 k N-12 k+5 N^2+16 N+12)}{(N+2) (k+N+2)}
  \Bigg],
\label{fivres3}
\eea}
together with the coefficients.
The last eigenvalue of (\ref{fivres3})
is the same as the one in (\ref{symmsymmbarEigen}).
Note that
the highest power terms of $(N,k)$ in the last term of
the eigenvalue of
higher spin $3$  current (\ref{antisymmantisymmbarlinear})
are exactly the same as the ones in (\ref{symmsymmbarlinearresult})
\footnote{For similar higher representation, one has
  \bea
{\bf \phi_2^{(1)}} (
\overline{\tiny\yng(1,1)};\tiny\yng(1,1))
& = & \phi_2^{(1)} (
\overline{\tiny\yng(1,1)};\tiny\yng(1,1)) + \Delta (
\overline{\tiny\yng(1,1)};0) +
 \Delta ( 0;\tiny\yng(1,1))
\nonu
\eea
plus the contributions from the following eigenvalues 
\bea
&& \Bigg[
  \frac{2 \left(4 k N^2+9 k N+9 N^2+24 N+12\right)}{(N+2) (k+N+2)^2},
  0,
  \frac{2 i \sqrt{\frac{N}{N+2}} (k+3 N+6)}{(k+N+2)},
  -\frac{2 i \sqrt{\frac{N}{N+2}} (3 k+N+2)}{(k+N+2)},
  \nonu \\
  &&
-\frac{2 \left(-5 k N-12 k+N^2+8 N+12\right)}{(N+2) (k+N+2)}
  \Bigg].
\nonu
\eea
The eigenvalue 
$\phi_2^{(1)} (
\overline{\tiny\yng(1,1)};\tiny\yng(1,1))$ can be obtained
from the footnote \ref{footlast},
the eigenvalue $\Delta (
\overline{\tiny\yng(1,1)};0)$ is given by  (\ref{Deltaantisymm0})
with minus sign,
and the eigenvalue $
\Delta ( 0;\tiny\yng(2))$ is given by
(\ref{Delta0symm}).
}.

In summary of this section,
the eigenvalues for the spin $2$ current and
the eigenvalues for the higher spin $3$ current are obtained
explicitly at finite $(N,k)$. It turns out that
they reproduce the ones in the nonlinear version described in section
$5$ under the large $(N,k)$ 't Hooft like limit. 

%%%%%%%%%%%%%%%%%%%%%%%%%%%%%%%%%%%%%%%%%%%%%%%%%%%%%%%%%%%%%%%%%%%%%%%%%
%%%%%%%%%%%%%%%%%%%%%%%%%%%%%%%%%%%%%%%%%%%%%%%%%%%%%%%%%%%%%%%%%%%%%%%%%%
\section{Conclusions and outlook }
%%section9%%%%%%%%%%%%%%%%%%%%%%%%%%%%%%%%%%%%%%%%%%%%%%%%%%%%%%%%%%%%%%%%%%%%%%%%%%%%%
%%%%%%%%%%%%%%%%%%%%%%%%%%%%%%%%%%%%%%%%%%%%%%%%%%%%%%%%%%%%%%%%%%%%%%%%%%%%%%%%

We present the first two Tables where
one can find the conformal dimensions for the spin $2$ current
acting on the various representations in the coset under the
large $(N,k)$ 't Hooft like limit (up to three boxes).
The other quantum numbers
$l^{\pm}$ and $\hat{u}$ can be made in Tables but
we did not do it.
We have found other quantum numbers associated with the higher spin
currents on some part of representations in the above Tables.
The explicit large $(N,k)$ behaviors of those eigenvalues
for the higher spin $3$ currents
are summarized in the last two Tables.
One realizes that although the conformal dimensions are equal to each
other for different two representations, the eigenvalues for the
higher spin $3$ current for those two
representations are different from each other under the large $(N,k)$ 't Hooft
like limit. 
For example, for the representations
$(\tiny\yng(1);\overline{\tiny\yng(1)})$ and
$(\overline{\tiny\yng(1)};\tiny\yng(1))$, the eigenvalues
$\phi_2^{(1)}$ are different but their conformal dimensions $h$
are the same under
the large $(N,k)$ 't Hooft
like limit. For the eigenvalue $h$, the corresponding conformal
dimension for the spin $2$ current is $2$ while
for the eigenvalue $\phi_2^{(1)}$, the corresponding
conformal dimension for the higher spin $3$ current is $3$.
Under the complex conjugation, the former remains the same (even spin)
but the latter changes the sign (odd spin). 

Let us present some related and open problems  
in the near future.

%%%%%%%%%%%%%%%%%%%%%%%%%%%%%%%%%%%%%%%%%%%%%%%%%%%%%%%%
$\bullet$ The other representations
%%%%%%%%%%%%%%%%%%%%%%%%%%%%%%%%%%%%%%%%%%%%%%%%%%%%%%%%

One can calculate the cases where
the higher representations have $\La_{+}=0$ and
$\La_{-} = \tiny\yng(1,1,1), \tiny\yng(2,1), \overline{\tiny\yng(1,1,1)}$
and $\overline{\tiny\yng(2,1)}$.
It is straightforward to calculate them because
one has the known higher spin currents for several $N$ values
and the OPEs between these currents and the product of
adjoint spin $\frac{1}{2}$ fermions can be obtained as before.

%%%%%%%%%%%%%%%%%%%%%%%%%%%%%%%%%%%%%%%%%%%%%%%%%%%%%%%%
$1)$ The $(0; \mbox{antisymm})$ representation with three boxes
%%%%%%%%%%%%%%%%%%%%%%%%%%%%%%%%%%%%%%%%%%%%%%%%%%%%%%%%

The relevant subsection is given by $2.2.7$. 
The four eigenvalues are given by 
\bea
\phi_0^{(1)} (0;\tiny\yng(1,1,1)) & = & -\frac{3k}{(N+k+2)},
\nonu \\
v^+ (0;\tiny\yng(1,1,1)) & = & \frac{216 k}{(k+N+2)^2},
\nonu \\
v^- (0;\tiny\yng(1,1,1)) & = & \frac{12 k (17 k+2 N+2)}{(k+N+2)^2},
\nonu \\
\phi_2^{(1)} (0;\tiny\yng(1,1,1)) & = &
\frac{4 k ({\bf 12 k^2 N}-8 k^2 {\bf +6 k N^2}+51 k N-22 k+5 N^2+64 N+12)}{
  (k+N+2)^2 (6 k N+5 k+5 N+4)}.
\label{zeroantisymmthree}
\eea
One can see that the first eigenvalue is {\bf three} times of the one in
(\ref{0feigenhigher}) even finite $(N,k)$.
The last eigenvalue is also {\bf three}
times of the one in (\ref{0feigenhigher})
under the large $(N,k)$ 't Hooft-like limit.
See also (\ref{zeroantisymmhigher}).

%%%%%%%%%%%%%%%%%%%%%%%%%%%%%%%%%%%%%%%%%%%%%%%%%%%%%%%%
$2)$ The $(0; \overline{\mbox{antisymm}})$ representation  with three boxes
%%%%%%%%%%%%%%%%%%%%%%%%%%%%%%%%%%%%%%%%%%%%%%%%%%%%%%%%

The four eigenvalues are
\bea
\phi_0^{(1)} (0;\overline{\tiny\yng(1,1,1)}) & = & \frac{3k}{(N+k+2)},
\nonu \\
v^+ (0;\overline{\tiny\yng(1,1,1)}) & = & \frac{216 k}{(k+N+2)^2},
\nonu \\
v^- (0;\overline{\tiny\yng(1,1,1)}) & = & \frac{12 k (17 k+2 N+2)}{(k+N+2)^2},
\nonu \\
\phi_2^{(1)} (0;\overline{\tiny\yng(1,1,1)}) & = &
-
\frac{4 k ({\bf 12 k^2 N}-8 k^2 {\bf +6 k N^2}
  +51 k N-22 k+5 N^2+64 N+12)}{
  (k+N+2)^2 (6 k N+5 k+5 N+4)}.
\label{zeroantisymmbarthree}
\eea
The first eigenvalue is {\bf three} times of the one in
(\ref{zerofbareigenhigher}) even finite $(N,k)$.
The last eigenvalue also is {\bf three}
times of the one in (\ref{zerofbareigenhigher})
under the large $(N,k)$ 't Hooft-like limit.
See also the footnote \ref{zeroantisymmbarexpression}.

%%%%%%%%%%%%%%%%%%%%%%%%%%%%%%%%%%%%%%%%%%%%%%%%%%%%%%%%
$3)$ The $(0; \mbox{mixed})$ representation  with three boxes
%%%%%%%%%%%%%%%%%%%%%%%%%%%%%%%%%%%%%%%%%%%%%%%%%%%%%%%%

The relevant subsection is given by $2.2.8$.
The four eigenvalues are described by 
\bea
\phi_0^{(1)} (0;\tiny\yng(2,1)) & = & -\frac{3k}{(N+k+2)},
\nonu \\
v^+ (0;\tiny\yng(2,1)) & = & \frac{216 k}{(k+N+2)^2},
\nonu \\
v^- (0;\tiny\yng(2,1)) & = & \frac{12 k (13 k+12 N-14)}{(k+N+2)^2},
\nonu \\
\phi_2^{(1)} (0;\tiny\yng(2,1)) & = &
\frac{4 k ({\bf 12 k^2 N}-8 k^2 {\bf +6 k N^2}+15 k N-16 k+5 N^2-2 N-12)}{
  (k+N+2)^2 (6 k N+5 k+5 N+4)}.
\nonu
\eea
The first two eigenvalues are the same as
the ones in (\ref{zeroantisymmthree}) while the last eigenvalue
is the same as the one in (\ref{zeroantisymmthree}) under the large
$(N,k)$ 't Hooft-like limit.

%%%%%%%%%%%%%%%%%%%%%%%%%%%%%%%%%%%%%%%%%%%%%%%%%%%%%%%%
$4)$ The $(0; \overline{\mbox{mixed}})$ representation  with three boxes
%%%%%%%%%%%%%%%%%%%%%%%%%%%%%%%%%%%%%%%%%%%%%%%%%%%%%%%%

The four eigenvalues are summarized by
\bea
\phi_0^{(1)} (0;\overline{\tiny\yng(2,1)}) & = & \frac{3k}{(N+k+2)},
\nonu \\
v^+ (0;\overline{\tiny\yng(2,1)}) & = & \frac{216 k}{(k+N+2)^2},
\nonu \\
v^- (0;\overline{\tiny\yng(2,1)}) & = & \frac{12 k (13 k+12 N-14)}{(k+N+2)^2},
\nonu \\
\phi_2^{(1)} (0;\overline{\tiny\yng(2,1)}) & = &
-
\frac{4 k ( {\bf 12 k^2 N}-8 k^2 {\bf +6 k N^2}+15 k N-16 k+5 N^2-2 N-12)}{
  (k+N+2)^2 (6 k N+5 k+5 N+4)}.
\nonu
\eea
The first two eigenvalues are the same as
the ones in (\ref{zeroantisymmbarthree}) while the last eigenvalue
is the same as the one in (\ref{zeroantisymmbarthree}) under the large
$(N,k)$ 't Hooft-like limit.

Then one considers the following higher representations
\bea
\La_{+} = \tiny\yng(1), \overline{\tiny\yng(1)}, \tiny\yng(2), \tiny\yng(1,1),
\overline{\tiny\yng(2)}, \overline{\tiny\yng(1,1)},
\qquad
\La_{-} = \tiny\yng(1,1,1), \overline{\tiny\yng(1,1,1)},
\tiny\yng(2,1), \overline{\tiny\yng(2,1)},
\nonu
\eea
which appears in the product of $(\La_{+};0)$ and $(0;\La_{-})$
with $\La_{+} \neq \La_{-}$.
One expects that the eigenvalues are summarized in the following
two Tables.

%%%%%%%%%%%%%%%%%%%%%%%%%%%%%%%%%%%%%%%%%%%%%%%%%%%%%%%%%%%%%%%%%%%
\begin{table}[ht]
\centering % used for centering table
{\small
\begin{tabular}{|c||c|c|c|c| } % centered columns (4 columns)
  \hline %inserts double horizontal lines
  %$(\La_+;\La_-)$
  &
$ \tiny\yng(1,1,1)$ &$\tiny\yng(2,1)$ & $\overline{ \tiny\yng(1,1,1)}$
& $\overline{\tiny\yng(2,1)}$
\\ [0.5ex] % inserts table
%heading
\hline \hline % inserts single horizontal line
$0$   & $3 \phi_0^{(1)}(0;\tiny\yng(1))$ &
$3 \phi_0^{(1)}(0;\tiny\yng(1))$ & $-3 \phi_0^{(1)}(0;\tiny\yng(1))$  &
 $-3 \phi_0^{(1)}(0;\tiny\yng(1))$
%\\ % inserting body of the table
%& $=\frac{8}{3} (1-\lambda ) (2-\lambda )$ &$=\frac{8}{3} (1-\lambda ) (2-\lambd%a )$ &$=-\frac{8}{3} (1-\lambda ) (2-\lambda )$ & $=-\frac{8}{3} (1-\lambda ) (2%-\lambda )$
\\
\hline
$\tiny\yng(1)$  &$ \phi_0^{(1)}[(\tiny\yng(1);0) +
3 (0;\tiny\yng(1))]$ & $ \phi_0^{(1)}[(\tiny\yng(1);0) +
3 (0;\tiny\yng(1))]$ &$ \phi_0^{(1)}[(\tiny\yng(1);0) -
3 (0;\tiny\yng(1))]$ & $ \phi_0^{(1)}[(\tiny\yng(1);0) -
3 (0;\tiny\yng(1))]$  \\
%& $=\frac{4}{3} (\la^2-7\la +4)$ & $=\frac{4}{3} (\la^2-7\la +4)$ &
%& $=-\frac{4}{3} (3\la^2-5\la +4)$ &  $=-\frac{4}{3} (3\la^2-5\la +4)$ \\
\hline
$\overline{\tiny\yng(1)}$    &$ \phi_0^{(1)}[-(\tiny\yng(1);0) +
3 (0;\tiny\yng(1))]$ & $ \phi_0^{(1)}[-(\tiny\yng(1);0) +
  3 (0;\tiny\yng(1))]$ &$ \phi_0^{(1)}[-(\tiny\yng(1);0) -
3 (0;\tiny\yng(1))]$ & $ \phi_0^{(1)}[-(\tiny\yng(1);0) -
3 (0;\tiny\yng(1))]$ \\
%& $=\frac{4}{3} (3\la^2-5\la +4)$ &  $=\frac{4}{3} (3\la^2-5\la +4)$  &
%$=-\frac{4}{3} (\la^2-7\la +4)$& $=-\frac{4}{3} (\la^2-7\la +4)$ \\
\hline
$ \tiny\yng(2)$
& $ \phi_0^{(1)}[2(\tiny\yng(1);0) +
3(0;\tiny\yng(1))] $ &
$ \phi_0^{(1)}[2(\tiny\yng(1);0) +
3(0;\tiny\yng(1))]$ &
$ \phi_0^{(1)}[2(\tiny\yng(1);0) -
 3 (0;\tiny\yng(1) )]$ & $ \phi_0^{(1)}[2(\tiny\yng(1);0) -
  3(0;\tiny\yng(1) )]$ \\
% &$=-\frac{16}{3 N} \lambda  (2 \lambda -1)$ & $=-\frac{16}{3}  (2 \lambda -1)$ %& $=-\frac{16}{3}(\lambda ^2-\lambda +1)$ &  $=-\frac{16}{3}(\lambda ^2-\lambda %+1)$\\
\hline
$\tiny\yng(1,1)$ & $ \phi_0^{(1)}[2(\tiny\yng(1);0) +
3(0;\tiny\yng(1))]$ &
$\phi_0^{(1)}[2(\tiny\yng(1);0) +
3(0;\tiny\yng(1))] $ &
 $
 \phi_0^{(1)}[2(\tiny\yng(1);0) -3(0;\tiny\yng(1) )] $
&   $
 \phi_0^{(1)}[2(\tiny\yng(1);0) -3(0;\tiny\yng(1) )]
$  \\
%& $=-\frac{16}{3}  (2 \lambda -1)$ & $=-\frac{16}{3 N} \lambda  (2 \lambda -1)$ %&$=-\frac{16}{3}(\lambda ^2-\lambda +1)$ & $=-\frac{16}{3}(\lambda ^2-\lambda +1%)$ \\
\hline
$\overline{\tiny\yng(2)}$  &
$ \phi_0^{(1)}[-2 (\tiny\yng(1);0) +3
  (0;\tiny\yng(1) )]$ &$ \phi_0^{(1)}[-2 (\tiny\yng(1);0) +3
  (0;\tiny\yng(1) )]$ &$\phi_0^{(1)}[-2(\tiny\yng(1);0) -3
(0;\tiny\yng(1))]$ &
$ \phi_0^{(1)}[-2(\tiny\yng(1);0) -3
(0;\tiny\yng(1))]$ \\ 
%&$=\frac{16}{3}(\lambda ^2-\lambda +1)$ &$=\frac{16}{3}(\lambda ^2-\lambda +1)$ %&$=\frac{16}{3 N} \lambda  (2 \lambda -1)$ &
%$=\frac{16}{3}  (2 \lambda -1)$
%\\
\hline
$\overline{\tiny\yng(1,1)}$  &$ \phi_0^{(1)}[-2(\tiny\yng(1);0) +3
  (0;\tiny\yng(1) )]$ &$ \phi_0^{(1)}[-2(\tiny\yng(1);0) +3
  (0;\tiny\yng(1) )]$ &$ \phi_0^{(1)}[-2(\tiny\yng(1);0) -3
(0;\tiny\yng(1))]$ & $ \phi_0^{(1)}[-2(\tiny\yng(1);0) -3
(0;\tiny\yng(1))]$  \\ 
%&$=\frac{16}{3}(\lambda ^2-\lambda +1)$ &$=\frac{16}{3}(\lambda ^2-\lambda +1)$ %&$=\frac{16}{3}  (2 \lambda -1)$ &$=\frac{16}{3 N} \lambda  (2 \lambda -1)$
%\\
[1ex] % [1ex] adds vertical space
\hline %inserts single line
\end{tabular}}
%\label{tableone} % is used to refer this table in the text
\caption{Expectation of the eigenvalue $\phi_0^{(1)}$
at finite  $(N,k)$.
Due the  fact that this higher spin current
has conformal dimension $1$,
there are no contributions from the commutator between the
zero mode of this higher spin current and the corresponding mode
$Q_{-\frac{1}{2}}^{13} Q_{-\frac{1}{2}}^{14} Q_{-\frac{1}{2}}^{15}$,
$Q_{-\frac{1}{2}}^{14} Q_{-\frac{1}{2}}^{16} Q_{-\frac{1}{2}}^{13}$
(and its complex conjugated ones) from the subsections $2.2.7$ and
$2.2.8$.
 The eigenvalues are  given by the linear combinations
  of the one of $(0;f)$ (or $(0;\overline{f})$)
  and the one of $(f;0)$ (or $(\overline{f};0)$).
  Then each coefficient depends on the  number of boxes in $\La_{+}$
  and $\La_{-}$.
  %Also one has plus sign for the fundamental representation
  %while minus sign for the complex conjugated (anti fundamental)
  %representation.
  The first row can be seen from the previous description.
  In order to express these eigenvalues in terms of $(N,k)$ dependence
  explicitly, the relations (\ref{0feigenhigher}) and
  (\ref{f0eigenhigher}) are needed.
} % title of Table
\end{table}
%%%%%%%%%%%%%%%%%%%%%%%%%%%%%%%%%%%%%%%%%%%%%%%%%%%%%%%%%%%%%%%%%%%%%%

%%%%%%%%%%%%%%%%%%%%%%%%%%%%%%%%%%%%%%%%%%%%%%%%%%%%%%%%%%%%%%%%%%%
\begin{table}[ht]
\centering % used for centering table
{\small
\begin{tabular}{|c||c|c|c|c| } % centered columns (4 columns)
  \hline %inserts double horizontal lines
  %$(\La_+;\La_-)$
  &
$ \tiny\yng(1,1,1)$ &$\tiny\yng(2,1)$ & $\overline{ \tiny\yng(1,1,1)}$
& $\overline{\tiny\yng(2,1)}$
\\ [0.5ex] % inserts table
%heading
\hline \hline % inserts single horizontal line
$0$   & $3 \phi_2^{(1)}(0;\tiny\yng(1))$ &
$3 \phi_2^{(1)}(0;\tiny\yng(1))$ & $-3 \phi_2^{(1)}(0;\tiny\yng(1))$  &
 $-3 \phi_2^{(1)}(0;\tiny\yng(1))$
%\\ % inserting body of the table
%& $=\frac{8}{3} (1-\lambda ) (2-\lambda )$ &$=\frac{8}{3} (1-\lambda ) (2-\lambd%a )$ &$=-\frac{8}{3} (1-\lambda ) (2-\lambda )$ & $=-\frac{8}{3} (1-\lambda ) (2%-\lambda )$
\\
\hline
$\tiny\yng(1)$  &$ \phi_2^{(1)}[(\tiny\yng(1);0) +
3 (0;\tiny\yng(1))]$ & $ \phi_2^{(1)}[(\tiny\yng(1);0) +
3 (0;\tiny\yng(1))]$ &$ \phi_2^{(1)}[(\tiny\yng(1);0) -
3 (0;\tiny\yng(1))]$ & $ \phi_2^{(1)}[(\tiny\yng(1);0) -
3 (0;\tiny\yng(1))]$  \\
%& $=\frac{4}{3} (\la^2-7\la +4)$ & $=\frac{4}{3} (\la^2-7\la +4)$ &
%& $=-\frac{4}{3} (3\la^2-5\la +4)$ &  $=-\frac{4}{3} (3\la^2-5\la +4)$ \\
\hline
$\overline{\tiny\yng(1)}$    &$ \phi_2^{(1)}[-(\tiny\yng(1);0) +
3 (0;\tiny\yng(1))]$ & $ \phi_2^{(1)}[-(\tiny\yng(1);0) +
  3 (0;\tiny\yng(1))]$ &$ \phi_2^{(1)}[-(\tiny\yng(1);0) -
3 (0;\tiny\yng(1))]$ & $ \phi_2^{(1)}[-(\tiny\yng(1);0) -
3 (0;\tiny\yng(1))]$ \\
%& $=\frac{4}{3} (3\la^2-5\la +4)$ &  $=\frac{4}{3} (3\la^2-5\la +4)$  &
%$=-\frac{4}{3} (\la^2-7\la +4)$& $=-\frac{4}{3} (\la^2-7\la +4)$ \\
\hline
$ \tiny\yng(2)$
& $ \phi_2^{(1)}[2(\tiny\yng(1);0) +
3(0;\tiny\yng(1))] $ &
$ \phi_2^{(1)}[2(\tiny\yng(1);0) +
3(0;\tiny\yng(1))]$ &
$ \phi_2^{(1)}[2(\tiny\yng(1);0) -
 3 (0;\tiny\yng(1) )]$ & $ \phi_2^{(1)}[2(\tiny\yng(1);0) -
  3(0;\tiny\yng(1) )]$ \\
% &$=-\frac{16}{3 N} \lambda  (2 \lambda -1)$ & $=-\frac{16}{3}  (2 \lambda -1)$ %& $=-\frac{16}{3}(\lambda ^2-\lambda +1)$ &  $=-\frac{16}{3}(\lambda ^2-\lambda %+1)$\\
\hline
$\tiny\yng(1,1)$ & $ \phi_2^{(1)}[2(\tiny\yng(1);0) +
3(0;\tiny\yng(1))]$ &
$\phi_2^{(1)}[2(\tiny\yng(1);0) +
3(0;\tiny\yng(1))] $ &
 $
 \phi_2^{(1)}[2(\tiny\yng(1);0) -3(0;\tiny\yng(1) )] $
&   $
 \phi_2^{(1)}[2(\tiny\yng(1);0) -3(0;\tiny\yng(1) )]
$  \\
%& $=-\frac{16}{3}  (2 \lambda -1)$ & $=-\frac{16}{3 N} \lambda  (2 \lambda -1)$ %&$=-\frac{16}{3}(\lambda ^2-\lambda +1)$ & $=-\frac{16}{3}(\lambda ^2-\lambda +1%)$ \\
\hline
$\overline{\tiny\yng(2)}$  &
$ \phi_2^{(1)}[-2 (\tiny\yng(1);0) +3
  (0;\tiny\yng(1) )]$ &$ \phi_2^{(1)}[-2 (\tiny\yng(1);0) +3
  (0;\tiny\yng(1) )]$ &$\phi_2^{(1)}[-2(\tiny\yng(1);0) -3
(0;\tiny\yng(1))]$ &
$ \phi_2^{(1)}[-2(\tiny\yng(1);0) -3
(0;\tiny\yng(1))]$ \\ 
%&$=\frac{16}{3}(\lambda ^2-\lambda +1)$ &$=\frac{16}{3}(\lambda ^2-\lambda +1)$ %&$=\frac{16}{3 N} \lambda  (2 \lambda -1)$ &
%$=\frac{16}{3}  (2 \lambda -1)$
%\\
\hline
$\overline{\tiny\yng(1,1)}$  &$ \phi_2^{(1)}[-2(\tiny\yng(1);0) +3
  (0;\tiny\yng(1) )]$ &$ \phi_2^{(1)}[-2(\tiny\yng(1);0) +3
  (0;\tiny\yng(1) )]$ &$ \phi_2^{(1)}[-2(\tiny\yng(1);0) -3
(0;\tiny\yng(1))]$ & $ \phi_2^{(1)}[-2(\tiny\yng(1);0) -3
(0;\tiny\yng(1))]$  \\ 
%&$=\frac{16}{3}(\lambda ^2-\lambda +1)$ &$=\frac{16}{3}(\lambda ^2-\lambda +1)$ %&$=\frac{16}{3}  (2 \lambda -1)$ &$=\frac{16}{3 N} \lambda  (2 \lambda -1)$
%\\
[1ex] % [1ex] adds vertical space
\hline %inserts single line
\end{tabular}}
%\label{tableone} % is used to refer this table in the text
\caption{Expectation of the eigenvalue $\phi_2^{(1)}$
  under the large $(N,k)$ 't Hooft-like limit (\ref{largenk}).
  The eigenvalues are  given by the linear combinations
  of the one of $(0;f)$ (or $(0;\overline{f})$)
  and the one of $(f;0)$ (or $(\overline{f};0)$).
  Then each coefficient depends on the the number of boxes in $\La_{+}$
  and $\La_{-}$.
  %Also one has plus sign for the fundamental representation
  %while minus sign for the complex conjugated (anti fundamental)
  %representation.
  The first row can be seen from the previous description.
  In order to express these eigenvalues in terms of $\la$ dependence
  explicitly, the relations (\ref{0feigenhigher}) and
  (\ref{f0eigenhigher}) with the large $(N,k)$ 't Hooft-like
  limit (\ref{largenk})
  are needed.
} % title of Table
\end{table}
%%%%%%%%%%%%%%%%%%%%%%%%%%%%%%%%%%%%%%%%%%%%%%%%%%%%%%%%%%%%%%%%%%%%%%

%%%%%%%%%%%%%%%%%%%%%%%%%%%%%%%%%%%%%%%%%%%%%%%%%%%%%%%%
$\bullet$ The three-point functions
%%%%%%%%%%%%%%%%%%%%%%%%%%%%%%%%%%%%%%%%%%%%%%%%%%%%%%%%

As mentioned in the abstract,
one can determine the three-point functions \cite{CY} of the higher
spin currents with two scalar operators at finite $(N,k)$.
From (\ref{symmzerohigher}),
\bea
&& <\overline{\cal O}_{(\overline{\tiny\yng(2)};0)}
    {\cal O}_{(\tiny\yng(2);0)} \Phi_2^{(1)} > \nonu \\
&& =  \Bigg[-\frac{8 N
  ({\bf 6 k^2 N}+5 k^2 {\bf +12 k N^2}+45 k N+43 k-2 N^2-N+12)}
{3 (k+N+2)^2 (6 k N+5 k+5 N+4)} \Bigg]
    <\overline{\cal O}_{(\overline{\tiny\yng(2)};0)}  {\cal O}_{(\tiny\yng(2);0)}>
    \nonu \\
    && \longrightarrow -\frac{8}{3} \la (\la+1)
     <\overline{\cal O}_{(\overline{\tiny\yng(2)};0)}  {\cal O}_{(\tiny\yng(2);0)}>.
\nonu
\eea
The large $(N,k)$ 't Hooft like limit (\ref{largenk}) in the final expression
is taken.
It is straightforward to write down all the three-point functions
we have found in this paper.

%%%%%%%%%%%%%%%%%%%%%%%%%%%%%%%%%%%%%%%%%%%%%%%%%%%%%%%%$$$$$$$$
$\bullet$ The higher spin $3$ current in terms of adjoint spin $1$ and
spin $\frac{1}{2}$ currents
%%%%%%%%%%%%%%%%%%%%%%%%%%%%%%%%%%%%%%%%%%%%%%%%%%%%%%%%%%%%%%%%%

In this paper, we used the higher spin currents
for several $N$ values which are written in terms of
adjoint spin $1$ and
spin $\frac{1}{2}$ currents.
In principle, one can find the explicit expression
for the higher spin $3$ current (by hand) as described in section $4$.
Although it will be rather complicated to obtain this form
because that all the calculations on the OPEs should be checked
step by step, it will be worthwhile to determine this full expression.
Once this will be found, then it will be an open problem to
obtain the corresponding eigenvalues associated with any representations.  

%%%%%%%%%%%%%%%%%%%%%%%%%%%%%%%%%%%%%%%%%%%%%%%%%%%%%%%%
$\bullet$ The spectrum for the higher spin $4$ current 
%%%%%%%%%%%%%%%%%%%%%%%%%%%%%%%%%%%%%%%%%%%%%%%%%%%%%%%%

What happens for the higher spin current with different spin?
For example, the next $16$ higher spin current contains
the higher spin $4$ current $\Phi_2^{(s=2)}(z)$.
One expects that the behavior of large $(N,k)$ 't Hooft-like limit
in the eigenvalues on this higher spin $4$ current looks similar to
the results of this paper.
 The general structure for the eigenvalue with $\La_{+} \neq \La_{-}$
  in the product of $(\La_{+};0)$ and $(0;\La_{-})$
  is given by the linear combinations
  of the one of $(0;\tiny\yng(1))$ (or $(0;\overline{\tiny\yng(1)})$)
  and the one of $(\tiny\yng(1);0)$ (or $(\overline{\tiny\yng(1)};0)$).
  Then each coefficient depends on the the number of boxes in $\La_{+}$
  and $\La_{-}$. Also one has plus sign for the fundamental representation
  while minus sign for the complex conjugated (anti fundamental)
  representation.
  The corresponding basic eigenvalues with an appropriate
  normalization are found in \cite{AKK1703} 
  \bea
\phi_2^{(2)} (\tiny\yng(1);0) = \frac{12}{5} \la (1+\la) (2+\la), \qquad
\phi_2^{(2)} (0;\tiny\yng(1)) = -\frac{12}{5} (1-\la)(2-\la)(3-\la).
\nonu
\eea 
Their complex conjugated ones
remain the same because this higher spin current has the conformal
spin $4$.
When the representation $\La_{-}$ appears in the branching of $\La_{+}$,
  one expects that
  the eigenvalue leads to the representation $(|\La_+|-|\La_{-}|;0)$
  where $|\La_{\pm}|$ denotes the number of boxes.
  The eigenvalue with $\La_{+} = \La_{-}$
  can be written in terms of the multiple of the eigenvalue of
  $(\tiny\yng(1);\tiny\yng(1))$
  or $(\overline{\tiny\yng(1)};\overline{\tiny\yng(1)})$. 
  It would be interesting to observe whether these behaviors
  occur.
  Similarly, it is an open problem to obtain the eigenvalues
  for the higher spin $2$ current $\Phi_0^{(s=2)}(z)$.
  
%%%%%%%%%%%%%%%%%%%%%%%%%%%%%%%%%%%%%%%%%%%%%%%%%%%%%%%%
$\bullet$ The three boxes in $\La_+$ 
%%%%%%%%%%%%%%%%%%%%%%%%%%%%%%%%%%%%%%%%%%%%%%%%%%%%%%%%

In this paper, the boxes for $\La_{+}$ in the eigenvalues of
the higher spin currents are limited to $2$.
One considers the case where $\La_{+}$ contains the three boxes:
$\tiny\yng(3)$,$\tiny\yng(2,1)$, and $\tiny\yng(1,1,1)$ (and its
conjugated ones).
At least, one needs to have the $SU(N+2)$ generators
with $N=3,5,7,9,11$ in these higher representations
in order to extract the eigenvalues.
It is known that the dimensions for the above higher representations
are given by $\frac{1}{6} (N+2)(N+3)(N+4)$, $\frac{1}{3}(N+2)(N+3)(N+1)$,
and $\frac{1}{6}(N+2)(N+1)N$ respectively.
In particular, for $N=11$, these become
$455$, $728$, and $286$. This implies that
the $84$ generators of $SU(13)$ should be written in terms of
$455\times 455$ matrices, $728\times 728$ matrices, and
$286\times 286$ matrices respectively.
It is rather difficult to obtain $84$ generators in
the mixed representation by using the methods in Appendix $D$.
It is an open problem to find out the systematic way to
read off the complete $84$ generators which are $728 \times 728$
matrices by using the general formula described in the subsection $3.5$.

%%%%%%%%%%%%%%%%%%%%%%%%%%%%%%%%%%%%%%%%%%%%%%%%%%%%%%%%
$\bullet$ The three-point functions from the decomposition
of the four-point functions of scalar operators with Virasoro
conformal blocks
%%%%%%%%%%%%%%%%%%%%%%%%%%%%%%%%%%%%%%%%%%%%%%%%%%%%%%%%

Recently \cite{HU},
using the decomposition of the scalar four-point functions
by Virasoro conformal blocks, the three-point functions including
$\frac{1}{N}$ corrections in the two dimensional (bosonic)
$W_N$ minimal model
were obtained using the result of
\cite{PR} (see also the works of \cite{CY1,CY2}).
The $\frac{1}{N}$ corrections for the conformal dimension
$6,7,8$ were new.
It would be interesting to obtain the
eigenvalues for the higher spin currents in the higher
representations of the $W_N$ minimal model. 
As observed in \cite{HU}, it is an open problem to
obtain the three-point functions from the decomposition of
four-point functions in the large ${
\cal N }=4$ holography.

%%%%%%%%%%%%%%%%%%%%%%%%%%%%%%%%%%%%%%%%%%%%%%%%%%%%%%%%
$\bullet$ The orthogonal Wolf space coset spectrum
%%%%%%%%%%%%%%%%%%%%%%%%%%%%%%%%%%%%%%%%%%%%%%%%%%%%%%%%

One can ask what happens for the orthogonal Wolf space coset spectrum.
The relevant previous works are given by \cite{AP1410,AKP1510}. It is
an open problem to obtain the eigenvalues for the
higher spin currents in the higher representations.
One should obtain the generators of $SO(N+4)$ in various higher
representations explicitly and obtain the higher spin currents
(where the spins are $2$, $3$ or $4$)
for several $N$ values.

\vspace{.7cm}

%%%%%%%%%%%%%%%%%%%%%%%%%%%%%%%%%%%%%%%%%%%%%%%%%%%%%%%%%%%%%%
%%%%%%%%%%%%%%%%%%%%%%%%%%%%%%%%%%%%%%%%%%%%%%%%%%%%%%%%%%%%%%%
\centerline{\bf Acknowledgments}
%%%%%%%%%%%%%%%%%%%%%%%%%%%%%%%%%%%%%%%%%%%%%%%%%%%%%%%%%%%%%%%
%%%%%%%%%%%%%%%%%%%%%%%%%%%%%%%%%%%%%%%%%%%%%%%%%%%%%%%%%%%%%%%

%CA acknowledges warm hospitality from 
%the School of  Liberal Arts (and Institute of Convergence Fundamental
%Studies), Seoul National University of Science and Technology.
This research was supported by Kyungpook National University Research Fund,
2017.
CA acknowledges warm hospitality from
C.N. Yang Institute for Theoretical Physics,
Stony Brook University.
CA would like to thank M.H. Kim for discussions and T. Kephart for pointing
out the reference \cite{Cvitanovic}. 
%and R. Gopakumar and H. Kim for discussions. 
%CA would like to thank his previous collaborators K. Schoutens and 
%A. Sevrin for earlier works.
%This research was supported by Basic Science Research Program through
%the National Research Foundation of Korea  
%funded by the Ministry of Education  
%(No. 2016R1D1A1B03931786).
%CA would like to thank the participants of the focus program of 
%Asia Pacific Center
%for Theoretical Physics (APCTP) 
%on
%``Liouville, Integrability and Branes (10) Focus Program at Asia-Pacific
%Center for Theoretical Physics'',
%Sept. 03-14, 2014 for their feedbacks.
%We thank the Galileo Galilei Institute for Theoretical Physics 
%for the hospitality and the INFN for partial support during 
%the completion of this work. 
%CA appreciates APCTP for its hospitality during completion of this work.

\newpage

\appendix

\renewcommand{\theequation}{\Alph{section}\mbox{.}\arabic{equation}}

%%%%%%%%%%%%%%%%%%%%%%%%%%%%%%%%%%%%%%%%%%%%%%%%%%%%%%%%%%%%%%%%%%%%%
%%%%%%%%%%%%%%%%%%%%%%%%%%%%%%%%%%%%%%%%%%%%%%%%%%%%%%%%%%%%%%%%%%%%%
\section{ The $24$ $SU(5)$ generators in the symmetric $\bf 15$
  representation of $SU(5)$ with two boxes }
%appendixA%%%%%%%%%%%%%%%%%%%%%%%%%%%%%%%%%%%%%%%%%%%%%%%%%%%%%%%%%%%%%%%%%%%
%%%%%%%%%%%%%%%%%%%%%%%%%%%%%%%%%%%%%%%%%%%%%%%%%%%%%%%%%%%%%%%%%%%%

In order to determine the eigenvalue equations for the zero modes
of (higher spin) currents in the higher representations,
one should obtain the $SU(N+2=5)$
generators in the higher representations.
The $24$ generators are given by $\frac{1}{2} (N+2)(N+3) \times
\frac{1}{2} (N+2)(N+3) = 15 \times 15 $ matrix
for the symmetric representation
$\tiny\yng(2)$.

Let us introduce the basis (of the five dimensional vector space $V$)
for the fundamental representation
${\bf 5}$ as follows \cite{Beveren}:
\bea
\hat{e}_1 & = & 
\left(
\begin{array}{c}
 1  \\
 0 \\
 0  \\
 0 \\
 0 \\
\end{array}
\right), \qquad
\hat{e}_2  =  
\left(
\begin{array}{c}
 0  \\
 1 \\
 0  \\
 0 \\
 0 \\
\end{array}
\right), \qquad
\hat{e}_3  =  
\left(
\begin{array}{c}
 0  \\
 0 \\
 1  \\
 0 \\
 0 \\
\end{array}
\right), \qquad
\hat{e}_4  =  
\left(
\begin{array}{c}
 0  \\
 0 \\
 0  \\
 1 \\
 0 \\
\end{array}
\right), \qquad
\hat{e}_5  =  
\left(
\begin{array}{c}
 0  \\
 0 \\
 0  \\
 0 \\
 1 \\
\end{array}
\right).
\nonu 
\eea
For given $24$ $SU(5)$ generators
which are $5 \times 5 $ matrices, one can calculate the following
matrix multiplications
\bea
T_1 \, \hat{e}_1 & = & \hat{e}_4, \qquad
T_2 \, \hat{e}_2 = \hat{e}_4, \qquad
T_3 \, \hat{e}_3 = \hat{e}_4, \qquad
T_4 \, \hat{e}_1 = \hat{e}_5, 
\nonu \\
T_5 \, \hat{e}_2  & = & \hat{e}_5, \qquad
T_6 \, \hat{e}_3 = \hat{e}_5, \qquad
T_7 \, \hat{e}_4 = \hat{e}_5, \qquad
T_8 \, \hat{e}_1 = \hat{e}_2, 
\nonu \\
T_9 \, \hat{e}_1  & = & \hat{e}_3, \qquad
T_{10} \, \hat{e}_2 = \hat{e}_3, \qquad
T_{11} \, \hat{e}_1 = (\frac{i}{2} + \frac{1}{2\sqrt{3}})\hat{e}_1, \qquad
T_{11} \, \hat{e}_2 =  (-\frac{i}{2} + \frac{1}{2\sqrt{3}}) \hat{e}_2, 
\nonu \\
T_{11} \, \hat{e}_3 & = & -\frac{1}{\sqrt{3}} \hat{e}_3
\qquad
T_{12} \, \hat{e}_1 = (\frac{i}{2\sqrt{6}} + \frac{1}{2\sqrt{10}})\hat{e}_1,
\qquad
T_{12} \, \hat{e}_2 =  (\frac{i}{2\sqrt{6}} + \frac{1}{2\sqrt{10}}) \hat{e}_2, 
\nonu \\
T_{12} \, \hat{e}_3  & = &
(\frac{i}{2\sqrt{6}} + \frac{1}{2\sqrt{10}})\hat{e}_3,
\qquad
T_{12} \, \hat{e}_4 =  (-\frac{i}{2} \sqrt{\frac{3}{2}} +
\frac{1}{2\sqrt{10}}) \hat{e}_4, 
\nonu \\
T_{12} \, \hat{e}_5 & = & -\sqrt{\frac{2}{5}} \hat{e}_5.
\label{fund5by5}
\eea
The remaining half of the generators can be obtained by taking
the transpose and the complex conjugate on these generators.
The coefficients of the right hand side of (\ref{fund5by5})
give us the nonzero matrix elements of these generators.
For example, the nonzero component of $T_1$ is given by the $(4,1)$
matrix element and the numerical value is equal to $1$.
Similarly, one of nonzero components of $T_{11}$ is given by the
$(1,1)$
matrix element and the numerical value is equal to
$(\frac{i}{2} + \frac{1}{2\sqrt{3}})$. 

Then the basis vectors of direct product space $V \otimes V$
can be obtained from $E_{ij} \equiv \hat{e}_i \otimes \hat{e}_j$
where $i, j =1, 2, \cdots, 5$.
The basis for the symmetric tensors can be realized by
\bea
\hat{u}_1 & = & E_{11}, \qquad
\hat{u}_2  =  \frac{1}{\sqrt{2}} (E_{12}+E_{21}), \qquad
\hat{u}_3  =  \frac{1}{\sqrt{2}} (E_{13}+E_{31}), \qquad
\hat{u}_4  =  E_{22},
\nonu \\
\hat{u}_5 & = &  \frac{1}{\sqrt{2}} (E_{23}+E_{32}), \qquad
\hat{u}_6  =  E_{33}, \qquad
\hat{u}_7  =   \frac{1}{\sqrt{2}} (E_{14}+E_{41}),
\qquad
\hat{u}_8  =   \frac{1}{\sqrt{2}} (E_{15}+E_{51}),
\nonu \\
\hat{u}_9 & = &  \frac{1}{\sqrt{2}} (E_{24}+E_{42}), \qquad
\hat{u}_{10}  =  \frac{1}{\sqrt{2}} (E_{25}+E_{52}), \qquad
\hat{u}_{11}  =   \frac{1}{\sqrt{2}} (E_{34}+E_{43}),
\nonu \\
\hat{u}_{12}   & = &   \frac{1}{\sqrt{2}} (E_{35}+E_{53}),
\qquad
\hat{u}_{13}  =   E_{44}, \qquad
\hat{u}_{14}  =  \frac{1}{\sqrt{2}} (E_{45}+E_{54}), \qquad
\hat{u}_{15}  =  E_{55}.
\nonu
\eea
Then it is straightforward to calculate the following quantities
based on (\ref{fund5by5})
{\small
  \bea
T_1 \, \hat{u}_1  & = & \sqrt{2} \hat{u}_7, \qquad
T_1 \, \hat{u}_2 =  \hat{u}_9,\qquad
T_1 \, \hat{u}_3 =  \hat{u}_{11}, \qquad
T_1 \, \hat{u}_7 = \sqrt{2} \hat{u}_{13},
\qquad
T_1 \, \hat{u}_8 =  \hat{u}_{14}, 
\nonu \\
T_2 \, \hat{u}_2  & = &  \hat{u}_7, \qquad
T_2 \, \hat{u}_4 =  \sqrt{2} \hat{u}_9,\qquad
T_2 \, \hat{u}_5 =  \hat{u}_{11}, \qquad
T_2 \, \hat{u}_9 = \sqrt{2} \hat{u}_{13},
\qquad
T_2 \, \hat{u}_{10} =  \hat{u}_{14}, 
\nonu \\
T_3 \, \hat{u}_3  & = &  \hat{u}_7, \qquad
T_3 \, \hat{u}_5 =   \hat{u}_9,\qquad
T_3 \, \hat{u}_6 =  \sqrt{2} \hat{u}_{11}, \qquad
T_3 \, \hat{u}_{11} = \sqrt{2} \hat{u}_{13},
\qquad
T_3 \, \hat{u}_{12} =  \hat{u}_{14}, 
\nonu \\
T_4 \, \hat{u}_1  & = &  \sqrt{2} \hat{u}_8, \qquad
T_4 \, \hat{u}_2 =   \hat{u}_{10},\qquad
T_4 \, \hat{u}_3 =   \hat{u}_{12}, \qquad
T_4 \, \hat{u}_{7} =  \hat{u}_{14},
\qquad
T_4 \, \hat{u}_{8} =  \sqrt{2} \hat{u}_{15}, 
\nonu \\
T_5 \, \hat{u}_2  & = &   \hat{u}_8, \qquad
T_5 \, \hat{u}_4 =   \sqrt{2} \hat{u}_{10},\qquad
T_5 \, \hat{u}_5 =   \hat{u}_{12}, \qquad
T_5 \, \hat{u}_{9} =  \hat{u}_{14},
\qquad
T_5 \, \hat{u}_{10} =  \sqrt{2} \hat{u}_{15}, 
\nonu \\
T_6 \, \hat{u}_3  & = &   \hat{u}_8, \qquad
T_6 \, \hat{u}_5 =   \hat{u}_{10},\qquad
T_6 \, \hat{u}_6 =   \sqrt{2} \hat{u}_{12}, \qquad
T_6 \, \hat{u}_{11} =  \hat{u}_{14},
\qquad
T_6 \, \hat{u}_{12} =  \sqrt{2} \hat{u}_{15}, 
\nonu \\
T_7 \, \hat{u}_7  & = &   \hat{u}_8, \qquad
T_7 \, \hat{u}_9 =   \hat{u}_{10},\qquad
T_7 \, \hat{u}_{11} =   \hat{u}_{12}, \qquad
T_7 \, \hat{u}_{13} =  \sqrt{2} \hat{u}_{14},
\qquad
T_7 \, \hat{u}_{14} =  \sqrt{2} \hat{u}_{15}, 
\nonu \\
T_8 \, \hat{u}_1  & = & \sqrt{2}  \hat{u}_2, \qquad
T_8 \, \hat{u}_2 =   \sqrt{2} \hat{u}_{4},\qquad
T_8 \, \hat{u}_{3} =   \hat{u}_{5}, \qquad
T_8 \, \hat{u}_{7} =   \hat{u}_{9},
\qquad
T_8 \, \hat{u}_{8} =   \hat{u}_{10}, 
\nonu \\
T_9 \, \hat{u}_1  & = & \sqrt{2}  \hat{u}_3, \qquad
T_9 \, \hat{u}_2 =    \hat{u}_{5},\qquad
T_9 \, \hat{u}_{3} = \sqrt{2}  \hat{u}_{6}, \qquad
T_9 \, \hat{u}_{7} =   \hat{u}_{11},
\qquad
T_9 \, \hat{u}_{8} =   \hat{u}_{12}, 
\nonu \\
T_{10} \, \hat{u}_2  & = &   \hat{u}_3, \qquad
T_{10} \, \hat{u}_4 =  \sqrt{2}  \hat{u}_{5},\qquad
T_{10} \, \hat{u}_{5} = \sqrt{2}  \hat{u}_{6}, \qquad
T_{10} \, \hat{u}_{9} =   \hat{u}_{11},
\qquad
T_{10} \, \hat{u}_{10} =   \hat{u}_{12}, 
\nonu \\
T_{11} \, \hat{u}_1  & = &   (i + \frac{1}{\sqrt{3}})\hat{u}_1, \qquad
T_{11} \, \hat{u}_2 =   \frac{1}{\sqrt{3}}
\hat{u}_{2},\qquad
T_{11} \, \hat{u}_{3} =  ( \frac{i}{2} - \frac{1}{2\sqrt{3}}) \hat{u}_{3},
\nonu \\
T_{11} \, \hat{u}_{4}  & = &   (-i + \frac{1}{\sqrt{3}}) \hat{u}_{4},
\qquad
T_{11} \, \hat{u}_{5}   =    ( -\frac{i}{2} - \frac{1}{2\sqrt{3}}) \hat{u}_{5}, 
\qquad
T_{11} \, \hat{u}_6   =    -\frac{2}{\sqrt{3}} \hat{u}_6,
\nonu \\
T_{11} \, \hat{u}_7  & = &   (\frac{i}{2} + \frac{1}{2\sqrt{3}})\hat{u}_7, \qquad
T_{11} \, \hat{u}_8 =  ( \frac{i}{2} + \frac{1}{2\sqrt{3}})
\hat{u}_{8},\qquad
T_{11} \, \hat{u}_{9} =  ( -\frac{i}{2} + \frac{1}{2\sqrt{3}}) \hat{u}_{9},
\nonu \\
T_{11} \, \hat{u}_{10}  & = &   (-\frac{i}{2} + \frac{1}{2\sqrt{3}}) \hat{u}_{10},
\qquad
T_{11} \, \hat{u}_{11}   =     - \frac{1}{\sqrt{3}} \hat{u}_{11}, 
\qquad
T_{11} \, \hat{u}_{12}   =    -\frac{1}{\sqrt{3}} \hat{u}_{12},
\nonu \\
T_{12} \, \hat{u}_1  & = &   (\frac{i}{\sqrt{6}} + \frac{1}{\sqrt{10}})\hat{u}_1, \qquad
T_{12} \, \hat{u}_2 =  ( \frac{i}{\sqrt{6}} + \frac{1}{\sqrt{10}})
\hat{u}_{2},\qquad
T_{12} \, \hat{u}_{3} =  ( \frac{i}{\sqrt{6}} + \frac{1}{\sqrt{10}})
\hat{u}_{3},
\nonu \\
T_{12} \, \hat{u}_{4}  & = &   (\frac{i}{\sqrt{6}} + \frac{1}{\sqrt{10}}) \hat{u}_{4},
\qquad
T_{12} \, \hat{u}_{5}   =    ( \frac{i}{\sqrt{6}} + \frac{1}{\sqrt{10}}) \hat{u}_{5}, 
\qquad
T_{12} \, \hat{u}_6   = ( \frac{i}{\sqrt{6}} + \frac{1}{\sqrt{10}})     \hat{u}_6,
\nonu \\
T_{12} \, \hat{u}_7  & = &   (-\frac{i}{\sqrt{6}} + \frac{1}{\sqrt{10}})\hat{u}_7, \qquad
T_{12} \, \hat{u}_8 =  ( \frac{5i\sqrt{6}}{60} - \frac{9\sqrt{10}}{60})
\hat{u}_{8},\qquad
T_{12} \, \hat{u}_{9} =  ( -\frac{i}{\sqrt{6}} + \frac{1}{\sqrt{10}})
\hat{u}_{9},
\nonu \\
T_{12} \, \hat{u}_{10}  & = &
 ( \frac{5i\sqrt{6}}{60} - \frac{9\sqrt{10}}{60})
\hat{u}_{10},
\qquad
T_{12} \, \hat{u}_{11}   =   (-\frac{i}{\sqrt{6}} + \frac{1}{\sqrt{10}})    \hat{u}_{11}, 
T_{12} \, \hat{u}_{12}   =    ( \frac{5i\sqrt{6}}{60} - \frac{9\sqrt{10}}{60})
 \hat{u}_{12},
 \nonu \\
 T_{12} \, \hat{u}_{13}  & = &   (-i \sqrt{\frac{3}{2}} + \frac{1}{\sqrt{10}})
\hat{u}_{13},
\qquad
T_{12} \, \hat{u}_{14}   =    (-\frac{i\sqrt{6}}{4} - \frac{3\sqrt{10}}{20})      \hat{u}_{14}, 
\qquad
T_{12} \, \hat{u}_{15}   =   
-2 \sqrt{\frac{2}{5}}  \hat{u}_{15}.
 \label{symm15by15}
\eea}
For example, one can write down $T_1 \, \hat{u}_1 = T_1 \, \hat{e}_1
\otimes \hat{e}_1$ which can be written as
$T_1 \hat{e}_1 \otimes \hat{e}_1 + \hat{e}_1 \otimes T_1 \hat{e}_1$.
By using (\ref{fund5by5}), this can be written as
$\hat{e}_4 \otimes \hat{e}_1 +  \hat{e}_1 \otimes \hat{e}_4$.
Then this is equal to $\sqrt{2} \hat{u}_7$ as in (\ref{symm15by15}).

From (\ref{symm15by15}),
the $24$ generators in terms of  $15 \times 15$
unitary matrix are given as follows:
{\small
\bea
  T_1 & = &
\left(
\begin{array}{ccccccccccccccc}
 0 & 0 & 0 & 0 & 0 & 0 & 0 & 0 & 0 & 0 & 0 & 0 & 0 & 0 & 0 \\
 0 & 0 & 0 & 0 & 0 & 0 & 0 & 0 & 0 & 0 & 0 & 0 & 0 & 0 & 0 \\
 0 & 0 & 0 & 0 & 0 & 0 & 0 & 0 & 0 & 0 & 0 & 0 & 0 & 0 & 0 \\
 0 & 0 & 0 & 0 & 0 & 0 & 0 & 0 & 0 & 0 & 0 & 0 & 0 & 0 & 0 \\
 0 & 0 & 0 & 0 & 0 & 0 & 0 & 0 & 0 & 0 & 0 & 0 & 0 & 0 & 0 \\
 0 & 0 & 0 & 0 & 0 & 0 & 0 & 0 & 0 & 0 & 0 & 0 & 0 & 0 & 0 \\
 \sqrt{2} & 0 & 0 & 0 & 0 & 0 & 0 & 0 & 0 & 0 & 0 & 0 & 0 & 0 & 0 \\
 0 & 0 & 0 & 0 & 0 & 0 & 0 & 0 & 0 & 0 & 0 & 0 & 0 & 0 & 0 \\
 0 & 1 & 0 & 0 & 0 & 0 & 0 & 0 & 0 & 0 & 0 & 0 & 0 & 0 & 0 \\
 0 & 0 & 0 & 0 & 0 & 0 & 0 & 0 & 0 & 0 & 0 & 0 & 0 & 0 & 0 \\
 0 & 0 & 1 & 0 & 0 & 0 & 0 & 0 & 0 & 0 & 0 & 0 & 0 & 0 & 0 \\
 0 & 0 & 0 & 0 & 0 & 0 & 0 & 0 & 0 & 0 & 0 & 0 & 0 & 0 & 0 \\
 0 & 0 & 0 & 0 & 0 & 0 & \sqrt{2} & 0 & 0 & 0 & 0 & 0 & 0 & 0 & 0 \\
 0 & 0 & 0 & 0 & 0 & 0 & 0 & 1 & 0 & 0 & 0 & 0 & 0 & 0 & 0 \\
 0 & 0 & 0 & 0 & 0 & 0 & 0 & 0 & 0 & 0 & 0 & 0 & 0 & 0 & 0 \\
\end{array}
\right),
\nonu \\
T_2 & = & \left(
\begin{array}{ccccccccccccccc}
 0 & 0 & 0 & 0 & 0 & 0 & 0 & 0 & 0 & 0 & 0 & 0 & 0 & 0 & 0 \\
 0 & 0 & 0 & 0 & 0 & 0 & 0 & 0 & 0 & 0 & 0 & 0 & 0 & 0 & 0 \\
 0 & 0 & 0 & 0 & 0 & 0 & 0 & 0 & 0 & 0 & 0 & 0 & 0 & 0 & 0 \\
 0 & 0 & 0 & 0 & 0 & 0 & 0 & 0 & 0 & 0 & 0 & 0 & 0 & 0 & 0 \\
 0 & 0 & 0 & 0 & 0 & 0 & 0 & 0 & 0 & 0 & 0 & 0 & 0 & 0 & 0 \\
 0 & 0 & 0 & 0 & 0 & 0 & 0 & 0 & 0 & 0 & 0 & 0 & 0 & 0 & 0 \\
 0 & 1 & 0 & 0 & 0 & 0 & 0 & 0 & 0 & 0 & 0 & 0 & 0 & 0 & 0 \\
 0 & 0 & 0 & 0 & 0 & 0 & 0 & 0 & 0 & 0 & 0 & 0 & 0 & 0 & 0 \\
 0 & 0 & 0 & \sqrt{2} & 0 & 0 & 0 & 0 & 0 & 0 & 0 & 0 & 0 & 0 & 0 \\
 0 & 0 & 0 & 0 & 0 & 0 & 0 & 0 & 0 & 0 & 0 & 0 & 0 & 0 & 0 \\
 0 & 0 & 0 & 0 & 1 & 0 & 0 & 0 & 0 & 0 & 0 & 0 & 0 & 0 & 0 \\
 0 & 0 & 0 & 0 & 0 & 0 & 0 & 0 & 0 & 0 & 0 & 0 & 0 & 0 & 0 \\
 0 & 0 & 0 & 0 & 0 & 0 & 0 & 0 & \sqrt{2} & 0 & 0 & 0 & 0 & 0 & 0 \\
 0 & 0 & 0 & 0 & 0 & 0 & 0 & 0 & 0 & 1 & 0 & 0 & 0 & 0 & 0 \\
 0 & 0 & 0 & 0 & 0 & 0 & 0 & 0 & 0 & 0 & 0 & 0 & 0 & 0 & 0 \\
\end{array}
\right),
\nonu \\
T_3  & = & \left(
\begin{array}{ccccccccccccccc}
 0 & 0 & 0 & 0 & 0 & 0 & 0 & 0 & 0 & 0 & 0 & 0 & 0 & 0 & 0 \\
 0 & 0 & 0 & 0 & 0 & 0 & 0 & 0 & 0 & 0 & 0 & 0 & 0 & 0 & 0 \\
 0 & 0 & 0 & 0 & 0 & 0 & 0 & 0 & 0 & 0 & 0 & 0 & 0 & 0 & 0 \\
 0 & 0 & 0 & 0 & 0 & 0 & 0 & 0 & 0 & 0 & 0 & 0 & 0 & 0 & 0 \\
 0 & 0 & 0 & 0 & 0 & 0 & 0 & 0 & 0 & 0 & 0 & 0 & 0 & 0 & 0 \\
 0 & 0 & 0 & 0 & 0 & 0 & 0 & 0 & 0 & 0 & 0 & 0 & 0 & 0 & 0 \\
 0 & 0 & 1 & 0 & 0 & 0 & 0 & 0 & 0 & 0 & 0 & 0 & 0 & 0 & 0 \\
 0 & 0 & 0 & 0 & 0 & 0 & 0 & 0 & 0 & 0 & 0 & 0 & 0 & 0 & 0 \\
 0 & 0 & 0 & 0 & 1 & 0 & 0 & 0 & 0 & 0 & 0 & 0 & 0 & 0 & 0 \\
 0 & 0 & 0 & 0 & 0 & 0 & 0 & 0 & 0 & 0 & 0 & 0 & 0 & 0 & 0 \\
 0 & 0 & 0 & 0 & 0 & \sqrt{2} & 0 & 0 & 0 & 0 & 0 & 0 & 0 & 0 & 0 \\
 0 & 0 & 0 & 0 & 0 & 0 & 0 & 0 & 0 & 0 & 0 & 0 & 0 & 0 & 0 \\
 0 & 0 & 0 & 0 & 0 & 0 & 0 & 0 & 0 & 0 & \sqrt{2} & 0 & 0 & 0 & 0 \\
 0 & 0 & 0 & 0 & 0 & 0 & 0 & 0 & 0 & 0 & 0 & 1 & 0 & 0 & 0 \\
 0 & 0 & 0 & 0 & 0 & 0 & 0 & 0 & 0 & 0 & 0 & 0 & 0 & 0 & 0 \\
\end{array}
\right),
\nonu \\
T_4  & = & \left(
\begin{array}{ccccccccccccccc}
 0 & 0 & 0 & 0 & 0 & 0 & 0 & 0 & 0 & 0 & 0 & 0 & 0 & 0 & 0 \\
 0 & 0 & 0 & 0 & 0 & 0 & 0 & 0 & 0 & 0 & 0 & 0 & 0 & 0 & 0 \\
 0 & 0 & 0 & 0 & 0 & 0 & 0 & 0 & 0 & 0 & 0 & 0 & 0 & 0 & 0 \\
 0 & 0 & 0 & 0 & 0 & 0 & 0 & 0 & 0 & 0 & 0 & 0 & 0 & 0 & 0 \\
 0 & 0 & 0 & 0 & 0 & 0 & 0 & 0 & 0 & 0 & 0 & 0 & 0 & 0 & 0 \\
 0 & 0 & 0 & 0 & 0 & 0 & 0 & 0 & 0 & 0 & 0 & 0 & 0 & 0 & 0 \\
 0 & 0 & 0 & 0 & 0 & 0 & 0 & 0 & 0 & 0 & 0 & 0 & 0 & 0 & 0 \\
 \sqrt{2} & 0 & 0 & 0 & 0 & 0 & 0 & 0 & 0 & 0 & 0 & 0 & 0 & 0 & 0 \\
 0 & 0 & 0 & 0 & 0 & 0 & 0 & 0 & 0 & 0 & 0 & 0 & 0 & 0 & 0 \\
 0 & 1 & 0 & 0 & 0 & 0 & 0 & 0 & 0 & 0 & 0 & 0 & 0 & 0 & 0 \\
 0 & 0 & 0 & 0 & 0 & 0 & 0 & 0 & 0 & 0 & 0 & 0 & 0 & 0 & 0 \\
 0 & 0 & 1 & 0 & 0 & 0 & 0 & 0 & 0 & 0 & 0 & 0 & 0 & 0 & 0 \\
 0 & 0 & 0 & 0 & 0 & 0 & 0 & 0 & 0 & 0 & 0 & 0 & 0 & 0 & 0 \\
 0 & 0 & 0 & 0 & 0 & 0 & 1 & 0 & 0 & 0 & 0 & 0 & 0 & 0 & 0 \\
 0 & 0 & 0 & 0 & 0 & 0 & 0 & \sqrt{2} & 0 & 0 & 0 & 0 & 0 & 0 & 0 \\
\end{array}
\right), 
\nonu \\
T_5 & = & \left(
\begin{array}{ccccccccccccccc}
 0 & 0 & 0 & 0 & 0 & 0 & 0 & 0 & 0 & 0 & 0 & 0 & 0 & 0 & 0 \\
 0 & 0 & 0 & 0 & 0 & 0 & 0 & 0 & 0 & 0 & 0 & 0 & 0 & 0 & 0 \\
 0 & 0 & 0 & 0 & 0 & 0 & 0 & 0 & 0 & 0 & 0 & 0 & 0 & 0 & 0 \\
 0 & 0 & 0 & 0 & 0 & 0 & 0 & 0 & 0 & 0 & 0 & 0 & 0 & 0 & 0 \\
 0 & 0 & 0 & 0 & 0 & 0 & 0 & 0 & 0 & 0 & 0 & 0 & 0 & 0 & 0 \\
 0 & 0 & 0 & 0 & 0 & 0 & 0 & 0 & 0 & 0 & 0 & 0 & 0 & 0 & 0 \\
 0 & 0 & 0 & 0 & 0 & 0 & 0 & 0 & 0 & 0 & 0 & 0 & 0 & 0 & 0 \\
 0 & 1 & 0 & 0 & 0 & 0 & 0 & 0 & 0 & 0 & 0 & 0 & 0 & 0 & 0 \\
 0 & 0 & 0 & 0 & 0 & 0 & 0 & 0 & 0 & 0 & 0 & 0 & 0 & 0 & 0 \\
 0 & 0 & 0 & \sqrt{2} & 0 & 0 & 0 & 0 & 0 & 0 & 0 & 0 & 0 & 0 & 0 \\
 0 & 0 & 0 & 0 & 0 & 0 & 0 & 0 & 0 & 0 & 0 & 0 & 0 & 0 & 0 \\
 0 & 0 & 0 & 0 & 1 & 0 & 0 & 0 & 0 & 0 & 0 & 0 & 0 & 0 & 0 \\
 0 & 0 & 0 & 0 & 0 & 0 & 0 & 0 & 0 & 0 & 0 & 0 & 0 & 0 & 0 \\
 0 & 0 & 0 & 0 & 0 & 0 & 0 & 0 & 1 & 0 & 0 & 0 & 0 & 0 & 0 \\
 0 & 0 & 0 & 0 & 0 & 0 & 0 & 0 & 0 & \sqrt{2} & 0 & 0 & 0 & 0 & 0 \\
\end{array}
\right),
\nonu \\
T_6 & = & \left(
\begin{array}{ccccccccccccccc}
 0 & 0 & 0 & 0 & 0 & 0 & 0 & 0 & 0 & 0 & 0 & 0 & 0 & 0 & 0 \\
 0 & 0 & 0 & 0 & 0 & 0 & 0 & 0 & 0 & 0 & 0 & 0 & 0 & 0 & 0 \\
 0 & 0 & 0 & 0 & 0 & 0 & 0 & 0 & 0 & 0 & 0 & 0 & 0 & 0 & 0 \\
 0 & 0 & 0 & 0 & 0 & 0 & 0 & 0 & 0 & 0 & 0 & 0 & 0 & 0 & 0 \\
 0 & 0 & 0 & 0 & 0 & 0 & 0 & 0 & 0 & 0 & 0 & 0 & 0 & 0 & 0 \\
 0 & 0 & 0 & 0 & 0 & 0 & 0 & 0 & 0 & 0 & 0 & 0 & 0 & 0 & 0 \\
 0 & 0 & 0 & 0 & 0 & 0 & 0 & 0 & 0 & 0 & 0 & 0 & 0 & 0 & 0 \\
 0 & 0 & 1 & 0 & 0 & 0 & 0 & 0 & 0 & 0 & 0 & 0 & 0 & 0 & 0 \\
 0 & 0 & 0 & 0 & 0 & 0 & 0 & 0 & 0 & 0 & 0 & 0 & 0 & 0 & 0 \\
 0 & 0 & 0 & 0 & 1 & 0 & 0 & 0 & 0 & 0 & 0 & 0 & 0 & 0 & 0 \\
 0 & 0 & 0 & 0 & 0 & 0 & 0 & 0 & 0 & 0 & 0 & 0 & 0 & 0 & 0 \\
 0 & 0 & 0 & 0 & 0 & \sqrt{2} & 0 & 0 & 0 & 0 & 0 & 0 & 0 & 0 & 0 \\
 0 & 0 & 0 & 0 & 0 & 0 & 0 & 0 & 0 & 0 & 0 & 0 & 0 & 0 & 0 \\
 0 & 0 & 0 & 0 & 0 & 0 & 0 & 0 & 0 & 0 & 1 & 0 & 0 & 0 & 0 \\
 0 & 0 & 0 & 0 & 0 & 0 & 0 & 0 & 0 & 0 & 0 & \sqrt{2} & 0 & 0 & 0 \\
\end{array}
\right),
\nonu \\
T_7 & = &\left(
\begin{array}{ccccccccccccccc}
 0 & 0 & 0 & 0 & 0 & 0 & 0 & 0 & 0 & 0 & 0 & 0 & 0 & 0 & 0 \\
 0 & 0 & 0 & 0 & 0 & 0 & 0 & 0 & 0 & 0 & 0 & 0 & 0 & 0 & 0 \\
 0 & 0 & 0 & 0 & 0 & 0 & 0 & 0 & 0 & 0 & 0 & 0 & 0 & 0 & 0 \\
 0 & 0 & 0 & 0 & 0 & 0 & 0 & 0 & 0 & 0 & 0 & 0 & 0 & 0 & 0 \\
 0 & 0 & 0 & 0 & 0 & 0 & 0 & 0 & 0 & 0 & 0 & 0 & 0 & 0 & 0 \\
 0 & 0 & 0 & 0 & 0 & 0 & 0 & 0 & 0 & 0 & 0 & 0 & 0 & 0 & 0 \\
 0 & 0 & 0 & 0 & 0 & 0 & 0 & 0 & 0 & 0 & 0 & 0 & 0 & 0 & 0 \\
 0 & 0 & 0 & 0 & 0 & 0 & 1 & 0 & 0 & 0 & 0 & 0 & 0 & 0 & 0 \\
 0 & 0 & 0 & 0 & 0 & 0 & 0 & 0 & 0 & 0 & 0 & 0 & 0 & 0 & 0 \\
 0 & 0 & 0 & 0 & 0 & 0 & 0 & 0 & 1 & 0 & 0 & 0 & 0 & 0 & 0 \\
 0 & 0 & 0 & 0 & 0 & 0 & 0 & 0 & 0 & 0 & 0 & 0 & 0 & 0 & 0 \\
 0 & 0 & 0 & 0 & 0 & 0 & 0 & 0 & 0 & 0 & 1 & 0 & 0 & 0 & 0 \\
 0 & 0 & 0 & 0 & 0 & 0 & 0 & 0 & 0 & 0 & 0 & 0 & 0 & 0 & 0 \\
 0 & 0 & 0 & 0 & 0 & 0 & 0 & 0 & 0 & 0 & 0 & 0 & \sqrt{2} & 0 & 0 \\
 0 & 0 & 0 & 0 & 0 & 0 & 0 & 0 & 0 & 0 & 0 & 0 & 0 & \sqrt{2} & 0 \\
\end{array}
\right),
\nonu \\
T_8 & = & \left(
\begin{array}{ccccccccccccccc}
 0 & 0 & 0 & 0 & 0 & 0 & 0 & 0 & 0 & 0 & 0 & 0 & 0 & 0 & 0 \\
 \sqrt{2} & 0 & 0 & 0 & 0 & 0 & 0 & 0 & 0 & 0 & 0 & 0 & 0 & 0 & 0 \\
 0 & 0 & 0 & 0 & 0 & 0 & 0 & 0 & 0 & 0 & 0 & 0 & 0 & 0 & 0 \\
 0 & \sqrt{2} & 0 & 0 & 0 & 0 & 0 & 0 & 0 & 0 & 0 & 0 & 0 & 0 & 0 \\
 0 & 0 & 1 & 0 & 0 & 0 & 0 & 0 & 0 & 0 & 0 & 0 & 0 & 0 & 0 \\
 0 & 0 & 0 & 0 & 0 & 0 & 0 & 0 & 0 & 0 & 0 & 0 & 0 & 0 & 0 \\
 0 & 0 & 0 & 0 & 0 & 0 & 0 & 0 & 0 & 0 & 0 & 0 & 0 & 0 & 0 \\
 0 & 0 & 0 & 0 & 0 & 0 & 0 & 0 & 0 & 0 & 0 & 0 & 0 & 0 & 0 \\
 0 & 0 & 0 & 0 & 0 & 0 & 1 & 0 & 0 & 0 & 0 & 0 & 0 & 0 & 0 \\
 0 & 0 & 0 & 0 & 0 & 0 & 0 & 1 & 0 & 0 & 0 & 0 & 0 & 0 & 0 \\
 0 & 0 & 0 & 0 & 0 & 0 & 0 & 0 & 0 & 0 & 0 & 0 & 0 & 0 & 0 \\
 0 & 0 & 0 & 0 & 0 & 0 & 0 & 0 & 0 & 0 & 0 & 0 & 0 & 0 & 0 \\
 0 & 0 & 0 & 0 & 0 & 0 & 0 & 0 & 0 & 0 & 0 & 0 & 0 & 0 & 0 \\
 0 & 0 & 0 & 0 & 0 & 0 & 0 & 0 & 0 & 0 & 0 & 0 & 0 & 0 & 0 \\
 0 & 0 & 0 & 0 & 0 & 0 & 0 & 0 & 0 & 0 & 0 & 0 & 0 & 0 & 0 \\
\end{array}
\right),
\nonu \\
T_9 & = &
\left(
\begin{array}{ccccccccccccccc}
 0 & 0 & 0 & 0 & 0 & 0 & 0 & 0 & 0 & 0 & 0 & 0 & 0 & 0 & 0 \\
 0 & 0 & 0 & 0 & 0 & 0 & 0 & 0 & 0 & 0 & 0 & 0 & 0 & 0 & 0 \\
 \sqrt{2} & 0 & 0 & 0 & 0 & 0 & 0 & 0 & 0 & 0 & 0 & 0 & 0 & 0 & 0 \\
 0 & 0 & 0 & 0 & 0 & 0 & 0 & 0 & 0 & 0 & 0 & 0 & 0 & 0 & 0 \\
 0 & 1 & 0 & 0 & 0 & 0 & 0 & 0 & 0 & 0 & 0 & 0 & 0 & 0 & 0 \\
 0 & 0 & \sqrt{2} & 0 & 0 & 0 & 0 & 0 & 0 & 0 & 0 & 0 & 0 & 0 & 0 \\
 0 & 0 & 0 & 0 & 0 & 0 & 0 & 0 & 0 & 0 & 0 & 0 & 0 & 0 & 0 \\
 0 & 0 & 0 & 0 & 0 & 0 & 0 & 0 & 0 & 0 & 0 & 0 & 0 & 0 & 0 \\
 0 & 0 & 0 & 0 & 0 & 0 & 0 & 0 & 0 & 0 & 0 & 0 & 0 & 0 & 0 \\
 0 & 0 & 0 & 0 & 0 & 0 & 0 & 0 & 0 & 0 & 0 & 0 & 0 & 0 & 0 \\
 0 & 0 & 0 & 0 & 0 & 0 & 1 & 0 & 0 & 0 & 0 & 0 & 0 & 0 & 0 \\
 0 & 0 & 0 & 0 & 0 & 0 & 0 & 1 & 0 & 0 & 0 & 0 & 0 & 0 & 0 \\
 0 & 0 & 0 & 0 & 0 & 0 & 0 & 0 & 0 & 0 & 0 & 0 & 0 & 0 & 0 \\
 0 & 0 & 0 & 0 & 0 & 0 & 0 & 0 & 0 & 0 & 0 & 0 & 0 & 0 & 0 \\
 0 & 0 & 0 & 0 & 0 & 0 & 0 & 0 & 0 & 0 & 0 & 0 & 0 & 0 & 0 \\
\end{array}
\right),
\nonu \\
T_{10} & = &
\left(
\begin{array}{ccccccccccccccc}
 0 & 0 & 0 & 0 & 0 & 0 & 0 & 0 & 0 & 0 & 0 & 0 & 0 & 0 & 0 \\
 0 & 0 & 0 & 0 & 0 & 0 & 0 & 0 & 0 & 0 & 0 & 0 & 0 & 0 & 0 \\
 0 & 1 & 0 & 0 & 0 & 0 & 0 & 0 & 0 & 0 & 0 & 0 & 0 & 0 & 0 \\
 0 & 0 & 0 & 0 & 0 & 0 & 0 & 0 & 0 & 0 & 0 & 0 & 0 & 0 & 0 \\
 0 & 0 & 0 & \sqrt{2} & 0 & 0 & 0 & 0 & 0 & 0 & 0 & 0 & 0 & 0 & 0 \\
 0 & 0 & 0 & 0 & \sqrt{2} & 0 & 0 & 0 & 0 & 0 & 0 & 0 & 0 & 0 & 0 \\
 0 & 0 & 0 & 0 & 0 & 0 & 0 & 0 & 0 & 0 & 0 & 0 & 0 & 0 & 0 \\
 0 & 0 & 0 & 0 & 0 & 0 & 0 & 0 & 0 & 0 & 0 & 0 & 0 & 0 & 0 \\
 0 & 0 & 0 & 0 & 0 & 0 & 0 & 0 & 0 & 0 & 0 & 0 & 0 & 0 & 0 \\
 0 & 0 & 0 & 0 & 0 & 0 & 0 & 0 & 0 & 0 & 0 & 0 & 0 & 0 & 0 \\
 0 & 0 & 0 & 0 & 0 & 0 & 0 & 0 & 1 & 0 & 0 & 0 & 0 & 0 & 0 \\
 0 & 0 & 0 & 0 & 0 & 0 & 0 & 0 & 0 & 1 & 0 & 0 & 0 & 0 & 0 \\
 0 & 0 & 0 & 0 & 0 & 0 & 0 & 0 & 0 & 0 & 0 & 0 & 0 & 0 & 0 \\
 0 & 0 & 0 & 0 & 0 & 0 & 0 & 0 & 0 & 0 & 0 & 0 & 0 & 0 & 0 \\
 0 & 0 & 0 & 0 & 0 & 0 & 0 & 0 & 0 & 0 & 0 & 0 & 0 & 0 & 0 \\
\end{array}
\right),
\nonu \\
T_{11} & = &
\mbox{diag} \left(
i+\frac{1}{\sqrt{3}},
\frac{1}{\sqrt{3}}, 
\frac{1}{6} (3 i-\sqrt{3}),
- i+\frac{1}{\sqrt{3}},  
\frac{1}{6} (-3 i-\sqrt{3}),
\right.
\nonu \\
&- & \left. \frac{2}{\sqrt{3}}, 
 \frac{1}{6} (3 i+\sqrt{3}), 
\frac{1}{6} (3 i+\sqrt{3}), 
\frac{1}{6} (-3 i+\sqrt{3}),
\frac{1}{6} (-3 i+\sqrt{3}),
-  \frac{1}{\sqrt{3}},
-\frac{1}{\sqrt{3}},
 0, 0,  0 \right),
\nonu \\
T_{12} & = &
\mbox{diag} \left(
\frac{i}{\sqrt{6}}+\frac{1}{\sqrt{10}},
\frac{i}{\sqrt{6}}+\frac{1}{\sqrt{10}},
\frac{i}{\sqrt{6}}+\frac{1}{\sqrt{10}},
\frac{i}{\sqrt{6}}+\frac{1}{\sqrt{10}},
\frac{i}{\sqrt{6}}+\frac{1}{\sqrt{10}}, \right.
\nonu \\
& + &
\frac{i}{\sqrt{6}}+\frac{1}{\sqrt{10}},
-\frac{i}{\sqrt{6}}+\frac{1}{\sqrt{10}},
\frac{1}{60} (5 i \sqrt{6}-9 \sqrt{10}),
 -\frac{i}{\sqrt{6}}+\frac{1}{\sqrt{10}},
 \frac{1}{60} (5 i \sqrt{6}-9 \sqrt{10}),
 \nonu \\
 & - & \left. \frac{i}{\sqrt{6}}+\frac{1}{\sqrt{10}},
\frac{1}{60} (5 i \sqrt{6}-9 \sqrt{10}),
-i \sqrt{\frac{3}{2}}+\frac{1}{\sqrt{10}},
\frac{1}{20} (-5 i \sqrt{6}-3 \sqrt{10}),
 -2 \sqrt{\frac{2}{5}}\right).
\label{symmgen}
\eea}
For example, from the first line of (\ref{symm15by15}),
the nonzero components of the generator $T_1$
are given by
$(7,1)$, $(9,2)$, $(11,3)$, $(13,7)$ and $(14,8)$
and their numerical values are
$\sqrt{2}$, $1$, $1$, $\sqrt{2}$, and $1$ respectively as in (\ref{symmgen}).
Then the remaining $12$ generators can be obtained from these
by taking the transpose and complex conjugate.
One has $\mbox{Tr} (T_a T_{a^{\ast}})=(N+2+2)=7$ with $N=3$.
The number $7$ is the index $l$ of the representation ${\bf 15}$
of $SU(5)$ \cite{Feger}.

%%%%%%%%%%%%%%%%%%%%%%%%%%%%%%%%%%%%%%%%%%%%%%%%%%%%%%%%%%%%%%%%%%%%%
%%%%%%%%%%%%%%%%%%%%%%%%%%%%%%%%%%%%%%%%%%%%%%%%%%%%%%%%%%%%%%%%%%%%%
\section{ The $24$ $SU(5)$ generators in the antisymmetric $\bf 10$
  representation of $SU(5)$ with two boxes}
%appendixB%%%%%%%%%%%%%%%%%%%%%%%%%%%%%%%%%%%%%%%%%%%%%%%%%%%%%%%%%%%%%%%%%%%
%%%%%%%%%%%%%%%%%%%%%%%%%%%%%%%%%%%%%%%%%%%%%%%%%%%%%%%%%%%%%%%%%%%%

The $24$ generators are given by $\frac{1}{2} (N+2)(N+1) \times
\frac{1}{2} (N+2)(N+1)=10 \times 10$ matrix
for the antisymmetric representation
$\tiny\yng(1,1)$.

The basis for the antisymmetric tensors can be realized by
\bea
\hat{u}_1   & = &   \frac{1}{\sqrt{2}} (E_{12}-E_{21}), \qquad
\hat{u}_2  =  \frac{1}{\sqrt{2}} (E_{13}-E_{31}), 
\nonu \\
\hat{u}_3   & = &   \frac{1}{\sqrt{2}} (E_{23}-E_{32}), \qquad
\hat{u}_4  =  \frac{1}{\sqrt{2}} (E_{14}-E_{41}), 
\nonu \\
\hat{u}_5   & = &   \frac{1}{\sqrt{2}} (E_{15}-E_{51}), \qquad
\hat{u}_6  =  \frac{1}{\sqrt{2}} (E_{24}-E_{42}), 
\nonu \\
\hat{u}_7   & = &   \frac{1}{\sqrt{2}} (E_{25}-E_{52}), \qquad
\hat{u}_8  =  \frac{1}{\sqrt{2}} (E_{34}-E_{43}), 
\nonu \\
\hat{u}_9   & = &   \frac{1}{\sqrt{2}} (E_{35}-E_{53}), \qquad
\hat{u}_{10}  =  \frac{1}{\sqrt{2}} (E_{45}-E_{54}). 
\label{basisantisymm}
\eea

After applying the $24$ generators which are $5 \times 5$  matrices in
(\ref{fund5by5}) to
(\ref{basisantisymm}),
the $24$ generators in terms of  $10 \times 10$
unitary matrix are given as follows as in Appendix $A$:
{\small
\bea
T_1 &= &
\left(
\begin{array}{cccccccccc}
 0 & 0 & 0 & 0 & 0 & 0 & 0 & 0 & 0 & 0 \\
 0 & 0 & 0 & 0 & 0 & 0 & 0 & 0 & 0 & 0 \\
 0 & 0 & 0 & 0 & 0 & 0 & 0 & 0 & 0 & 0 \\
 0 & 0 & 0 & 0 & 0 & 0 & 0 & 0 & 0 & 0 \\
 0 & 0 & 0 & 0 & 0 & 0 & 0 & 0 & 0 & 0 \\
 -1 & 0 & 0 & 0 & 0 & 0 & 0 & 0 & 0 & 0 \\
 0 & 0 & 0 & 0 & 0 & 0 & 0 & 0 & 0 & 0 \\
 0 & -1 & 0 & 0 & 0 & 0 & 0 & 0 & 0 & 0 \\
 0 & 0 & 0 & 0 & 0 & 0 & 0 & 0 & 0 & 0 \\
 0 & 0 & 0 & 0 & 1 & 0 & 0 & 0 & 0 & 0 \\
\end{array}
\right),
T_2=\left(
\begin{array}{cccccccccc}
 0 & 0 & 0 & 0 & 0 & 0 & 0 & 0 & 0 & 0 \\
 0 & 0 & 0 & 0 & 0 & 0 & 0 & 0 & 0 & 0 \\
 0 & 0 & 0 & 0 & 0 & 0 & 0 & 0 & 0 & 0 \\
 1 & 0 & 0 & 0 & 0 & 0 & 0 & 0 & 0 & 0 \\
 0 & 0 & 0 & 0 & 0 & 0 & 0 & 0 & 0 & 0 \\
 0 & 0 & 0 & 0 & 0 & 0 & 0 & 0 & 0 & 0 \\
 0 & 0 & 0 & 0 & 0 & 0 & 0 & 0 & 0 & 0 \\
 0 & 0 & -1 & 0 & 0 & 0 & 0 & 0 & 0 & 0 \\
 0 & 0 & 0 & 0 & 0 & 0 & 0 & 0 & 0 & 0 \\
 0 & 0 & 0 & 0 & 0 & 0 & 1 & 0 & 0 & 0 \\
\end{array}
\right),
\nonu \\
T_3 &=&
\left(
\begin{array}{cccccccccc}
 0 & 0 & 0 & 0 & 0 & 0 & 0 & 0 & 0 & 0 \\
 0 & 0 & 0 & 0 & 0 & 0 & 0 & 0 & 0 & 0 \\
 0 & 0 & 0 & 0 & 0 & 0 & 0 & 0 & 0 & 0 \\
 0 & 1 & 0 & 0 & 0 & 0 & 0 & 0 & 0 & 0 \\
 0 & 0 & 0 & 0 & 0 & 0 & 0 & 0 & 0 & 0 \\
 0 & 0 & 1 & 0 & 0 & 0 & 0 & 0 & 0 & 0 \\
 0 & 0 & 0 & 0 & 0 & 0 & 0 & 0 & 0 & 0 \\
 0 & 0 & 0 & 0 & 0 & 0 & 0 & 0 & 0 & 0 \\
 0 & 0 & 0 & 0 & 0 & 0 & 0 & 0 & 0 & 0 \\
 0 & 0 & 0 & 0 & 0 & 0 & 0 & 0 & 1 & 0 \\
\end{array}
\right),
T_4 =
\left(
\begin{array}{cccccccccc}
 0 & 0 & 0 & 0 & 0 & 0 & 0 & 0 & 0 & 0 \\
 0 & 0 & 0 & 0 & 0 & 0 & 0 & 0 & 0 & 0 \\
 0 & 0 & 0 & 0 & 0 & 0 & 0 & 0 & 0 & 0 \\
 0 & 0 & 0 & 0 & 0 & 0 & 0 & 0 & 0 & 0 \\
 0 & 0 & 0 & 0 & 0 & 0 & 0 & 0 & 0 & 0 \\
 0 & 0 & 0 & 0 & 0 & 0 & 0 & 0 & 0 & 0 \\
 -1 & 0 & 0 & 0 & 0 & 0 & 0 & 0 & 0 & 0 \\
 0 & 0 & 0 & 0 & 0 & 0 & 0 & 0 & 0 & 0 \\
 0 & -1 & 0 & 0 & 0 & 0 & 0 & 0 & 0 & 0 \\
 0 & 0 & 0 & -1 & 0 & 0 & 0 & 0 & 0 & 0 \\
\end{array}
\right),
\nonu \\
T_5 & = &
\left(
\begin{array}{cccccccccc}
 0 & 0 & 0 & 0 & 0 & 0 & 0 & 0 & 0 & 0 \\
 0 & 0 & 0 & 0 & 0 & 0 & 0 & 0 & 0 & 0 \\
 0 & 0 & 0 & 0 & 0 & 0 & 0 & 0 & 0 & 0 \\
 0 & 0 & 0 & 0 & 0 & 0 & 0 & 0 & 0 & 0 \\
 1 & 0 & 0 & 0 & 0 & 0 & 0 & 0 & 0 & 0 \\
 0 & 0 & 0 & 0 & 0 & 0 & 0 & 0 & 0 & 0 \\
 0 & 0 & 0 & 0 & 0 & 0 & 0 & 0 & 0 & 0 \\
 0 & 0 & 0 & 0 & 0 & 0 & 0 & 0 & 0 & 0 \\
 0 & 0 & -1 & 0 & 0 & 0 & 0 & 0 & 0 & 0 \\
 0 & 0 & 0 & 0 & 0 & -1 & 0 & 0 & 0 & 0 \\
\end{array}
\right),
T_6 = \left(
\begin{array}{cccccccccc}
 0 & 0 & 0 & 0 & 0 & 0 & 0 & 0 & 0 & 0 \\
 0 & 0 & 0 & 0 & 0 & 0 & 0 & 0 & 0 & 0 \\
 0 & 0 & 0 & 0 & 0 & 0 & 0 & 0 & 0 & 0 \\
 0 & 0 & 0 & 0 & 0 & 0 & 0 & 0 & 0 & 0 \\
 0 & 1 & 0 & 0 & 0 & 0 & 0 & 0 & 0 & 0 \\
 0 & 0 & 0 & 0 & 0 & 0 & 0 & 0 & 0 & 0 \\
 0 & 0 & 1 & 0 & 0 & 0 & 0 & 0 & 0 & 0 \\
 0 & 0 & 0 & 0 & 0 & 0 & 0 & 0 & 0 & 0 \\
 0 & 0 & 0 & 0 & 0 & 0 & 0 & 0 & 0 & 0 \\
 0 & 0 & 0 & 0 & 0 & 0 & 0 & -1 & 0 & 0 \\
\end{array}
\right),
\nonu \\
T_7 &= &
\left(
\begin{array}{cccccccccc}
 0 & 0 & 0 & 0 & 0 & 0 & 0 & 0 & 0 & 0 \\
 0 & 0 & 0 & 0 & 0 & 0 & 0 & 0 & 0 & 0 \\
 0 & 0 & 0 & 0 & 0 & 0 & 0 & 0 & 0 & 0 \\
 0 & 0 & 0 & 0 & 0 & 0 & 0 & 0 & 0 & 0 \\
 0 & 0 & 0 & 1 & 0 & 0 & 0 & 0 & 0 & 0 \\
 0 & 0 & 0 & 0 & 0 & 0 & 0 & 0 & 0 & 0 \\
 0 & 0 & 0 & 0 & 0 & 1 & 0 & 0 & 0 & 0 \\
 0 & 0 & 0 & 0 & 0 & 0 & 0 & 0 & 0 & 0 \\
 0 & 0 & 0 & 0 & 0 & 0 & 0 & 1 & 0 & 0 \\
 0 & 0 & 0 & 0 & 0 & 0 & 0 & 0 & 0 & 0 \\
\end{array}
\right),
T_8 =\left(
\begin{array}{cccccccccc}
 0 & 0 & 0 & 0 & 0 & 0 & 0 & 0 & 0 & 0 \\
 0 & 0 & 0 & 0 & 0 & 0 & 0 & 0 & 0 & 0 \\
 0 & 1 & 0 & 0 & 0 & 0 & 0 & 0 & 0 & 0 \\
 0 & 0 & 0 & 0 & 0 & 0 & 0 & 0 & 0 & 0 \\
 0 & 0 & 0 & 0 & 0 & 0 & 0 & 0 & 0 & 0 \\
 0 & 0 & 0 & 1 & 0 & 0 & 0 & 0 & 0 & 0 \\
 0 & 0 & 0 & 0 & 1 & 0 & 0 & 0 & 0 & 0 \\
 0 & 0 & 0 & 0 & 0 & 0 & 0 & 0 & 0 & 0 \\
 0 & 0 & 0 & 0 & 0 & 0 & 0 & 0 & 0 & 0 \\
 0 & 0 & 0 & 0 & 0 & 0 & 0 & 0 & 0 & 0 \\
\end{array}
\right),
\nonu \\
T_9 & = & \left(
\begin{array}{cccccccccc}
 0 & 0 & 0 & 0 & 0 & 0 & 0 & 0 & 0 & 0 \\
 0 & 0 & 0 & 0 & 0 & 0 & 0 & 0 & 0 & 0 \\
 -1 & 0 & 0 & 0 & 0 & 0 & 0 & 0 & 0 & 0 \\
 0 & 0 & 0 & 0 & 0 & 0 & 0 & 0 & 0 & 0 \\
 0 & 0 & 0 & 0 & 0 & 0 & 0 & 0 & 0 & 0 \\
 0 & 0 & 0 & 0 & 0 & 0 & 0 & 0 & 0 & 0 \\
 0 & 0 & 0 & 0 & 0 & 0 & 0 & 0 & 0 & 0 \\
 0 & 0 & 0 & 1 & 0 & 0 & 0 & 0 & 0 & 0 \\
 0 & 0 & 0 & 0 & 1 & 0 & 0 & 0 & 0 & 0 \\
 0 & 0 & 0 & 0 & 0 & 0 & 0 & 0 & 0 & 0 \\
\end{array}
\right),
T_{10} =
\left(
\begin{array}{cccccccccc}
 0 & 0 & 0 & 0 & 0 & 0 & 0 & 0 & 0 & 0 \\
 1 & 0 & 0 & 0 & 0 & 0 & 0 & 0 & 0 & 0 \\
 0 & 0 & 0 & 0 & 0 & 0 & 0 & 0 & 0 & 0 \\
 0 & 0 & 0 & 0 & 0 & 0 & 0 & 0 & 0 & 0 \\
 0 & 0 & 0 & 0 & 0 & 0 & 0 & 0 & 0 & 0 \\
 0 & 0 & 0 & 0 & 0 & 0 & 0 & 0 & 0 & 0 \\
 0 & 0 & 0 & 0 & 0 & 0 & 0 & 0 & 0 & 0 \\
 0 & 0 & 0 & 0 & 0 & 1 & 0 & 0 & 0 & 0 \\
 0 & 0 & 0 & 0 & 0 & 0 & 1 & 0 & 0 & 0 \\
 0 & 0 & 0 & 0 & 0 & 0 & 0 & 0 & 0 & 0 \\
\end{array}
\right),
\nonu \\
T_{11} & = &
\mbox{daig} \left(
\frac{1}{\sqrt{3}},
\frac{1}{6} (3 i-\sqrt{3}),
\frac{1}{6} (-3 i-\sqrt{3}),
\frac{1}{6} (3 i+\sqrt{3}),
\frac{1}{6} (3 i+\sqrt{3})
, \right.
\nonu \\
& + & \left.
\frac{1}{6} (-3 i+\sqrt{3}),
\frac{1}{6} (-3 i+\sqrt{3}),
-\frac{1}{\sqrt{3}},
-\frac{1}{\sqrt{3}},0
\right),
\nonu \\
T_{12} & = &
\mbox{diag}
\left(
\frac{i}{\sqrt{6}}+\frac{1}{\sqrt{10}},
\frac{i}{\sqrt{6}}+\frac{1}{\sqrt{10}},
\frac{i}{\sqrt{6}}+\frac{1}{\sqrt{10}},
-\frac{i}{\sqrt{6}}+\frac{1}{\sqrt{10}},
\frac{1}{60} (5 i \sqrt{6}-9 \sqrt{10}),
\right. \nonu \\
& - & \left. \frac{i}{\sqrt{6}}+\frac{1}{\sqrt{10}},
\frac{1}{60} (5 i \sqrt{6}-9 \sqrt{10}),
-\frac{i}{\sqrt{6}}+\frac{1}{\sqrt{10}},
\frac{1}{60} (5 i \sqrt{6}-9 \sqrt{10}),
\frac{1}{20} (-5 i \sqrt{6}-3 \sqrt{10})
\right).
\nonu
\eea}
Again the half of generators can be obtained from these 
by taking the transpose and complex conjugate. 
One has $\mbox{Tr} (T_a T_{a^{\ast}})=(N+2-2)=3$ which 
 is the index $l$ of the representation ${\bf 10}$
of $SU(5)$ \cite{Feger}.
For the generators of $SU(N+2)$ with $N=5,7,9,11$, the similar
constructions can be done.

%%%%%%%%%%%%%%%%%%%%%%%%%%%%%%%%%%%%%%%%%%%%%%%%%%%%%%%%%%%%%%%%%%%%%
%%%%%%%%%%%%%%%%%%%%%%%%%%%%%%%%%%%%%%%%%%%%%%%%%%%%%%%%%%%%%%%%%%%%%
\section{ The $24$ $SU(5)$ generators in the symmetric $\overline{\bf 35}$
  representation of $SU(5)$ with three boxes }
%appendixC%%%%%%%%%%%%%%%%%%%%%%%%%%%%%%%%%%%%%%%%%%%%%%%%%%%%%%%%%%%%%%%%%%%
%%%%%%%%%%%%%%%%%%%%%%%%%%%%%%%%%%%%%%%%%%%%%%%%%%%%%%%%%%%%%%%%%%%%

The basis vectors of direct product space $V \otimes V \otimes V$
can be obtained from $E_{ijk} \equiv \hat{e}_i \otimes \hat{e}_j \otimes
\hat{e}_k$
where $i, j, k =1, 2, \cdots, 5$.
The basis for the symmetric ($\tiny\yng(3)$) tensors can be realized by
\footnote{In the notation of \cite{Feger}, their $\hat{u}$ charge
  is twice of the one of this paper. For example, the branching rule
(Table $A.71$ of \cite{Feger})
   of $SU(5) \rightarrow SU(3) \times SU(2) \times U(1)$ gives
  ${\bf \overline{35}} =\tiny\yng(3)= ({\bf 10},{\bf 1})_{6} +
   ({\bf 6},{\bf 2})_{1} + ({\bf 3},{\bf 3})_{-4}+ ({\bf 1},{\bf 4})_{-9}$
   by taking the complex conjugation of ${\bf 35}$. One should take
   the complex conjugation of $SU(3)$ representation and
   take the minus sign of $\hat{u}$ charge.
   This can be compared to the relation in (\ref{transtranstrans}).
   For $N \geq 7$, the three box does not contain the complex conjugated
   notation.
   Note that
   the corresponding relation for the $SU(11)$ branching in ${\bf 286}=
   \tiny\yng(3)$ is given by
   Table $A.77$ of \cite{Feger}. }
\bea
\hat{u}_1   & = & E_{111}, \qquad
\hat{u}_2    =  \frac{1}{\sqrt{3}} (E_{112} + E_{211} + E_{121}),
\nonu \\
\hat{u}_3   &  = &  \frac{1}{\sqrt{3}} (E_{113} + E_{311} + E_{131}),
\qquad
\hat{u}_4    =  \frac{1}{\sqrt{3}} (E_{122} + E_{221} + E_{212}),
\nonu \\
\hat{u}_5   & = &   \frac{1}{\sqrt{6}} (E_{123}+E_{132}+E_{312}+E_{213}+E_{231}+
E_{321}),
\qquad
\hat{u}_6     =   \frac{1}{\sqrt{3}} (E_{133} + E_{331} + E_{313}),
\nonu \\
\hat{u}_7   & = & E_{222}, \qquad
\hat{u}_8    =  \frac{1}{\sqrt{3}} (E_{223} + E_{232} + E_{322}),
\nonu \\
\hat{u}_9   & = &  \frac{1}{\sqrt{3}} (E_{233} + E_{332} + E_{323}),
\qquad
\hat{u}_{10}    =  E_{333},
\nonu \\
\hat{u}_{11}   &  = &  \frac{1}{\sqrt{3}} (E_{114} + E_{411} + E_{141}),
\qquad
\hat{u}_{12}    =  \frac{1}{\sqrt{3}} (E_{115} + E_{511} + E_{151}),
\nonu \\
\hat{u}_{13}   & = &  \frac{1}{\sqrt{6}} (E_{124}+E_{142}+E_{412}+E_{214}+E_{241}+
E_{421}), 
\nonu \\
\hat{u}_{14}   & = &   \frac{1}{\sqrt{6}} (E_{125}+E_{152}+E_{512}+E_{215}+E_{251}+
E_{521}),
\nonu \\
\hat{u}_{15}   & = &  \frac{1}{\sqrt{6}} (E_{134}+E_{143}+E_{413}+E_{314}+E_{341}+
E_{431}), 
\nonu \\
\hat{u}_{16}   & = &   \frac{1}{\sqrt{6}} (E_{135}+E_{153}+E_{513}+E_{315}+E_{351}+
E_{531}),
\qquad
\hat{u}_{17}    =   \frac{1}{\sqrt{3}} (E_{144} + E_{414} + E_{441}),
\nonu \\
\hat{u}_{18}   & = &  \frac{1}{\sqrt{6}} (E_{145}+E_{154}+E_{514}+E_{415}+E_{451}+
E_{541}), 
\nonu \\
\hat{u}_{19}  &  =  &  \frac{1}{\sqrt{3}} (E_{155} + E_{515} + E_{551}),
\qquad
\hat{u}_{20}    =   \frac{1}{\sqrt{3}} (E_{224} + E_{422} + E_{242}),
\nonu \\
\hat{u}_{21}   & = &   \frac{1}{\sqrt{3}} (E_{225} + E_{522} + E_{252}),
\qquad
\hat{u}_{22}    =    \frac{1}{\sqrt{6}} (E_{234}+E_{243}+E_{423}+E_{324}+E_{342}+
E_{432}),
\nonu \\
\hat{u}_{23}   & = &  \frac{1}{\sqrt{6}} (E_{235}+E_{253}+E_{523}+E_{325}+E_{352}+
E_{532}),
\qquad
\hat{u}_{24}    =    \frac{1}{\sqrt{3}} (E_{244} + E_{424} + E_{442}),
\nonu \\
\hat{u}_{25}   & = &   \frac{1}{\sqrt{6}} (E_{245}+E_{254}+E_{524}+E_{425}+E_{452}+
E_{542}),
\qquad
\hat{u}_{26}    =    \frac{1}{\sqrt{3}} (E_{255} + E_{525} + E_{552}),
\nonu \\
\hat{u}_{27}  &  = &    \frac{1}{\sqrt{3}} (E_{334} + E_{343} + E_{433}),
\qquad
\hat{u}_{28}    =    \frac{1}{\sqrt{3}} (E_{335} + E_{533} + E_{353}),
\nonu \\
\hat{u}_{29}   &  =  &   \frac{1}{\sqrt{3}} (E_{344} + E_{434} + E_{443}),
\qquad
\hat{u}_{30}    =   \frac{1}{\sqrt{6}} (E_{345}+E_{354}+E_{534}+E_{435}+E_{453}+
E_{543}), 
\nonu \\
\hat{u}_{31}   & = &    \frac{1}{\sqrt{3}} (E_{355} + E_{535} + E_{553}),
\qquad
\hat{u}_{32}    =  E_{444},
\nonu \\
\hat{u}_{33}   & = &    \frac{1}{\sqrt{3}} (E_{445} + E_{544} + E_{454}),
\qquad
\hat{u}_{34}    =     \frac{1}{\sqrt{3}} (E_{455} + E_{545} + E_{554}),
\nonu \\
\hat{u}_{35}   & = &  E_{555}.
\label{basissymm3}
\eea
After applying the $24$ generators which are $5 \times 5$  matrices in
(\ref{fund5by5}) to
(\ref{basissymm3}),
the $24$ generators in terms of  $35 \times 35$
unitary matrix are given as follows
as in Appendices $A$ and $B$. We present only the nonzero elements. 
{\small 
{\small
  \bea
  T_1  & : &
 (20,4), (22,5), (25,14), (27,6), (30,16), (34,19) \, \mbox{has} \, 1,
      \qquad
        (17,11) \, \mbox{has} \, 2,
      \nonu \\
      & : &  (13,2), (15,3), (18,12), (24,13), (29,15), (33,18) \, \mbox{has} \, \sqrt{2},
      \qquad
        (11,1), (32,17) \, \mbox{has} \, \sqrt{3},
      \nonu \\
  T_2  & : &
 (11,2), (15,5), (18,14), (27,9), (30,23), (34,26) \, \mbox{has} \, 1,
      \qquad
        (24,20) \, \mbox{has} \, 2,
      \nonu \\
      & : &  (13,4), (17,13), (22,8), (25,21), (29,22), (33,25) \, \mbox{has} \, \sqrt{2},
      \qquad
        (20,7), (32,24) \, \mbox{has} \, \sqrt{3},
      \nonu \\
        T_3  & : &
 (11,3), (13,5), (18,16), (20,8), (25,23), (34,31) \, \mbox{has} \, 1,
      \qquad
        (29,27) \, \mbox{has} \, 2,
      \nonu \\
      & : &  (15,6), (17,15), (22,9), (24,22), (30,28), (33,30) \, \mbox{has} \, \sqrt{2},
      \qquad
        (27,10), (32,29) \, \mbox{has} \, \sqrt{3},
      \nonu \\
 T_4  & : &
 (21,4), (23,5), (25,13), (28,6), (30,15), (33,17) \, \mbox{has} \, 1,
      \qquad
        (19,12) \, \mbox{has} \, 2,
      \nonu \\
      & : &  (14,2), (16,3), (18,11), (26,14), (31,16), (34,18)
      \, \mbox{has} \, \sqrt{2},
      \qquad
        (12,1), (35,19) \, \mbox{has} \, \sqrt{3},
      \nonu \\
 T_5  & : &
 (12,2), (16,5), (18,13), (28,9), (30,22), (33,24) \, \mbox{has} \, 1,
      \qquad
        (26,21) \, \mbox{has} \, 2,
      \nonu \\
      & : &  (14,4), (19,14), (23,8), (25,20), (31,23), (34,25)
      \, \mbox{has} \, \sqrt{2},
      \qquad
        (21,7), (35,26) \, \mbox{has} \, \sqrt{3},
      \nonu \\
 T_6  & : &
 (12,3), (14,5), (18,15), (21,8), (25,22), (33,29) \, \mbox{has} \, 1,
      \qquad
        (31,28) \, \mbox{has} \, 2,
      \nonu \\
      & : &  (16,6), (19,16), (23,9), (26,23), (30,27), (34,30)
      \, \mbox{has} \, \sqrt{2},
      \qquad
        (28,10), (35,31) \, \mbox{has} \, \sqrt{3},
      \nonu \\
 T_7  & : &
 (12,11), (14,13), (16,15), (21,20), (23,22), (28,27) \, \mbox{has} \, 1,
      \qquad
        (34,33) \, \mbox{has} \, 2,
      \nonu \\
      & : &  (18,17), (19,18), (25,24), (26,25), (30,29), (31,30)
      \, \mbox{has} \, \sqrt{2},
      \qquad
        (33,32), (35,34) \, \mbox{has} \, \sqrt{3},
      \nonu \\
T_8  & : &
 (9,6), (22,15), (23,16), (24,17), (25,18), (26,19) \, \mbox{has} \, 1,
      \qquad
        (4,2) \, \mbox{has} \, 2,
      \nonu \\
      & : &  (5,3), (8,5), (13,11), (14,12), (20,13), (21,14)
      \, \mbox{has} \, \sqrt{2},
      \qquad
        (2,1), (7,4) \, \mbox{has} \, \sqrt{3},
      \nonu \\
T_9  & : &
 (8,4), (22,13), (23,14), (29,17), (30,18), (31,19) \, \mbox{has} \, 1,
      \qquad
        (6,3) \, \mbox{has} \, 2,
      \nonu \\
      & : &  (5,2), (9,5), (15,11), (16,12), (27,15), (28,16)
      \, \mbox{has} \, \sqrt{2},
      \qquad
        (3,1), (10,6) \, \mbox{has} \, \sqrt{3},
      \nonu \\
T_{10}  & : &
 (3,2), (15,13), (16,14), (29,24), (30,25), (31,26) \, \mbox{has} \, 1,
      \qquad
        (9,8) \, \mbox{has} \, 2,
      \nonu \\
      & : &  (5,4), (6,5), (22,20), (23,21), (27,22), (28,23)
      \, \mbox{has} \, \sqrt{2},
      \qquad
        (8,7), (10,9) \, \mbox{has} \, \sqrt{3},
      \nonu \\
      T_{11}  & : &
 (8,8) \, \mbox{has} \, -i,
      \qquad
        (3,3) \, \mbox{has} \, i,
      \nonu \\
      & : &  (27,27), (28,28)
      \, \mbox{has} \, -\frac{2}{\sqrt{3}},
      \qquad
        (29,29), (30,30), (31,31) \, \mbox{has} \, -\frac{1}{\sqrt{3}},
      \nonu \\
   & : &  (13,13), (14,14) \, \mbox{has} \, \frac{1}{\sqrt{3}},
      \qquad
     (10,10) \, \mbox{has} \, -\sqrt{3},
      \nonu \\
        & : &  (9,9) \, \mbox{has} \,  \frac{1}{2} (-\sqrt{3}-i),
      \qquad
    (6,6) \, \mbox{has} \,  \frac{1}{2} (-\sqrt{3}+i),
      \nonu     \\
      & : &  (22,22), (23,23) \,
      \mbox{has} \, \frac{1}{6} (-\sqrt{3}-3 i),
      \qquad
      (15,15), (16,16) \,
      \mbox{has} \, \frac{1}{6} (-\sqrt{3}+3 i),
      \nonu \\
   & : &  (4,4) \,
      \mbox{has} \, \frac{1}{2} (\sqrt{3}-i),
      \qquad
      (2,2) \,
      \mbox{has} \, \frac{1}{2} (\sqrt{3}+i),
      \nonu    \\
    & : &  (25,25) \,
      \mbox{has} \, \frac{1}{6} (\sqrt{3}-3 i),
      \qquad
       (20,20), (21,21) \,
      \mbox{has} \, \frac{1}{3} (\sqrt{3}-3 i),
      \nonu  \\
    & : &  (7,7) \,
      \mbox{has} \, \frac{1}{2} (\sqrt{3}-3 i),
      \qquad
          (18,18) \,
      \mbox{has} \, \frac{1}{6} (\sqrt{3}+3 i),
      \nonu  \\
     & : &  (11,11), (12,12) \,
      \mbox{has} \, \frac{1}{3} (\sqrt{3}+3 i),
      \qquad
        (1,1) \,
      \mbox{has} \, \frac{1}{2} (\sqrt{3}+3 i),
      \nonu \\
      T_{12}  & : &
 (35,35) \, \mbox{has} \, -3 \sqrt{\frac{2}{5}},
      \qquad
        (18,18), (25,25), (30,30) \, \mbox{has} \,
      -\frac{i (5 \sqrt{3}-3 i \sqrt{5})}{15 \sqrt{2}},
      \nonu \\
      & : &  (14,14), (16,16), (23,23)
      \, \mbox{has} \, \frac{i
        (5 \sqrt{3}+3 i \sqrt{5})}{15 \sqrt{2}},
      \qquad
        (34,34) \, \mbox{has} \,
      -\frac{i (5 \sqrt{3}-7 i \sqrt{5})}{10 \sqrt{2}},
      \nonu \\
     & : &  (19,19), (26,26), (31,31) \, \mbox{has} \,
     \frac{i (5 \sqrt{3}+21 i \sqrt{5})}{30 \sqrt{2}},
      \qquad
        (32,32)
      \, \mbox{has} \,
      \frac{3 (\sqrt{5}-5 i \sqrt{3})}{10 \sqrt{2}},
      \nonu \\
      & : &  (33,33) \, \mbox{has} \,
   -\frac{(\sqrt{5}+5 i \sqrt{3})}{5 \sqrt{2}},
   \qquad
      (12,12), (21,21), (28,28) \, \mbox{has} \,
    -\frac{(3 \sqrt{5}-5 i \sqrt{3})}{15 \sqrt{2}},
      \nonu \\
      & : &  (1,1), (2,2), (3,3), (4,4), (5,5), (6,6), (7,7),
      (8,8),(9,9),(10,10)
      \, \mbox{has} \,
  \frac{(3 \sqrt{5}+5 i \sqrt{3})}{10 \sqrt{2}} ,
      \nonu \\
      & : &  (11,11), (13,13), (15,15), (20,20), (22,22), (27,27)
      \, \mbox{has} \,
 \frac{(9 \sqrt{5}-5 i \sqrt{3})}{30 \sqrt{2}},
 \nonu \\
 & : &  (17,17), (24,24), (29,29)
      \, \mbox{has} \,
 \frac{(9 \sqrt{5}-25 i \sqrt{3})}{30 \sqrt{2}}.
 \nonu
      \eea}
The remaining generators can be obtained from these
by taking the transpose and the complex conjugation.
One has $\mbox{Tr} (T_a T_{a^{\ast}})=\frac{1}{2} (N+2+2)(N+2+3)= 28$
which 
 is the index $l$ of the representation ${\bf \overline{35}}$
of $SU(5)$ \cite{Feger}.

%%%%%%%%%%%%%%%%%%%%%%%%%%%%%%%%%%%%%%%%%%%%%%%%%%%%%%%%%%%%%%%%%%%%%
%%%%%%%%%%%%%%%%%%%%%%%%%%%%%%%%%%%%%%%%%%%%%%%%%%%%%%%%%%%%%%%%%%%%%
\section{ The $24$ $SU(5)$ generators in the mixed $\overline{\bf 40}$
  representation of $SU(5)$ with three boxes}
%appendixD%%%%%%%%%%%%%%%%%%%%%%%%%%%%%%%%%%%%%%%%%%%%%%%%%%%%%%%%%%%%%%%%%%%
%%%%%%%%%%%%%%%%%%%%%%%%%%%%%%%%%%%%%%%%%%%%%%%%%%%%%%%%%%%%%%%%%%%%

The basis for the mixed ($\tiny\yng(2,1)$) tensors can be realized by
\footnote{ The branching rule (Table $A.71$ of \cite{Feger})
  of $SU(5) \rightarrow SU(3) \times SU(2) \times U(1)$ gives
  ${\bf \overline{40}} =\tiny\yng(2,1)= ({\bf 8},{\bf 1})_{6} +
  ({\bf \overline{3}},{\bf 2})_{1}+  ({\bf 6},{\bf 2})_{1}
  + ({\bf 3},{\bf 3})_{-4}+ ({\bf 3}, {\bf 1})_{-4} + ({\bf 1},{\bf 2})_{-9}$
  which can be compared to the relation in (\ref{transtranstrans}).
 For $N \geq 7$, the three box does not contain the complex conjugated
   notation.
   Note that
   the corresponding relation for the $SU(11)$ branching in ${\bf 440}=
   \tiny\yng(2,1)$ is given by
   Table $A.77$ of \cite{Feger}.}
{\small
\bea
\hat{u}_1 & = & \frac{1}{\sqrt{6}} ( 2 E_{112}-E_{121}-E_{211}),
\qquad
\hat{u}_2  =  \frac{1}{\sqrt{6}} ( 2 E_{113}-E_{131}-E_{311}),
\nonu \\
\hat{u}_3 & = & \frac{1}{\sqrt{6}} ( 2 E_{114}-E_{141}-E_{411}),
\qquad
\hat{u}_4  =  \frac{1}{\sqrt{6}} ( 2 E_{115}-E_{151}-E_{511}),
\nonu \\
\hat{u}_5 & = & \frac{1}{\sqrt{6}} ( E_{122}+E_{212}-2 E_{221}),
\qquad
\hat{u}_6  =  \frac{1}{\sqrt{6}} ( E_{133}+E_{313}-2 E_{331}),
\nonu \\
\hat{u}_7 & = & \frac{1}{\sqrt{6}} (  E_{144}+E_{414}-2 E_{441}),
\qquad
\hat{u}_8  =  \frac{1}{\sqrt{6}} (  E_{155}+E_{515}-2 E_{551}),
\nonu \\
\hat{u}_9 & = & \frac{1}{\sqrt{6}} ( 2 E_{223}-E_{232}-E_{322}),
\qquad
\hat{u}_{10}  =  \frac{1}{\sqrt{6}} ( 2 E_{224}-E_{242}-E_{422}),
\nonu \\
\hat{u}_{11} & = & \frac{1}{\sqrt{6}} ( 2 E_{225}-E_{252}-E_{522}),
\qquad
\hat{u}_{12}  =  \frac{1}{\sqrt{6}} ( E_{233}+E_{323}-2 E_{332}),
\nonu \\
\hat{u}_{13} & = & \frac{1}{\sqrt{6}} (  E_{244}+E_{424}-2 E_{442}),
\qquad
\hat{u}_{14}  =  \frac{1}{\sqrt{6}} (  E_{255}+E_{525}-2 E_{552}),
\nonu \\
\hat{u}_{15} & = & \frac{1}{\sqrt{6}} ( 2 E_{334}-E_{343}-E_{433}),
\qquad
\hat{u}_{16}  =  \frac{1}{\sqrt{6}} ( 2 E_{335}-E_{353}-E_{533}),
\nonu \\
\hat{u}_{17} & = & \frac{1}{\sqrt{6}} (  E_{344}+E_{434}-2 E_{443}),
\qquad
\hat{u}_{18}  =  \frac{1}{\sqrt{6}} (  E_{355}+E_{535}-2 E_{553}),
\nonu \\
\hat{u}_{19} & = & \frac{1}{\sqrt{6}} ( 2 E_{445}-E_{454}-E_{544}),
\qquad
\hat{u}_{20}  =  \frac{1}{\sqrt{6}} (  E_{455}+E_{545}-2 E_{554}),
\nonu \\
\hat{u}_{21} & = & \frac{1}{\sqrt{12}} ( 2 E_{123}-E_{132}+2 E_{213}-E_{231}-
E_{312}-E_{321}),
\nonu \\
\hat{u}_{22} & = &  \frac{1}{\sqrt{12}} ( - E_{123}+2 E_{132}-E_{213}-E_{231}+
2 E_{312}-E_{321}),
\label{basismixed} \\
\hat{u}_{23} & = & \frac{1}{\sqrt{12}} ( 2 E_{124}-E_{142}+2 E_{214}-E_{241}-
E_{412}-E_{421}),
\hat{u}_{24}  =  \frac{1}{2} (  E_{142}-E_{241}+E_{412}-E_{421}),
\nonu \\
\hat{u}_{25} & = & \frac{1}{\sqrt{12}} ( 2 E_{125}-E_{152}+2 E_{215}-E_{251}-
E_{512}-E_{521}),
\hat{u}_{26}  =  \frac{1}{2} (  E_{152}-E_{251}+E_{512}-E_{521}),
\nonu \\
\hat{u}_{27} & = & \frac{1}{\sqrt{12}} ( 2 E_{134}-E_{143}+2 E_{314}-E_{341}-
E_{413}-E_{431}),
\hat{u}_{28}  =  \frac{1}{2} (  E_{143}-E_{341}+E_{413}-E_{431}),
\nonu \\
\hat{u}_{29} & = & \frac{1}{\sqrt{12}} ( 2 E_{135}-E_{153}+2 E_{315}-E_{351}-
E_{513}-E_{531}),
\hat{u}_{30}  =  \frac{1}{2} (  E_{153}-E_{351}+E_{513}-E_{531}),
\nonu \\
\hat{u}_{31} & = & \frac{1}{\sqrt{12}} ( 2 E_{145}-E_{154}+2 E_{415}-E_{451}-
E_{514}-E_{541}),
\hat{u}_{32}  =  \frac{1}{2} (  E_{154}-E_{451}+E_{514}-E_{541}),
\nonu \\
\hat{u}_{33} & = & \frac{1}{\sqrt{12}} ( 2 E_{245}-E_{254}+2 E_{425}-E_{452}-
E_{524}-E_{542}),
\hat{u}_{34}  =  \frac{1}{2} (  E_{254}-E_{452}+E_{524}-E_{542}),
\nonu \\
\hat{u}_{35} & = & \frac{1}{\sqrt{12}} ( - E_{234}-E_{243}- E_{324}+2 E_{342}-
E_{423}+ 2 E_{432}),
\hat{u}_{36}  =  \frac{1}{2} (  E_{423}-E_{324}-E_{234}+E_{243}),
\nonu \\
\hat{u}_{37} & = & \frac{1}{\sqrt{12}} ( - E_{235}-E_{253}- E_{325}+ 2 E_{352}-
E_{523}+2 E_{532}),
\hat{u}_{38}  =  \frac{1}{2} (  E_{523}-E_{325}+E_{253}-E_{235}),
\nonu \\
\hat{u}_{39} & = & \frac{1}{\sqrt{12}} ( - E_{345}-E_{354}- E_{435}+ 2 E_{453}-
E_{534}+ 2 E_{543}),
\hat{u}_{40}  =  \frac{1}{2} (  E_{534}-E_{435}+E_{354}-E_{345}).
\nonu
\eea}
It is not obvious to see these
basis at first sight. There is other way to obtain
these basis by using the
roots and weights \cite{Semenoff}.
After applying the $24$ generators which are $5 \times 5$  matrices
in (\ref{fund5by5}) to
(\ref{basismixed}),
the $24$ generators in terms of  $40 \times 40$
unitary matrix are given as follows. We present only the nonzero elements. 
{\small
  \bea
  T_1  & : &
 (10,5), (15,6), (34,26) \, \mbox{has} \, -1,
      \qquad
      (35,21),(39,29), (40,30) \, \mbox{has} \, -\frac{1}{2},
      \nonu \\
      & : &  (36,22) \, \mbox{has} \, \frac{1}{2},
      \qquad
        (7,3), (20,8), (33,25) \, \mbox{has} \, 1,
      \nonu \\
      & : &  (13,24), (17,28)
      \, \mbox{has} \, -\sqrt{\frac{3}{2}},
      \qquad
         (24,1),(28,2) \, \mbox{has} \, \sqrt{\frac{3}{2}},
      \nonu \\
      & : &  (23,1), (27,2) \, \mbox{has} \, -\frac{1}{\sqrt{2}},
      \qquad
        (13,23), (17,27) \, \mbox{has} \, \frac{1}{\sqrt{2}},
      \nonu \\
      & : &  (19,31), (31,4) \, \mbox{has} \, \sqrt{2},
      \qquad
         (40,29) \, \mbox{has} \,
      -\frac{\sqrt{3}}{2},
  \nonu \\
      & : &  (35,22),(36,21),(39,30) \, \mbox{has} \,
      \frac{\sqrt{3}}{2},
  \nonu \\
   T_2  & : &
 (15,12) \, \mbox{has} \, -1,
      \qquad
        (27,21),(39,37) \, \mbox{has} \, -\frac{1}{2},
      \nonu \\
      & : &  (28,22), (40,38) \, \mbox{has} \, \frac{1}{2},
      \qquad
       (3,1), (13,10), (20,14), (31,25), (32,26) \, \mbox{has} \, 1,
      \nonu \\
      & : &  (17,36)
      \, \mbox{has} \, -\sqrt{\frac{3}{2}},
      \qquad
         (7,24),(24,5),(36,9) \, \mbox{has} \, \sqrt{\frac{3}{2}},
      \nonu \\
      & : &  (35,9) \, \mbox{has} \, -\frac{1}{\sqrt{2}},
      \qquad
        (7,23), (17,35), (23,5) \, \mbox{has} \, \frac{1}{\sqrt{2}},
      \nonu \\
      & : &  (19,33), (33,11) \, \mbox{has} \, \sqrt{2},
      \qquad
        (27,22),(28,21),(39,38),(40,37) \, \mbox{has} \,
      \frac{\sqrt{3}}{2},
   \nonu \\
   T_3  & : &
    (33,37) \, \mbox{has} \, -\frac{1}{2},
      \qquad
        (34,38) \, \mbox{has} \, \frac{1}{2},
      \nonu \\
      & : &  (3,2), (10,9), (17,15), (20,18), (23,21), (24,22),
      (31,29), (32,30)\, \mbox{has} \, 1,
      \nonu \\
           & : &  (19,40), (40,16)
      \, \mbox{has} \, -\sqrt{\frac{3}{2}},
      \qquad
         (7,28),(28,6) \, \mbox{has} \, \sqrt{\frac{3}{2}},
      \nonu \\
      & : &  (19,39), (39,16) \, \mbox{has} \, -\frac{1}{\sqrt{2}},
      \qquad
       (7,27), (27,6) \, \mbox{has} \, \frac{1}{\sqrt{2}},
      \nonu \\
 & : &  (13,35), (35,12) \, \mbox{has} \, -\sqrt{2},
      \qquad
        (33,38), (34,37) \, \mbox{has} \, -\frac{\sqrt{3}}{2},
      \nonu \\
           T_4  & : &
   (11,5), (16,6), (19,7) \, \mbox{has} \, -1,
      \qquad
        (33,23),(34,24),(37,21),(39,27) \, \mbox{has} \, -\frac{1}{2},
      \nonu \\
      & : &  (38,22), (40,28) \, \mbox{has} \, \frac{1}{2},
      \qquad
        (8,4) \, \mbox{has} \, 1,
      \nonu \\
      & : &  (14,26), (18,30), (20,32)
      \, \mbox{has} \, -\sqrt{\frac{3}{2}},
      \qquad
         (26,1),(30,2),(32,3) \, \mbox{has} \, \sqrt{\frac{3}{2}},
      \nonu \\
      & : &  (25,1), (29,2), (31,3) \, \mbox{has} \, -\frac{1}{\sqrt{2}},
      \qquad
        (14,25), (18,29), (20,31) \, \mbox{has} \, \frac{1}{\sqrt{2}},
      \nonu \\
      & : &  (33,24) \, \mbox{has} \, -\frac{\sqrt{3}}{2},
      \qquad
        (34,23),(37,22),(38,21),(39,28),(40,27) \, \mbox{has} \, \frac{\sqrt{3}}{2},
     \nonu \\
     T_5  
      & : &  (16,12), (19,13) \, \mbox{has} \, -1,
      \qquad
        (29,21),(31,23),(39,35),(40,36) \, \mbox{has} \, -\frac{1}{2},
      \nonu \\
      & : &  (30,22), (32,24) \, \mbox{has} \, \frac{1}{2},
      \qquad
        (4,1), (14,11) \, \mbox{has} \, 1,
      \nonu \\
      & : &  (18,38), (20,34) \, \mbox{has} \, -\sqrt{\frac{3}{2}},
      \qquad
        (8,26),(26,5),(34,10),(38,9) \, \mbox{has} \, \sqrt{\frac{3}{2}},
      \nonu \\
      & : &  (33,10), (37,9) \, \mbox{has} \, -\frac{1}{\sqrt{2}},
      \qquad
        (8,25), (18,37), (20,33), (25,5) \, \mbox{has} \, \frac{1}{\sqrt{2}},
      \nonu \\
      & : &  (40,35) \, \mbox{has} \, -\frac{\sqrt{3}}{2},
      \qquad
        (29,22),(30,21),(31,24),(32,23),(39,36) \, \mbox{has} \, \frac{\sqrt{3}}{2},
      \nonu \\
       T_6  & : &  (19,17) \, \mbox{has} \, -1,
       \qquad
          (31,27), (33,35), (34,36) \, \mbox{has} \, -\frac{1}{2},
       \nonu \\
  & : &  (32,28) \, \mbox{has} \, \frac{1}{2},
       \qquad
  (4,2), (11,9), (18,16), (25,21), (26,22) \, \mbox{has} \, 1,
       \nonu \\
  & : &  (20,40) \, \mbox{has} \,  -\sqrt{\frac{3}{2}},
       \qquad
   (8,30), (30,6), (40,15) \, \mbox{has} \,  \sqrt{\frac{3}{2}},
       \nonu \\
 & : &  (39,15) \, \mbox{has} \,  -\frac{1}{\sqrt{2}},
       \qquad
   (8,29), (20,39), (29,6) \, \mbox{has} \,  \frac{1}{\sqrt{2}},
       \nonu \\
 & : &  (14,37), (37,12) \, \mbox{has} \,  -\sqrt{2},
       \qquad
   (34,35) \, \mbox{has} \,  -\frac{\sqrt{3}}{2},
       \nonu \\
 & : &  (31,28), (32,27), (33,36) \, \mbox{has} \,  \frac{\sqrt{3}}{2},
       \nonu \\
  T_7  & : &  (4,3), (11,10), (16,15), (20,19), (25,23), (26,24),
  (29,27), (30,28), (37,35), (38,36) \, \mbox{has} \, 1,
  \nonu \\
  & : &  (8,32), (14,34), (32,7), (34,13)
   \, \mbox{has} \, \sqrt{\frac{3}{2}},
   \qquad
    (8,31), (14,33), (31,7), (33,13)
   \, \mbox{has} \, \frac{1}{\sqrt{2}},
   \nonu \\
   & : &  (18,39), (39,17)
   \, \mbox{has} \, -\sqrt{2},
   \nonu \\
  T_8  & : &  (35,27), (37,29) \, \mbox{has} \, -\frac{1}{2},
  \qquad
   (36,28), (38,30) \, \mbox{has} \, \frac{1}{2},
  \nonu \\
  & : &  (5,1), (12,6), (13,7), (14,8), (33,31), (34,32) \, \mbox{has} \,
  1,
  \nonu \\
 & : &  (9,21), (10,23), (11,25), (21,2), (23,3), (25,4) \, \mbox{has} \,
  \sqrt{2},
  \nonu \\
  & : &  (35,28), (36,27), (37,30), (38,29)
   \, \mbox{has} \,
  -\frac{\sqrt{3}}{2},
  \nonu \\
T_9  & : &  (9,5) \, \mbox{has} \, -1,
\qquad
  (35,23), (36,24), (37,25), (38,26), (39,31) \, \mbox{has} \,
-\frac{1}{2},
\nonu \\
 & : &  (40,32) \, \mbox{has} \, \frac{1}{2},
\qquad
   (6,2), (17,7), (18,8) \, \mbox{has} \, 1,
\nonu \\
& : &  (12,22) \, \mbox{has} \, -\sqrt{\frac{3}{2}},
\qquad
  (22,1) \, \mbox{has} \, \sqrt{\frac{3}{2}},
\nonu \\
& : &  (21,1) \, \mbox{has} \, -\frac{1}{\sqrt{2}},
\qquad
  (12,21) \, \mbox{has} \, \frac{1}{\sqrt{2}},
\nonu \\
& : &  (15,27), (16,29), (27,3), (29,4)
\, \mbox{has} \, \sqrt{2},
\nonu \\
& : &  (36,23), (38,25), (39,32), (40,31)
\, \mbox{has} \, -\frac{\sqrt{3}}{2},
\qquad
  (35,24), (37,26)
\, \mbox{has} \, \frac{\sqrt{3}}{2},
\nonu \\
  T_{10} & : &  (39,33) \, \mbox{has} \, -\frac{1}{2},
\qquad
  (40,34) \, \mbox{has} \, \frac{1}{2},
\nonu \\
& :  & (2,1), (12,9),(17,13),(18,14),(27,23),(28,24),(29,25),(30,26)
 \, \mbox{has} \, 1,
 \nonu \\
& : & (15,36),(16,38),(36,10),(38,11) \, \mbox{has} \, -\sqrt{\frac{3}{2}},
 \nonu \\
& : & (6,22),(22,5) \, \mbox{has} \, \sqrt{\frac{3}{2}},
 \qquad
 (6,21),(21,5) \, \mbox{has} \, \frac{1}{\sqrt{2}},
 \nonu \\
 & : & (15,35),(16,37), (35,10), (37,11) \,
 \mbox{has} \, -\frac{1}{\sqrt{2}},
 \nonu \\
& : & (39,34),(40,33) \, \mbox{has} \, -\frac{\sqrt{3}}{2},
 \nonu \\
  T_{11} & : & (9,9) \, \mbox{has} \, -i,
\qquad
(2,2) 
\, \mbox{has} \, i,
   \nonu \\
& : & (15,15),(16,16) \,
   \mbox{has} \,
   -\frac{2}{\sqrt{3}},
  \qquad
   (17,17), (18,18), (39,39), (40,40) \, \mbox{has} \,
-\frac{1}{\sqrt{3}},
  \nonu \\
   & : & (23,23), (24,24), (25,25), (26,26) \, \mbox{has} \,
\frac{1}{\sqrt{3}},
 \nonu \\
 & : & (12,12) \, \mbox{has} \,
\frac{1}{2} (-\sqrt{3}-i),
\qquad
(6,6) \, \mbox{has} \,
\frac{1}{2} (-\sqrt{3}+i),
 \nonu \\
 & : & (35,35), (36,36), (37,37), (38,38)
  \, \mbox{has} \,
\frac{1}{6} (-\sqrt{3}-3 i),
\nonu \\
 & : & (27,27), (28,28), (29,29), (30,30)
  \, \mbox{has} \,
\frac{1}{6} \left(-\sqrt{3}+3 i\right),
\nonu \\
 & : & (5,5)
\, \mbox{has} \,
\frac{1}{2} (\sqrt{3}-i),
\qquad
  (1,1)
\, \mbox{has} \,
\frac{1}{2} (\sqrt{3}+i),
\nonu \\
 & : & (13,13), (14,14), (33,33), (34,34)
\, \mbox{has} \,
\frac{1}{6} (\sqrt{3}-3 i),
\qquad
 (10,10), (11,11)
\, \mbox{has} \,
\frac{1}{3} (\sqrt{3}-3 i),
\nonu \\
 & : & (7,7), (8,8), (31,31), (32,32)
\, \mbox{has} \,
\frac{1}{6} (\sqrt{3}+3 i),
\qquad
  (3,3), (4,4)
\, \mbox{has} \,
\frac{1}{3} (\sqrt{3}+3 i),
\nonu \\
T_{12} & : & (19,19) \, \mbox{has} \,
-\frac{i (5 \sqrt{3}-i \sqrt{5})}{5 \sqrt{2}},
\nonu \\
& : & (31,31),(32,32),(33,33),(34,34),(39,39),(40,40) 
\, \mbox{has} \,
-\frac{i (5 \sqrt{3}-3 i \sqrt{5})}{15 \sqrt{2}},
   \nonu \\
& : & (4,4),(11,11),(16,16),(25,25),(26,26),(29,29) 
   (30,30),(37,37), (38,38) \,
   \nonu \\
   & & \mbox{has} \, 
\frac{i (5 \sqrt{3}+3 i \sqrt{5})}{15 \sqrt{2}},
   \nonu   
  \\
  & : & (20,20) \, \mbox{has} \,
 -\frac{i (5 \sqrt{3}-7 i \sqrt{5})}{10 \sqrt{2}},
  \qquad
    (8,8), (14,14), (18,18) \, \mbox{has} \,
\frac{i (5 \sqrt{3}+21 i \sqrt{5})}{30 \sqrt{2}},
 \nonu \\
 & : & (1,1), (2,2), (5,5), (6,6), (9,9), (12,12), (21,21),
 (22,22) \, \mbox{has} \,
\frac{(3 \sqrt{5}+5 i \sqrt{3})}{10 \sqrt{2}},
 \nonu \\
 & : & (3,3), (10,10), (15,15), (23,23), (24,24), (27,27),
 (28,28),
 (35,35), (36,36) \,
 \nonu \\
 && \mbox{has} \,
\frac{(9 \sqrt{5}-5 i \sqrt{3})}{30 \sqrt{2}},
 \nonu \\
 & : & (7,7), (13,13), (17,17)
  \, \mbox{has} \,
\frac{(9 \sqrt{5}-25 i \sqrt{3})}{30 \sqrt{2}}.
 \nonu
  \eea}
The remaining generators can be obtained similarly. 
One has $\mbox{Tr} (T_a T_{a^{\ast}})=(N+2)^2-3=22$
which 
 is the index $l$ of the representation ${\bf \overline{40}}$
of $SU(5)$ \cite{Feger}.

%%%%%%%%%%%%%%%%%%%%%%%%%%%%%%%%%%%%%%%%%%%%%%%%%%%%%%%%%%%%%%%%%%%%%
%%%%%%%%%%%%%%%%%%%%%%%%%%%%%%%%%%%%%%%%%%%%%%%%%%%%%%%%%%%%%%%%%%%%%
\section{ The $24$ $SU(5)$ generators in the antisymmetric
  $\overline{\bf 10} $
  representation of $SU(5)$ with three boxes }
%appendixE%%%%%%%%%%%%%%%%%%%%%%%%%%%%%%%%%%%%%%%%%%%%%%%%%%%%%%%%%%%%%%%%%%%
%%%%%%%%%%%%%%%%%%%%%%%%%%%%%%%%%%%%%%%%%%%%%%%%%%%%%%%%%%%%%%%%%%%%

%%%
%%%%%%%%%%%%%%%%%%%%%%%%%%%%%%%%%%%%%%%%%%%%%%%%%%%%%%%%%%%%%%%%%%%%%%%%%%

The basis for the mixed ($\tiny\yng(1,1,1)$) tensors can be realized by
\footnote{ The branching rule
  (Table $A.71$ of \cite{Feger})
  of $SU(5) \rightarrow SU(3) \times SU(2) \times U(1)$ gives
  ${\bf \overline{10}} =\tiny\yng(1,1,1)= ({\bf 1},{\bf 1})_{6} +
  ({\bf \overline{3}},{\bf 2})_{1}+  ({\bf 3},{\bf 1})_{-4}
 $
  which can be compared to the relation in (\ref{transtranstrans}).
 For $N \geq 7$, the three box does not contain the complex conjugated
   notation.
   Note that
   the corresponding relation for the $SU(11)$ branching in ${\bf 165}=
   \tiny\yng(1,1,1)$ is given by
   Table $A.77$ of \cite{Feger}.}
\bea
\hat{u}_1   & = &   \frac{1}{\sqrt{6}} (E_{123}-E_{132}+E_{312}-E_{213}+E_{231}-
E_{321}),
\nonu \\
\hat{u}_2   & = &  \frac{1}{\sqrt{6}} (E_{124}-E_{142}+E_{412}-E_{214}+E_{241}-
E_{421}), 
\nonu \\
\hat{u}_3   & = &   \frac{1}{\sqrt{6}} (E_{125}-E_{152}+E_{512}-E_{215}+E_{251}-
E_{521}),
\nonu \\
\hat{u}_4   & = &  \frac{1}{\sqrt{6}} (E_{134}-E_{143}+E_{413}-E_{314}+E_{341}-
E_{431}), 
\nonu \\
\hat{u}_5   & = &   \frac{1}{\sqrt{6}} (E_{135}-E_{153}+E_{513}-E_{315}+E_{351}-
E_{531}),
\nonu \\
\hat{u}_6   & = &  \frac{1}{\sqrt{6}} (E_{145}-E_{154}+E_{514}-E_{415}+E_{451}-
E_{541}), 
\nonu \\
\hat{u}_7   & = &   \frac{1}{\sqrt{6}} (E_{234}-E_{243}+E_{423}-E_{324}+E_{342}-
E_{432}),
\nonu \\
\hat{u}_8   & = &  \frac{1}{\sqrt{6}} (E_{235}-E_{253}+E_{523}-E_{325}+E_{352}-
E_{532}), 
\nonu \\
\hat{u}_9   & = &   \frac{1}{\sqrt{6}} (E_{245}-E_{254}+E_{524}-E_{425}+E_{452}-
E_{542}),
\nonu \\
\hat{u}_{10}   & = &  \frac{1}{\sqrt{6}} (E_{345}-E_{354}+E_{534}-E_{435}+E_{453}-
E_{543}). 
\label{basisantisymm3}
\eea
After applying the $24$ generators which are $5 \times 5$  matrices
in (\ref{fund5by5}) to
(\ref{basisantisymm3}),
the $24$ generators in terms of  $10 \times 10$
unitary matrix are given as follows as done in previous Appendices:
{\small
\bea
T_1 & = & \left(
\begin{array}{cccccccccc}
 0 & 0 & 0 & 0 & 0 & 0 & 0 & 0 & 0 & 0 \\
 0 & 0 & 0 & 0 & 0 & 0 & 0 & 0 & 0 & 0 \\
 0 & 0 & 0 & 0 & 0 & 0 & 0 & 0 & 0 & 0 \\
 0 & 0 & 0 & 0 & 0 & 0 & 0 & 0 & 0 & 0 \\
 0 & 0 & 0 & 0 & 0 & 0 & 0 & 0 & 0 & 0 \\
 0 & 0 & 0 & 0 & 0 & 0 & 0 & 0 & 0 & 0 \\
 1 & 0 & 0 & 0 & 0 & 0 & 0 & 0 & 0 & 0 \\
 0 & 0 & 0 & 0 & 0 & 0 & 0 & 0 & 0 & 0 \\
 0 & 0 & -1 & 0 & 0 & 0 & 0 & 0 & 0 & 0 \\
 0 & 0 & 0 & 0 & -1 & 0 & 0 & 0 & 0 & 0 \\
\end{array}
\right), 
T_2=\left(
\begin{array}{cccccccccc}
 0 & 0 & 0 & 0 & 0 & 0 & 0 & 0 & 0 & 0 \\
 0 & 0 & 0 & 0 & 0 & 0 & 0 & 0 & 0 & 0 \\
 0 & 0 & 0 & 0 & 0 & 0 & 0 & 0 & 0 & 0 \\
 -1 & 0 & 0 & 0 & 0 & 0 & 0 & 0 & 0 & 0 \\
 0 & 0 & 0 & 0 & 0 & 0 & 0 & 0 & 0 & 0 \\
 0 & 0 & 1 & 0 & 0 & 0 & 0 & 0 & 0 & 0 \\
 0 & 0 & 0 & 0 & 0 & 0 & 0 & 0 & 0 & 0 \\
 0 & 0 & 0 & 0 & 0 & 0 & 0 & 0 & 0 & 0 \\
 0 & 0 & 0 & 0 & 0 & 0 & 0 & 0 & 0 & 0 \\
 0 & 0 & 0 & 0 & 0 & 0 & 0 & -1 & 0 & 0 \\
\end{array}
\right),
\nonu \\
T_3 & = & \left(
\begin{array}{cccccccccc}
 0 & 0 & 0 & 0 & 0 & 0 & 0 & 0 & 0 & 0 \\
 1 & 0 & 0 & 0 & 0 & 0 & 0 & 0 & 0 & 0 \\
 0 & 0 & 0 & 0 & 0 & 0 & 0 & 0 & 0 & 0 \\
 0 & 0 & 0 & 0 & 0 & 0 & 0 & 0 & 0 & 0 \\
 0 & 0 & 0 & 0 & 0 & 0 & 0 & 0 & 0 & 0 \\
 0 & 0 & 0 & 0 & 1 & 0 & 0 & 0 & 0 & 0 \\
 0 & 0 & 0 & 0 & 0 & 0 & 0 & 0 & 0 & 0 \\
 0 & 0 & 0 & 0 & 0 & 0 & 0 & 0 & 0 & 0 \\
 0 & 0 & 0 & 0 & 0 & 0 & 0 & 1 & 0 & 0 \\
 0 & 0 & 0 & 0 & 0 & 0 & 0 & 0 & 0 & 0 \\
\end{array}
\right),
T_4 = \left(
\begin{array}{cccccccccc}
 0 & 0 & 0 & 0 & 0 & 0 & 0 & 0 & 0 & 0 \\
 0 & 0 & 0 & 0 & 0 & 0 & 0 & 0 & 0 & 0 \\
 0 & 0 & 0 & 0 & 0 & 0 & 0 & 0 & 0 & 0 \\
 0 & 0 & 0 & 0 & 0 & 0 & 0 & 0 & 0 & 0 \\
 0 & 0 & 0 & 0 & 0 & 0 & 0 & 0 & 0 & 0 \\
 0 & 0 & 0 & 0 & 0 & 0 & 0 & 0 & 0 & 0 \\
 0 & 0 & 0 & 0 & 0 & 0 & 0 & 0 & 0 & 0 \\
 1 & 0 & 0 & 0 & 0 & 0 & 0 & 0 & 0 & 0 \\
 0 & 1 & 0 & 0 & 0 & 0 & 0 & 0 & 0 & 0 \\
 0 & 0 & 0 & 1 & 0 & 0 & 0 & 0 & 0 & 0 \\
\end{array}
\right),
\nonu \\
T_5 & = &\left(
\begin{array}{cccccccccc}
 0 & 0 & 0 & 0 & 0 & 0 & 0 & 0 & 0 & 0 \\
 0 & 0 & 0 & 0 & 0 & 0 & 0 & 0 & 0 & 0 \\
 0 & 0 & 0 & 0 & 0 & 0 & 0 & 0 & 0 & 0 \\
 0 & 0 & 0 & 0 & 0 & 0 & 0 & 0 & 0 & 0 \\
 -1 & 0 & 0 & 0 & 0 & 0 & 0 & 0 & 0 & 0 \\
 0 & -1 & 0 & 0 & 0 & 0 & 0 & 0 & 0 & 0 \\
 0 & 0 & 0 & 0 & 0 & 0 & 0 & 0 & 0 & 0 \\
 0 & 0 & 0 & 0 & 0 & 0 & 0 & 0 & 0 & 0 \\
 0 & 0 & 0 & 0 & 0 & 0 & 0 & 0 & 0 & 0 \\
 0 & 0 & 0 & 0 & 0 & 0 & 1 & 0 & 0 & 0 \\
\end{array}
\right), 
T_6  = 
\left(
\begin{array}{cccccccccc}
 0 & 0 & 0 & 0 & 0 & 0 & 0 & 0 & 0 & 0 \\
 0 & 0 & 0 & 0 & 0 & 0 & 0 & 0 & 0 & 0 \\
 1 & 0 & 0 & 0 & 0 & 0 & 0 & 0 & 0 & 0 \\
 0 & 0 & 0 & 0 & 0 & 0 & 0 & 0 & 0 & 0 \\
 0 & 0 & 0 & 0 & 0 & 0 & 0 & 0 & 0 & 0 \\
 0 & 0 & 0 & -1 & 0 & 0 & 0 & 0 & 0 & 0 \\
 0 & 0 & 0 & 0 & 0 & 0 & 0 & 0 & 0 & 0 \\
 0 & 0 & 0 & 0 & 0 & 0 & 0 & 0 & 0 & 0 \\
 0 & 0 & 0 & 0 & 0 & 0 & -1 & 0 & 0 & 0 \\
 0 & 0 & 0 & 0 & 0 & 0 & 0 & 0 & 0 & 0 \\
\end{array}
\right), 
\nonu \\
T_7 & = &
\left(
\begin{array}{cccccccccc}
 0 & 0 & 0 & 0 & 0 & 0 & 0 & 0 & 0 & 0 \\
 0 & 0 & 0 & 0 & 0 & 0 & 0 & 0 & 0 & 0 \\
 0 & 1 & 0 & 0 & 0 & 0 & 0 & 0 & 0 & 0 \\
 0 & 0 & 0 & 0 & 0 & 0 & 0 & 0 & 0 & 0 \\
 0 & 0 & 0 & 1 & 0 & 0 & 0 & 0 & 0 & 0 \\
 0 & 0 & 0 & 0 & 0 & 0 & 0 & 0 & 0 & 0 \\
 0 & 0 & 0 & 0 & 0 & 0 & 0 & 0 & 0 & 0 \\
 0 & 0 & 0 & 0 & 0 & 0 & 1 & 0 & 0 & 0 \\
 0 & 0 & 0 & 0 & 0 & 0 & 0 & 0 & 0 & 0 \\
 0 & 0 & 0 & 0 & 0 & 0 & 0 & 0 & 0 & 0 \\
\end{array}
\right),
T_8 = \left(
\begin{array}{cccccccccc}
 0 & 0 & 0 & 0 & 0 & 0 & 0 & 0 & 0 & 0 \\
 0 & 0 & 0 & 0 & 0 & 0 & 0 & 0 & 0 & 0 \\
 0 & 0 & 0 & 0 & 0 & 0 & 0 & 0 & 0 & 0 \\
 0 & 0 & 0 & 0 & 0 & 0 & 0 & 0 & 0 & 0 \\
 0 & 0 & 0 & 0 & 0 & 0 & 0 & 0 & 0 & 0 \\
 0 & 0 & 0 & 0 & 0 & 0 & 0 & 0 & 0 & 0 \\
 0 & 0 & 0 & 1 & 0 & 0 & 0 & 0 & 0 & 0 \\
 0 & 0 & 0 & 0 & 1 & 0 & 0 & 0 & 0 & 0 \\
 0 & 0 & 0 & 0 & 0 & 1 & 0 & 0 & 0 & 0 \\
 0 & 0 & 0 & 0 & 0 & 0 & 0 & 0 & 0 & 0 \\
\end{array}
\right),
\nonu \\
T_9 & = &
\left(
\begin{array}{cccccccccc}
 0 & 0 & 0 & 0 & 0 & 0 & 0 & 0 & 0 & 0 \\
 0 & 0 & 0 & 0 & 0 & 0 & 0 & 0 & 0 & 0 \\
 0 & 0 & 0 & 0 & 0 & 0 & 0 & 0 & 0 & 0 \\
 0 & 0 & 0 & 0 & 0 & 0 & 0 & 0 & 0 & 0 \\
 0 & 0 & 0 & 0 & 0 & 0 & 0 & 0 & 0 & 0 \\
 0 & 0 & 0 & 0 & 0 & 0 & 0 & 0 & 0 & 0 \\
 0 & -1 & 0 & 0 & 0 & 0 & 0 & 0 & 0 & 0 \\
 0 & 0 & -1 & 0 & 0 & 0 & 0 & 0 & 0 & 0 \\
 0 & 0 & 0 & 0 & 0 & 0 & 0 & 0 & 0 & 0 \\
 0 & 0 & 0 & 0 & 0 & 1 & 0 & 0 & 0 & 0 \\
\end{array}
\right), 
T_{10} =\left(
\begin{array}{cccccccccc}
 0 & 0 & 0 & 0 & 0 & 0 & 0 & 0 & 0 & 0 \\
 0 & 0 & 0 & 0 & 0 & 0 & 0 & 0 & 0 & 0 \\
 0 & 0 & 0 & 0 & 0 & 0 & 0 & 0 & 0 & 0 \\
 0 & 1 & 0 & 0 & 0 & 0 & 0 & 0 & 0 & 0 \\
 0 & 0 & 1 & 0 & 0 & 0 & 0 & 0 & 0 & 0 \\
 0 & 0 & 0 & 0 & 0 & 0 & 0 & 0 & 0 & 0 \\
 0 & 0 & 0 & 0 & 0 & 0 & 0 & 0 & 0 & 0 \\
 0 & 0 & 0 & 0 & 0 & 0 & 0 & 0 & 0 & 0 \\
 0 & 0 & 0 & 0 & 0 & 0 & 0 & 0 & 0 & 0 \\
 0 & 0 & 0 & 0 & 0 & 0 & 0 & 0 & 1 & 0 \\
\end{array}
\right),
\nonu \\
T_{11}  & = & \mbox{diag} \left( 0,\frac{1}{\sqrt{3}} ,
\frac{1}{\sqrt{3}},
\frac{1}{6} (-\sqrt{3}+3 i),
\frac{1}{6} (-\sqrt{3}+3 i),
\right.
\nonu \\
&& \left. \frac{1}{6} (\sqrt{3}+3 i),
\frac{1}{6} (-\sqrt{3}-3 i),
\frac{1}{6} (-\sqrt{3}-3 i),
\frac{1}{6} (\sqrt{3}-3 i),
-\frac{1}{\sqrt{3}}\right),
\nonu \\
T_{12} & = & \mbox{diag} \left(\frac{1}{20}
(3 \sqrt{10}+5 i \sqrt{6}),
\frac{1}{60} (9 \sqrt{10}-5 i \sqrt{6}),
-\frac{1}{\sqrt{10}}+\frac{i}{\sqrt{6}},
\frac{1}{60} (9 \sqrt{10}-5 i \sqrt{6}),
\right.
\nonu \\
& - & \left. \frac{1}{\sqrt{10}}+\frac{i}{\sqrt{6}},
- (\frac{1}{\sqrt{10}}+\frac{i}{\sqrt{6}}),
\frac{1}{60} (9 \sqrt{10}-5 i \sqrt{6}),
-\frac{1}{\sqrt{10}}+\frac{i}{\sqrt{6}},
-\frac{1}{\sqrt{10}}-\frac{i}{\sqrt{6}},
-\frac{1}{\sqrt{10}}-\frac{i}{\sqrt{6}}\right).
\nonu 
\eea}
The remaining generators can be obtained
from these generators by taking the transpose and the complex
conjugation.
One has $\mbox{Tr} (T_a T_{a^{\ast}})=\frac{1}{2}(N+2-2)(N+2-3)=3$
which 
 is the index $l$ of the representation ${\bf \overline{10}}$
of $SU(5)$ \cite{Feger}.
For the generators of $SU(N+2)$ with $N=5,7,9,11$,
the corresponding ones in the higher representations more than three
boxes are rather involved.

%%%%%%%%%%%%%%%%%%%%%%%%%%%%%%%%%%%%%%%%%%%%%%%%%%%%%%%%%%%%%%%%%%%%%
%%%%%%%%%%%%%%%%%%%%%%%%%%%%%%%%%%%%%%%%%%%%%%%%%%%%%%%%%%%%%%%%%%%%%%


\begin{thebibliography}{99}

%\cite{Gaberdiel:2013vva}
\bibitem{GG1305} 
  M.~R.~Gaberdiel and R.~Gopakumar,
  ``Large N=4 Holography,''
  JHEP {\bf 1309}, 036 (2013)
  doi:10.1007/JHEP09(2013)036
  [arXiv:1305.4181 [hep-th]].
  %%CITATION = doi:10.1007/JHEP09(2013)036;%%
  %74 citations counted in INSPIRE as of 03 Nov 2017
  
%\cite{Sevrin:1988ew}
\bibitem{STVplb} 
  A.~Sevrin, W.~Troost and A.~Van Proeyen,
  ``Superconformal Algebras in Two-Dimensions with N=4,''
  Phys.\ Lett.\ B {\bf 208}, 447 (1988).
  doi:10.1016/0370-2693(88)90645-4
  %%CITATION = doi:10.1016/0370-2693(88)90645-4;%%
  %162 citations counted in INSPIRE as of 28 Feb 2017

%\cite{Sevrin:1988ps}
\bibitem{npb1988} 
  A.~Sevrin, W.~Troost, A.~Van Proeyen and P.~Spindel,
  ``EXTENDED SUPERSYMMETRIC sigma MODELS ON GROUP MANIFOLDS. 2. CURRENT ALGEBRAS,''
  Nucl.\ Phys.\ B {\bf 311}, 465 (1988).
  doi:10.1016/0550-3213(88)90070-3
  %%CITATION = doi:10.1016/0550-3213(88)90070-3;%%
  %103 citations counted in INSPIRE as of 28 Feb 2017

%\cite{Schoutens:1988ig}
\bibitem{Schoutens88} 
  K.~Schoutens,
  ``O(n) Extended Superconformal Field Theory in Superspace,''
  Nucl.\ Phys.\ B {\bf 295}, 634 (1988).
  doi:10.1016/0550-3213(88)90539-1
  %%CITATION = doi:10.1016/0550-3213(88)90539-1;%%
  %125 citations counted in INSPIRE as of 28 Feb 2017

%\cite{Sevrin:1989ce}
\bibitem{ST} 
  A.~Sevrin and G.~Theodoridis,
  ``N=4 Superconformal Coset Theories,''
  Nucl.\ Phys.\ B {\bf 332}, 380 (1990).
  doi:10.1016/0550-3213(90)90100-R
  %%CITATION = doi:10.1016/0550-3213(90)90100-R;%%
  %39 citations counted in INSPIRE as of 28 Feb 2017

%\cite{Saulina:2004su}
\bibitem{Saulina} 
  N.~Saulina,
  ``Geometric interpretation of the large N=4 index,''
  Nucl.\ Phys.\ B {\bf 706}, 491 (2005)
  doi:10.1016/j.nuclphysb.2004.11.049
  [hep-th/0409175].
  %%CITATION = doi:10.1016/j.nuclphysb.2004.11.049;%%
  %5 citations counted in INSPIRE as of 28 Feb 2017

%\cite{Gunaydin:1988re}
\bibitem{GPTV} 
  M.~Gunaydin, J.~L.~Petersen, A.~Taormina and A.~Van Proeyen,
  ``On the Unitary Representations of a Class of $N=4$ Superconformal Algebras,''
  Nucl.\ Phys.\ B {\bf 322}, 402 (1989).
  doi:10.1016/0550-3213(89)90421-5
  %%CITATION = doi:10.1016/0550-3213(89)90421-5;%%
  %51 citations counted in INSPIRE as of 10 Nov 2017

%\cite{Petersen:1989zz}
\bibitem{PT1} 
  J.~L.~Petersen and A.~Taormina,
  ``Characters of the $N=4$ Superconformal Algebra With Two Central Extensions,''
  Nucl.\ Phys.\ B {\bf 331}, 556 (1990).
  doi:10.1016/0550-3213(90)90084-Q
  %%CITATION = doi:10.1016/0550-3213(90)90084-Q;%%
  %20 citations counted in INSPIRE as of 10 Nov 2017

%\cite{Petersen:1989pp}
\bibitem{PT2} 
  J.~L.~Petersen and A.~Taormina,
  ``Characters of the $N=4$ Superconformal Algebra With Two Central Extensions: 2. Massless Representations,''
  Nucl.\ Phys.\ B {\bf 333}, 833 (1990).
  doi:10.1016/0550-3213(90)90141-Y
  %%CITATION = doi:10.1016/0550-3213(90)90141-Y;%%
  %26 citations counted in INSPIRE as of 10 Nov 2017

%\cite{Goddard:1988wv}
\bibitem{GS} 
  P.~Goddard and A.~Schwimmer,
  ``Factoring Out Free Fermions and Superconformal Algebras,''
  Phys.\ Lett.\ B {\bf 214}, 209 (1988).
  doi:10.1016/0370-2693(88)91470-0
  %%CITATION = doi:10.1016/0370-2693(88)91470-0;%%
  %80 citations counted in INSPIRE as of 28 Feb 2017
  
%\cite{VanProeyen:1989me}
\bibitem{VP} 
  A.~Van Proeyen,
  ``Realizations of $N=4$ Superconformal Algebras on Wolf Spaces,''
  Class.\ Quant.\ Grav.\  {\bf 6}, 1501 (1989).
  doi:10.1088/0264-9381/6/10/018
  %%CITATION = doi:10.1088/0264-9381/6/10/018;%%
  %45 citations counted in INSPIRE as of 28 Feb 2017

%\cite{Gates:1995ip}
\bibitem{GK} 
  S.~J.~Gates, Jr. and S.~V.~Ketov,
  ``No N=4 strings on wolf spaces,''
  Phys.\ Rev.\ D {\bf 52}, 2278 (1995)
  doi:10.1103/PhysRevD.52.2278
  [hep-th/9501140].
  %%CITATION = doi:10.1103/PhysRevD.52.2278;%%
  %14 citations counted in INSPIRE as of 28 Feb 2017
  
\bibitem{Wolf}
J.~A.~Wolf,
``Complex Homogeneous Contact Manifolds and Quaternionic Symmetric Spaces,''
J. \ Math. \ Mech. {\bf 14}, 1033 (1965).

\bibitem{Alek}
D.~V.~Alekseevskii,
``Classification of Quarternionic Spaces with a Transitive Solvable Group of
Motions,''
\ Math. \ USSR \ Izv. {\bf 9}, 297 (1975).

\bibitem{Salamon}
S.~Salamon,
``Quaternionic Kahler Manifolds,''
\ Invent. \ Math. {\bf 67}, 143 (1982).

\bibitem{BW} 
  J.~Bagger and E.~Witten,
  ``Matter Couplings in N=2 Supergravity,''  Nucl.\ Phys.\ B {\bf 222}, 1 (1983).  %%CITATION = NUPHA,B222,1;%%  %365 citations counted in INSPIRE as of 11 Nov 2013

%\cite{Ahn:2015rma}
\bibitem{AK1506} 
  C.~Ahn and H.~Kim,
  ``Three point functions in the large $ \mathcal{N}=4 $ holography,''
  JHEP {\bf 1510}, 111 (2015)
  doi:10.1007/JHEP10(2015)111
  [arXiv:1506.00357 [hep-th]].
  %%CITATION = doi:10.1007/JHEP10(2015)111;%%
  %7 citations counted in INSPIRE as of 03 Nov 2017

%\cite{Eberhardt:2017fsi}
\bibitem{EGGL} 
  L.~Eberhardt, M.~R.~Gaberdiel, R.~Gopakumar and W.~Li,
  ``BPS spectrum on AdS$_3\times $S$^3 \times $S$^3 \times $S$^1$,''
  JHEP {\bf 1703}, 124 (2017)
  doi:10.1007/JHEP03(2017)124
  [arXiv:1701.03552 [hep-th]].
  %%CITATION = doi:10.1007/JHEP03(2017)124;%%
  %8 citations counted in INSPIRE as of 03 Nov 2017  

  %\cite{Ferreira:2017pgt}
\bibitem{FGJ} 
  K.~Ferreira, M.~R.~Gaberdiel and J.~I.~Jottar,
  ``Higher spins on AdS$_{3}$ from the worldsheet,''
  JHEP {\bf 1707}, 131 (2017)
  doi:10.1007/JHEP07(2017)131
  [arXiv:1704.08667 [hep-th]].
  %%CITATION = doi:10.1007/JHEP07(2017)131;%%
  %4 citations counted in INSPIRE as of 13 Nov 2017
  
 %\cite{Eberhardt:2017pty}
\bibitem{EGL} 
  L.~Eberhardt, M.~R.~Gaberdiel and W.~Li,
  ``A holographic dual for string theory on
  AdS$_{3}\times$ S$^{3} \times$ S$^{3} \times$ S$^{1}$,''
  JHEP {\bf 1708}, 111 (2017)
  doi:10.1007/JHEP08(2017)111
  [arXiv:1707.02705 [hep-th]].
  %%CITATION = doi:10.1007/JHEP08(2017)111;%%
  %3 citations counted in INSPIRE as of 03 Nov 2017  

   \bibitem{Gopakumar17} 
   R. Gopakumar's talk
   at the conference,
``20 Years Later: The Many Faces of AdS/CFT'',
   October 31- November 3, 2017, PCTS, Princeton.
   
%\cite{Ahn:2017dqo}
\bibitem{AKK1703} 
  C.~Ahn, D.~g.~Kim and M.~H.~Kim,
  ``The next 16 higher spin currents and three-point functions in the large $\mathcal{N}=4$ holography,''
  Eur.\ Phys.\ J.\ C {\bf 77}, no. 8, 523 (2017)
  doi:10.1140/epjc/s10052-017-5064-6
  [arXiv:1703.01744 [hep-th]].
  %%CITATION = doi:10.1140/epjc/s10052-017-5064-6;%%
  %1 citations counted in INSPIRE as of 03 Nov 2017

\bibitem{mathematica}
  Wolfram Research, Inc., Mathematica, Version 11.0, Champaign, IL (2016).    
%\cite{Thielemans:1991uw}
\bibitem{Thielemans} 
  K.~Thielemans,
  ``A Mathematica package for computing operator product expansions,''
  Int.\ J.\ Mod.\ Phys.\ C {\bf 2}, 787 (1991).
  doi:10.1142/S0129183191001001
  %%CITATION = doi:10.1142/S0129183191001001;%%
  %117 citations counted in INSPIRE as of 21 Mar 2016

%\cite{Kac:1986dv}
\bibitem{KT1985} 
  V.~G.~Kac and I.~T.~Todorov,
  ``Superconformal Current Algebras And Their Unitary Representations,''
  Commun.\ Math.\ Phys.\  {\bf 102}, 337 (1985).
  doi:10.1007/BF01229384
  %%CITATION = doi:10.1007/BF01229384;%%
  %109 citations counted in INSPIRE as of 28 Feb 2017

%\cite{Slansky:1981yr}
\bibitem{Slansky} 
  R.~Slansky,
  ``Group Theory for Unified Model Building,''
  Phys.\ Rept.\  {\bf 79}, 1 (1981).
  doi:10.1016/0370-1573(81)90092-2
  %%CITATION = doi:10.1016/0370-1573(81)90092-2;%%
  %1080 citations counted in INSPIRE as of 10 Nov 2017

%\cite{Feger:2012bs}
\bibitem{Feger} 
  R.~Feger and T.~W.~Kephart,
  ``LieART-A Mathematica application for
  Lie algebras and representation theory,''
  Comput.\ Phys.\ Commun.\  {\bf 192}, 166 (2015)
  doi:10.1016/j.cpc.2014.12.023
  [arXiv:1206.6379 [math-ph]].
  %%CITATION = doi:10.1016/j.cpc.2014.12.023;%%
  %59 citations counted in INSPIRE as of 24 Aug 2017

  %\cite{Gaberdiel:2010pz}
\bibitem{GG1011} 
  M.~R.~Gaberdiel and R.~Gopakumar,
  ``An $AdS_3$ Dual for Minimal Model CFTs,''
  Phys.\ Rev.\ D {\bf 83}, 066007 (2011).
  doi:10.1103/PhysRevD.83.066007
  [arXiv:1011.2986 [hep-th]].
  %%CITATION = doi:10.1103/PhysRevD.83.066007;%%
  %247 citations counted in INSPIRE as of 15 Mar 2016 
  
  %\cite{Cvitanovic:1976am}
\bibitem{Cvitanovic} 
  P.~Cvitanovic,
  ``Group theory for Feynman diagrams in non-Abelian gauge theories,''
  Phys.\ Rev.\ D {\bf 14}, 1536 (1976).
  doi:10.1103/PhysRevD.14.1536
  %%CITATION = doi:10.1103/PhysRevD.14.1536;%%
  %200 citations counted in INSPIRE as of 24 Aug 2017

%\cite{Joseph:2014bwa}
\bibitem{Joseph} 
  A.~Joseph,
  ``Two-dimensional $ \mathcal{N} $ = (2, 2) lattice gauge theories with matter in higher representations,''
  JHEP {\bf 1407}, 067 (2014)
  doi:10.1007/JHEP07(2014)067
  [arXiv:1403.4390 [hep-lat]].
  %%CITATION = doi:10.1007/JHEP07(2014)067;%%
  %6 citations counted in INSPIRE as of 24 Aug 2017
  
%\cite{Cvitanovic:2008zz}
\bibitem{Cvitanovicbook} 
  P.~Cvitanovic,
  ``Group theory: Birdtracks, Lie's and exceptional groups,''
  Princeton, USA: Univ. Pr. (2008) 273 p
  %31 citations counted in INSPIRE as of 10 Nov 2017  

  \bibitem{Eichmann}
  G. Eichmann, Lecture note of Hadron Physics,
  http://physik.uni-graz.at/~gxe/2013-hadron-physics/hadron-app-3.pdf. 
  
  %\cite{Gaberdiel:2011wb}
\bibitem{GH} 
  M.~R.~Gaberdiel and T.~Hartman,
  ``Symmetries of Holographic Minimal Models,''
  JHEP {\bf 1105}, 031 (2011)
  doi:10.1007/JHEP05(2011)031
  [arXiv:1101.2910 [hep-th]].
  %%CITATION = doi:10.1007/JHEP05(2011)031;%%
  %162 citations counted in INSPIRE as of 01 Jan 2017

%\cite{Gaberdiel:2011zw}
\bibitem{GGHR} 
  M.~R.~Gaberdiel, R.~Gopakumar, T.~Hartman and S.~Raju,
  ``Partition Functions of Holographic Minimal Models,''
  JHEP {\bf 1108}, 077 (2011)
  doi:10.1007/JHEP08(2011)077
  [arXiv:1106.1897 [hep-th]].
  %%CITATION = doi:10.1007/JHEP08(2011)077;%%
  %122 citations counted in INSPIRE as of 03 Jan 2017
  
%\cite{Ahn:2014nia}
\bibitem{AK1411} 
  C.~Ahn and H.~Kim,
  ``Higher Spin Currents in Wolf Space for Generic N,''
  JHEP {\bf 1412}, 109 (2014),
  doi:10.1007/JHEP12(2014)109
  [arXiv:1411.0356 [hep-th]].
  %%CITATION = doi:10.1007/JHEP12(2014)109;%%
  %4 citations counted in INSPIRE as of 31 mars 2016  

   %\cite{Ahn:2013oya}
\bibitem{Ahn1311} 
  C.~Ahn,
  ``Higher Spin Currents in Wolf Space. Part I,''
  JHEP {\bf 1403}, 091 (2014),
  doi:10.1007/JHEP03(2014)091
  [arXiv:1311.6205 [hep-th]].
  %%CITATION = doi:10.1007/JHEP03(2014)091;%%
  %13 citations counted in INSPIRE as of 31 mars 2016

   %\cite{Ahn:2014via}
\bibitem{Ahn1408} 
  C.~Ahn,
  ``Higher Spin Currents in Wolf Space: Part II,''
  Class.\ Quant.\ Grav.\  {\bf 32}, no. 1, 015023 (2015),
  doi:10.1088/0264-9381/32/1/015023
  [arXiv:1408.0655 [hep-th]].
  %%CITATION = doi:10.1088/0264-9381/32/1/015023;%%
  %7 citations counted in INSPIRE as of 31 mars 2016

   %\cite{Ahn:2015gba}
\bibitem{Ahn1504} 
  C.~Ahn,
  ``Higher spin currents in Wolf space: III,''
  Class.\ Quant.\ Grav.\  {\bf 32}, no. 18, 185001 (2015),
  doi:10.1088/0264-9381/32/18/185001
  [arXiv:1504.00070 [hep-th]].
  %%CITATION = doi:10.1088/0264-9381/32/18/185001;%%
  %3 citations counted in INSPIRE as of 31 mars 2016
  
%\cite{Beccaria:2014jra}
\bibitem{BCG} 
  M.~Beccaria, C.~Candu and M.~R.~Gaberdiel,
  ``The large N = 4 superconformal $W_{\infty}$ algebra,''
  JHEP {\bf 1406}, 117 (2014)
  doi:10.1007/JHEP06(2014)117
  [arXiv:1404.1694 [hep-th]].
  %%CITATION = doi:10.1007/JHEP06(2014)117;%%
  %19 citations counted in INSPIRE as of 28 Feb 2017  
  
%\cite{Ahn:2015nga}
\bibitem{AK1509} 
  C.~Ahn and M.~H.~Kim,
  ``The operator product expansion between the 16 lowest higher spin
  currents in the $\mathcal{N}=4$ superspace,''
  Eur.\ Phys.\ J.\ C {\bf 76}, no. 7, 389 (2016).
  doi:10.1140/epjc/s10052-016-4234-2
  [arXiv:1509.01908 [hep-th]].
  %%CITATION = doi:10.1140/epjc/s10052-016-4234-2;%%
  %3 citations counted in INSPIRE as of 18 Nov 2016      

  %\cite{Chang:2011mz}
\bibitem{CY} 
  C.~M.~Chang and X.~Yin,
  ``Higher Spin Gravity with Matter in $AdS_3$ and Its CFT Dual,''
  JHEP {\bf 1210}, 024 (2012)
  doi:10.1007/JHEP10(2012)024
  [arXiv:1106.2580 [hep-th]].
  %%CITATION = doi:10.1007/JHEP10(2012)024;%%
  %98 citations counted in INSPIRE as of 24 Dec 2016
  
%\cite{Hikida:2017byl}
\bibitem{HU} 
  Y.~Hikida and T.~Uetoko,
  ``Three point functions in higher spin $AdS_3$ holography with 1/N corrections,''
  Universe {\bf 3}, no. 4, 70 (2017)
  doi:10.3390/universe3040070
  [arXiv:1708.02017 [hep-th]].
  %%CITATION = doi:10.3390/universe3040070;%%
  %1 citations counted in INSPIRE as of 03 Nov 2017

   %\cite{Papadodimas:2011pf}
\bibitem{PR} 
  K.~Papadodimas and S.~Raju,
  ``Correlation Functions in Holographic Minimal Models,''
  Nucl.\ Phys.\ B {\bf 856}, 607 (2012)
  doi:10.1016/j.nuclphysb.2011.11.006
  [arXiv:1108.3077 [hep-th]].
  %%CITATION = doi:10.1016/j.nuclphysb.2011.11.006;%%
  %54 citations counted in INSPIRE as of 13 Nov 2017

  %\cite{Chang:2011vka}
\bibitem{CY1} 
  C.~M.~Chang and X.~Yin,
  ``Correlators in $W_N$ Minimal Model Revisited,''
  JHEP {\bf 1210}, 050 (2012)
  doi:10.1007/JHEP10(2012)050
  [arXiv:1112.5459 [hep-th]].
  %%CITATION = doi:10.1007/JHEP10(2012)050;%%
  %42 citations counted in INSPIRE as of 13 Nov 2017

  %\cite{Chang:2013izp}
\bibitem{CY2} 
  C.~M.~Chang and X.~Yin,
  ``A semilocal holographic minimal model,''
  Phys.\ Rev.\ D {\bf 88}, no. 10, 106002 (2013)
  doi:10.1103/PhysRevD.88.106002
  [arXiv:1302.4420 [hep-th]].
  %%CITATION = doi:10.1103/PhysRevD.88.106002;%%
  %14 citations counted in INSPIRE as of 13 Nov 2017

  %\cite{Ahn:2014fea}
\bibitem{AP1410} 
  C.~Ahn and J.~Paeng,
  ``Higher Spin Currents in Orthogonal Wolf Space,''
  Class.\ Quant.\ Grav.\  {\bf 32}, no. 4, 045011 (2015),
  doi:10.1088/0264-9381/32/4/045011
  [arXiv:1410.0080 [hep-th]].
  %%CITATION = doi:10.1088/0264-9381/32/4/045011;%%
  %5 citations counted in INSPIRE as of 31 Mar 2016

 %\cite{Ahn:2015gxa}
  \bibitem{AKP1510} 
  C.~Ahn, H.~Kim and J.~Paeng,
  ``Three-point functions in the N = 4 orthogonal coset theory,''
  Int.\ J.\ Mod.\ Phys.\ A {\bf 31}, no. 16, 1650090 (2016).
  doi:10.1142/S0217751X16500901
  [arXiv:1510.03139 [hep-th]].
  %%CITATION = doi:10.1142/S0217751X16500901;%%
  %2 citations counted in INSPIRE as of 18 Nov 2016
  
\bibitem{Beveren}
  E. van Beveren, Lecture note of Group Theory,
  http://cft.fis.uc.pt/eef/evbgroups.pdf.
    
\bibitem{Semenoff}
  G. Semenoff, Lecture note of Group Theory,
  http://www.phas.ubc.ca/~gordonws/521/chapter9a.pdf.
  
\end{thebibliography}
\end{document}